\documentclass[useAMS,usenatbib]{mn2e}

\usepackage[active]{srcltx}
\usepackage{graphicx}
\usepackage{txfonts}
\usepackage{natbib}
\usepackage{multirow}
\usepackage{longtable}
\usepackage[francais]{babel}
\usepackage[applemac]{inputenc}
\usepackage{rotating}
\usepackage{array}

\newcommand{\kms}{km.s$^{-1}$}

\newcommand{\vsini}{$v\sin i$}
\newcommand{\vsinis}{$v\sin i\;$}
\newcommand{\vrad}{$v_{\rm rad}$}
\newcommand{\vrads}{$v_{\rm rad}\;$}
\newcommand{\angs}{\AA$\;$}
\newcommand{\msun}{M$_{\odot}$}
\newcommand{\bz}{\ensuremath{\langle B_z\rangle}}
\def\gtrsim{\mathrel{\hbox{\rlap{\hbox{\lower4pt\hbox{$\sim$}}}\hbox{$>$}}}}
\def\ltsim{\mathrel{\hbox{\rlap{\hbox{\lower4pt\hbox{$\sim$}}}\hbox{$<$}}}}

\title[A spectropolarimetric survey of HAeBe stars - I. Observations and measurements]{A high-resolution spectropolarimetric survey of Herbig Ae/Be stars\thanks{Based on observations obtained at the Canada-France-Hawaii Telescope (CFHT) which is operated by the National Research Council of Canada, the Institut National des Sciences de l'Univers of the Centre National de la Recherche Scientifique of France, and the University of Hawaii}\\I. Observations and measurements}
\author[E. Alecian et al.]
{E.~Alecian$^{1,2}$\thanks{E-mail: evelyne.alecian@obspm.fr},
 G.A.~Wade$^2$,
 C.~Catala$^1$,
 J.H.~Grunhut $^{2,3}$,
 J.D.~Landstreet$^{4,5}$,
 S. Bagnulo$^5$,
 \newauthor
 T.~B\"ohm$^{6,7}$,
 C.P.~Folsom$^5$,
 S.~Marsden$^{8,9}$,
 I.~Waite$^{9}$ \\
 $^1$LESIA-Observatoire de Paris, CNRS, UPMC Univ., Univ. Paris-Diderot, 5 place Jules Janssen, F-92195 Meudon Principal Cedex, France, \\
 $^2$Dept. of Physics, Royal Military College of Canada, PO Box 17000, Stn Forces, Kingston K7K 7B4, Canada \\
 $^3$Department of Physics, Queen's University, Kingston, Canada \\
 $^4$Dept. of Physics \& Astronomy, University of Western Ontario, London N6A 3K7, Canada \\
 $^5$Armagh Observatory, College Hill, Armagh BT61 9DG, Northern Ireland, UK\\
 $^6$ Universit\'e de Toulouse; UPS-OMP; IRAP; Toulouse, France \\
 $^7$CNRS; IRAP; 14, avenue Edouard Belin, F-31400 Toulouse, France \\
 $^8$Centre for Astronomy, School of Engineering and Physical Sciences, James Cook University, Townsville, 4811, Australia \\
 $^{9}$Faculty of Sciences, University of Southern Queensland, Toowoomba, 4350, Australia
}

\begin{document}

\date{Accepted . Received ; in original form }

\pagerange{\pageref{firstpage}--\pageref{lastpage}} \pubyear{2002}

\maketitle

\label{firstpage}

\begin{abstract}
This is the first in a series of papers in which we describe and report the analysis of a large survey of Herbig Ae/Be stars in circular spectropolarimetry. Using the ESPaDOnS and Narval high-resolution spectropolarimeters at the Canada-France-Hawaii and Bernard Lyot Telescopes, respectively, we have acquired 132 circularly-polarised spectra of 70 Herbig Ae/Be stars and Herbig candidates. The large majority of these spectra are characterised by a resolving power of about 65,000, and a spectral coverage from about 3700~\angs to 1~$\mu$m. The peak signal-to-noise ratio per CCD pixel ranges from below 100 (for the faintest targets) to over 1000 (for the brightest). The observations were acquired with the primary aim of searching for magnetic fields in these objects. However, our spectra are suitable for a variety of other important measurements, including rotational properties, variability, binarity, chemical abundances, circumstellar environment conditions and structure, etc. In this first paper, we describe the sample selection, the observations and their reduction, and the measurements that will comprise the basis of much of our following analysis. We describe the determination of fundamental parameters for each target. We detail the Least-Squares Deconvolution that we have applied to each of our spectra, including the selection, editing and tuning of the LSD line masks. We describe the fitting of the LSD Stokes $I$ profiles using a multi-component model that yields the rotationally-broadened photospheric profile (providing the projected rotational velocity and radial velocity for each observation) as well as circumstellar emission and absorption components. Finally, we diagnose the longitudinal Zeeman effect via the measured circular polarisation, and report the longitudinal magnetic field and Stokes $V$ Zeeman signature detection probability. As an appendix, we provide a detailed review of each star observed.\end{abstract}


%
%

\section{Introduction}

\citet{herbig60} was the first to perform a systematic study of a certain class of stars that we call now Herbig Ae/Be (HAeBe) stars, and whose observational parameters are as follows: 

\begin{enumerate}
\item the spectral type is A or earlier, with emission lines,
\item the star lies in an obscured region of space, and
\item the star illuminates fairly bright nebulosity in its immediate vicinity.
\end{enumerate}

Herbig selected these characteristics following the observational properties of the lower-mass counterparts of HAeBe stars - T Tauri stars - and built a list of 26 HAeBe stars satisfying these criteria. This list, as well as the Herbig characteristics, have been extended since this original study (e.g. Herbig \& Bell 1988, Thé et al. 1994, Vieira et al. 2003), and we now know of more than a hundred HAeBe stars of spectral type earlier than F5. All of them do not necessarily show all Herbig characteristics, but all of them have infrared excess with an abnormal extinction law { (compared to classical Be stars)}.

The characteristics enumerated above suggest that HAeBe stars are very young, still surrounded by dust and gas in an envelope or disk. However, it was the spectroscopic study of Str\"om et al. (1972) that brought the first solid evidence that Herbig Ae/Be stars are in the pre-main sequence (PMS) phase of quasi-static contraction, by showing that their surface gravities are systematically lower than those of their main sequence (MS) counterparts. Herbig Ae/Be stars are therefore generally believed to be the evolutionary progenitors of main sequence (MS) intermediate mass (A/B) stars. 

Before the work of Palla \& Stahler in the early 90's, intermediate-mass stars with masses above 3~M$_{\odot}$ were believed to not experience a pre-main sequence phase similar to that of lower-mass stars \citep{larson72}. More detailed calculations performed by \citet{palla90,palla91,palla92,palla93}, including deuterium burning during the protostellar collapse and the pre-main sequence phase, show that optically-visible pre-main sequence stars could be observed, up to masses of about 8~M$_{\odot}$. There calculations considered a constant mass accretion rate during the protostellar collapse, and from the upper envelope of the distribution of Herbig Ae/Be stars in the HR diagram, they concluded that all stars with masses lower than 8~M$_{\odot}$ are formed with a similar mass accretion rate of the order of $10^{-5}$~M$_{\odot}$.yr$^{-1}$.

However, following this work, many Herbig Be stars with masses larger than 8~M$_{\odot}$ have been discovered in the field of the Galaxy (see e.g. Fig. \ref{fig:hr} of this paper), as well as in very young clusters \citep[e.g. ][]{martayan08}, implying that the simplified model of Palla \& Stahler - with a constant mass accretion rate at all masses - may be insufficient to explain the observations. In fact, Palla \& Stahler themselves proposed that the mass accretion rate should be time-dependent. \citet[][NM00 hereinafter]{norberg00}, then \citet[][BM01 hereinafter]{behrend01} proposed that the mass accretion rate depends on the mass or the luminosity of the growing star. As a result, the mass accretion rate should increase while the star is growing and gaining in mass and luminosity, until the circumstellar (CS) matter become sufficiently rare and the massive accretion phase stops. Whereas a unique accretion rate of $10^{-5}$~M$_{\odot}$.yr$^{-1}$ as proposed by Palla \& Stahler (1993) results in a maximum PMS star mass of around 8~M$_{\odot}$, a modulated mass accretion rate, as proposed by NM00 and BM01, allows the birthline to reach the zero-age main sequence (ZAMS) at much higher masses (above 20~M$_{\odot}$).

Herbig Ae/Be stars are indeed observed with masses as large as 20~M$_{\odot}$, with a distribution more concentrated between 1.5 and 3~M$_{\odot}$ (see Table \ref{tab:fp} and Figure \ref{fig:hr} of this paper). Their spectral types are found between F5 and B2 \citep{vieira03}, and many of them show some spectroscopic and photometric activity, reflective of their young age. Many types of activity can be found among Herbig Ae/Be stars, but their origins are not well understood. Among them we find the UX Ori stars, with the Herbig Ae star UX Ori as the prototype. These stars are characterised by a very strong photometric variability (up to 3 magnitudes in the $V$ band), and by the presence of transient absorption features in their spectra that may be due to episodic accretion events \citep[e.g.][]{mora02}. While some authors think that these characteristics are created by the infall of cometary bodies onto the star \citep[e.g.][]{grady00b}, others are more convinced by the theory of accretion from a disk, either via the intermediary of a magnetic field, or not \citep{natta00,mora04}. The presence of winds is detected in many HAeBe stars through P Cygni profiles observed in H$\alpha$ and sometimes in metallic and He lines \citep[e.g. ][]{finkenzeller84,bouret97,bouret98}. Some authors have proposed that these winds have a stellar origin \citep[e.g.][]{bohm94}, while others believe that a disc wind is present \citep[e.g. ][]{corcoran98,vieira03}. However, strong variability in H$\alpha$ emission profiles is observed in a few HAeBe stars, some of them at times showing double-peaked profiles, and sometimes P Cygni profiles \citep[e.g.][]{the85a,catala86a}. These stars show periodic cyclical modulations of their H$\alpha$ emission, but also of metallic lines such as the UV Mg~{\sc ii} h \& k doublet \citep[e.g.][]{catala89}, as well as of their X-ray emission \citep[e.g.][]{testa08}. Their spectra also show UV emission lines of highly-ionised species such as N~{\sc v} and O~{\sc vi} \citep[e.g.][]{bouret97}. \citet{bouret97} proposed that these characteristics are due to the presence of a non-axisymetric wind controlled by a stellar magnetic field. Various non-photospheric spectral features, in addition to those discussed above, are observed in the spectra of HAeBe stars, some with variability and others without. However, the interpretation of each one of these features, as well as their diversity, is not understood at all. The fact that HAeBe stars cover such a large range of mass, temperatures, age and evolutionary state, as well as the fact that these stars evolve at a variety of rates, certainly must be connected with the large variety of observed HAeBe activity phenomena and our difficulties to interpret them.

HAeBe stars are important astrophysical objects because they represent the late formative stages of {intermediate mass stars}. They are therefore significant for understanding general and specific phenomena involved in star formation. Moreover, HAeBe stars can help us to understand a number of perplexing properties of their main sequence (MS) descendants: in particular chemical peculiarities, very slow rotation, and magnetic fields, observed in individually or in combination in a significant fraction of MS A/B stars.

Among the MS A/B stars, a significant fraction shows photospheric abundance anomalies (as compared to solar abundances, and to the abundances of the majority of MS A/B stars). These anomalies are believed to result from atomic diffusion within their surface layers due to the competition between radiative levitation and gravitational settling \citep[e.g.][]{michaud70}. One important condition necessary to allow this phenomenon to occur is the absence of strong deep mixing in those layers, which would tend to overwhelm these separation processes. As rotation-driven circulation is an important source of such mixing, this condition implies that such {chemically peculiar} stars should be slow rotators. It has been observed that nearly all chemically peculiar Am, Ap/Bp and HgMn stars are characterised by slow rotation (rotation periods longer than $\sim 1$ day) compared to the "normal" (non-peculiar) A/B stars \citep{abt95}. The origin of this slow rotation is not well understood. In the case of Am and HgMn stars, slow rotation might be the result of tidal interaction occurring in close binary systems \citep[i.e. those with orbital periods shorter than 100 days ; e.g.][]{abt09}. The mechanism responsible for the slow rotation of Ap/Bp stars is likely related to their strong magnetic fields. \citet{stepien00} discussed different theories aimed at explaining this slow rotation, and he concluded that magnetic braking must occur during the PMS phase in order to reproduce the rotational angular momenta of MS A/B stars. Stepien demonstrated that magnetic coupling of a PMS star with its accretion disk would slow the rotation of the star and increase its rotation period to a few days. In order to produce the very slowest rotators - those with observed rotation periods greater than about one month - the disk must disappear sufficiently early during the PMS phase to allow strong magnetised winds to carry away a large quantity of angular momentum before the star reaches the ZAMS.

Until recently we had very few observational constraints on the magnetic fields and the rotation of Herbig Ae/Be stars. To our knowledge, only two thorough observational studies of the evolution of the angular momentum of intermediate-mass stars during the PMS phase have been undertaken. {\citet{bohm95} concluded that if these stars rotate as solid bodies, the evolution of the angular momentum must depend on stellar mass, while if the internal rotation varies as (radius)$^{-2}$, the observations of HAeBe and MS A/B stars in young clusters are consistent with conservation of total angular momentum at all masses. \citet{wolff04} concluded that PMS intermediate-mass stars lose angular momentum before they start the PMS phase, while angular momentum is conserved during the radiative phase of PMS evolution.} Both of these analyses provided very interesting results that should be discussed in the framework of a scenario of angular momentum evolution that includes magnetic fields.

A number of studies have been attempted to detect magnetic fields in Herbig Ae/Be stars, without much success \citep[e.g. ][]{catala93,hubrig04}. {Apart from a marginal detection in HD~104237 reported by \citep{donati97}, and a possible detection in HD~101412 proposed by \citep{wade07} (both being now firmly confirmed magnetic stars: Alecian et al. in prep.), no other convincing magnetic detections have been reported before the present survey. The reason is likely limited precision and an insufficiently large stellar sample as a consequence of limited observational capabilities. Fortunately, many of these limitations are overcome by today's spectropolarimetric facilities: telescopes with large collecting area, high-efficiency instruments, large spectral range, and high spectral resolution.}

In order to thoroughly investigate magnetism and rotation in HAeBe stars, we have performed a large survey of 70 stars using the newest high-resolution spectropolarimetric instruments: ESPaDOnS (at the Canada-France-Hawaii telescope, CFHT, USA) and Narval (at the T\'elescope Bernard Lyot, TBL, France). Within the context of this survey we have detected a small number of new magnetic stars and confirmed the presence of a magnetic field already discovered during a parallel ESPaDOnS program focused on massive stars in Orion \citep[LP Ori,][]{petit08}. Those  discoveries (HD 190073, HD 200775, HD 72106, V380 Ori and LP Ori), and the analysis we performed to characterise their magnetic fields and related properties, have already or will be described in other papers \citep[][Petit et al. in prep.]{wade05,catala07,alecian08a,folsom08,petit08,alecian09b}. While this survey is focusing on HAeBe stars in the field of the Galaxy, we have also performed a similar survey of HAeBe stars in three young clusters and detected three more magnetic stars: NGC~6611~601, NGC~2244~201 and NGC~2264~83 \citep{alecian08b,alecian09a}. The description of this cluster survey will be presented in an upcoming paper (Alecian et al. in prep.).

We are now publishing a series of papers describing the complete sample of observed field stars, discussing the observations and their analysis (this paper, paper I), an analysis of their rotation velocities \citep[][paper II]{paperii}, an analysis of their magnetic properties (Wade et al. in prep., paper III), and the characterisation of the CS contributions to the spectra of the sample (Alecian et al. in prep., paper IV).

This paper is organised as follows. In Sect. 2 we review the sample selection, and in Sect. 3 the observational procedure and data reduction, and summarise the characteristics and quality of the reduced spectra. In Sect. 4 we determine fundamental parameters for the stars of the sample, and in Sect. 5 describe the extraction and fitting of the Least-Squares Deconvolved profiles that we use for the majority of our analysis. In Sect. 6 we discuss the magnetic field diagnosis carried out in a number of different ways. Sect. 7 provides a discussion of the results and conclusions relevant to the analysis to be reported in papers II, III and IV.

%
%

\begin{figure*}
\centering
\includegraphics[width=6.5cm,angle=-90]{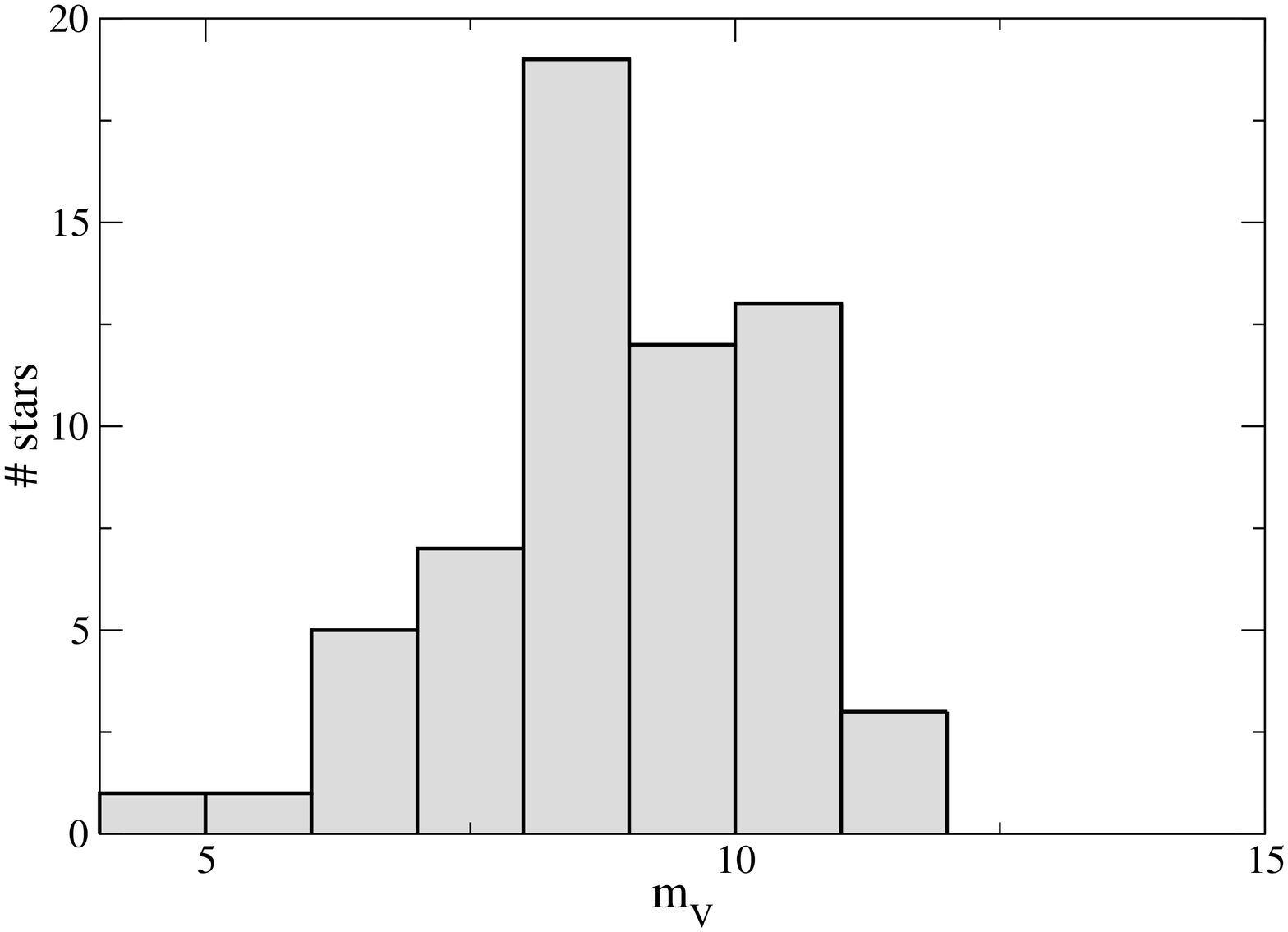}
\includegraphics[width=6.5cm,angle=-90]{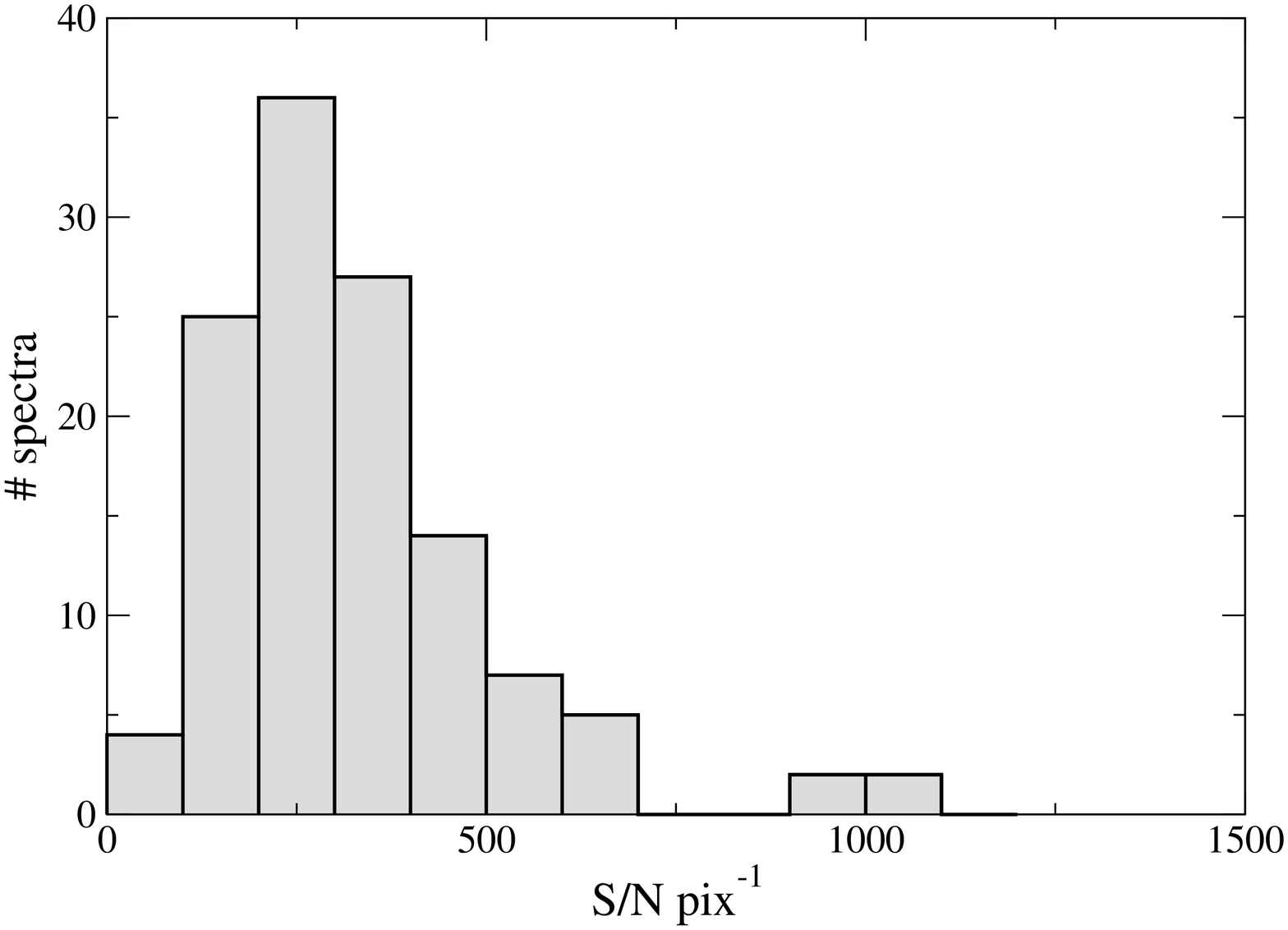}
\caption{Distributions of $m_V$ (left) and SNR (right) of the sample stars.}
\label{SNRs}
\end{figure*}

\section{Sample selection}

Our study required the selection of a relatively large number of HAeBe stars to allow us to derive statistically meaningful conclusions about the presence of magnetic fields in these stars. Various literature sources were used for target selection, primarily the catalogues of HAeBe stars and HAeBe candidates by Th\'e et al. (1994) and by Vieira et al. (2003). 

The catalogue of Th\'e et al. (1994) contains six categories of stars; the stars selected for our study were obtained only from the first category, which contains stars historically known as HAeBe stars, or strong candidates of the group. According to the authors, all of these stars possess near- or far-infrared excess and emission lines, associated with the presence of CS dust, {discs} and energetic ouflows which are usually found in HAeBe stellar environments. On the other hand, \citet{vieira03} produced a catalogue of HAeBe stars and probable candidates from an initial search for new T Tauri stars (pre-main sequence stars of lower mass) using the Infrared Astronomical Satellite (\textit{IRAS}) point source catalogue\footnote{\textit{http://irsa.ipac.caltech.edu/}}. Because the initial search was based on CS dust properties, it included HAeBe stars along with T Tauri stars. {Vieira et al.} extracted the HAeBe stars by filtering the data using specific requirements such as a spectral type earlier than F5, emission at H$\alpha$, and a minimum level of infrared emission. The majority of the stars were associated by the authors with a star forming region. Based on the quality of these two literature sources and the arguments presented by their authors, we conclude that all of the stars in our sample are {\em bona fide} HAeBe stars. In total, 70 HAeBe stars have been selected with visual magnitudes lower than 12, spanning in spectral type from F5 to B0.

{ Measurements with high-resolution spectropolarimeters (such as ESPaDOnS@CFHT or Narval@TBL) have a high enough resolving power to take advantage of the information contained in the line profiles of metallic lines, as has already been demonstrated in earlier studies with the MuSiCoS spectropolarimeter (e.g. Wade et al. 2000, Petit et al. 2004)}. ESPaDOnS magnetic field measurements have standard errors which decrease strongly with decreasing $v\sin i$ and with increasing richness and strength of the metallic line spectrum (c.f. Landstreet 1982; Shorlin et al. 2002). To fully exploit this dependence, and thus to obtain the most precise measurements possible, we have preferentially selected our targets for low $v\sin i$ ($\ltsim 100$~\kms) where available $v\sin i$ data allowed us to perform such a selection. However, because accurate measurements of $v\sin i$ are not available for many HAeBe stars, a significant
fraction of our targets (about one-third) turn out to be relatively rapid rotators.

\section{Observations and data reduction}

{The 132 program star observations reported here were obtained between 2004 and 2010 using two high-resolution spectropolarimeters: the ESPaDOnS spectropolarimeter at the Canada-France-Hawaii Telescope (80 spectra), and the Narval spectropolarimeter at the T\'elescope Bernard Lyot (52 spectra). The ESPaDOnS observations were obtained during 6 observing runs in 2004 (technical and commissioning runs), 2005 and 2006 (competitively-allocated PI time), including the first scientific ESPaDOnS run. The Narval observations were obtained during 7 observing runs between 2007 and 2010 (competitively-allocated PI time).} 

The basic technical characteristics of ESPaDOnS and Narval are nearly identical. The polarisation analysis unit is located at the Cassegrain focus of the telescope. The stellar image is formed on an aperture followed by a collimating lens. The beam then passes through a rotatable $\lambda$/2 retarder, a fixed $\lambda$/4 retarder, a second rotatable $\lambda$/2 retarder, and finally a small-angle Wollaston prism, followed by a lens which refocuses the (now double) star image on the inputs of two optical fibres. This relatively complex polarisation analyser is necessary because one of the fundamental design parameters for ESPaDOnS/Narval was very wide wavelength coverage (approximately 3700~\AA\ to 1.04 $\mu$m). To have retarders which are approximately achromatic over this wide range, ESPaDOnS uses Fresnel rhombs. A single Fresnel rhomb acts as a $\lambda$/4 retarder, but deviates the beam, while two Fresnel rhombs in series form a $\lambda$/2 plate without beam deviation. To minimise mechanical complications, only the double (non-deviating) Fresnel rhombs are allowed to rotate; the configuration chosen is the minimum which allows one to analyse all of the Stokes polarisation components ($Q, U, V$) by appropriate orientation of the axes of the successive retarders. 

The two output beams from the Wollaston prism, which have been { split} into the two components of circular polarisation (for this study) by appropriate retarder orientations, are then carried by the pair of optical fibres to a stationary and temperature-controlled cross-dispersed spectrograph where two interleaved spectra are formed, covering virtually the entire desired wavelength range with a resolving power of $R\simeq 65 000$. The $I$ component of the stellar Stokes vector is formed by adding the two corresponding spectra, while the desired polarisation component ($Q, U, V$) is obtained essentially from the difference of the two spectra. To minimise systematic errors due to small misalignments, differences in transmission, effects of seeing, etc., one complete observation of a star consists of four successive sub exposures; for the second and third, the retarder orientations are changed so as to exchange the beam paths of the two analysed spectra (see Donati et al. 1997).

The actual reduction of observations is carried out at the observatory using the dedicated software package Libre-ESpRIT. Libre-Esprit subtracts bias, locates the various spectral orders on the CCD image, measures the shape of each order and models the (varying) slit geometry, identifies comparison lines for each order and computes a global wavelength model of all orders, performs an optimal extraction of each order, and combines the resulting spectra to obtain one-dimensional intensity (Stokes $I$) and circular polarisation (Stokes $V$) spectra. The Stokes $V$ spectrum normally has the continuum polarisation removed, as this arises mainly from instrumental effects and carries little information about the star. Each spectrum is corrected to the heliocentric frame of reference, and may optionally be divided by a flat field and be approximately normalised (see Donati et al. 1997, and ESPaDOnS www pages\footnote{http://www.cfht.hawaii.edu/Instruments/Spectroscopy/Espadons}). Due to the presence of strong emission lines in the spectra of many of our targets, the automatic Libre-Esprit normalisation fails to achieve a satisfactory rectification of the continuum.  We have therefore turned off this option in Libre-Esprit, and normalised the final reduced one-dimensional spectra manually, order-by-order.

Diagnostic { null} spectra called $N$ spectra, computed by combining the four successive sub exposures of polarisation in such a way as to have the real polarisation cancel out { \citep{donati97}}, are also calculated by Libre-Esprit. The $N$ spectra test the system for spurious polarisation signals. In all of our observations, the $N$ spectra are quite featureless, as expected. The final spectra consist of ascii files tabulating $I/I_{\rm c}$, $V/I_{\rm c}$, $N/I_{\rm c}$, and estimated uncertainty per pixel as a function of wavelength, order by order. 

The log of spectropolarimetric observations is reported in Table 1. The 132 observations include 112 observations of 64 apparently non-magnetic program stars, 9 observations of 5 magnetic program stars (the discovery and/or confirmation observations of each), and 11 observations of one possibly newly-detected magnetic program star (HD 35929).

Because the program stars observed in this study are characterised by a large range of visual magnitudes (reflecting their diverse luminosities, distances and extinctions due to their surrounding environments), the distribution of their apparent magnitudes $m_{\rm V}$ (shown in Fig.~\ref{SNRs}, left panel) is rather broad (with a mean of 8.9, a minimum of 4.2 and a maximum of 11.9). As a consequence, our data yield a broad distribution of signal-to-noise ratios (SNRs, illustrated in Fig.~\ref{SNRs}, right panel), ranging from below 100 per CCD pixel to over 1000.

\begin{table*}
\caption{Log of observations of the HAeBe program stars. Columns 1 and 2 give the designations of the stars. The date, Universal Time (UT), and Heliocentric Julian Date (HJD) of the start of the observation are given in columns 3 and 4. The total exposure time is given in column 5. Columns 6 gives the peak SNR per CCD pixel at the wavelength indicated in column 7. Columns 8 to 11 give the number of lines used to compute the LSD profiles with the full and cleaned masks and the SNR in the LSD $V$ profile. The final column indicates the instrument associated to the observation.}
\centering
\begin{tabular}{llllllllllll}
\hline
     &     &      &      &      &       &     &     \multicolumn{2}{c}{Full mask} & \multicolumn{2}{c}{Cleaned mask} & \\
HD or BD & Other & Date (d/m/y) & HJD -     & Total exp. & Peak & $\lambda$ (nm) & $\#$ LSD & LSD & $\#$ LSD & LSD & Instrument \\
number    & name & UT time      & 2450000& time (s)    & SNR   &                            & lines         & SNR & lines         & SNR &                    \\
\hline
BD-06 1259 & BF Ori & 21/02/05 09:17 & 3422.88915 & 4800 & 192 & 515 & 2401 & 1986 & 466 & 1426 & ESPaDOnS \\
 &  & 12/03/09 19:41 & 4903.32074 & 4640 & 83 & 567 & \multirow{2}{*}{2398} & \multirow{2}{*}{3224} & \multirow{2}{*}{397} & \multirow{2}{*}{ 874} & Narval \\
 &  & 12/03/09 21:03 & 4903.37805 & 4640 & 88 & 731 &  &  &  &  & Narval \\
BD-06 1253 & V380 Ori & 20/02/05 09:32 & 3421.90001 & 4800 & 144 & 781 &  &  &  &  & ESPaDOnS \\
BD-05 1329 & T Ori & 24/08/05 14:53 & 3607.11832 & 3600 & 245 & 731 & 1487 & 2890 & 662 & 2201 & ESPaDOnS \\
BD-05 1324 & NV Ori & 12/01/06 04:56 & 3747.71083 & 3200 & 163 & 708 & 5896 & 2759 & 368 &  860 & ESPaDOnS \\
BD+41 3731 &  & 26/08/05 09:03 & 3608.88285 & 4000 & 309 & 527 & 362 & 1564 & 320 & 2552 & ESPaDOnS \\
 &  & 06/11/07 21:54 & 4411.41380 & 5800 & 178 & 552 & 382 & 2675 & 310 & 1444 & Narval \\
BD+46 3471 & V1578 Cyg & 26/08/05 10:59 & 3608.96354 & 4800 & 304 & 708 & 1274 & 3540 & 586 & 2811 & ESPaDOnS \\
BD+61 154 & V594 Cas & 22/02/05 05:56 & 3423.74367 & 3600 & 144 & 730 & 550 & 1332 & 12 &  148 & ESPaDOnS \\
 &  & 24/08/05 11:01 & 3606.96314 & 5600 & 208 & 515 & 570 &  870 & 12 &  355 & ESPaDOnS \\
BD+65 1637 & V361 Cep & 11/06/06 14:49 & 3898.11845 & 2400 & 237 & 730 & 371 & 1728 & 86 &  738 & ESPaDOnS \\
 &  & 24/09/09 21:43 & 5099.40934 & 8400 & 276 & 731 & 343 & 2151 & 73 &  911 & Narval \\
BD+72 1031 & SV Cep & 12/06/06 15:00 & 3899.12535 & 1600 & 159 & 730 & 967 & 1547 & 561 & 1301 & ESPaDOnS \\
 &  & 11/11/07 21:46 & 4416.40939 & 6400 & 139 & 731 & 1025 & 1394 & 543 & 1228 & Narval \\
HD 9672 & 49 Cet & 25/08/05 11:40 & 3607.98968 &  800 & 910 & 515 & 2079 & 17572 & 2079 & 17572 & ESPaDOnS \\
HD 17081 & $\pi$ Cet & 20/02/05 05:29 & 3421.72749 &  480 & 925 & 515 & 518 & 8006 & 234 & 4885 & ESPaDOnS \\
 &  & 21/02/05 05:31 & 3422.72864 &  480 & 1049 & 515 & 517 & 9158 & 234 & 5837 & ESPaDOnS \\
HD 31293 & AB Aur &  &  &  &  &  & 1525 & 5293 & 590 & 6641 & ESPaDOnS \\
 &  & 20/02/05 05:53 & 3421.74513 & 1200 & 395 & 527 & 1536 & 8391 & 559 & 3500 & ESPaDOnS \\
 &  & 22/02/05 08:26 & 3423.85121 & 2400 & 547 & 527 & 1604 & 9895 & 559 & 5575 & ESPaDOnS \\
HD 31648 & MWC 480 & 22/02/05 09:13 & 3423.88401 & 2400 & 389 & 527 & 3411 & 9335 & 1073 & 6583 & ESPaDOnS \\
 &  & 25/08/05 12:45 & 3608.03208 & 2000 & 435 & 708 & 3420 & 9671 & 1067 & 6934 & ESPaDOnS \\
HD 34282 &  & 25/08/05 13:57 & 3608.07957 & 4000 & 246 & 708 & 2924 & 4904 & 2924 & 4904 & ESPaDOnS \\
HD 35187 B &  & 26/08/05 13:13 & 3609.05072 & 2000 & 360 & 708 & 2104 & 5967 & 1383 & 5466 & ESPaDOnS \\
HD 35929 &  & 13/11/07 00:50 & 4417.53905 & 4000 & 415 & 566 & 4853 & 12302 & 3055 & 10076 & Narval \\
 &  & 14/11/07 00:29 & 4418.52488 & 2000 & 213 & 708 & 4962 & 11800 & 3050 & 4999 & Narval \\
 &  & 20/02/09 19:33 & 4883.31686 & 2000 & 335 & 731 & \multirow{3}{*}{4951} & \multirow{3}{*}{5830} & \multirow{3}{*}{3010} & \multirow{3}{*}{14214} & Narval \\
 &  & 20/02/09 20:12 & 4883.34386 & 2000 & 341 & 731 &  &  &  &  & Narval \\
 &  & 20/02/09 20:49 & 4883.36939 & 2000 & 323 & 731 &  &  &  &  & Narval \\
 &  & 21/02/09 19:13 & 4884.30273 & 2000 & 309 & 731 & \multirow{3}{*}{4852} & \multirow{3}{*}{16589} & \multirow{3}{*}{3007} & \multirow{3}{*}{12287} & Narval \\
 &  & 21/02/09 19:49 & 4884.32826 & 2000 & 298 & 731 &  &  &  &  & Narval \\
 &  & 21/02/09 20:26 & 4884.35380 & 2000 & 281 & 731 &  &  &  &  & Narval \\
 &  & 11/03/09 19:31 & 4902.31377 & 2000 & 284 & 731 & \multirow{3}{*}{4851} & \multirow{3}{*}{14318} & \multirow{3}{*}{3006} & \multirow{3}{*}{10574} & Narval \\
 &  & 11/03/09 20:08 & 4902.33930 & 2000 & 250 & 731 &  &  &  &  & Narval \\
 &  & 11/03/09 20:44 & 4902.36482 & 2000 & 265 & 731 &  &  &  &  & Narval \\
HD 36112 & MWC 758 &  &  &  &  &  & 3758 & 7123 & 284 & 2792 & ESPaDOnS \\
 &  & 20/02/05 06:30 & 3421.77182 & 2400 & 322 & 708 & 4185 & 9217 & 271 & 2236 & ESPaDOnS \\
HD 36910 & CQ Tau & 04/04/08 20:12 & 4561.33902 & 6000 & 198 & 731 & 5671 & 5037 & 1219 & 3209 & Narval \\
HD 36917 & V372 Ori & 08/11/07 23:48 & 4413.49636 & 4000 & 208 & 552 & 1064 & 2477 & 1064 & 2477 & Narval \\
HD 36982 & LP Ori & 09/11/07 01:05 & 4413.54984 & 4000 & 136 & 552 & 614 & 3430 & 187 &  937 & Narval \\
 &  & 10/11/07 00:56 & 4414.54299 & 6000 & 306 & 552 & 615 & 3270 & 187 & 2178 & Narval \\
 &  & 11/11/07 00:03 & 4415.50631 & 4400 & 264 & 552 & \multirow{2}{*}{609} & \multirow{2}{*}{4838} & \multirow{2}{*}{187} & \multirow{2}{*}{2088} & Narval \\
 &  & 11/11/07 01:21 & 4415.56085 & 4400 & 314 & 552 &  &  &  &  & Narval \\
 &  & 12/11/07 01:05 & 4416.54958 & 6000 & 426 & 552 & 579 & 1447 & 187 & 3091 & Narval \\
HD 37258 & V586 Ori & 24/02/09 19:15 & 4887.30404 & 6600 & 292 & 553 & 1872 & 5262 & 559 & 3215 & Narval \\
HD 37357 &  & 24/02/09 22:15 & 4887.42954 & 4440 & 270 & 553 & 1965 & 4795 & 723 & 3550 & Narval \\
HD 37806 & MWC 120 & 25/08/05 15:10 & 3608.13042 & 2000 & 468 & 515 & 577 & 4293 & 53 & 1793 & ESPaDOnS \\
HD 38120 &  & 13/03/09 22:23 & 4904.43316 & 3600 & 230 & 553 & 1400 & 3152 & 1436 & 3021 & Narval \\
HD 38238 & V351 Ori & 16/03/07 20:16 & 4176.34501 & 4680 & 247 & 731 & 4790 & 6822 & 3356 & 6190 & Narval \\
HD 50083 & V742 Mon & 13/11/07 01:50 & 4417.57965 & 2000 & 487 & 553 & 584 & 5699 & 157 & 2508 & Narval \\
 &  & 03/04/08 20:58 & 4560.37363 & 2000 & 500 & 553 & 628 & 7080 & 147 & 2287 & Narval \\
HD 52721 &  & 07/11/07 03:33 & 4411.65007 & 2000 & 523 & 553 & 628 & 5350 & 273 & 4090 & Narval \\
 &  & 03/04/08 20:14 & 4560.34420 & 2000 & 467 & 553 & 662 & 6253 & 250 & 3590 & Narval \\
HD 53367 &  & 20/02/05 10:31 & 3421.94295 & 1200 & 363 & 708 & 545 & 2646 & 59 &  977 & ESPaDOnS \\
 &  & 21/02/05 10:25 & 3422.93832 & 2400 & 505 & 566 & 548 & 4289 & 59 & 1621 & ESPaDOnS \\
HD 68695 &  & 22/02/05 10:53 & 3423.95853 & 2400 & 128 & 527 & 1550 & 1579 & 1550 & 1575 & ESPaDOnS \\
HD 72106 &  & 22/02/05 10:04 & 3423.92478 & 2400 & 236 & 515 &  &  &  &  & ESPaDOnS \\
HD 76534 A &  & 22/02/05 11:40 & 3423.99189 & 1800 & 221 & 708 & 436 & 1578 & 11 &  353 & ESPaDOnS \\
\hline
\end{tabular}
\end{table*}

\begin{table*}
\contcaption{}
\centering
\begin{tabular}{llllllllllll}
\hline
     &     &      &      &      &       &     &     \multicolumn{2}{c}{Full mask} & \multicolumn{2}{c}{Cleaned mask} & \\
HD or BD & Other & Date (d/m/y) & HJD -     & Total exp. & Peak & $\lambda$ (nm) & $\#$ LSD & LSD & $\#$ LSD & LSD & Instrument \\
number    & name & UT time      & 2450000& time (s)    & SNR   &                            & lines         & SNR & lines         & SNR &                    \\
\hline
HD 98922 &  & 21/02/05 11:54 & 3423.00115 & 1600 & 451 & 527 & 683 & 3089 & 578 & 2872 & ESPaDOnS \\
HD 114981 & V958 Cen & 20/02/05 12:13 & 3422.01396 & 1600 & 329 & 515 & 518 & 6680 & 210 & 1756 & ESPaDOnS \\
 &  & 12/01/06 15:01 & 3748.12665 & 2400 & 633 & 515 & 531 & 3018 & 211 & 4060 & ESPaDOnS \\
HD 135344 &  & 10/01/06 15:40 & 3746.15200 & 2400 & 128 & 731 & 6631 & 2555 & 6631 & 2555 & ESPaDOnS \\
HD 139614 &  & 20/02/05 13:50 & 3422.07904 & 3600 & 298 & 708 & 3495 & 5853 & 7023 & 9542 & ESPaDOnS \\
 &  & 21/02/05 13:46 & 3423.07633 & 2800 & 274 & 708 & 3519 & 5587 & 7004 & 9242 & ESPaDOnS \\
 &  & 22/02/05 14:14 & 3424.09588 & 2400 & 294 & 708 & 3513 & 6592 & 7020 & 10579 & ESPaDOnS \\
HD 141569 &  & 13/02/06 12:29 & 3780.02056 & 4000 & 301 & 708 & 1496 & 2821 & 1418 & 2723 & ESPaDOnS \\
 &  & 07/03/07 13:20 & 4167.05806 & 5400 & 1053 & 566 & 1478 & 16345 & 1395 & 15985 & ESPaDOnS \\
HD 142666 & V1026 Sco & 20/02/05 12:54 & 3422.03917 & 2400 & 237 & 708 & 3855 & 4123 & 2496 & 3711 & ESPaDOnS \\
 &  & 22/02/05 13:16 & 3424.05441 & 3600 & 338 & 708 & 3885 & 7184 & 2549 & 6402 & ESPaDOnS \\
 &  & 22/05/05 07:50 & 3512.83235 & 3600 & 292 & 708 & 3866 & 6005 & 2543 & 5381 & ESPaDOnS \\
 &  & 22/05/05 08:55 & 3512.87760 & 3600 & 293 & 708 & 3867 & 5141 & 2506 & 5365 & ESPaDOnS \\
 &  & 23/05/05 08:25 & 3513.85638 & 3600 & 256 & 708 & 3920 & 6425 & 2507 & 4593 & ESPaDOnS \\
 &  & 24/05/05 07:54 & 3514.83487 & 3600 & 308 & 708 & 3893 & 5863 & 2530 & 5725 & ESPaDOnS \\
 &  & 25/05/05 08:02 & 3515.84083 & 3600 & 282 & 708 & 4775 & 7989 & 2507 & 5209 & ESPaDOnS \\
HD 144432 &  & 20/02/05 14:47 & 3422.11732 & 2400 & 323 & 708 & 4792 & 9430 & 1751 & 5893 & ESPaDOnS \\
 &  & 21/02/05 14:46 & 3423.11688 & 3200 & 368 & 708 & 3426 & 12788 & 1742 & 7118 & ESPaDOnS \\
HD 144668 & HR 5999 & 24/08/05 05:38 & 3606.73345 & 1200 & 569 & 708 & 3863 & 8503 & 2821 & 12340 & ESPaDOnS \\
HD 145718 & V718 Sco & 26/08/05 05:32 & 3608.72972 & 2800 & 403 & 730 & 1760 & 6149 & 2284 & 7274 & ESPaDOnS \\
HD 150193 & V2307 Oph & 24/08/05 06:20 & 3606.76406 & 2800 & 453 & 730 & 1768 & 9845 & 1541 & 6043 & ESPaDOnS \\
HD 152404 & AK Sco & 15/02/06 14:46 & 3782.11538 & 3600 & 393 & 708 & 6684 & 10308 & 2634 & 7454 & ESPaDOnS \\
HD 163296 &  & 22/05/05 09:54 & 3512.91769 & 2400 & 588 & 527 & 1764 & 7008 & 1123 & 8391 & ESPaDOnS \\
 &  & 23/05/05 10:02 & 3513.92318 & 3600 & 460 & 515 & 1801 & 10426 & 1112 & 5924 & ESPaDOnS \\
 &  & 24/05/05 09:49 & 3514.91391 & 3600 & 615 & 515 & 1798 & 7439 & 1157 & 8981 & ESPaDOnS \\
 &  & 24/05/05 14:48 & 3515.12170 & 2400 & 448 & 515 & 1714 & 3678 & 1121 & 6362 & ESPaDOnS \\
 &  & 25/05/05 09:52 & 3515.91662 & 3600 & 617 & 527 & 1791 & 10394 & 1097 & 8744 & ESPaDOnS \\
 &  & 25/05/05 14:53 & 3516.12540 & 2400 & 436 & 566 & 1753 & 11221 & 1056 & 6092 & ESPaDOnS \\
 &  & 25/08/05 05:39 & 3607.73704 & 1200 & 641 & 515 & 5421 & 6488 & 1104 & 9561 & ESPaDOnS \\
HD 169142 &  & 20/02/05 15:34 & 3422.14750 & 2400 & 270 & 708 & 5450 & 5421 & 3718 & 5865 & ESPaDOnS \\
 &  & 22/02/05 15:01 & 3424.12494 & 2400 & 208 & 708 & 5502 & 8648 & 3788 & 4861 & ESPaDOnS \\
 &  & 22/05/05 10:41 & 3512.95012 & 2400 & 311 & 708 & 5479 & 14187 & 3737 & 7728 & ESPaDOnS \\
 &  & 24/08/05 07:10 & 3606.80061 & 2000 & 473 & 708 & 554 & 1926 & 3751 & 12614 & ESPaDOnS \\
HD 174571 & MWC 610 & 17/03/07 03:51 & 4176.65909 & 3600 & 283 & 731 & 586 & 2557 & 343 & 2086 & Narval \\
 &  & 16/04/08 02:35 & 4572.60859 & 3900 & 245 & 731 & 626 & 5359 & 342 & 1621 & Narval \\
HD 176386 &  & 25/08/05 06:09 & 3607.75872 & 1600 & 573 & 708 & 820 & 9459 & 601 & 5297 & ESPaDOnS \\
HD 179218 &  & 21/02/05 15:30 & 3423.14213 & 1200 & 298 & 527 & 849 & 2949 & 271 & 2117 & ESPaDOnS \\
 &  & 26/08/05 08:10 & 3608.84465 & 1600 & 631 & 515 & 841 & 6965 & 271 & 4945 & ESPaDOnS \\
 &  & 03/10/09 20:53 & 5108.37217 & 7200 & 866 & 553 & 1768 & 5059 & 255 & 6667 & Narval \\
HD 190073 & V1295 Aql & 22/05/05 11:40 & 3512.98887 & 3290 & 411 & 527 &  &  &  &  & ESPaDOnS \\
HD 200775 & MWC 361 & 22/05/05 14:35 & 3513.10716 & 3600 & 555 & 731 &  &  &  &  & ESPaDOnS \\
HD 203024 &  & 24/08/05 09:38 & 3606.90562 & 2800 & 365 & 515 & 1765 & 5954 & 1238 & 5567 & ESPaDOnS \\
 &  & 07/11/07 22:17 & 4412.43070 & 4800 & 307 & 552 & 1799 & 2353 & 1238 & 4811 & Narval \\
HD 216629 & IL Cep & 10/06/06 15:06 & 3897.12935 & 1200 & 301 & 730 & 437 & 1773 & 90 &  910 & ESPaDOnS \\
 &  & 08/12/06 07:09 & 4077.79899 & 1200 & 227 & 708 & 438 & 2115 & 81 &  758 & ESPaDOnS \\
 &  & 05/11/07 21:18 & 4410.39095 & 5700 & 339 & 731 & 2871 & 3856 & 90 & 1112 & Narval \\
HD 244314 & V1409 Ori & 05/11/07 23:19 & 4410.47647 & 6000 & 158 & 552 & 3449 & 7804 & 1799 & 2363 & Narval \\
HD 244604 & V1410 Ori & 24/08/05 13:47 & 3607.07333 & 3600 & 329 & 527 & 1736 & 2631 & 2016 & 6964 & ESPaDOnS \\
HD 245185 & V1271 Ori & 20/02/05 07:39 & 3421.82069 & 4800 & 192 & 515 & 1971 & 2213 & 1736 & 2631 & ESPaDOnS \\
HD 249879 &  & 05/04/08 21:03 & 4562.37504 & 6000 & 133 & 553 & 492 & 1726 & 982 & 1951 & Narval \\
HD 250550 & V1307 Ori & 08/11/07 00:23 & 4412.52023 & 5360 & 202 & 553 & 416 & 2282 & 387 & 1403 & Narval \\
HD 259431 & V700 Mon & 17/03/07 22:40 & 4177.44524 & 5100 & 302 & 731 & 402 & 1528 & 261 & 1583 & Narval \\
 &  & 24/02/09 23:31 & 4887.48247 & 2400 & 274 & 731 & 404 & 2078 & 253 & 1473 & Narval \\
 &  & 17/03/10 20:23 & 5273.35057 & 2400 & 199 & 731 & 1100 & 1072 & 258 & 1178 & Narval \\
HD 275877 & XY Per & 11/12/06 06:24 & 4080.77111 & 3600 & 348 & 708 & 2897 & 6912 & 453 & 3058 & ESPaDOnS \\
 &  & 25/09/09 00:46 & 5099.53643 & 7200 & 299 & 731 & 2715 & 5390 & 412 & 2419 & Narval \\
HD 278937 & IP Per & 21/02/05 06:26 & 3422.76644 & 4800 & 195 & 666 & 2892 & 3084 & 2871 & 3856 & ESPaDOnS \\
 &  & 21/02/05 07:50 & 3422.82473 & 4800 & 171 & 527 & 2924 & 2735 & 2892 & 3084 & ESPaDOnS \\
 &  & 22/02/05 07:14 & 3423.79948 & 4800 & 172 & 708 & 395 & 30034 & 2924 & 2735 & ESPaDOnS \\
HD 287823 &  & 17/03/07 21:16 & 4177.38573 & 3120 & 112 & 552 & 3671 & 2836 & 1100 & 1072 & Narval \\
HD 287841 & V346 Ori & 20/02/09 22:23 & 4883.43440 & 7740 & 151 & 731 & 2112 & 2979 & 3671 & 2836 & Narval \\
HD 290409 &  & 07/11/07 00:17 & 4411.51658 & 6000 & 171 & 552 & 1983 & 1315 & 2112 & 2979 & Narval \\
\hline
\end{tabular}
\end{table*}

\begin{table*}
\contcaption{}
\centering
\begin{tabular}{llllllllllll}
\hline
     &     &      &      &      &       &     &     \multicolumn{2}{c}{Full mask} & \multicolumn{2}{c}{Cleaned mask} & \\
HD or BD & Other & Date (d/m/y) & HJD -     & Total exp. & Peak & $\lambda$ (nm) & $\#$ LSD & LSD & $\#$ LSD & LSD & Instrument \\
number    & name & UT time      & 2450000& time (s)    & SNR   &                            & lines         & SNR & lines         & SNR &                    \\
\hline
HD 290500 &  & 21/02/09 21:49 & 4884.41137 & 6180 & 92 & 731 & \multirow{2}{*}{484} & \multirow{2}{*}{2468} & \multirow{2}{*}{360} & \multirow{2}{*}{ 629} & Narval \\
 &  & 21/02/09 23:38 & 4884.48650 & 6180 & 82 & 731 &  &  &  &  & Narval \\
HD 290770 &  & 24/02/09 20:53 & 4887.37237 & 4200 & 280 & 553 & 423 & 2513 & 172 & 1878 & Narval \\
HD 293782 & UX Ori & 11/01/06 04:55 & 3746.70929 & 3200 & 174 & 708 & 1521 & 2375 & 1521 & 2375 & ESPaDOnS \\
HD 344361 & WW Vul & 24/08/05 08:20 & 3606.85206 & 5200 & 274 & 708 & 1477 & 2060 & 764 & 3036 & ESPaDOnS \\
 &  & 06/11/07 18:59 & 4411.29021 & 6800 & 152 & 552 & 1502 & 3660 & 756 & 1763 & Narval \\
 & LkHa 215 & 15/04/08 20:49 & 4572.36556 & 6000 & 90 & 731 & 370 & 32681 & 204 & 26607 & Narval \\
 &  & 11/03/09 22:14 & 4902.42813 & 5820 & 176 & 731 & 615 & 1476 & 233 & 26231 & Narval \\
 & MWC 1080 & 25/08/05 08:10 & 3607.84527 & 6400 & 320 & 839 &  &  &  &  & ESPaDOnS \\
 & VV Ser & 26/08/05 06:57 & 3608.79270 & 6400 & 241 & 809 & 491 & 1418 & 226 &  856 & ESPaDOnS \\
 & VX Cas & 25/08/05 10:05 & 3607.92466 & 6400 & 146 & 515 & 1230 & 1499 & 1230 & 1499 & ESPaDOnS \\
\hline
\end{tabular}
\end{table*}

%
%

\section{Fundamental parameters}

\subsection{Effective temperature and surface gravity determination}

The temperature and gravity{, as well as their errors,} of each star was first taken from the literature, and has then been compared to our data as follows.  For effective temperatures below 15000~K, we have calculated synthetic spectra in the local thermodynamic equilibrium (LTE) approximation, using the code SYNTH of \citet{piskunov92}. SYNTH requires, as input, atmosphere models, obtained using the ATLAS~9 code \citep{kurucz93}, and a list of spectral line data obtained from the Vienna Atomic Line Database\footnotemark\footnotetext{http://ams.astro.univie.ac.at/$\sim$vald/} (VALD, Piskunov et al. 1995; Kupka et al. 1999; Ryabchikova et al. 1999). Above 15000~K we used TLUSTY non-LTE atmosphere models and the SYNSPEC code (Hubeny 1988; Hubeny \& Lanz 1992, 1995), to calculate synthetic spectra. { At all temperatures the synthetic spectra have been computed with a solar metallicity (see Sc. 4.2).} Then we compared, by eye, the observed to the synthetic spectra, and if necessary adjusted the temperature (holding $\log g = 4.0$ constant) until a best fit was achieved. For some stars, the temperatures found in the literature were not able to reproduce our spectra, and therefore we give here new determinations of $T_{\rm eff}$. 

{ In this procedure we fixed the surface gravity $\log g=4.0$ because for most of the stars of our sample, the determination of $\log g$ using our data is not possible for the following reasons. First, the continuum level is very difficult to determine in echelle spectra and most of the Balmer lines are spread over two orders, making the determination of $\log g$ from the wings of the Balmer lines very imprecise. Then, the spectrum of many of our targets are heavily contaminated with CS emission/absorption and especially in the spectral lines of Fe, Ti, Si, Cr, which makes impossible the determination of surface gravity from the ionisation equilibrium of abundant species. Typical values of $\log g$ in HAeBe stars are comprised between 3.5 and 4.5 \citep[e.g.][]{folsom12}. We have therefore adopted a value of 4.0 for all stars for which a determination from our observations was not possible.}

For a few Herbig stars whose metallic spectral lines are only very faintly contaminated with CS emission/absorption, we obtained very high quality, high resolution spectra. To determine accurate effective temperature and gravity in these few cases, we developed an automatic procedure, based on a comparison of the observed spectrum to a grid of model spectra. The grid is composed of LTE SYNTH3 models (Kochukhov 2007), computed using Kurucz's ATLAS~9 atmospheres. The models assume solar abundances and no macroturbulence, while the grid varies the microturbulence between 0, 1, 2, 3, 4, and 5~km~s$^{-1}$. The atomic line lists were extracted from VALD, for all lines with a predicted line depth greater than 0.01 times the continuum. The models in our grid range from an effective temperature of 6500~K to 15000~K, in steps of 100~K. In addition, the models range from 3.0 to 5.0~dex in $\log(g)$ in 0.5 dex steps. The rotational broadening and disk integration of the SYNTH3 models used for comparison with the observed spectra was carried out using the code S3DIV (Kochukhov 2007).

For this procedure the spectral region between 420~nm and 520~nm was modeled, as lines in this region show the strongest sensitivity to temperature variations for stars with temperatures within our grid range. We used a brute-force search for the lowest $\chi^2$ by comparing the observation to each model within a pre-selected parameter space (corresponding to a predefined temperature range, a predefined range in $\log(g)$, and each value of microturbulence) of the grid. The initial search parameters were chosen based on photometric or spectroscopic literature estimates. For each model, we fit the $v \sin i$ to the observations using a $\chi^2$ minimisation routine as well. The Balmer line regions were ignored due to imprecise continuum normalisation of the short echelle orders, and strong local emission. 

\begin{figure*}
\centering
\includegraphics[width=6cm,angle=90]{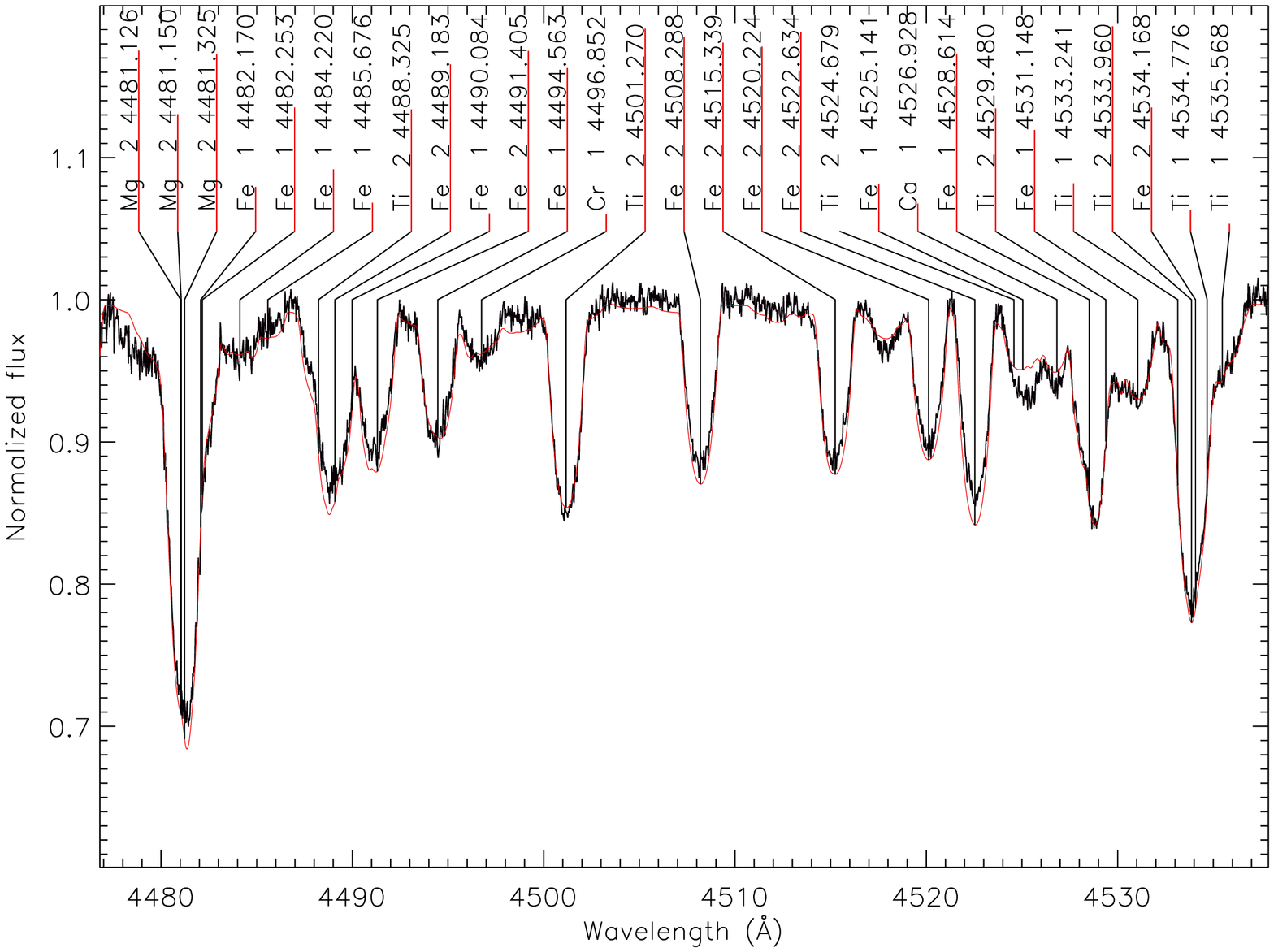}
\includegraphics[width=6cm,angle=90]{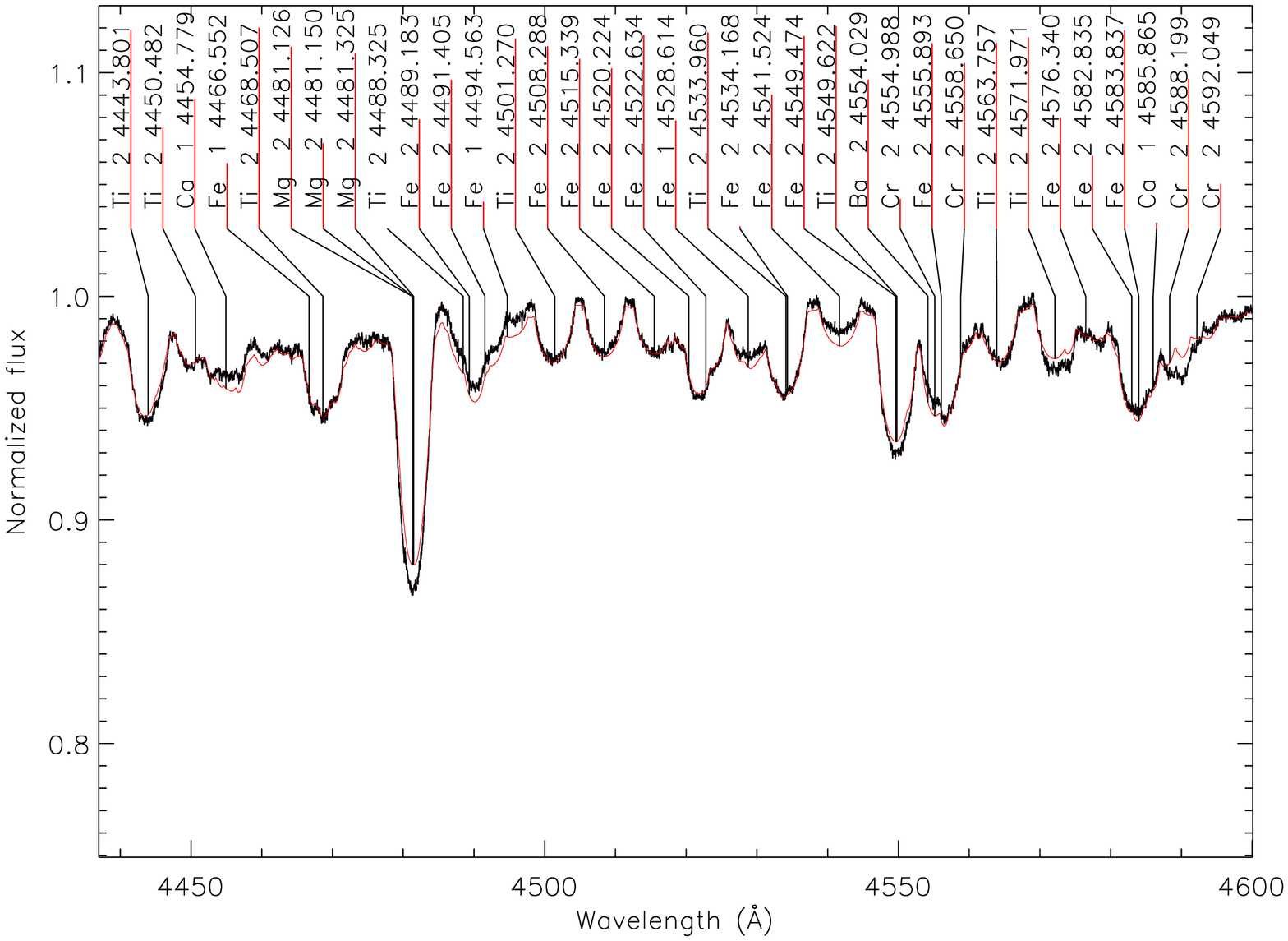}
\caption{Comparison of our best-fit model (smooth red) to an observation (noisy black) of HD~142666 (left panel) with a best fit $v \sin i$ of 67~km~s$^{-1}$, and HD~9672 (right panel) with a best fit $v \sin i$ of 191~km~s$^{-1}$ from this method. Identification of the ions with the strongest contribution to the line are indicated above.}
\label{spec_fit_examp}
\end{figure*}

Once the best-fit model was identified, we carried out a visual comparison between the model and the observation to determine which regions (if any) were poorly fit, likely due to contamination from CS material. We then re-ran our spectrum fitting procedure with these regions ignored to improve our fits. If an observation showed strong emission in metallic lines, this procedure would not provide a realistic estimate on the stellar parameters. Examples demonstrating the quality of fit we are able to achieve, for both a low $v\sin i$ and high $v\sin i$ star, are shown in Fig.~\ref{spec_fit_examp}. The parameters derived in this way were used to check those obtained for the same stars from visual comparison and to refine the visual matching process. 

\begin{figure}
\centering
\includegraphics[width=7.0cm,angle=-90]{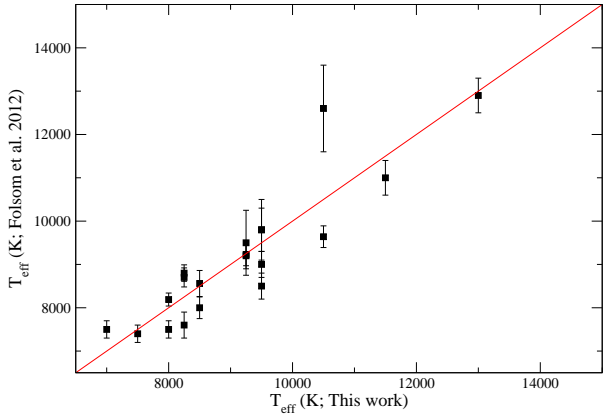}
\caption{Comparison of effective temperature $T_{\rm eff}$ derived by Folsom et al. (2012) for a sample of $\sim 20$ HAeBe stars, and those derived in this study. The solid line indicates a perfect agreement.}
\label{comp_teff}
\end{figure}

We note that the effective temperatures derived from the visual comparison method depend on the assumed (solar) abundance and fixed $\log g=4.0$. To check the sensitivity of our atmospheric parameter determinations to these assumptions, we compare our $T_{\rm eff}$ values with the results of Folsom et al. (2012), who derive $T_{\rm eff}$ and $\log g$ from detailed spectrum synthesis of a sample of approximately 20 HAeBe stars, simultaneously determining the abundance table and microturbulence parameter. As is illustrated in Fig.~\ref{comp_teff}, the results of our more approximate procedures are in reasonable agreement with those from detailed fitting. Ultimately, the results of the LSD procedure, which is used in particular to determine $v \sin i$, to identify circumstellar and interstellar line features, and to diagnose the magnetic field, are only weakly sensitive to the details of the line mask, and errors in the adopted parameters of up to about $\pm 10-20$\% have little impact on the results.

The adopted effective temperatures and surface gravities are summarised in Table 2. The stars whose effective temperature and surface gravity have been determined { using a visual inspection or the automatic procedure are labelled with a dagger ($\dag$) or a double dagger ($\ddag$) respectively. In these two cases no uncertainties have been determined in $\log g$, and the indicated values are estimated at the model grid precision, i.e. $\pm$0.5 dex.}


{ 
\subsection{Metallicities}

In this work we have assumed a solar metallicity for all our objects (except one, HD 34282, that shows a very low metallicity, Merin et al. 2004) for the following reasons. The spectra of most of the stars of our sample are heavily contaminated with CS emission and/or absorption, making difficult a reliable abundance analysis, and therefore a metallicity determination. However, Folsom et al. (2012), using our data, have isolated the 20 stars of our sample showing only faint CS contamination, and have determined their abundances. They find that about half of them display $\lambda$ Boo peculiarities, one of them is a magnetic Bp star (V380 Ori A), and all others are chemically normal with solar abundances. We know that magnetic Bp stars have peculiar abundances due to gravitational settling in their atmosphere \citep{michaud81}. These peculiarities do not reflect the global chemical composition of the star. The $\lambda$ Boo peculiarities are not yet fully understood, but are also very likely due to a surface effect \citep[e.g.][]{folsom12}. Among these 20 stars, there is therefore no evidence at all that their compositions deviate significantly from from that of the sun. While the Folsom et al. sample concerns only about 30\% of our sample, by extrapolation it is reasonable to assume that a large fraction of our targets have a metallicity similar to solar. \citet{acke04b} have analysed the spectra of 24 HAeBe and Vega-type stars and have found solar metallicities in 21 of them and $\lambda$ Boo patterns in one of them. Following these works, it is therefore reasonable to assume a solar metallicity in all the stars of our sample (except HD 34282).
}


\subsection{Luminosity determination}

The photometric data employed to determine luminosities were taken from the Hipparcos and Tycho catalogues \citep{perryman97}, when available, and from \citet{herbst99} and \citet{vieira03} otherwise. When a strong photometric variability ($\Delta V > 0.6$ mag) has been observed by Hipparcos, we assumed that the reduced brightness of the star was due to variable occultation by CS matter situated around the star. In these few cases, we adopted the brightest Hipparcos magnitude $H_{\rm P}$ value as the intrinsic (unocculted) magnitude of the star and we converted it to the Johnson system using the $(V-I)$ color of Hipparcos, and the conversion table of the Hipparcos and Tycho catalogues \citep[][p. 59]{perryman97}. We used the visual magnitude ($V$) corrected for reddening ($A_v$) obtained from the colour excess $E(B-V)$ and a total-to-selective extinction $R_v (A_v = Rv\times E(B-V))$ of 5.0 \citep{hernandez04}, the bolometric corrections, and the distances of the stars, to estimate their luminosity. The intrinsic $(B-V)$ and bolometric correction of all stars have been obtained from the effective temperature reported in Table 2 and the calibration of \citet{kenyon95}. { The errors on the luminosities have been determined by propagating the error on the distances. For the double-lined spectroscopic binaries (SB2) we have attempted a visual estimation of their luminosity ratio using the observed ratio of the individual spectral line depths of our data. However, in all but one of the systems we also had to adjust the temperatures of both components at the same time. The results being highly inaccurate we have not attempted an estimation of their error bars and therefore no errors on the luminosity of the individual components are reported for these systems. In the case of AK Sco, well constrained temperatures of both components are been obtained by \citet{alencar03}. It was therefore possible to derive an error bare on the luminosity ratio of the system, and therefore on the individual luminosities (see Appendix A and Table2).}

All the quantities are summarised in Table 2, while the sources of the data are detailed, star by star, in Appendix A. For three stars (HD 50083, HD 52721, HD 174571) no reliable distance could be found in the literature. We have therefore estimated their luminosity from their effective temperature and surface gravity by comparing their position  in a $\log g - T_{\rm eff}$ diagram with theoretical evolutionary tracks (described in Sec. 4.4).

\subsection{Mass and radius determination}

\begin{figure}
\centering
\includegraphics[width=8.5cm]{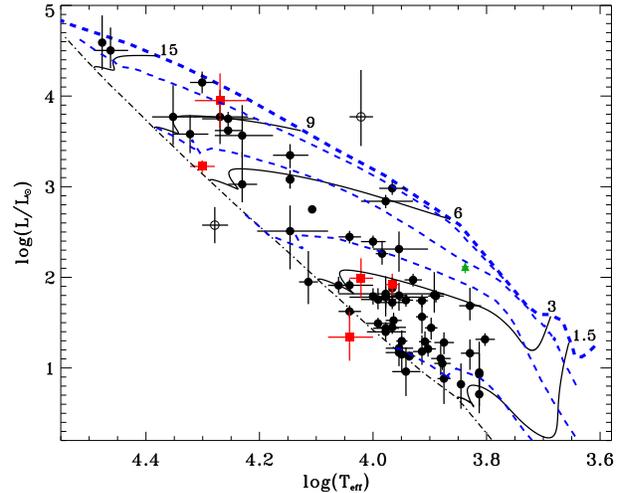}
\caption{Magnetic (red squares) and non-magnetic (black circle) Herbig Ae/Be stars plotted in an HR diagram. The green triangle is the candidate magnetic star HD 35929. Open circles correspond to HD~98922 (above the birthline) and IL~Cep (below the ZAMS) that fall outside of the PMS region of the HR diagram, whose positions cannot be reproduced with the theoretical evolutionary tracks considered in this paper. The CESAM PMS evolutionary tracks for 1.5, 3, 6, 9 and 15~M$_{\odot}$ (black full lines), 0.01, 0.1, 1 and 10~Myr isochrones (blue thin dashed lines), and ZAMS (black dot-dashed line) are also plotted. The birthline taken from \citet{behrend01} is plotted with a blue thick dashed line.}
\label{fig:hr}
\end{figure}

We placed all the stars in an HR diagram (Fig. \ref{fig:hr}), { with the error bars when available}, and compared their positions with evolutionary tracks calculated with the CESAM stellar evolutionary code \citep{morel97} version 2K. Using a 2D-linear interpolation and a grid of 120 evolutionary tracks with masses from 1 to 20 M$_{\odot}$, and mass steps varying between 0.01 and 1 M$_{\odot}$ (depending on the mass and the position of the stars in the HR diagram), we determined the mass, radius and age of each star. 

{ The errors have been determined using the intersection of the evolutionary tracks with the error ellipses, as defined by the errors in effective temperature, luminosity or surface gravity. When the ellipses are intersecting the ZAMS or the birthline, only the portion of the ellipse between the birthline and the ZAMS was considered.}

The ages have been measured from the birthline, i.e the locus in the HR diagram where the newly-formed stars become observable at optical wavelength, meaning that the CS matter in which the stars were buried during the proto-stellar phase becomes optically thin. We used the birthline of \citet{behrend01} that has been computed with a mass accretion rate increasing with the luminosity of the growing star. We favoured a birthline calculated with a modulated accretion rate \citep[instead of a constant accretion rate as computed by][]{palla93} as it better fits the upper envelope of the distribution of massive Herbig Be stars in the HR diagram (see Fig. \ref{fig:hr}). Furthermore, \citet[][PS93 hereinafter]{palla93} argue that a constant accretion rate of $10^{-5}$~M$_{\odot}$.yr$^{-1}$ during the proto-stellar phase is a good approximation as it fits well the upper envelope of the known Herbig Ae/Be stars. { However their work was only including HAeBe stars of masses lower than 6\msun. Since their work, more massive stars} have been identified as Herbig Be \citep[e.g.][]{vieira03,martayan08}, while intermediate-mass T Tauri stars that are cooler and younger than Herbig Ae/Be star and that are identified as the progenitors of the Herbig Ae/Be phases, have also been found \citep[e.g.][]{wolff04,hussain09}. The latter are filling the right part of the HR diagram (with $\log T_{\rm eff} \le 3.8$) and are evolving along the Hayashi track up to the radiative phase of the PMS evolution. We are therefore convinced that the BM01 birthline is a reasonable assumption for the start of the PMS phase at all masses.

\begin{table*}
{
\setlength{\extrarowheight}{3pt}
\centering
\caption{{ Photometric and fundamental parameters. Columns 1 and 2 give the stars designations. Columns 3 to 6 give the effective temperature, the surface gravity and their origin. The Johnson $V$ magnitude and $(B-V)$ color and the reference are indicated in columns 7 to 9. The visual extinction and magnitude corrected from extinction are given in columns 10 and 11. The distance and its reference are given in column 12 and 13. The luminosity is given in column 14. Columns 11, 12 and 13 give the mass, radius and age, while the PMS duration and the predicted radius on the ZAMS are indicated in columns 14 and 15. The measured \vsini\ and \vrad\ are given in columns 16 and 17. In the column 14 to 21, a reference is given when it was not determined in this work. An asterisk ($*$) indicates a note at the end of the table. All references are indicated at the end of the table.}}
\begin{tabular}{l l         r@{$\pm$}l                                       @{}r         r@{$\pm$}l                               @{}r         p{0.1mm}@{}          r                                     @{$\;\;$}r                                 @{$\;\;$}r       r                                                   @{$\;\;$}r                                           r@{}l                                  @{$\;\;$}r     @{$\;\;$}p{0.1mm}}
\hline
HD or BD & Other      & \multicolumn{2}{c}{$T_{\rm eff}$} & Ref        & \multicolumn{2}{c}{$\log g$} & Ref        &                             & \multicolumn{1}{c}{$V$}   & \multicolumn{1}{c}{$(B-V)$} & Ref            & \multicolumn{1}{c}{$A_{\rm V}$} & \multicolumn{1}{c}{$V_{\rm 0}$} &  \multicolumn{2}{c}{$d$} &  Ref          &      \\
number   & Name      & \multicolumn{2}{c}{(K)}                  &              & \multicolumn{2}{c}{(cgs)}      &              &                             & \multicolumn{1}{c}{(mag)} & \multicolumn{1}{c}{(mag)}   &                  & \multicolumn{1}{c}{(mag)}           & \multicolumn{1}{c}{(mag)}           &  \multicolumn{2}{c}{(pc)} &                 &      \\
(1)           & (2)           & \multicolumn{2}{c}{(3)}                  & (4)         & \multicolumn{2}{c}{(5)}         & (6)         &                             & \multicolumn{1}{c}{(7)}      & \multicolumn{1}{c}{(8)}        & (9)            & \multicolumn{1}{c}{(10)}              & \multicolumn{1}{c}{(11)}              & \multicolumn{2}{c}{(12)}   & (13)          &      \\
\hline
BD-06 1259 & BF Ori & 8750 & 250 & ar & 4.0 & 0.5 & ar & & 7.85 & -0.028 & j & -0.57 &  8.42 &  375 & $_{-30}^{+30}$ & d &  \\
BD-06 1253 A & V380 Ori A & 10500 & 500 & b & \multicolumn{2}{c}{4.0} &  & & 10.34 & 0.689 & j &  3.74 &  6.59 &  400 & $_{-40}^{+40}$ & b &  \\
BD-05 1329 & T Ori & 8500 & 300 & ao & 4.2 & 0.3 & ao & & 10.63 & 0.53 & r &  2.64 &  7.99 &  375 & $_{-30}^{+30}$ & d &  \\
BD-05 1324 & NV Ori & 6350 & 250 & ay & \multicolumn{2}{c}{4.0} & & & 9.90 & 0.46 & r &  0.45 &  9.45 &  375 & $_{-30}^{+30}$ & d &  \\
BD+41 3731 &  & 17000 & 1000 & \dag & \multicolumn{2}{c}{4.0} & & & 9.90 & 0.052 & j &  1.06 &  8.84 &  980 & $_{-200}^{+500}$ & ai &  \\
BD+46 3471 & V1578 Cyg & 9500 & 1000 & aq & \multicolumn{2}{c}{4.0} & & & 10.14 & 0.436 & j &  2.25 &  7.89 &  950 & $_{-80}^{+80}$ & p &  \\
BD+61 154 & V594 Cas & 13000 & 500 & \dag & \multicolumn{2}{c}{4.0} & & & 10.51 & 0.566 & j &  3.29 &  7.22 &  202 & $_{-49}^{+97}$ & al &  \\
BD+65 1637 & V361 Cep & 18000 & 1000 & \dag & \multicolumn{2}{c}{4.0} & & & 10.83 & 0.429 & j &  2.94 &  7.89 & 1250 & $_{-50}^{+50}$ & ah &  \\
BD+72 1031 & SV Cep & 9500 & 2000 & aq & \multicolumn{2}{c}{4.0} & & & 10.48 & 0.344 & ag &  1.87 &  8.61 &  400 & $_{-100}^{+100}$ & v &  \\
HD 9672 & 49 Cet & 8900 & 200 & \ddag & \multicolumn{2}{c}{4.5} & \ddag & & 5.62 & 0.066 & j &  0.05 &  5.57 &   59 & $_{-1.0}^{+1.0}$ & al &  \\
HD 17081 & $\pi$ Cet & 12800 & 200 & as & 3.77 & 0.15 & as & & 4.24 & -0.122 & j & -0.06 &  4.30 &  120 & $_{-3}^{+3}$ & al &  \\
HD 31293 & AB Aur & 9800 & 700 & ao & 3.9 & 0.3 & ao & & 7.03 & 0.132 & j &  0.65 &  6.38 &  139 & $_{-16}^{+21}$ & al &  \\
HD 31648 & MWC 480 & 8200 & 300 & \ddag & \multicolumn{2}{c}{4.0} & \ddag & & 7.73 & 0.160 & j &  0.10 &  7.63 &  137 & $_{-21}^{+31}$ & al &  \\
HD 34282 &  & 8625 & 200 & x & 4.2 & 0.20 & x & & 9.92 & 0.299 & j &  1.00 &  8.93 &  191 & $_{-46}^{+89}$ & al &  \\
HD 35187 B &  & 8900 & 200 & \ddag & \multicolumn{2}{c}{4.0} & \ddag & & 8.169 & 0.218 & j &  0.81 &  7.36 &  114 & $_{-24}^{+41}$ & al &  \\
HD 35929 &  & 6800 & 100 & at & 3.3 & 0.1 & at & & 8.11 & 0.438 & j &  0.69 &  7.42 &  375 & $_{-30}^{+30}$ & d &  \\
HD 36112 & MWC 758 & 7800 & 150 & \ddag & \multicolumn{2}{c}{4.0} & \ddag & & 8.27 & 0.317 & j &  0.69 &  7.59 &  279 & $_{-56}^{+94}$ & al &  \\
HD 36910 & CQ Tau & 6750 & 300 & aq & \multicolumn{2}{c}{4.0} &  & & 8.77 & 0.94 & j &  2.85 &  5.92 &  113 & $_{-19}^{+29}$ & al &  \\
HD 36917 & V372 Ori & 10000 & 500 & \dag & \multicolumn{2}{c}{4.0} &  & & 8.03 & 0.17 & an &  1.00 &  7.03 &  375 & $_{-30}^{+30}$ & d &  \\
HD 36982 & LP Ori & 20000 & 1000 & \dag & \multicolumn{2}{c}{4.0} &  & & 8.46 & 0.09 & an &  1.55 &  6.91 &  375 & $_{-30}^{+30}$ & d &  \\
HD 37258 & V586 Ori & 9500 & 500 & \dag & \multicolumn{2}{c}{4.0} &  & & 9.64 & 0.140 & g &  0.41 &  9.23 &  375 & $_{-30}^{+30}$ & d &  \\
HD 37357 &  & 9250 & 500 & am & \multicolumn{2}{c}{4.0} &  & & 8.88 & 0.13 & am &  0.37 &  8.52 &  375 & $_{-30}^{+30}$ & d &  \\
HD 37806 & MWC 120 & 11000 & 500 & \dag & \multicolumn{2}{c}{4.0} &  & & 7.91 & 0.025 & j &  0.58 &  7.32 &  375 & $_{-30}^{+30}$ & d &  \\
HD 38120 &  & 11000 & 500 & \dag & \multicolumn{2}{c}{4.0} &  & & 9.07 & 0.044 & j &  0.21 &  8.86 &  375 & $_{-30}^{+30}$ & d &  \\
HD 38238 & V351 Ori & 7750 & 250 & \dag & \multicolumn{2}{c}{4.0} &  & & 8.89 & 0.381 & j &  0.65 &  8.23 &  375 & $_{-30}^{+30}$ & d &  \\
HD 50083 & V742 Mon & 20000 & 1000 & m & 3.43 & 0.15 & m & & 6.91 & 0.008 & j &  1.09 &  5.82 & 1000 & $_{-100}^{+100}$  &  & \\
HD 52721 &  & 22500 & 2000 & m & 3.99 & 0.20 & m & & 6.54 & 0.016 & j &  1.28 &  5.26 & 670 & $_{-110}^{140}$ &   & \\
HD 53367 &  & 29000 & 2000 & \dag & \multicolumn{2}{c}{4.0} &  & & 6.97 & 0.357 & j &  3.29 &  3.68 &  255 & $_{-51}^{+86}$ & al &  \\
HD 68695 &  & 9000 & 300 & ao & 4.3 & 0.3 & ao & & 9.82 & 0.10 & am &  0.49 &  9.33 &  570 & $_{-100}^{+100}$ & i &  \\
HD 72106 A &  & 11000 & 1000 & l & 4.0 & 0.5 & l & & 9.00 & -0.090 & k & -0.09 &  9.10 &  289 & $_{-85}^{+204}$ & j &  \\
HD 72106 B &  & 8750 & 500 & l & 4.0 & 0.5 & l & & 9.62 & 0.20 & k &  0.10 &  9.52 &  289 & $_{-85}^{+204}$ & j &  \\
HD 76534 A &  & 18000 & 2000 & \dag & \multicolumn{2}{c}{4.0} & & & 8.35 & 0.107 & j &  1.43 &  6.91 &  870 & $_{-80}^{+80}$ & q &  \\
HD 98922 &  & 10500 & 500 & am & \multicolumn{2}{c}{4.0} &  & & 6.77 & 0.037 & j &  0.54 &  6.24 & 1150 & $_{-360}^{+930}$ & al &  \\
HD 114981 & V958 Cen & 17000 & 2000 & \dag & \multicolumn{2}{c}{4.0} &  & & 7.16 & -0.098 & j &  0.51 &  6.65 &  550 & $_{-130}^{+260}$ & al &  \\
HD 135344 &  & 6750 & 250 & \dag & \multicolumn{2}{c}{4.0} &  & & 8.70 & 0.60 & ae &  0.96 &  7.74 &  142 & $_{-27}^{+27}$ & aa &  \\
HD 139614 &  & 7600 & 300 & ao & 3.9 & 0.3 & ao & & 8.40 & 0.24 & am &  0.50 &  7.90 &  142 & $_{-27}^{+27}$ & aa &  \\
HD 141569 &  & 9800 & 500 & ao & 4.2 & 0.4 & ao & & 7.11 & 0.095 & j &  0.46 &  6.65 &   116 & $_{-8}^{+9}$ & al &  \\
HD 142666 & V1026 Sco & 7900 & 200 & \ddag & \multicolumn{2}{c}{4.0} & \ddag & & 8.67 & 0.50 & am &  1.60 &  7.07 &  145 & $_{-20}^{+20}$ & af &  \\
HD 144432 &  & 7500 & 300 & \ddag & \multicolumn{2}{c}{3.5} & \ddag & & 8.19 & 0.397 & j &  0.74 &  7.45 &  145 & $_{-20}^{+20}$ & af &  \\
HD 144668 & HR 5999 & 8200 & 200 & \ddag & \multicolumn{2}{c}{3.5} & \ddag & & 7.00 & 0.190 & j &  0.25 &  6.75 &  142 & $_{-27}^{+27}$ & aa &  \\
HD 145718 & V718 Sco & 8100 & 200 & \ddag & \multicolumn{2}{c}{4.0} & \ddag & & 8.83 & 0.456 & j &  1.38 &  7.45 &  145 & $_{-20}^{+20}$ & af &  \\
HD 150193 & V2307 Oph & 9500 & 500 & \dag & \multicolumn{2}{c}{4.0} &  & & 8.79 & 0.522 & j &  2.47 &  6.32 &  145 & $_{-20}^{+20}$ & af &  \\
\hline
\end{tabular}
\label{tab:fp}
}
\end{table*}

\begin{table*}
{
\centering
\setlength{\extrarowheight}{3pt}
\contcaption{}
\begin{tabular}{l    r@{}l                                                          p{2.5mm}         @{}r@{}l                                             r@{}l                                                  r@{}l                                    r@{}l                                                  r@{}l                                                       r@{$\pm$}l                              r@{$\pm$}l     }
\hline
ID                      &  \multicolumn{2}{c}{$\log(L/L_{\odot})$} &                       & \multicolumn{2}{c}{$M/M_{\odot}$} & \multicolumn{2}{c}{$R/R_{\odot}$} & \multicolumn{2}{c}{age}   & \multicolumn{2}{c}{$t_{\rm PMS}$} & \multicolumn{2}{c}{$R_{\rm ZAMS}$} & \multicolumn{2}{c}{$v\sin i$} & \multicolumn{2}{c}{$v_{\rm rad}$} \\
                         & &                                                             &                       & &                                                      & &                                                    & \multicolumn{2}{c}{(Myr)} & \multicolumn{2}{c}{(Myr)}                & \multicolumn{2}{c}{$R_{\odot}$}         & \multicolumn{2}{c}{(\kms)}    & \multicolumn{2}{c}{(\kms)}        \\
(1) or (2)           &  \multicolumn{2}{c}{(14)}                        &                       &  \multicolumn{2}{c}{(15)}                  & \multicolumn{2}{c}{(16)}                & \multicolumn{2}{c}{(17)}    & \multicolumn{2}{c}{(18)}                 & \multicolumn{2}{c}{(19)}                      & \multicolumn{2}{c}{(20)}        & \multicolumn{2}{c}{(21)} \\
\hline
BD-06 1259 &  1.75 & $_{-0.7}^{+0.7}$$^{}$ &  &  2.58 & $_{-0.14}^{+0.14}$ & 3.26 & $_{-0.31}^{+0.31}$ & 3.15 & $_{-0.44}^{+0.58}$ & 5.1 & $_{-0.8}^{+1.0}$ & 1.88 & $_{-0.06}^{+0.06}$ & 39 & 9$^{}$ & 22 & 6$^{}$ \\
BD-06 1253 A & 1.99 & $_{-0.22}^{+0.22}$$^{b}$ &  &  2.87 & $_{-0.32}^{+0.52}$$^{b}$ &  3.00 & $_{-0.8}^{+1.1}$$^{b}$ &  2.5 & $_{-1.0}^{+1.0}$$^{}$ &  3.56 & $_{-1.5}^{+1.3}$ &  1.99 & $_{-0.10}^{+0.10}$ & 6.7 & 1.1$^{b}$ & \multicolumn{2}{c}{[27.3,28.2]$^{b}$} \\
BD-05 1329 &  1.97 & $_{-0.07}^{+0.07}$$^{}$ &  &  3.13 & $_{-0.19}^{+0.19}$ & 4.47 & $_{-0.46}^{+0.46}$ & 1.77 & $_{-0.32}^{+0.38}$ & 2.66 & $_{-0.49}^{+0.60}$ & 2.10 & $_{-0.07}^{+0.07}$ & 147 & 9$^{}$ & 29 & 8$^{}$ \\
BD-05 1324 &  1.32 & $_{-0.07}^{+0.07}$$^{}$ &  &  2.28 & $_{-0.16}^{+0.18}$ & 3.77 & $_{-0.41}^{+0.41}$ & 3.7 & $_{-0.9}^{+1.0}$ & 7.6 & $_{-1.7}^{+2.0}$ & 1.75 & $_{-0.07}^{+0.07}$ & 74 & 7$^{}$ & 30 & 5$^{}$ \\
BD+41 3731 &  3.03 & $_{-0.20}^{+0.36}$$^{}$ &  &  5.50 & $_{-0.38}^{+1.37}$ & 3.8 & $_{-0.8}^{+0.8}$ & 0.24 & $_{-0.15}^{+0.18}$ & 0.344 & $_{-0.119}^{+0.042}$ & 2.89 & $_{-0.11}^{+0.11}$ & 345 & 27$^{}$ & -14 & 22$^{}$ \\
BD+46 3471 &  2.84 & $_{-0.08}^{+0.07}$$^{}$ &  &  5.9 & $_{-0.5}^{+0.6}$ & 9.7 & $_{-1.9}^{+1.9}$ & 0.06 & $_{-0.06}^{+0.06}$ & 0.31 & $_{-0.06}^{+0.05}$ & 3.00 & $_{-0.15}^{+0.15}$ & 199 & 11$^{}$ & -3 & 9$^{}$ \\
BD+61 154 &  1.95 & $_{-24}^{+34}$$^{}$ &  &  3.41 & $_{-0.38}^{+0.38}$ & 2.42 & $_{-0.35}^{+0.35}$ & 2.2 & $_{-0.9}^{+0.9}$ & 2.2 & $_{-0.9}^{+0.9}$ & 2.20 & $_{-0.14}^{+0.14}$ & 112 & 24$^{}$ & -16 & 18$^{}$ \\
BD+65 1637 &  3.620 & $_{-0.035}^{+0.034}$$^{}$ &  &  8.11 & $_{-0.23}^{+0.24}$ & 6.7 & $_{-0.7}^{+0.7}$ & 0.035 & $_{-0.010}^{+0.012}$ & 0.153 & $_{-0.010}^{+0.012}$ & 3.56 & $_{-0.05}^{+0.05}$ & 278 & 27$^{}$ & -26 & 20$^{}$ \\
BD+72 1031 &  1.82 & $_{-0.25}^{+0.19}$$^{}$ &  &  2.62 & $_{-0.34}^{+0.59}$ & 3.0 & $_{-1.1}^{+1.1}$ & 3.2 & $_{-1.6}^{+1.9}$ & 4.8 & $_{-2.4}^{+2.8}$ & 1.89 & $_{-0.14}^{+0.14}$ & 180 & 15$^{}$ & -9 & 11$^{}$ \\
HD 9672 &  1.297 & $_{-0.014}^{+0.015}$$^{}$ &  &  2.13 & $_{-0.07}^{+0.08}$ & 1.88 & $_{-0.09}^{+0.09}$ & 7.0 & $_{-1.3}^{+1.1}$ & 9.0 & $_{-1.0}^{+1.0}$ & 1.690 & $_{-0.030}^{+0.030}$ & 195 & 6$^{}$ & 13.1 & 4.6$^{}$ \\
HD 17081 &  2.750 & $_{-0.022}^{+0.022}$$^{}$ &  &  4.65 & $_{-0.08}^{+0.08}$ & 4.84 & $_{-0.19}^{+0.19}$ & 0.279 & $_{-0.023}^{+0.012}$ & 0.469 & $_{-0.021}^{+0.012}$ & 2.630 & $_{-0.030}^{+0.030}$ & 19.9 & 0.9$^{}$ & \multicolumn{2}{c}{[11.0,12.7]$^{}$} \\
HD 31293 &  1.76 & $_{-0.11}^{+0.12}$$^{}$ &  &  2.50 & $_{-0.13}^{+0.29}$ & 2.62 & $_{-0.44}^{+0.44}$ & 3.7 & $_{-0.8}^{+0.6}$ & 5.6 & $_{-1.7}^{+1.1}$ & 1.84 & $_{-0.06}^{+0.06}$ & 116 & 6$^{}$ & 24.7 & 4.7$^{}$ \\
HD 31648 &  1.18 & $_{-0.15}^{+0.18}$$^{}$ &  &  1.93 & $_{-0.14}^{+0.09}$ & 1.93 & $_{-0.32}^{+0.32}$ & 7.8 & $_{-1.5}^{+4.5}$ & 12.9 & $_{-1.8}^{+3.3}$ & 1.60 & $_{-0.06}^{+0.06}$ & 97.5 & 4.7$^{}$ & 12.9 & 3.5$^{}$ \\
HD 34282 & 1.13 & $_{-0.22}^{+0.27}$$^{x}$ &  &  1.59 & $_{-0.07}^{+0.30}$$^{x}$ &  1.66 & $_{-0.37}^{+0.62}$$^{x}$ &  6.4 & $_{-1.9}^{+2.6}$$^{x}$ & & &  & & 105 & 6$^{}$ & 16.2 & 4.8$^{}$ \\
HD 35187 B &  1.15 & $_{-0.20}^{+0.27}$$^{}$ &  &  1.93 & $_{-0.04}^{+0.28}$ & 1.58 & $_{-0.02}^{+0.02}$ & 10.7 & $_{-5.2}^{+3.7}$ & 12.9 & $_{-4.5}^{+1.4}$ & 1.60 & $_{-0.04}^{+0.04}$ & 93.3 & 2.8$^{}$ & 27.0 & 2.1$^{}$ \\
HD 35929 &  2.12 & $_{-0.07}^{+0.07}$$^{}$ &  &  4.13 & $_{-0.24}^{+0.23}$ & 8.1 & $_{-0.7}^{+0.7}$ & 0.16 & $_{-0.08}^{+0.49}$ & 0.68 & $_{-0.13}^{+0.54}$ & 2.46 & $_{-0.08}^{+0.08}$ & 61.8 & 2.2$^{}$ & 21.1 & 1.8$^{}$ \\
HD 36112 &  1.81 & $_{-0.19}^{+0.25}$$^{}$ &  &  2.90 & $_{-0.43}^{+0.67}$ & 4.4 & $_{-0.9}^{+0.9}$ & 2.1 & $_{-1.1}^{+1.1}$ & 3.4 & $_{-1.8}^{+2.4}$ & 2.01 & $_{-0.17}^{+0.17}$ & 54.1 & 4.9$^{}$ & 17.8 & 3.7$^{}$ \\
HD 36910 &  1.69 & $_{-0.16}^{+0.20}$$^{}$ &  &  2.93 & $_{-0.37}^{+0.54}$ & 5.1 & $_{-0.9}^{+0.9}$ & 1.9 & $_{-0.8}^{+0.9}$ & 3.3 & $_{-1.5}^{+1.9}$ & 2.02 & $_{-0.15}^{+0.15}$ & 98 & 5$^{}$ & 35.7 & 4.5$^{}$ \\
HD 36917 &  2.39 & $_{-0.07}^{+0.07}$$^{}$ &  &  3.98 & $_{-0.24}^{+0.25}$ & 5.2 & $_{-0.6}^{+0.6}$ & 0.72 & $_{-0.42}^{+0.29}$ & 1.06 & $_{-0.45}^{+0.33}$ & 2.41 & $_{-0.09}^{+0.09}$ & 127.1 & 4.6$^{}$ & 26.3 & 3.6$^{}$ \\
HD 36982 &  3.22 & $_{-0.07}^{+0.07}$$^{}$ &  &  6.70 & $_{-0.37}^{+0.64}$ & 3.42 & $_{-0.30}^{+0.30}$ & 0.20 & $_{-0.07}^{+0.07}$ & 0.230 & $_{-0.030}^{+0.043}$ & 3.22 & $_{-0.10}^{+0.10}$ & 88 & 8$^{}$ & 30 & 6$^{}$ \\
HD 37258 &  1.44 & $_{-0.07}^{+0.07}$$^{}$ &  &  2.28 & $_{-0.16}^{+0.15}$ & 1.94 & $_{-0.24}^{+0.24}$ & 5.9 & $_{-1.5}^{+1.8}$ & 7.5 & $_{-1.4}^{+1.9}$ & 1.75 & $_{-0.07}^{+0.07}$ & 200 & 14$^{}$ & 31 & 12$^{}$ \\
HD 37357 &  1.72 & $_{-0.07}^{+0.07}$$^{}$ &  &  2.47 & $_{-0.11}^{+0.13}$ & 2.83 & $_{-0.35}^{+0.35}$ & 3.7 & $_{-0.5}^{+0.6}$ & 5.8 & $_{-0.9}^{+1.0}$ & 1.83 & $_{-0.05}^{+0.05}$ & 124 & 7$^{}$ & 21.4 & 4.7$^{}$ \\
HD 37806 &  2.45 & $_{-0.07}^{+0.07}$$^{}$ &  &  3.94 & $_{-0.23}^{+0.23}$ & 4.6 & $_{-0.5}^{+0.5}$ & 0.88 & $_{-0.51}^{+0.21}$ & 1.18 & $_{-0.53}^{+0.25}$ & 2.39 & $_{-0.08}^{+0.08}$ & 120 & 27$^{}$ & 47 & 21$^{}$ \\
HD 38120 &  1.62 & $_{-0.07}^{+0.07}$$^{}$ &  &  2.49 & $_{-0.09}^{+0.09}$ & 1.91 & $_{-0.11}^{+0.11}$ & 5.1 & $_{-0.5}^{+0.5}$ & 5.6 & $_{-0.8}^{+0.8}$ & 1.840 & $_{-0.040}^{+0.040}$ & 97 & 17$^{}$ & 28 & 12$^{}$ \\
HD 38238 &  1.79 & $_{-0.07}^{+0.07}$$^{}$ &  &  2.88 & $_{-0.18}^{+0.18}$ & 4.38 & $_{-0.44}^{+0.44}$ & 2.16 & $_{-0.38}^{+0.46}$ & 3.5 & $_{-0.6}^{+0.8}$ & 2.00 & $_{-0.07}^{+0.07}$ & 99.8 & 4.2$^{}$ & 15.0 & 2.9$^{}$ \\
HD 50083 & 4.15 & $_{-0.12}^{+0.12}$$^{}$ &  & 12.1 & $_{-1.1}^{+1.1}$ & 10.0 & $_{-1.0}^{+1.0}$ & 0.004 & $_{-0.006}^{+0.006}$ & 0.033 & $_{-0.035}^{+0.035}$ & 7.6 & $_{-3.4}^{+3.4}$ & 233 & 22$^{}$ & -0.4 & 1.2$^{}$ \\
HD 52721 & 3.77 & $_{-0.31}^{+0.35}$$^{}$ &  &  9.1 & $_{-1.4}^{+2.4}$ & 5.0 & $_{-1.2}^{+1.2}$ & 0.044 & $_{-0.030}^{+0.073}$ & 0.12 & $_{-0.06}^{+0.06}$ & 3.78 & $_{-0.33}^{+0.33}$ & 215 & 18$^{}$ & 21 & 14$^{}$ \\
HD 53367 &  4.50 & $_{-0.20}^{+0.25}$$^{}$ &  & 16.1 & $_{-1.6}^{+2.7}$ & 7.1 & $_{-1.6}^{+1.6}$ & 0.008 & $_{-0.008}^{+0.016}$ & 0.036 & $_{-0.035}^{+0.000}$ & 5.13 & $_{-0.29}^{+0.29}$ & 41 & 7$^{}$ & 47.2 & 4.8$^{}$ \\
HD 68695 &  1.80 & $_{-0.17}^{+0.14}$$^{}$ &  &  2.64 & $_{-0.30}^{+0.31}$ & 3.3 & $_{-0.6}^{+0.6}$ & 3.0 & $_{-0.8}^{+1.2}$ & 4.7 & $_{-1.4}^{+2.3}$ & 1.90 & $_{-0.13}^{+0.13}$ & 43.8 & 2.6$^{}$ & 20.3 & 1.7$^{}$ \\
HD 72106 A & 1.34 & $_{-0.26}^{+0.28}$$^{l}$ &  &  2.40 & $_{-0.3}^{+0.3}$$^{l}$ &  1.3 & $_{-0.5}^{+0.5}$$^{l}$ &  9.0 & $_{-3}^{+4}$$^{}$ &  9.0 & $_{-3}^{+4}$$^{}$ &  1.3 & $_{-0.5}^{+0.5}$ & 41.0 & 0.3$^{l}$ & 22 & 1$^{l}$ \\
HD 72106 B & 0.96 & $_{-0.27}^{+0.27}$$^{l}$ &  &  1.9 & $_{-0.2}^{+0.2}$$^{l}$ &  1.3 & $_{-0.5}^{+0.5}$$^{l}$ & 9.0 & $_{-3}^{+4}$$^{}$ &  9.0 & $_{-3}^{+4}$$^{}$ &  1.3 & $_{-0.5}^{+0.5}$ & 53.9 & 1.0$^{l}$ & 22 & 1$^{l}$ \\
HD 76534 A &  3.75 & $_{-0.08}^{+0.08}$$^{}$ &  &  9.0 & $_{-0.6}^{+0.6}$ & 7.7 & $_{-1.6}^{+1.6}$ & 0.021 & $_{-0.013}^{+0.018}$ & 0.122 & $_{-0.022}^{+0.019}$ & 3.76 & $_{-0.13}^{+0.13}$ & 68 & 30$^{}$ & 23 & 18$^{}$ \\
HD 98922 &  3.77 & $_{-0.32}^{+0.52}$$^{}$ &  & $^{}$ & $^{}$ & $^{}$ &  &  & & & & & & 50.0 & 3.0$^{}$ & 0.2 & 2.2$^{}$ \\
HD 114981 &  3.56 & $_{-0.24}^{+0.34}$$^{}$ &  &  7.9 & $_{-1.3}^{+2.4}$ & 7.0 & $_{-2.0}^{+2.0}$ & 0.038 & $_{-0.038}^{+0.064}$ & 0.16 & $_{-0.08}^{+0.08}$ & 3.51 & $_{-0.33}^{+0.33}$ & 239 & 13$^{}$ & -50 & 11$^{}$ \\
HD 135344 &  1.16 & $_{-0.18}^{+0.15}$$^{}$ &  &  1.90 & $_{-0.24}^{+0.25}$ & 2.8 & $_{-0.6}^{+0.6}$ & 6.6 & $_{-2.0}^{+3.4}$ & 13.6 & $_{-4.4}^{+6.9}$ & 1.58 & $_{-0.11}^{+0.11}$ & 82.4 & 2.0$^{}$ & -0.0011 & 0.0006$^{}$ \\
HD 139614 &  1.10 & $_{-0.18}^{+0.15}$$^{}$ &  &  1.76 & $_{-0.08}^{+0.15}$ & 2.06 & $_{-0.42}^{+0.42}$ & 8.8 & $_{-1.9}^{+4.5}$ & 17.2 & $_{-3.9}^{+2.7}$ & 1.520 & $_{-0.040}^{+0.040}$ & 24.1 & 3.0$^{}$ & 0.3 & 2.3$^{}$ \\
HD 141569 &  1.49 & $_{-0.06}^{+0.06}$$^{}$ &  &  2.33 & $_{-0.12}^{+0.20}$ & 1.94 & $_{-0.21}^{+0.21}$ & 5.7 & $_{-1.4}^{+1.3}$ & 7.1 & $_{-1.7}^{+1.4}$ & 1.77 & $_{-0.05}^{+0.05}$ & 228 & 10$^{}$ & -12 & 7$^{}$ \\
HD 142666 &  1.44 & $_{-0.13}^{+0.11}$$^{}$ &  &  2.15 & $_{-0.19}^{+0.20}$ & 2.82 & $_{-0.41}^{+0.41}$ & 5.0 & $_{-1.1}^{+1.6}$ & 9.2 & $_{-2.2}^{+3.1}$ & 1.70 & $_{-0.08}^{+0.08}$ & 65.3 & 3.1$^{}$ & -7.0 & 2.7$^{}$ \\
HD 144432 &  1.28 & $_{-0.13}^{+0.11}$$^{}$ &  &  1.95 & $_{-0.16}^{+0.18}$ & 2.59 & $_{-0.40}^{+0.40}$ & 6.4 & $_{-1.4}^{+1.8}$ & 12.4 & $_{-3.0}^{+3.7}$ & 1.61 & $_{-0.07}^{+0.07}$ & 78.8 & 4.2$^{}$ & -3.0 & 3.5$^{}$ \\
HD 144668 &  1.56 & $_{-0.18}^{+0.15}$$^{}$ &  &  2.31 & $_{-0.28}^{+0.29}$ & 3.0 & $_{-0.6}^{+0.6}$ & 4.2 & $_{-1.2}^{+2.0}$ & 7.2 & $_{-2.3}^{+3.7}$ & 1.77 & $_{-0.12}^{+0.12}$ & 199 & 11$^{}$ & -10 & 8$^{}$ \\
HD 145718 &  1.29 & $_{-0.13}^{+0.11}$$^{}$ &  &  1.93 & $_{-0.08}^{+0.14}$ & 2.25 & $_{-0.33}^{+0.33}$ & 7.4 & $_{-1.7}^{+0.7}$ & 12.8 & $_{-2.5}^{+1.8}$ & 1.60 & $_{-0.04}^{+0.04}$ & 113.4 & 3.4$^{}$ & -3.6 & 2.3$^{}$ \\
HD 150193 &  1.79 & $_{-0.13}^{+0.11}$$^{}$ &  &  2.56 & $_{-0.19}^{+0.22}$ & 2.89 & $_{-0.48}^{+0.48}$ & 3.3 & $_{-0.7}^{+0.9}$ & 5.2 & $_{-1.2}^{+1.5}$ & 1.87 & $_{-0.08}^{+0.08}$ & 108 & 5$^{}$ & -4.9 & 3.9$^{}$ \\
\hline
\end{tabular}
}
\end{table*}

\begin{table*}
{
\begin{minipage}{18cm}
\centering
\setlength{\extrarowheight}{3pt}
\contcaption{}
\begin{tabular}{l l         r@{$\pm$}l                                       @{}r         r@{$\pm$}l                               @{}r         p{0.1mm}@{}          r                                     @{$\;\;$}r                                 @{$\;\;$}r       r                                                   @{$\;\;$}r                                           r@{}l                                  @{$\;\;$}r     @{$\;\;$}p{0.1mm}}
\hline
HD or BD & Other      & \multicolumn{2}{c}{$T_{\rm eff}$} & Ref        & \multicolumn{2}{c}{$\log g$} & Ref        &                             & \multicolumn{1}{c}{$V$}   & \multicolumn{1}{c}{$(B-V)$} & Ref            & \multicolumn{1}{c}{$A_{\rm V}$} & \multicolumn{1}{c}{$V_{\rm 0}$} &  \multicolumn{2}{c}{$d$} &  Ref          &      \\
number   & Name      & \multicolumn{2}{c}{(K)}                  &              & \multicolumn{2}{c}{(cgs)}      &              &                             & \multicolumn{1}{c}{(mag)} & \multicolumn{1}{c}{(mag)}   &                  & \multicolumn{1}{c}{(mag)}           & \multicolumn{1}{c}{(mag)}           &  \multicolumn{2}{c}{(pc)} &                 &      \\
(1)           & (2)           & \multicolumn{2}{c}{(3)}                  & (4)         & \multicolumn{2}{c}{(5)}         & (6)         &                             & \multicolumn{1}{c}{(7)}      & \multicolumn{1}{c}{(8)}        & (9)            & \multicolumn{1}{c}{(10)}              & \multicolumn{1}{c}{(11)}              & \multicolumn{2}{c}{(12)}   & (13)          &      \\
\hline
HD 152404 A & AK Sco A & 6500 & 100 & c & \multicolumn{2}{c}{4.0} & & \multirow{2}{*}{$\big \rbrace$} & \multirow{2}{*}{8.839} & \multirow{2}{*}{0.622} & \multirow{2}{*}{w} & \multirow{2}{*}{ 1.06} & \multirow{2}{*}{ 7.78} & \multirow{2}{*}{ 103} & \multirow{2}{*}{ $_{-18}^{+27}$} & \multirow{2}{*}{al} & \multirow{2}{0.1mm}{$\big \lbrace$}  \\
HD 152404 B & AK Sco B & 6500 & 100 & c & \multicolumn{2}{c}{4.0} & & & & & & & &  & \\
HD 163296 &  & 9200 & 300 & ao & 4.2 & 0.3 & ao & & 6.86 & 0.092 & j &  0.32 &  6.54 &  119 & $_{-10}^{+12}$ & al &  \\
HD 169142 &  & 7500 & 200 & ao & 4.3 & 0.2 & ao & & 8.15 & 0.28 & am & -0.30 &  8.45 &  145 & $_{-40}^{+40}$ & ak &  \\
HD 174571 & MWC 610 & 21000 & 1500 & m & 4.00 & 0.10 & m & & 8.87 & 0.610 & j &  4.15 &  4.72 & 540 & $_{-70}^{+80}$ & &  \\
HD 176386 &  & 11500 & 350 & \ddag & \multicolumn{2}{c}{4.5} & \ddag & & 7.22 & 0.121 & j &  1.00 &  6.22 &  128 & $_{-12}^{+15}$ & al &  \\
HD 179218 &  & 9640 & 250 & ao & 3.9 & 0.2 & ao & & 7.40 & 0.094 & j &  0.77 &  6.63 &  254 & $_{-33}^{+45}$ & al &  \\
HD 190073 & V1295 Aql & 9250 & 250 & e & 3.5 & 0.5 & e & & 7.84 & 0.113 & j &  0.43 &  7.41 &      & &  \\
HD 200775 A & MWC 361 A & 18600 & 2000 & a & \multicolumn{2}{c}{3.5} & a & \multirow{2}{*}{$\big \rbrace$} & \multirow{2}{*}{7.34} & \multirow{2}{*}{0.306} & \multirow{2}{*}{j} & \multirow{2}{*}{ 2.43} & \multirow{2}{*}{ 4.91} & \multirow{2}{*}{ 429} & \multirow{2}{*}{$_{-90}^{+156}$} & \multirow{2}{*}{j} & \multirow{2}{0.1mm}{$\big \lbrace$}  \\
HD 200775 B & MWC 361 B & 18600 & 2000 & a & \multicolumn{2}{c}{3.6} & a & & & & & & &  & \\
HD 203024 A &  & \multicolumn{2}{c}{9250} & \dag & \multicolumn{2}{c}{4.0} &  & \multirow{2}{*}{$\big \rbrace$} & \multirow{2}{*}{8.80} & \multirow{2}{*}{0.40} & \multirow{2}{*}{aj} & \multirow{2}{*}{ 1.86} & \multirow{2}{*}{ 6.94} & \multirow{2}{*}{ 420} &  \multirow{2}{*}{$_{-50}^{+50}$} & \multirow{2}{*}{u} & \multirow{2}{0.1mm}{$\big \lbrace$}  \\
HD 203024 B &  & \multicolumn{2}{c}{6500} & \dag & \multicolumn{2}{c}{4.0} &  & & & & & & &  & \\
HD 216629 A & IL Cep A & \multicolumn{2}{c}{19000} & \dag & \multicolumn{2}{c}{4.0} &  & \multirow{2}{*}{$\big \rbrace$} & \multirow{2}{*}{9.34} & \multirow{2}{*}{-0.240} & \multirow{2}{*}{j} & \multirow{2}{*}{-0.25} & \multirow{2}{*}{ 9.59} & \multirow{2}{*}{ 720} & \multirow{2}{*}{$_{-150}^{+190}$} & \multirow{2}{*}{f} & \\
HD 216629 B & IL Cep B & \multicolumn{2}{c}{19000} & \dag & \multicolumn{2}{c}{4.0} &  & & & & & & & & \\
HD 244314 & V1409 Ori & 9250 & 500 & am & \multicolumn{2}{c}{4.0} &  & & 10.19 & 0.22 & z &  0.96 &  9.23 &  375 & $_{-30}^{+30}$ & d &  \\
HD 244604 & V1410 Ori & 8200 & 200 & \ddag & \multicolumn{2}{c}{4.0} & \ddag & & 8.99 & 0.255 & g &  0.57 &  8.41 &  375 & $_{-30}^{+30}$ & d &  \\
HD 245185 & V1271 Ori & 9500 & 750 & ao & 4.0 & 0.4 & ao & & 9.96 & 0.070 & s &  0.21 &  9.75 &  450 & $_{-50}^{+50}$ & ab &  \\
HD 249879 &  & 9000 & 1000 & \dag & \multicolumn{2}{c}{4.0} &  & & 10.64 & 0.05 & am & -0.04 & 10.68 & 2000 & $_{-500}^{+500}$ & am &  \\
HD 250550 & V1307 Ori & 12000 & 1500 & \dag & \multicolumn{2}{c}{4.0} &  & & 9.51 & 0.044 & j &  0.68 &  8.83 &  &  &  &  \\
HD 259431 & V700 Mon & 14000 & 1000 & \dag & \multicolumn{2}{c}{4.0} &  & & 8.71 & 0.274 & j &  2.02 &  6.69 &  660 & $_{-100}^{+100}$ & t &  \\
HD 275877 & XY Per & 9000 & 500 & ay & \multicolumn{2}{c}{4.0} &  & & 9.04 & 0.47 & ad &  1.75 &  7.29 &  120 & $_{-35}^{+87}$ & j &  \\
HD 278937 & IP Per & 8500 & 250 & ao & 4.1 & 0.2 & ao & & 10.36 & 0.31 & y &  0.95 &  9.41 &  320 & $_{-30}^{+30}$ & h &  \\
HD 287823 A &  & \multicolumn{2}{c}{10000} & \dag & \multicolumn{2}{c}{4.0} &  & \multirow{2}{*}{$\big \rbrace$} & \multirow{2}{*}{9.71} & \multirow{2}{*}{0.223} & \multirow{2}{*}{j} & \multirow{2}{*}{ 1.26} & \multirow{2}{*}{ 8.45} & \multirow{2}{*}{ 375} & \multirow{2}{*}{$_{-30}^{+30}$} & \multirow{2}{*}{d} & \multirow{2}{0.1mm}{$\big \lbrace$}  \\
HD 287823 B &  & \multicolumn{2}{c}{7000} & \dag & \multicolumn{2}{c}{4.0} &  & & & & & & & &  \\
HD 287841 & V346 Ori & 7550 & 250 & av & 3.5 & 0.4 & av & & 10.21 & 0.199 & j &  0.09 & 10.11 &  375 & $_{-30}^{+30}$ & d &  \\
HD 290409 A &  & 9000 & 500 & \dag & \multicolumn{2}{c}{4.0} &  & & 10.02 & 0.09 & am &  0.17 &  9.85 &  375 & $_{-30}^{+30}$ & o &  \\
HD 290500 &  & 9000 & 500 & \dag & \multicolumn{2}{c}{4.0} &  & & 11.04 & 0.31 & n &  1.26 &  9.77 &  375 & $_{-30}^{+30}$ & d &  \\
HD 290770 &  & 11000 & 1000 & \dag & \multicolumn{2}{c}{4.0} &  & & 9.27 & 0.03 & am &  0.61 &  8.66 &  375 & $_{-30}^{+30}$ & d &  \\
HD 293782 & UX Ori & 9250 & 500 & ax & \multicolumn{2}{c}{4.0} & ax & & 8.53 & 0.615 & j &  3.06 &  5.47 &  375 & $_{-30}^{+30}$ & d &  \\
HD 344361 & WW Vul & 9000 & 1000 & ar & \multicolumn{2}{c}{4.0} &  & & 10.74 & 0.41 & r &  2.04 &  8.70 &  700 & $_{-150}^{+260}$ & ap &  \\
 & LkHa 215 A & \multicolumn{2}{c}{14000} & \dag & \multicolumn{2}{c}{4.0} &  & \multirow{2}{*}{$\big \rbrace$} & \multirow{2}{*}{10.54} & \multirow{2}{*}{0.52} & \multirow{2}{*}{r} & \multirow{2}{*}{ 3.25} & \multirow{2}{*}{ 7.29} & \multirow{2}{*}{ 900} & \multirow{2}{*}{$_{-100}^{+100}$} & \multirow{2}{*}{ac} & \multirow{2}{0.1mm}{$\big \lbrace$}  \\
 & LkHa 215 B & \multicolumn{2}{c}{14000} & \dag & \multicolumn{2}{c}{4.0} &  & & & & & & & &  \\
 & MWC 1080 & \multicolumn{2}{c}{30000} & aw & \multicolumn{2}{c}{4.0} &  & & 11.58 & 1.197 & j &  7.09 &  4.49 & 2300 & $_{-600}^{+600}$ & &  \\
 & VV Ser & 14000 & 2000 & aq & \multicolumn{2}{c}{4.0} &  & & 11.92 & 0.96 & r &  5.35 &  6.57 &  260 & $_{-100}^{+100}$ & az &  \\
 & VX Cas & 9500 & 1500 & aq & \multicolumn{2}{c}{4.0} &  & & 11.28 & 0.32 & r &  1.67 &  9.61 &  620 & $_{-60}^{+60}$ & ap &  \\
\hline
\end{tabular}
\end{minipage}
}
\end{table*}

\begin{table*}
{ 
\begin{minipage}{18cm}
\centering
\setlength{\extrarowheight}{3pt}
\contcaption{}
\begin{tabular}{l    r@{}l                                                          p{2.5mm}         @{}r@{}l                                             r@{}l                                                  r@{}l                                    r@{}l                                                  r@{}l                                                       r@{$\pm$}l                              r@{$\pm$}l     }
\hline
ID                      &  \multicolumn{2}{c}{$\log(L/L_{\odot})$} &                       & \multicolumn{2}{c}{$M/M_{\odot}$} & \multicolumn{2}{c}{$R/R_{\odot}$} & \multicolumn{2}{c}{age}   & \multicolumn{2}{c}{$t_{\rm PMS}$} & \multicolumn{2}{c}{$R_{\rm ZAMS}$} & \multicolumn{2}{c}{$v\sin i$} & \multicolumn{2}{c}{$v_{\rm rad}$} \\
                         & &                                                             &                       & &                                                      & &                                                    & \multicolumn{2}{c}{(Myr)} & \multicolumn{2}{c}{(Myr)}                & \multicolumn{2}{c}{$R_{\odot}$}         & \multicolumn{2}{c}{(\kms)}    & \multicolumn{2}{c}{(\kms)}        \\
(1) or (2)           &  \multicolumn{2}{c}{(14)}                        &                       &  \multicolumn{2}{c}{(15)}                  & \multicolumn{2}{c}{(16)}                & \multicolumn{2}{c}{(17)}    & \multicolumn{2}{c}{(18)}                 & \multicolumn{2}{c}{(19)}                      & \multicolumn{2}{c}{(20)}        & \multicolumn{2}{c}{(21)} \\
\hline
HD 152404 A & 0.95 & $_{-0.21}^{+0.21}$$^{}$ &  &  1.66 & $_{-0.21}^{+0.29}$ & 2.4 & $_{-0.5}^{+0.5}$ & 9.3 & $_{-3.3}^{+3.8}$ & 20 & $_{-8}^{+10}$ & 1.48 & $_{-0.10}^{+0.10}$ & 18.2 & 1.7$^{}$ & -17.0 & 1.3$^{}$ \\
HD 152404 B & 0.71 & $_{-0.21}^{+0.21}$$^{}$ &  &  1.43 & $_{-0.09}^{+0.20}$ & 1.79 & $_{-0.38}^{+0.38}$ & 13.7 & $_{-4.3}^{+3.6}$ & 31 & $_{-10}^{+6}$ & 1.37 & $_{-0.04}^{+0.04}$ &  17.6 & 0.9$^{}$ & 14.3 & 0.9$^{}$ \\
HD 163296 &  1.52 & $_{-0.08}^{+0.08}$$^{}$ &  &  2.23 & $_{-0.07}^{+0.22}$ & 2.28 & $_{-0.23}^{+0.23}$ & 5.10 & $_{-0.77}^{+0.31}$ & 8.1 & $_{-2.1}^{+0.9}$ & 1.73 & $_{-0.03}^{+0.03}$ & 129 & 8$^{}$ & -9 & 6$^{}$ \\
HD 169142 &  0.88 & $_{-0.28}^{+0.21}$$^{}$ &  &  1.69 & $_{-0.14}^{+0.06}$ & 1.64 & $_{-0.20}^{+0.20}$ & 13.5 & $_{-4.7}^{+11.2}$ & 19.2 & $_{-1.9}^{+5.5}$ & 1.49 & $_{-0.06}^{+0.06}$ & 47.8 & 2.3$^{}$ & -0.4 & 2.0$^{}$ \\
HD 174571 & 3.58 &  $_{-0.21}^{+0.21}$$^{}$ &  &  8.0 & $_{-1.0}^{+1.2}$ & 4.7 & $_{-0.6}^{+0.6}$ & 0.065 & $_{-0.026}^{+0.050}$ & 0.161 & $_{-0.043}^{+0.060}$ & 3.53 & $_{-0.24}^{+0.24}$ & 219 & 31$^{}$ & 14 & 24$^{}$ \\
HD 176386 &  1.91 & $_{-0.09}^{+0.09}$$^{}$ &  &  3.02 & $_{-0.26}^{+0.23}$ & 2.28 & $_{-0.24}^{+0.24}$ & 2.8 & $_{-0.8}^{+1.0}$ & 3.0 & $_{-0.7}^{+1.1}$ & 2.05 & $_{-0.10}^{+0.10}$ & 175 & 6$^{}$ & -2 & 5$^{}$ \\
HD 179218 &  2.26 & $_{-0.12}^{+0.14}$$^{}$ &  &  3.66 & $_{-0.34}^{+0.44}$ & 4.8 & $_{-0.7}^{+0.7}$ & 1.08 & $_{-0.70}^{+0.48}$ & 1.5 & $_{-0.8}^{+0.6}$ & 2.29 & $_{-0.12}^{+0.12}$ & 68.8 & 2.9$^{}$ & 15.1 & 2.3$^{}$ \\
HD 190073 & 1.92 & $_{-0.12}^{+0.12}$$^{e}$ &  &  2.85 & $_{-0.25}^{+0.25}$$^{e}$ &  3.60 & $_{-0.5}^{+0.5}$$^{e}$ &  2.40 & $_{-0.6}^{+0.7}$$^{}$ &  3.6 & $_{-1.0}^{+1.3}$ &  1.99 & $_{-0.10}^{+0.10}$ & \multicolumn{2}{c}{[0-8.3]$^{d}$} & 0.21 & 0.10$^{d}$ \\
HD 200775 A & 3.95 & $_{-0.30}^{+0.30}$$^{a}$ &  & 10.7 & $_{-2.5}^{+2.5}$$^{a}$ & 10.4 & $_{-4.9}^{+4.9}$$^{a}$ &  0.016 & $_{-0.009}^{+0.009}$$^{}$ &  0.07 & $_{-0.07}^{+0.07}$ &  4.1 & $_{-0.5}^{+0.5}$ & 26 & 2$^{a}$ & \multicolumn{2}{c}{[-23.3,8.2]$^{a}$} \\
HD 200775 B & 3.77 & $_{-0.30}^{+0.30}$$^{a}$ &  &  9.3 & $_{-2.1}^{+2.1}$$^{a}$ &  8.3 & $_{-3.9}^{+3.9}$$^{a}$ & 0.016 & $_{-0.009}^{+0.009}$$^{}$ &  0.12 & $_{-0.06}^{+0.08}$ &  3.8 & $_{-0.4}^{+0.4}$ & 59 & 5$^{a}$ & \multicolumn{2}{c}{[-21.1,9.3]$^{a}$} \\
HD 203024 A & 1.88 & &  &  \multicolumn{2}{c}{2.8} & \multicolumn{2}{c}{3.4} & \multicolumn{2}{c}{2.7} & \multicolumn{2}{c}{4.0} & \multicolumn{2}{c}{1.9} & 162 & 11$^{}$ & -14 & 9$^{}$ \\
HD 203024 B & 0.93 & &  &  \multicolumn{2}{c}{1.6} & \multicolumn{2}{c}{2.3} & \multicolumn{2}{c}{9.3} & \multicolumn{2}{c}{21.1} & \multicolumn{2}{c}{1.5} & 57.1 & 3.8$^{}$ & \multicolumn{2}{c}{[-10.5,-5.3]$^{}$} \\
HD 216629 A & \multirow{2}{*}{ 2.58} & \multirow{2}{*}{$_{-0.20}^{+0.20}$$^{}$} & \multirow{2}{0.1cm}{$\big \lbrace$}  & $^{}$ & $^{}$ & $^{}$ &  &  & & & & & & 179 & 27$^{}$ & \multicolumn{2}{c}{[-39,31]$^{}$} \\
HD 216629 B &  &  & $^{}$&  & $^{}$ & $^{}$ &  & & & & & &  & 152 & 33$^{}$ & \multicolumn{2}{c}{[-87,-30]$^{}$} \\
HD 244314 &  1.45 & $_{-0.07}^{+0.07}$$^{}$ &  &  2.33 & $_{-0.23}^{+0.08}$ & 2.07 & $_{-0.26}^{+0.26}$ & 4.78 & $_{-0.19}^{+2.40}$ & 7.1 & $_{-0.7}^{+2.8}$ & 1.77 & $_{-0.10}^{+0.10}$ & 51.9 & 2.2$^{}$ & 22.5 & 1.8$^{}$ \\
HD 244604 &  1.74 & $_{-0.07}^{+0.07}$$^{}$ &  &  2.66 & $_{-0.15}^{+0.15}$ & 3.69 & $_{-0.34}^{+0.34}$ & 2.79 & $_{-0.41}^{+0.52}$ & 4.6 & $_{-0.8}^{+1.0}$ & 1.91 & $_{-0.06}^{+0.06}$ & 98.3 & 1.8$^{}$ & 26.8 & 1.6$^{}$ \\
HD 245185 &  1.40 & $_{-0.10}^{+0.09}$$^{}$ &  &  2.19 & $_{-0.12}^{+0.27}$ & 1.85 & $_{-0.20}^{+0.20}$ & 6.9 & $_{-2.5}^{+2.0}$ & 8.7 & $_{-2.7}^{+1.8}$ & 1.71 & $_{-0.06}^{+0.06}$ & 118 & 22$^{}$ & 16 & 16$^{}$ \\
HD 249879 &  2.31 & $_{-0.25}^{+0.19}$$^{}$ &  &  4.0 & $_{-0.8}^{+0.8}$ & 5.9 & $_{-1.8}^{+1.8}$ & 0.7 & $_{-0.5}^{+1.0}$ & 1.1 & $_{-0.6}^{+1.3}$ & 2.40 & $_{-0.27}^{+0.27}$ & 254 & 26$^{}$ & 11 & 20$^{}$ \\
HD 250550 &  & &  & & & & & &  & &  &  &  & 79 & 9$^{}$ & -22 & 8$^{}$ \\
HD 259431 &  3.35 & $_{-0.14}^{+0.12}$$^{}$ &  &  7.1 & $_{-0.8}^{+0.8}$ & 8.0 & $_{-1.6}^{+1.6}$ & 0.059 & $_{-0.041}^{+0.035}$ & 0.218 & $_{-0.053}^{+0.040}$ & 3.32 & $_{-0.20}^{+0.20}$ & 83 & 11$^{}$ & 26 & 8$^{}$ \\
HD 275877 &  1.21 & $_{-0.30}^{+0.47}$$^{}$ &  &  1.95 & $_{-0.09}^{+0.46}$ & 1.65 & $_{-0.11}^{+0.11}$ & 10 & $_{-6}^{+6}$ & 12.4 & $_{-6.2}^{+3.1}$ & 1.61 & $_{-0.06}^{+0.06}$ & 224 & 12$^{}$ & 2 & 10$^{}$ \\
HD 278937 &  1.21 & $_{-0.09}^{+0.08}$$^{}$ &  &  1.86 & $_{-0.06}^{+0.10}$ & 2.10 & $_{-0.23}^{+0.23}$ & 8.19 & $_{-1.18}^{+0.40}$ & 14.3 & $_{-2.2}^{+1.4}$ & 1.570 & $_{-0.030}^{+0.030}$ & 79.8 & 2.9$^{}$ & 13.7 & 2.1$^{}$ \\
HD 287823 A & 1.79 &  &  &  \multicolumn{2}{c}{2.5} & \multicolumn{2}{c}{2.6} & \multicolumn{2}{c}{3.5} & \multicolumn{2}{c}{5.2} & \multicolumn{2}{c}{1.9} & 10.3 & 1.5$^{}$ & -0.3 & 1.1$^{}$ \\
HD 287823 B & 0.82 &  &  &  \multicolumn{2}{c}{1.6} & \multicolumn{2}{c}{1.8} & \multicolumn{2}{c}{11.9} & \multicolumn{2}{c}{24.2} & \multicolumn{2}{c}{1.4} & 8.2 & 3.3$^{}$ & 54.0 & 1.6$^{}$ \\
HD 287841 &  1.05 & $_{-0.07}^{+0.07}$$^{}$ &  &  1.72 & $_{-0.05}^{+0.15}$ & 1.96 & $_{-0.20}^{+0.20}$ & 9.3 & $_{-0.8}^{+0.6}$ & 18.2 & $_{-4.2}^{+1.6}$ & 1.510 & $_{-0.020}^{+0.020}$ & 115.8 & 4.2$^{}$ & 20.0 & 3.6$^{}$ \\
HD 290409 A & 1.32 & $_{-0.06}^{+0.06}$$^{}$ &  &  2.04 & $_{-0.18}^{+0.18}$ & 1.75 & $_{-0.20}^{+0.20}$ & 10.5 & $_{-4.8}^{+4.8}$ & 11.8 & $_{-3.5}^{+3.5}$ & 1.64 & $_{-0.09}^{+0.09}$ & 250 & 120$^{}$ & 1 & 70$^{}$ \\
HD 290500 &  1.22 & $_{-0.07}^{+0.07}$$^{}$ &  &  1.96 & $_{-0.06}^{+0.21}$ & 1.68 & $_{-0.09}^{+0.09}$ & 9.8 & $_{-3.7}^{+2.9}$ & 12.3 & $_{-3.4}^{+1.3}$ & 1.61 & $_{-0.03}^{+0.03}$ & 85 & 15$^{}$ & 29 & 11$^{}$ \\
HD 290770 &  1.91 & $_{-0.07}^{+0.07}$$^{}$ &  &  2.86 & $_{-0.21}^{+0.27}$ & 2.49 & $_{-0.44}^{+0.44}$ & 2.76 & $_{-0.33}^{+0.86}$ & 3.6 & $_{-0.9}^{+1.0}$ & 1.99 & $_{-0.08}^{+0.08}$ & 240 & 100$^{}$ & 4 & 60$^{}$ \\
HD 293782 &  2.98 & $_{-0.07}^{+0.07}$$^{}$ &  &  6.72 & $_{-0.43}^{+0.42}$ & 12.1 & $_{-1.5}^{+1.5}$ & 0.009 & $_{-0.009}^{+0.027}$ & 0.229 & $_{-0.226}^{+0.034}$ & 3.22 & $_{-0.12}^{+0.12}$ & 221 & 13$^{}$ & 12 & 10$^{}$ \\
HD 344361 &  2.23 & $_{-0.21}^{+0.27}$$^{}$ &  &  3.7 & $_{-0.6}^{+1.0}$ & 5.4 & $_{-1.4}^{+1.4}$ & 0.9 & $_{-0.8}^{+0.9}$ & 1.0 & $_{-1.0}^{+1.0}$ & 2.31 & $_{-0.22}^{+0.22}$ & 196 & 8$^{}$ & -4 & 8$^{}$ \\
LkHa 215 A & 3.08 & $_{-0.10}^{+0.10}$$^{*}$ &  &  \multicolumn{2}{c}{5.8} & \multicolumn{2}{c}{5.9} & \multicolumn{2}{c}{0.1} & \multicolumn{2}{c}{0.3} & \multicolumn{2}{c}{3.0} & 210 & 70$^{}$ & 0 & 40$^{}$ \\
LkHa 215 B & 3.08 & $_{-0.10}^{+0.10}$$^{*}$ &  &  \multicolumn{2}{c}{5.8} & \multicolumn{2}{c}{5.9} & \multicolumn{2}{c}{0.1} & \multicolumn{2}{c}{0.3} & \multicolumn{2}{c}{3.0} & 11.7 & 4.6$^{}$ & \multicolumn{2}{c}{[12,22]$^{}$} \\
MWC 1080 & 5.77 & $_{-0.26}^{+0.20}$$^{}$ &  & \multicolumn{2}{c}{17.4} & \multicolumn{2}{c}{7.3} & \multicolumn{2}{c}{0.0028} & \multicolumn{2}{c}{0.033} & \multicolumn{2}{c}{5.4} & \multicolumn{2}{c}{$^{}$} & \multicolumn{2}{c}{$^{}$} \\
VV Ser  &  2.51 & $_{-0.42}^{+0.28}$$^{}$ &  &  4.0 & $_{-0.8}^{+0.8}$ & 3.1 & $_{-0.9}^{+0.9}$ & 0.64 & $_{-0.35}^{+1.91}$ & 0.76 & $_{-0.33}^{+1.78}$ & 2.43 & $_{-0.29}^{+0.29}$ & 124 & 24$^{}$ & 51 & 18$^{}$ \\
VX Cas &  1.78 & $_{-0.09}^{+0.08}$$^{}$ &  &  2.55 & $_{-0.14}^{+0.37}$ & 2.9 & $_{-0.8}^{+0.8}$ & 3.4 & $_{-1.0}^{+0.6}$ & 5.3 & $_{-1.9}^{+1.0}$ & 1.86 & $_{-0.06}^{+0.06}$ & 158 & 23$^{}$ & -9 & 18$^{}$ \\
\hline
\end{tabular}
\end{minipage}
\begin{minipage}{15cm}
{\footnotesize Notes: $^{*}$~Assuming that both components have the same temperature, and therefore the same luminosity ; References: $^{\dag}$~Visual temperature determination ; $^{\ddag}$~Automatic temperature determination ; $^{a}$~\citet{alecian08a}; $^{b}$~\citet{alecian09b}; $^{c}$~\citet{alencar03}; $^{d}$~\citet{brown94}; $^{e}$~\citet{catala07}; $^{f}$~\citet{crawford70}; $^{g}$~\citet{dewinter01}; $^{h}$~\citet{dezeeuw99}; $^{i}$~\citet{eggen86}; $^{j}$~\citet{esa97}; $^{k}$~\citet{fabricius00}; $^{l}$~\citet{folsom08}; $^{m}$~\citet{fremat06}; $^{n}$~\citet{guetter79}; $^{o}$~\citet{guetter81}; $^{p}$~\citet{harvey08}; $^{q}$~\citet{herbst75}; $^{r}$~\citet{herbst99}; $^{s}$~\citet{hog00}; $^{t}$~\citet{kharchenko05}; $^{u}$~\citet{kun00}; $^{v}$~\citet{kun98}; $^{w}$~\citet{manset05}; $^{x}$~\citet{merin04}; $^{y}$~\citet{miroshnichenko01}; $^{z}$~\citet{miroshnichenko99a}; $^{aa}$~\citet{muller11}; $^{ab}$~\citet{murdin77}; $^{ac}$~\citet{oliver96}; $^{ad}$~\citet{oudmaijer01}; $^{ae}$~\citet{oudmaijer92}; $^{af}$~\citet{preibisch08}; $^{ag}$~\citet{rostopchina00}; $^{ah}$~\citet{shevchenko89}; $^{ai}$~\citet{shevchenko91}; $^{aj}$~Simbad (http://simbad.u-strasbg.fr/simbad/); $^{ak}$~\citet{sylvester96}; $^{al}$~\citet{vanleeuwen07}; $^{am}$~\citet{vieira03}; $^{an}$~\citet{wolff04}; $^{ao}$~\citet{folsom12}; $^{ap}$~\citet{montesinos09}; $^{aq}$~\citet{hernandez04}; $^{ar}$~\citet{mora04}; $^{as}$~\citet{fossati09}; $^{at}$~\citet{miroshnichenko04}; $^{au}$~\citet{hernandez05}; $^{av}$~\citet{bernabei09}; $^{aw}$~\citet{hillenbrand92}; $^{ax}$~\citet{mora02}; $^{ay}$~\citet{mora01}; $^{az}$~\citet{straizys96}}
\end{minipage}
}
\end{table*}

In Fig. \ref{fig:hr} all the stars of our sample are plotted, with circles for non-magnetic stars, squares for magnetic stars, and a triangle for the candidate-magnetic star. The BM01 birthline and the CESAM zero-age-main-sequence (ZAMS) are also overplotted. We observe that two points (the open circles) are situated way outside of the theoretical limits of the PMS region (from the birthline to the ZAMS), { even taking into account their error bars}. For the two corresponding stars (HD 98922 and IL Cep) we are therefore not able to estimate their mass, radius and age using the CESAM theoretical tracks. { For stars situated just below the ZAMS, we have estimated the ranges of the parameters covered by the intersection area between the error ellipse and the HR diagram, and took the middle values}. For the four magnetic stars, HD~190073, HD~200775, HD~72106 and V380~Ori, we have adopted the masses and radii reported in the papers that describe their spectroscopic and magnetic analyses \citep{catala07,alecian08a,folsom08,alecian09b}. However we have redetermined their ages as different birthlines were used in these papers.

This method does not take into account the uncertainties { in the metallicity (however see Sec. 4.2), nor the choice of the birthline}. The inferred masses, radii and ages are therefore approximate, but will be useful when considering comparisons between the stars themselves. The masses, radii and ages are summarised in Table 2. In the same table we also give the PMS duration for each star, and the predicted radius that each star will have once it reaches the ZAMS. Both have been calculated by assuming a mass-constant evolution for each star. The PMS duration has been computed from the birthline. In the case of HD 34282, as we did not calculate the ZAMS radius and the PMS duration because our models are of solar metallicity. HD 98922 and IL Cep fall well outside of the HR diagram, even when the error on their temperatures and luminosities are taken into account. Furthermore, no distance, accurate enough to estimate a luminosity with reasonable error bars, could be found for HD 250550. Therefore no age, mass, radius, ZAMS radius and PMS duration could be estimated for these stars.

%
%

\section{The Least-Squares Deconvolved profile analysis}

\subsection{The LSD method}

\begin{figure}
\centering
\includegraphics[width=4.cm]{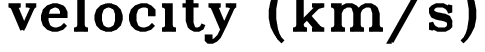}
\hfill
\includegraphics[width=4.cm]{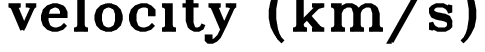}
\caption{LSD Stokes $I$ (bottom) and $V$ (top) profiles of the non-magnetic Herbig Ae star 49 Cet (left), and the magnetic Herbig Ae star V380 Ori (right). The diagnostic $N$ profile is also plotted in the middle. $V$ and $N$ have been shifted on the Y-axis, and magnified by a factor of 10 for 49~Cet, and 2 for V380~Ori, for display purposes.}
\label{fig:lsd}
\end{figure}

In order to increase the signal to noise ratio of our line profiles, we applied the Least Squares Deconvolution (LSD) procedure to all spectra \citep{donati97}. This procedure combines the information contained in many metal lines of the spectrum, in order to extract the mean intensity (Stokes $I$) and polarised (Stokes $V$) line profiles. In Stokes $I$, each line is weighted according to its central depth, while in Stokes $V$ the profiles are weighted according to the product of the central depth, wavelength and Land\'e factor. These parameters are contained in a "line mask" derived from a synthetic spectrum corresponding to the effective temperature and gravity of the star given in Table 2. The construction of the line mask for each star involved several steps. First, we used Kurucz ATLAS~9 models \citep{kurucz93} to obtain generic masks of solar abundances,  and of $T_{\rm eff}/\log g$ following the Kurucz models grid. Our masks contain only lines with intrinsic depths larger or equal to 0.1, which, according to the SNR of our data, is sufficient. We then excluded from the masks hydrogen Balmer lines, strong resonance lines, and lines whose Land\'e factor is unknown. At this stage, the mask contains all predicted lines satisfying the initial assumption of the LSD procedure, i.e. a similar shape for all spectral lines considered in the procedure. This mask will be called in the following the "full mask". Finally, each mask was carefully examined in order to exclude lines predicted by the models, but not appearing in the spectrum, as well as lines contaminated by non-photospheric features. This final "cleaning" procedure is explained in Sec. 5.2 and detailed for each star in Appendix A. Following this procedure, we executed Least-Squares Deconvolution using the full and cleaned masks and the observed spectra, obtaining for each star the mean intensity Stokes $I$ profile, the mean circularly-polarised Stokes $V$ profile, { and the null $N$ profile}. Fig \ref{fig:lsd} shows the LSD $I$, $V$, and $N$ profiles for two stars: one with a magnetic field detection, and one without. { In both cases - as well as in all of our observations, the $N$ profiles are null indicating the absence of spurious polarisation signals, and confirming that the Zeeman signatures detected in the magnetic stars are real.} The use of two separate masks per star is justified in the following sections. The analysis of the LSD $I$ profile and the rotation velocity measurements are described in Section~5.2, while the magnetic analysis performed using the $I$ and $V$ profiles is detailed in Section~6. 

{ The LSD method implies that all lines of the spectra have a similar shape, differing only in their relative strength. The strengths depend on the central depth when the method is applied to an $I$ spectrum, and on the central depth and Land\'e factor for a polarised $V$ spectrum. This hypothesis is reasonable for purely photospheric lines. However in the case of a spectrum contaminated with CS features, these hypotheses should be discussed. The CS features contaminating the photospheric lines of the spectra of the Herbig Ae/Be stars have the same shape (except in few lines like the Balmer, Ca II h\&k or Na D lines, which are removed from the mask). However, their relative strength is not dependent on the central depths of the photospheric lines, which are used to weight each line in the LSD procedure. Therefore, the LSD method, by averaging the contaminated lines, applies inappropriate weights to the CS contribution, while taking correctly into account the different weights of the photospheric lines. In the LSD procedure, using wrong weights does not change the global shape of the resulting profile, however its relative strength (with respect to the $V$ profile, for example) cannot be trusted. Therefore, the strength of the CS contribution of our LSD I profiles must be investigated in more detail before it can be reliably used to drawn any quantitative conclusions. However, its shape can be modelled and removed to be able to analyse the photospheric contribution of the $I$ profile, which is the one of interests for this paper. We describe in the following section how the CS contamination has been handled in this study.}

\subsection{Fitting of the LSD $I$ profiles}

The LSD $I$ profiles computed with the full mask reveal a rather complex average of the lines of the spectrum included in the line mask. Most HAeBe stars show CS emission and absorption in their spectrum; these effects are also reflected in the LSD profiles. Those lines that are the most strongly contaminated by CS emission can be easily identified directly in the spectrum itself, and excluded from the mask. For some of the stars, the resulting LSD profiles show a relatively clean rotational profile indicative of simple photospheric absorption. For others, the resulting profiles still show significant CS absorption and/or emission that is not possible to remove by further refinement of the line mask. However, the investigations of rotation and magnetic fields in papers II and III require that we are able to extract an approximation of the uncontaminated photospheric profiles in order to infer $v\sin i$ and to model the magnetic constraint imposed by Stokes $V$.

To characterise the various contributions to the LSD $I$ profiles, we have performed a least-squares fit to each of the LSD $I$ profiles using several models. In the first case we consider a simple photospheric profile modelled using the convolution of a rotation function (depending on the projected rotational velocity $v\sin i$ and the radial velocity of the star $v_{\rm rad}$), and a Gaussian (approximating the local photospheric profile) whose width is fixed and computed from the spectral resolution and the macroturbulent velocity (Gray 1992). This convolution will be called the photospheric function. The free parameters of the fitting procedure are the line depth, \vsini\ and $v_{\rm rad}$. In order to fit the wings of the observed LSD profiles of our sample, a macroturbulent velocity ($v_{\rm mac}$) is frequently required to be added to the model. {Only a few stars of our sample have LSD $I$ profiles suitable for estimating the value of $v_{\rm mac}$, and the typical value is found to be around 2~\kms. The other stars of our sample (i.e. most of them) either display too large a Doppler broadening, and/or the wings of the profile are sufficiently contaminated by CS features, that they do not allow us to obtain useful information about $v_{\rm mac}$. Nonetheless, we assumed that a macroturbulent velocity field is present near the surface of the star, and we adopted for all the stars an isotropic macroturbulent velocity of 2~\kms. However, taking into account a $v_{\rm mac}\sim2$~\kms\ seems to improve the fit, when fitting with the eye, only for stars with \vsinis lower than 40~\kms. In order to estimate the error on the \vsinis introduced by fixing $v_{\rm mac}$, we varied the value of $v_{\rm mac}$ between 0 and 4~\kms, and we find that it introduces significant variations of \vsini\ only if \vsini\ is lower than 10~\kms. For \vsini\ between 10 and 40 \kms, changing $v_{\rm mac}$ modifies the value of \vsini\ within the error bars. For \vsini\ larger than 40 \kms, changing $v_{\rm mac}$ has no impact on \vsini. The macroturbulent velocity is therefore not a significant parameter to be considered within this fitting procedure. It has been included in all the fitted models for consistency from one star to the other, but should only be considered with caution at very low \vsini\ (lower than 10~\kms).}

The second model considers one or more Gaussian functions meant to model the CS features present in the LSD $I$ profile. These functions are added to the photospheric function with the aim of providing the best reproduction of the observed profile. The parameters of each Gaussian contributing to the CS function are the full-width at half maximum (FWHM), the centroid position, and the amplitude. The amplitude may be either positive or negative, corresponding to emission or absorption contributions. Depending on the star analysed, we require between zero and four circumstellar functions to fit the Stokes $I$ profile. Hence, the number of fitting parameters required to reproduce the LSD $I$ profiles ranges from 3 up to 15. For a few stars for which this simple model was unable to reproduce features in the observed profile, an alternative method we applied was to suppress from the fit the regions contaminated with CS (or other) features. We accomplished this by assigning a null weight to all points inside the affected region, and fit the profile with only a photospheric function. 

{ The LSD profiles of our observations show systematically a continuum level lower than one (while the spectra are all normalised to one). We believe that this is due to spectral features not taken into account in the LSD procedure, and not due to poor continuum normalisation or to a reduction problem, for two reasons. First, multiple LSD profiles of a star obtained at different times show a continuum at the same level. However, multiple observations of different stars do not have the same continuum level, suggesting that the choice of the mask might determine the level of the continuum. Secondly, when we apply the LSD procedure on a simulated ESPaDOnS or Narval observation, in which the continuum is by construction perfectly normalised to one, the continuum level of the resulting $I$ profile is also lower  than one.

In order to simplify the fitting procedure and avoid additional parameters to fit, we have normalised each profile before fitting them. The continuum levels have been determined by fitting a line between two points chosen by eye on each side of the profiles. The profiles are then divided by the fitted continuum. We have checked that this normalisation process does not introduce significant errors by repeating the procedure many times and checking that the fitted parameters converge all towards the same value.
}

A remarkable result of this procedure is the conclusion that the mean line profiles of most HAeBe stars can be satisfactorily reproduced by a simple model consisting of the sum of a rotationally-broadened photospheric profile and a small number of local absorption/emission profiles assumed to be contributed by the CS environment. The analysis of the quantitative characteristics of the CS-contributed profiles will be described in a future paper.

\begin{figure}
\centering
\includegraphics[width=3cm,angle=90]{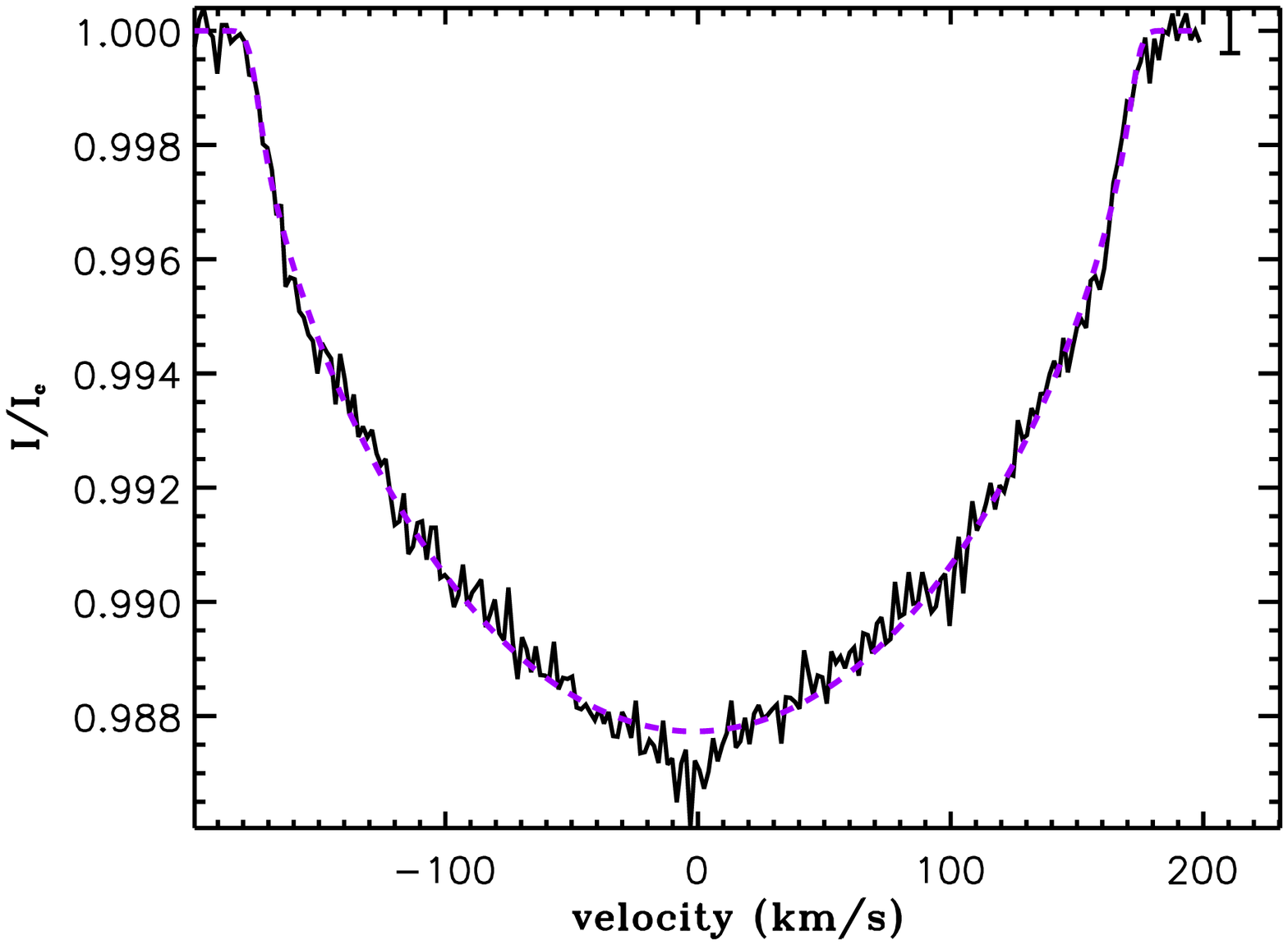}
\includegraphics[width=3cm,angle=90]{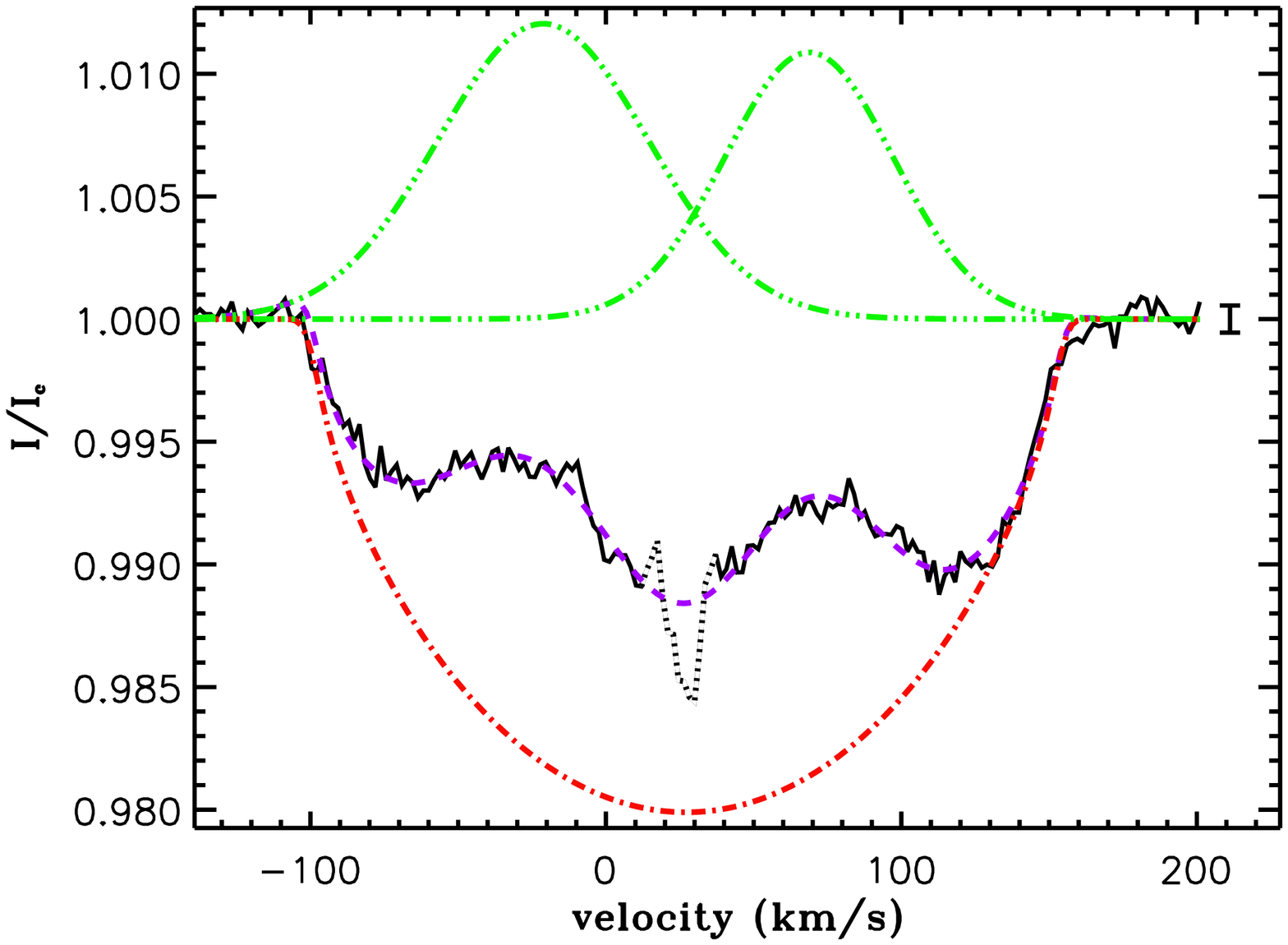}
\caption{Examples of LSD fits (dashed purple line) using only a photospheric (red dot-dashed line) function for HD~176386 (left) and using a photospheric + two Gaussian (green dot-dot-dot-dashed line) functions for HD~36917 (right) observed on Nov. 8th 2007. In the case of HD~36917, the narrow absorption in the core of the profile (dotted line) has been excluded from the fit.}
\label{fig:fitlsdI}
\end{figure}

As an additional complication, some stars in our sample are double-lined spectroscopic binaries (see Sec. 5.3). In these cases, we fitted the LSD $I$ profiles using the sum of two photospheric functions, each one having independent fitting parameters. Examples of the results of the fitting procedure are shown in Fig.~\ref{fig:fitlsdI}.

This fitting procedure is automatic and requires as input first guesses for each free parameter. We checked the uniqueness of the derived fit by modifying the first guesses, and checking that whatever the first guesses, the fits always converge towards a unique solution. This verification procedure was always successful for the photospheric parameters (the photospheric line depth, \vsinis and \vrad). However in some cases when one or more Gaussian CS functions was required, we could find multiple solutions (e.g. the addition of a Gaussian in emission or a Gaussians in absorption could give fits of comparable quality). In those cases we checked individual spectral lines in order to determine which of the solutions was the most consistent with the observed spectrum. Whenever multiple observations were obtained for the same star, we also chose the solution the most consistent over all observations.

When multiple observations were obtained for the same star with similar SNRs, we performed a simultaneous fit to the whole set of observations, by forcing the photospheric parameters (depth, $v\sin i$, $v_{\rm rad}$) to be the same for each of the observations. When necessary, one or more Gaussian functions were added to the profiles, with independent parameters for each observation.

We checked that changing the effective temperature and gravity of the masks, within the error bares, do not change our determination of our \vsini\ values. The fit performed to reproduce the shape of the LSD $I$ profile is dominating the uncertainties. The uncertainties on the adopted parameters (see Table 2) have therefore been determined by calculating the confidence intervals at a level of 99.73\%, as described in \citet[][p. 697]{press92}. 

After completing the analysis of the LSD profiles, we checked that the derived $v\sin i$ were consistent with individual spectral lines in the reduced spectra. The adopted values of $v\sin i$ and $v_{\rm rad}$ are summarised in Table 2, and the fitting procedure for each of the spectra is described in Appendix A. In the case of MWC 1080, no photospheric lines could be identified in the spectrum, therefore no \vsini\ values could be measured.

\subsection{The discovery of spectroscopic binaries}

The inspection of the LSD profiles allowed us to easily identify a number of spectroscopic binaries among our sample. Among the stars not detected as magnetic, a total of 5 SB2 systems were identified: 3 previously known or suspected (AK Sco, HD 287823, IL Cep), and two new discoveries previously unreported in the literature (HD 203024 and HD 290409). Based on the detection of a LiI lines at 6707 A, \citet{corporon99} claimed the presence of a low mass companion orbiting HD 203024. However this claimed companion cannot be the spectroscopic companion that we detect because its temperature is too high. We therefore cast some doubts on the stellar nature of the feature observed at $\sim 6708$~\AA, and leave open the interpretation of its origin. 

For each of the SB2 systems, we have attempted a determination of the effective temperature of both components, as well as of their luminosity ratio. The description of each system is detailed in Appendix A.

%
\section{Magnetic field diagnosis}

\subsection{Method}

Each of the spectra that we have acquired with ESPaDOnS and Narval was obtained in Stokes $V$ polarimetric mode, in order to allow us to measure the longitudinal Zeeman effect in spectral lines. We employ two methods to detect magnetic fields in our program and standard stars. First, we use the Stokes $V$ spectra to measure the mean longitudinal magnetic field strength \bz\ of each star at the time of observation. This is the conventional measure of field strength normally used for detection of magnetic fields in main sequence stars \citep[e.g.][]{landstreet82}. However, because of the high value of the resolving power, we can also examine spectral lines for the presence of circular polarisation signatures: Zeeman splitting combined with Doppler broadening of lines by rotation leads to non-zero values of $V$ within spectral lines even when the value of \bz\ is close to, or even equal to, zero. This possibility substantially increases the sensitivity of our measurements as a discriminant of whether a star is in fact a magnetic star or not, as discussed by \citet{shorlin02}, \citet{silvester09} and \citet{shultz11}.

Each set of LSD Stokes $I$ and $V$ profiles is therefore analysed in two ways. First, the value of \bz\ is determined by computing the first-order moment of Stokes $V$, normalised to the equivalent width of Stokes $I$ \citep{mathys91,donati97,wade00}:

\begin{equation}
\bz\ = -2.14 \times 10^{11} \frac{\int_{}^{}vV(v)dv}{\lambda {\bar g}c\int_{}^{}[I_c - I(v)]dv},
\label{bleqn}
\end{equation}

\noindent where \bz\ is in G, $\bar g$ is the mean Land\' e factor of the LSD weights in the line mask (typically $1.3\pm 0.1$), and $\lambda$ is the SNR-weighted mean wavelength of the LSD weights in the mask in~nm (typically $520\pm 40$~nm). The uncertainties $\sigma$ on \bz\ have been computed by propagating the error of each pixel within the Stokes $I$ and $V$ profiles through Eq. (1). The limits of integration are usually chosen for each star to coincide with the observed limits of the LSD $I$ and $V$ profiles; using a smaller window would neglect some of the signal coming from the limb of the star, while a window larger than the actual line would increase the noise without adding any further signal, thus degrading the SNR below the optimum value achievable \citep[see e.g.][]{neiner12}.

In addition, the LSD Stokes $V$ profile is itself examined. We evaluate the false alarm probability (FAP) of $V/I_{\rm c}$ inside the line according to:

\begin{equation}
{\rm FAP}\,(\chi_{\rm r}^2,\nu) = 1 - P\,({\nu\over 2}, {{\nu\chi_{\rm r}^2}\over 2}),
\label{fapeqn}
\end{equation}

\noindent where $P$ is the incomplete gamma function, $\nu$ is the number of spectral points inside the line, and $\chi_{\rm r}^2$ is the reduced chi-square ($\chi^2/ \nu$) computed across the $V$ profile relative to zero \citep[e.g.][]{donati92}. The FAP value gives the probability that the observed $V$ signal inside the spectral line could be produced by chance if there is actually no field present. Thus a very small value of the FAP implies that a field is actually present. We evaluate FAP using the detection thresholds of \citet{donati97}. We consider that an observation displays a ``definite detection" (DD) of Stokes $V$ Zeeman signature if the FAP is lower than 0.00001, a ``marginal detection" (MD) if it falls between 0.001 and 0.00001, and a ``null detection" (ND) otherwise. { As mentioned above a {\it significant signal} (i.e. with a MD or DD) may occur even if \bz\ is not significantly different from zero. On the contrary, a profile can give a \bz\ different from zero at a level of 3 $\sigma$ or lower without displaying a marginal or a definite detection (see Fig. 10 and Sec. 6.3). For these reasons, the FAP is the most sensitive diagnostic of the presence of a magnetic field, and will be the only one applied in this paper. Table 4 summarises our measurements of FAP and the magnetic diagnosis (DD, MD or ND) for all observations, except those already published in previous papers.}

If a significant signal (i.e. with a MD or DD) is detected within the line, while always remaining insignificant in the neighbouring continuum and in the $N$ profile, and is detected in multiple observations, { we conclude that the star is unambiguously magnetic.}

\subsection{The magnetic HAeBe stars detected within the survey}

\begin{table}
\caption{Log of our first observations of HAeBe stars in which we have detected a magnetic field. Column one gives the name of the star. Columns 2, 3 and 4 report the Heliocentric Julian Date, the type of detection (MD for marginal detection and DD for definite detection) and the longitudinal field measurement reported in the literature. In the final column we indicate the first refereed publication in which the field detection was reported. Discoveries by Wade et al. (2005), Catala et al. (2007) and Alecian et al. (2008a) are derived from this survey. LP Ori was reported to be magnetic by Petit et al. (2008) as part of a parallel program. However, as it was serendipitously detected within this survey as well, we include this star as a {\em bona fide} blind detection in our statistics.}
\begin{center}

\begin{tabular}{@{}lccr@{\,$\pm$\,}lc@{}}\hline\hline
Name & HJD         & Det & $\langle B_{\rm z}\rangle$ & $\sigma_{\rm B}$ & Discovery \\
           &-2450000 &        &                         \multicolumn{2}{c}{(G)}         & paper \\
\hline
\multicolumn{6}{c}{\it Confirmed magnetic HAeBe stars}\\
V380 Ori & 3421.900 & MD & -165 & 190 & Wade et al. (2005)\\ 
HD 36982 & 4416.549 & MD & -240 & 70 & Petit et al. (2008)\\                          
HD 72106 & 3423.924 & DD & 228 & 50 & Wade et al. (2005)\\     
HD 190073  & 3607.789  & DD & 111 & 13  &  Catala et al. (2007)\\
HD 200775 & 3608.920  & MD & 74 & 63  &  Alecian et al. (2008a)\\
\hline
\multicolumn{6}{c}{\it Suspected magnetic HAeBe stars}\\
HD 35929 & 4884.328 & DD & 74 & 19 &  This paper\\
\hline
\end{tabular}
\label{magtable}
\end{center}
\end{table}

Of the 70 stars observed with ESPaDOnS and Narval, 6 show Stokes $V$ Zeeman signatures in their LSD profiles. Four of these stars (V380~Ori, HD~72106, HD~190073, HD~200775) were unambiguously detected for the first time as a result of this study. First results for V380~Ori and HD~72106 were reported by \citet{wade05}; for HD~190073 by \citet{catala07}; for HD~200775 by \citet{alecian08a}.  In addition, one HAeBe star for which magnetic field detections were previously reported \citep[HD~36982=Par~1772=LP~Ori ;][]{petit08} is confirmed to be magnetic. Finally, one new suspected magnetic HAeBe star (HD~35929) is reported here. However as HD~35929 is a $\delta$-Scuti pulsating star \citep{marconi00}, more analysis and observations are required to verify that the signature detected in the $V$ profile is of magnetic origin.

In Table~\ref{magtable}, for each of the detected program stars, we list the observational details corresponding to the ESPaDOnS/Narval observations from which the presence of a magnetic field was first inferred. In addition, in Fig. 4 we show the positions on the HR diagram of the magnetic HAeBe stars of our sample. 

Each of the 5 detected stars has been or will be discussed in detail in a dedicated paper (e.g. HD 190073 by Catala et al. 2007, HD 200775 by Alecian et al. 2008a, HD 72106 by Folsom et al. 2008, V380 Ori by Alecian et al. 2009b). They will not be discussed further here.

Five other magnetic HAeBe stars have been discovered and confirmed during recent years, as part of parallel observational programs with ESPaDOnS, Narval, FORS1 on the Very-Large-Telescope (ESO, Chile), or the Semel Polarimeter \citep[SEMPOL][]{semel89,semel93,donati03} coupled with the spectrograph UCLES on the Anglo-Australian-Telescope (AAT, Australia). The first magnetic detections and their confirmations are reported or are to be reported in other publications: HD~101412 \citep[][Alecian et al. in prep.]{wade07,hubrig09}, HD~104237 \citep[][Alecian et al. in prep.]{donati97}, NGC~6611~601 \citep{alecian08b}, NGC~2244~201 \citep{alecian08b}, and NGC~6611~83 \citep{alecian09a}. We are mentioning them in this paper for the sake of completeness, however they won't be included in the statistical analyses that will be presented in paper II and III as they are not part of the survey presented in this series of papers. They will not be discussed any further here.

\subsection{Magnetic analysis of the remaining sample}

The polarised spectra of the undetected stars, i.e. displaying no magnetic signatures, contain a valuable information that we want to extract: the upper limits on admissible surface magnetic fields. In this section we first describe the problems that a typical spectrum of Herbig Ae/Be stars can bring in evaluating realistic values of such limits due to the CS contribution to the spectra. Then we propose the method that we adopted to solve the problems: the hybrid method.

\subsubsection{The CS contribution to the Stokes $I$ and $V$ spectra}

Before discussing the analysis of the stars in which no firm magnetic detections was obtained, it is instructive to consider the formation of the stellar spectrum, beginning in the photosphere of a magnetised HAeBe star. Upon exiting the "top" of the photosphere, the (absorption) lines will be partially circularly polarized due the magnetic field. As the flux propagates into the CS environment, it will undergo absorption or emission contributions due to the CS material. As observed in the spectra of real HAeBe stars, this can strongly modify the Stokes $I$ line profiles. However, we expect that the magnetic field strength will decrease rapidly with distance from the star: as $1/r^3$ for a dipole, and more rapidly for more complex fields. Therefore the contribution of the Zeeman effect to Stokes $V$ in the CS environment should be very small compared to the photospheric contribution. In other words, the CS contribution to the flux is expected to be negligibly circularly polarized. A consequence of this conclusion is that the observed Stokes $V/I_{\rm c}$ spectrum of a magnetic HAeBe star is expected to be reflective of the {\em photospheric} spectrum of the star, even if the Stokes $I$ spectrum is strongly modified by the CS environment. An important implication of this conclusion it that CS contamination of spectral lines cannot serve to "hide" the Zeeman signatures produced by a photospheric magnetic field. Note however, because the $V$ spectrum is normalised to the inferred continuum, this conclusion and its implication rely on the assumption that no significant veiling is present \citep{ghandour94,folsom12}. Although veiling does not modify the absolute amplitude or shape of the Zeeman signature, it serves to increase the noise, and could therefore render a Zeeman signature undetectable. 

\begin{figure}
\centering
\includegraphics[width=7cm,angle=-90]{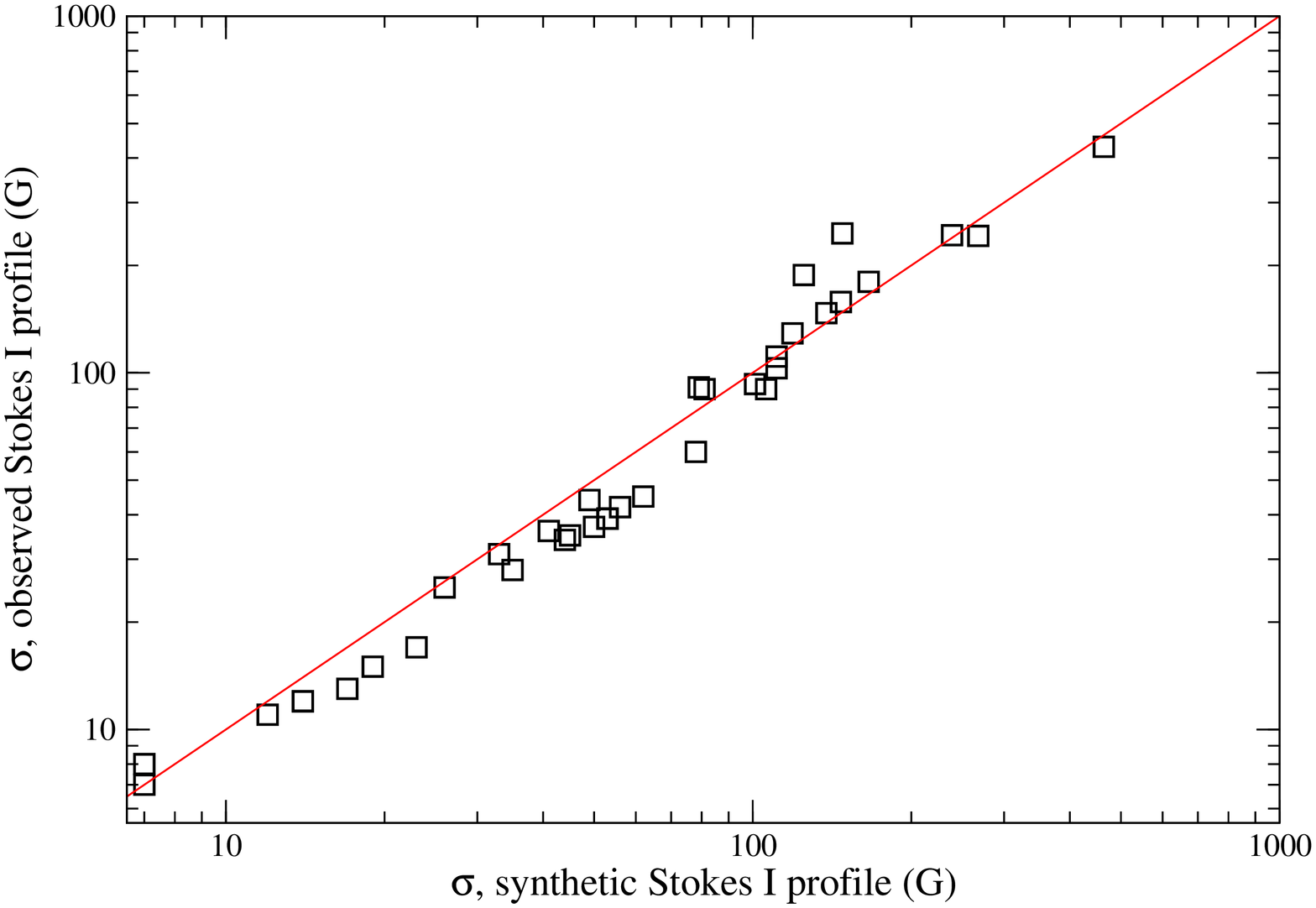}
\includegraphics[width=7cm,angle=-90]{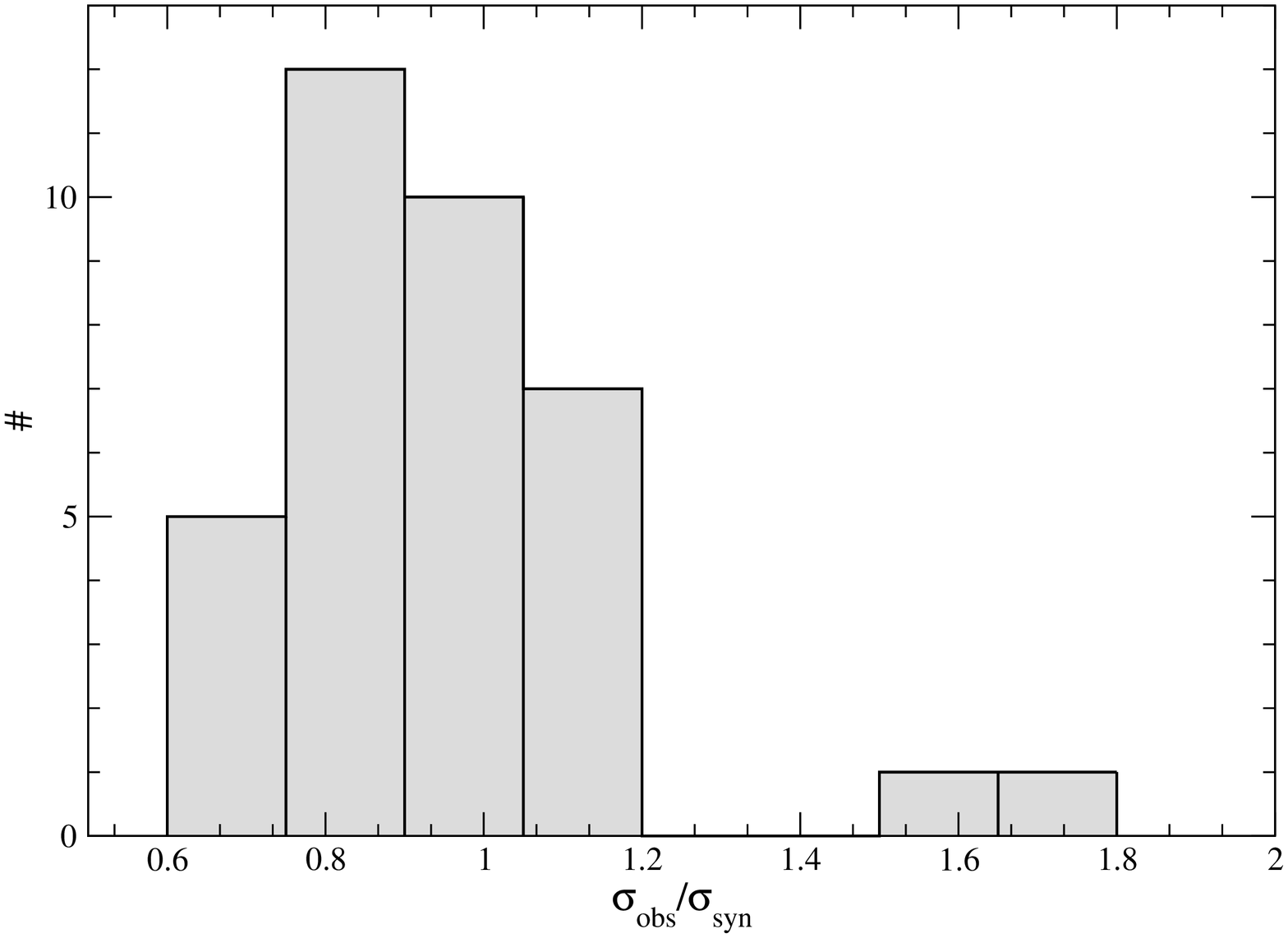}
\caption{Comparison of longitudinal field error bars for synthetic versus observed Stokes $I$ profiles. {\em Upper frame -}\ Scatter plot of longitudinal field error bars computed using synthetic Stokes $I$ versus error bars computed using the observed Stokes $I$ dominated by the photospheric component only. The solid (red) line corresponds to perfect agreement. {\em Lower frame -}\ Histogram of the ratio of error bars with synthetic Stokes $I$ to error bars with observed Stokes $I$, for observations with LSD profiles dominated by the photospheric component.}
\label{mag3}
\end{figure}

The LSD profiles produced for line profile analysis in Sect.~5.2 with the cleaned masks are heavily filtered: many lines have been removed from the line masks in order to reduce the CS contribution and to reveal the photospheric profile. In the case of the Stokes $V$ profile, this results in a relatively high noise level (because of the relatively small number of lines used in the deconvolution) and consequently low sensitivity to magnetic fields (see the SNR in $V$ obtained from both full and cleaned masks in Table 1). However, in contrast to the Stokes $I$ profile, we have concluded that Stokes $V$ is not strongly modified by the CS contribution to the line. Therefore the most sensitive magnetic diagnosis should be obtained by including as many lines as possible in the mask. However, as we have seen in Sect. 5, LSD Stokes $I$ profiles derived from such masks can be heavily modified by CS contributions. Even if the Stokes $V$ Zeeman signature is not modified significantly, using such contaminated $I$  profiles has two important consequences for our diagnosis of the magnetic field. First, uncertainty is introduced into the appropriate integration range to use to compute the longitudinal magnetic field in Eq. (1), and the reduced $\chi^2$ in Eq. (2). Secondly, the equivalent width of the Stokes $I$ profile (i.e. the denominator of Eq. (1)) is modified. Both of these consequences can change the inferred values of the longitudinal field and its error bar, while the first can influence the derived FAP. Of these, the impact on the longitudinal field is the most severe. 
For example, CS emission/absorption superimposed with the photospheric spectral lines can reduce/increase the equivalent with of the Stokes $I$ profile significantly, artificially increasing/decreasing the derived longitudinal field and its error.
In the absence of a magnetic detection, the longitudinal field error bar is the only important statistical quantity, as it provides an estimate of the upper limit on admissible fields. Because it is sensitive to CS contamination of Stokes $I$, it is important to understand, and potentially limit, the CS contribution to the $I$ profile (even if the $V$ profile is unmodified).

\subsubsection{An hybrid approch}

With these insights, we approached the problem of obtaining realistic quantitative longitudinal field measurements of the 65 program stars for which no significant magnetic field was detected. Our goal was to obtain measurements of the longitudinal field for which the error bars were simultaneously accurate and precise. The first option considered was to use the cleaned line masks described in Sect. 5. These have the advantage that they reveal, in many cases, the photospheric profile of the star. The disadvantage is that, in many cases, this is accomplished by excluding most of the lines - especially strong lines - that contribute significantly to reducing the Stokes $V$ noise level. The second option was to use the full line masks. This has the advantage of reducing the noise level of Stokes $V$ to the greatest extent, but the disadvantages of a strongly-contaminated $I$ profile (as described above). We considered using masks for which an intermediate level of cleaning had been applied, but it was not obvious to what extent to clean the masks, nor was it clear that we were not simply combining the uncertainties and disadvantages of both options 1 and 2.

\begin{figure}
\centering
\includegraphics[width=3cm,angle=-90]{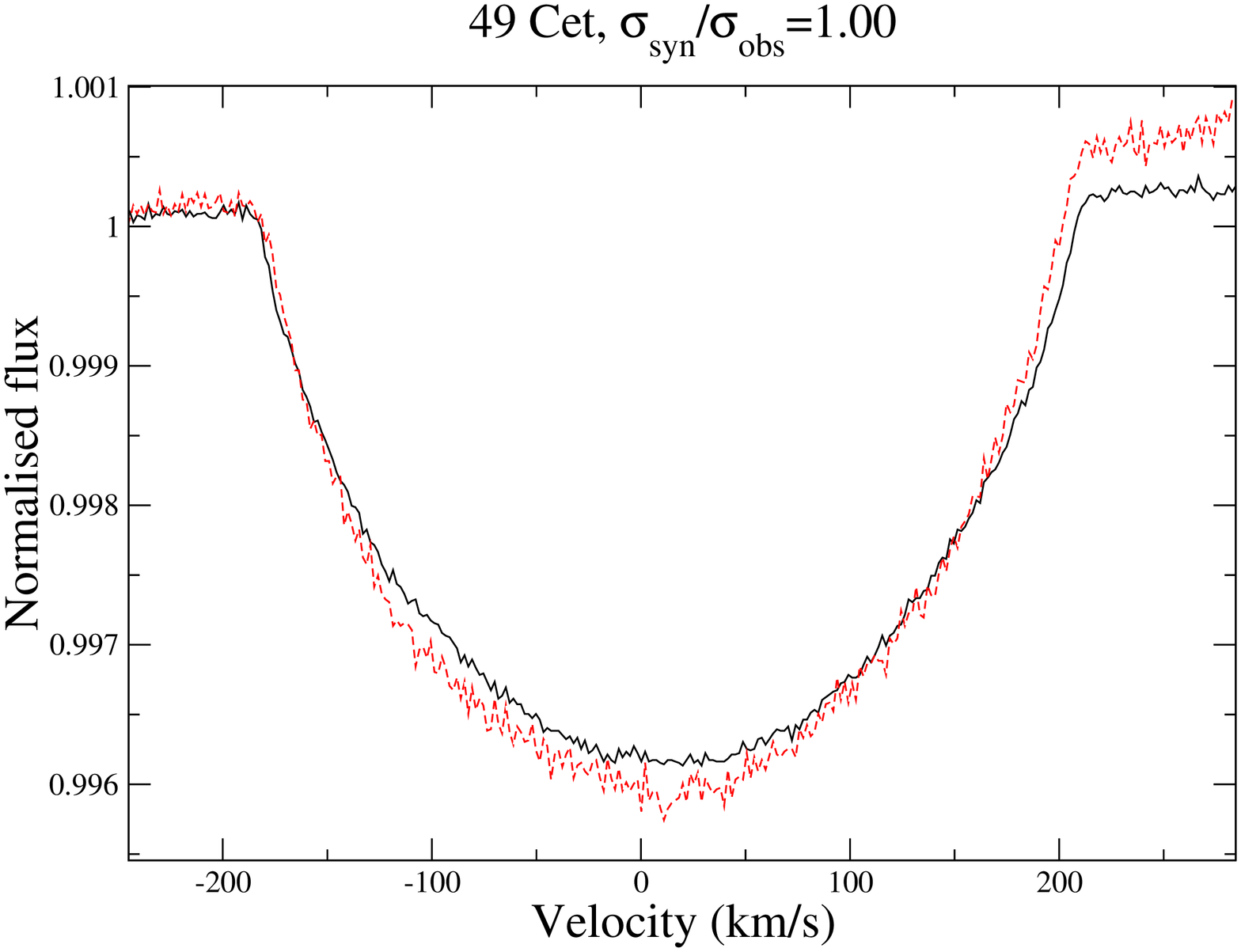}
\includegraphics[width=3cm,angle=-90]{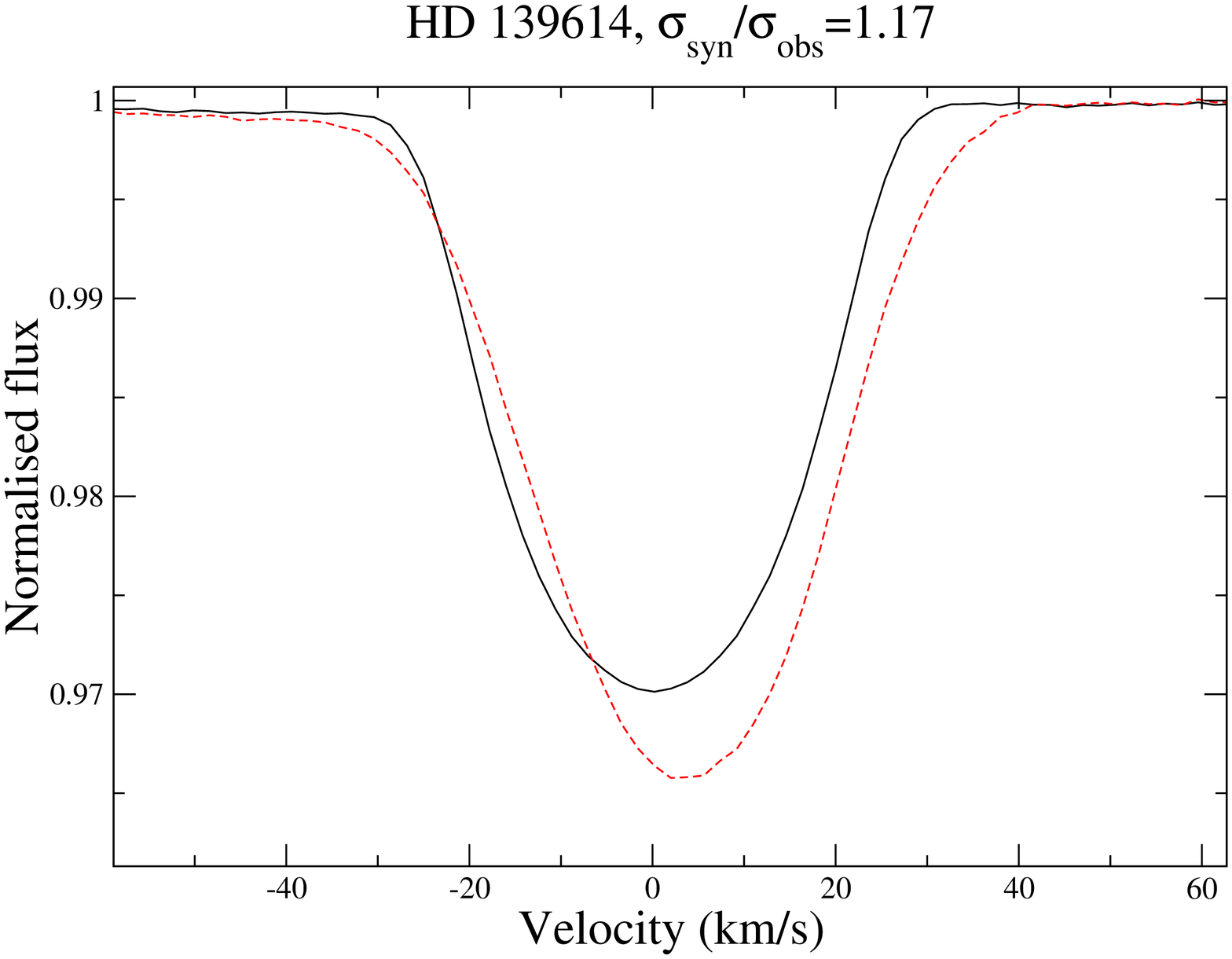}
\includegraphics[width=3cm,angle=-90]{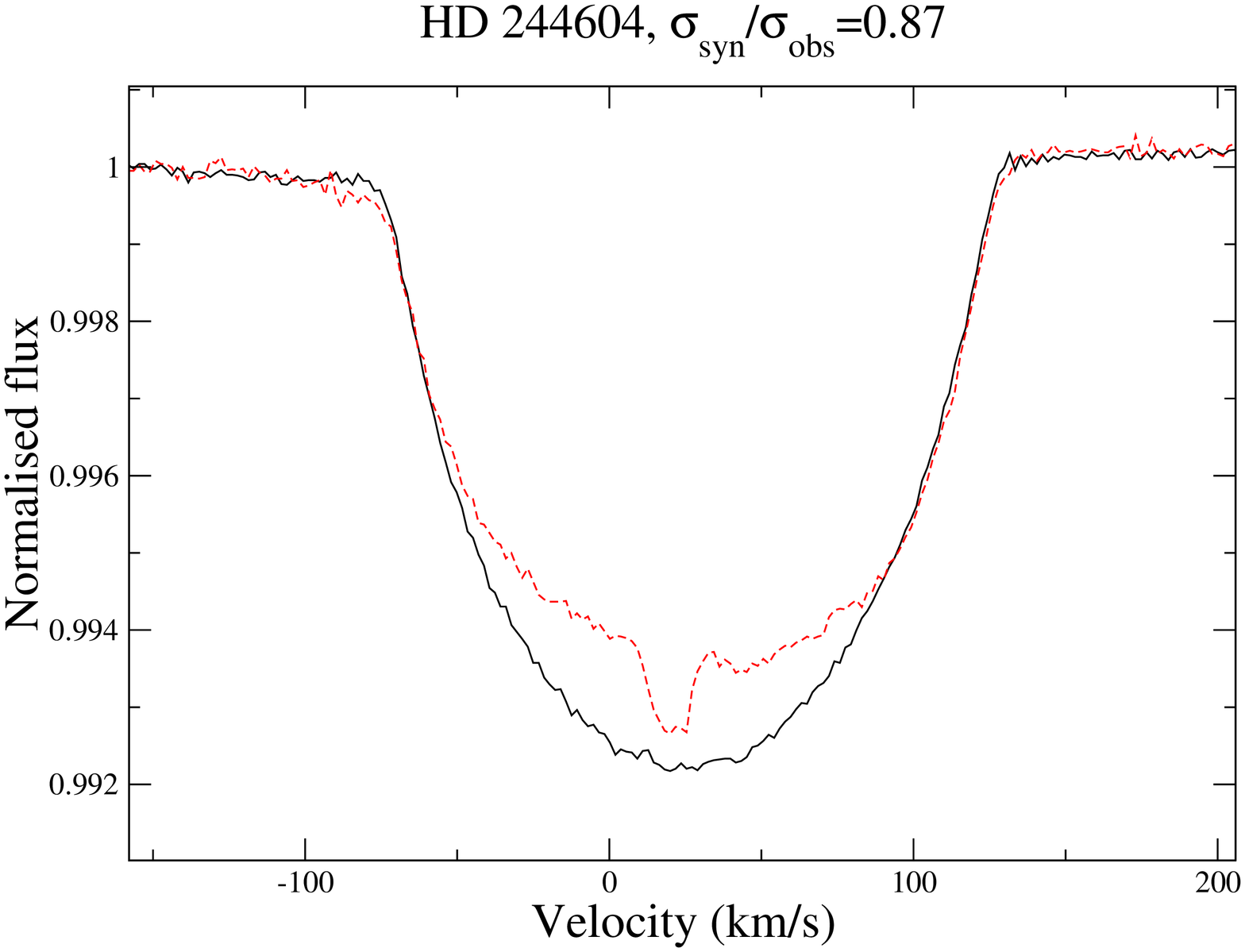}
\includegraphics[width=3cm,angle=-90]{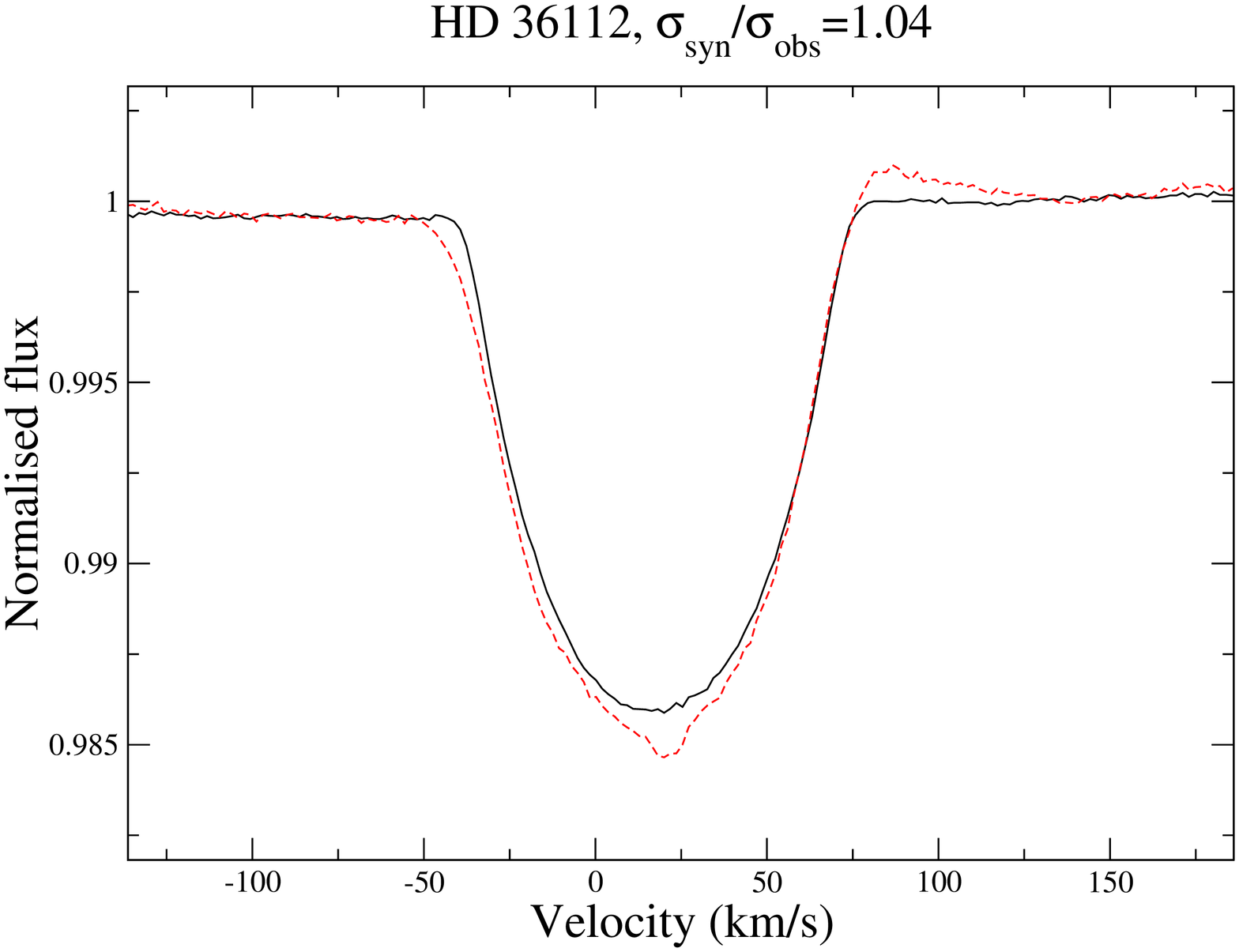}
\caption{Illustrations of observed versus synthetic LSD Stokes $I$ profiles for stars with profiles dominated by the photospheric component. The black full line is the synthetic, and the red dashed line is observed. Upper left: 49 Cet. Upper right: HD 139614. Lower left: HD 244604. Lower right: HD 36112.}
\label{mag4}
\end{figure}

In reality, we wished to combined the {\em advantages} of options 1 and 2, essentially by combining the more nearly photospheric Stokes $I$ profiles obtained from the cleaned masks and the high-SNR Stokes $V$ profiles from the full masks. While such an approach can solve the problem of determination of the integration range, it does not solve the problem of determination of the longitudinal field: the $I$  and $V$ profiles obtained from two different masks correspond to averages of different lines with different weights, and are therefore not directly comparable or quantitively compatible in Eq.~(\ref{bleqn}). 

As a solution to this problem, we decided to take advantage of the atmospheric and spectral parameters determined in Sects.~4 and 5 and to compute the approximate photospheric spectrum of each star using spectrum synthesis. We used the Synth3 LTE spectrum synthesis code \citep{kochukhov07} and effective temperature, surface gravity, $v\sin i$ and $v_{\rm rad}$ of each star (reported in Table \ref{tab:fp}) to compute its photospheric Stokes $I$ spectrum with the same spectral domain and resolution as ESPaDOnS/Narval. We assumed solar abundances. To each synthetic spectrum we added synthetic gaussian noise (calculated from the SNR of the spectra) that varied with wavelength in the same manner as in the observed spectrum. For each observed spectrum we then used the full line mask appropriate to the star to extract the Stokes $I$ LSD profile from the synthetic spectrum, and the Stokes $V$ LSD profiles from the observed spectrum. Combining the $I$ and $V$ profiles, this ultimately resulted in "hybrid" LSD profiles consisting of the real, observed Stokes $V$ profiles and a synthetic Stokes $I$ profile, both extracted using the same mask. The advantage of this approach is that we avoid the uncertainty related to CS contributions to the Stokes $I$ profile. On the other hand, we introduce uncertainty related to the compatibility of the synthetic photospheric spectrum with the real stellar photospheric spectrum.

\subsubsection{Tests and effectiveness of the hybrid method}

Using an integration range equal to the 1.2 times the measured $v\sin i$ of each star symmetric about the measured radial velocity, we evaluated Eqs. (\ref{bleqn}) and (\ref{fapeqn}) for each of the hydrid LSD profiles. For comparison, we also performed the same measurements, but using the original LSD profiles extracted using the full masks obtained only from the observed spectra. These measurements are listed in Table~\ref{bltable}.

\begin{figure}
\centering
\includegraphics[width=3cm,angle=-90]{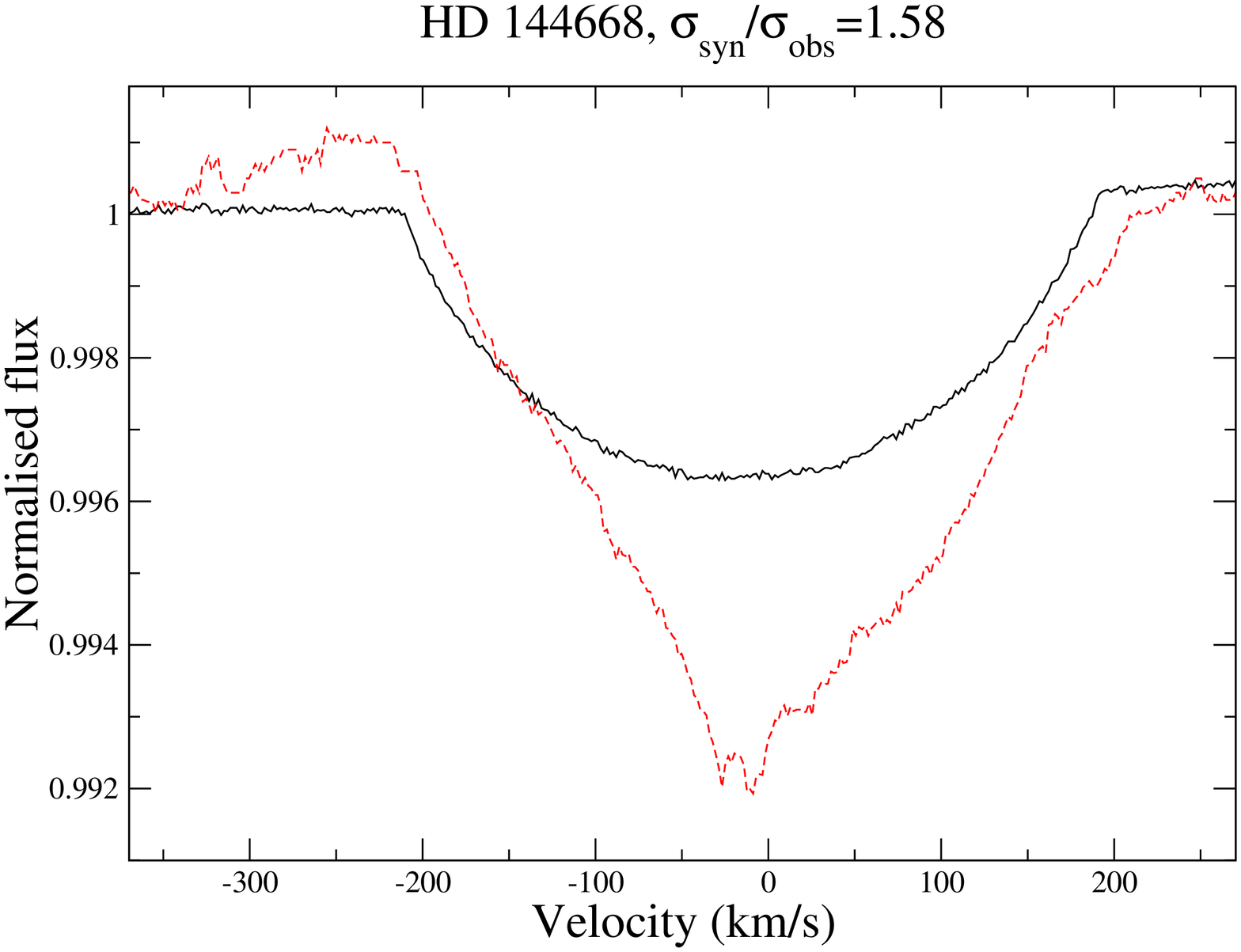}
\includegraphics[width=3cm,angle=-90]{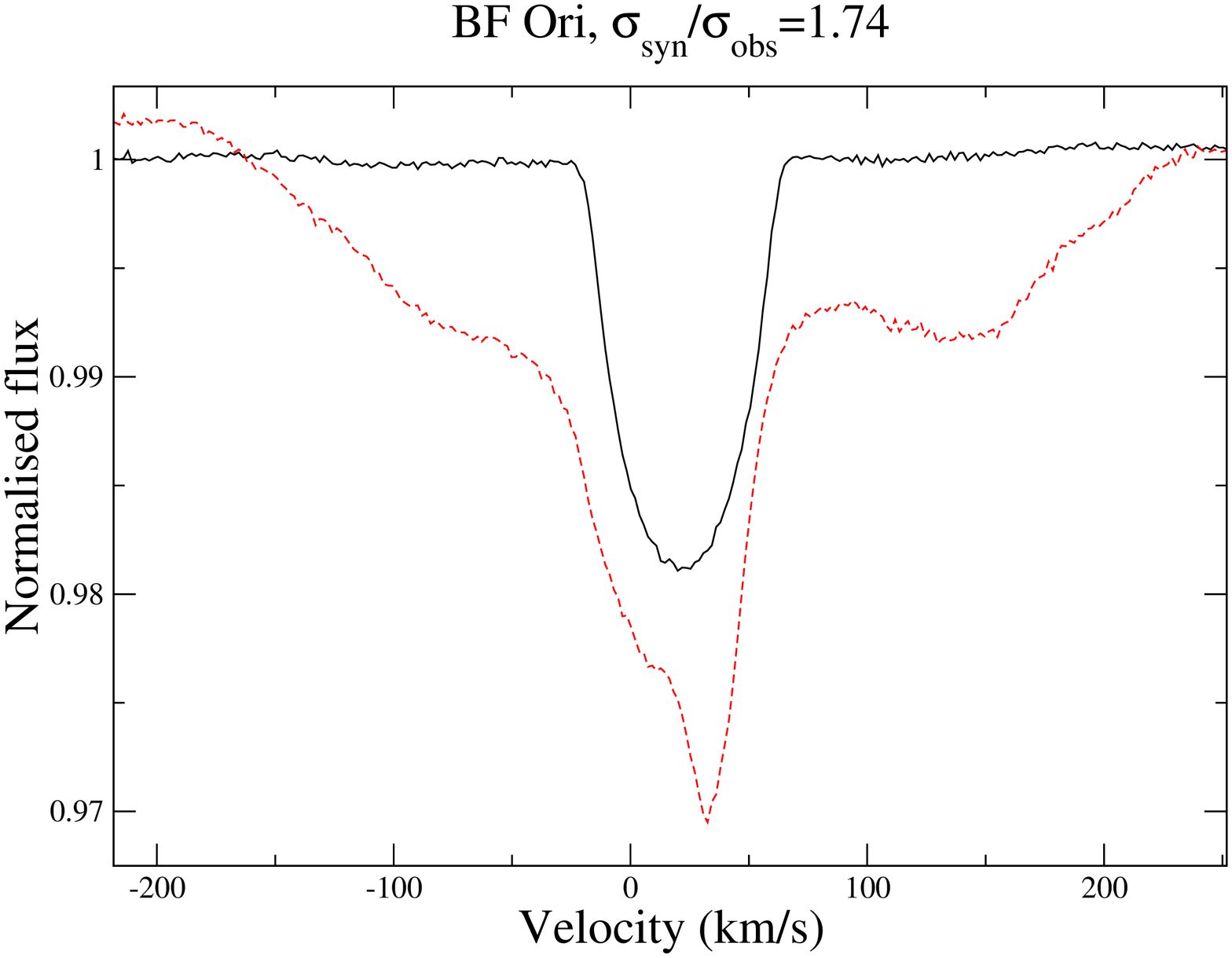}
\includegraphics[width=3cm,angle=-90]{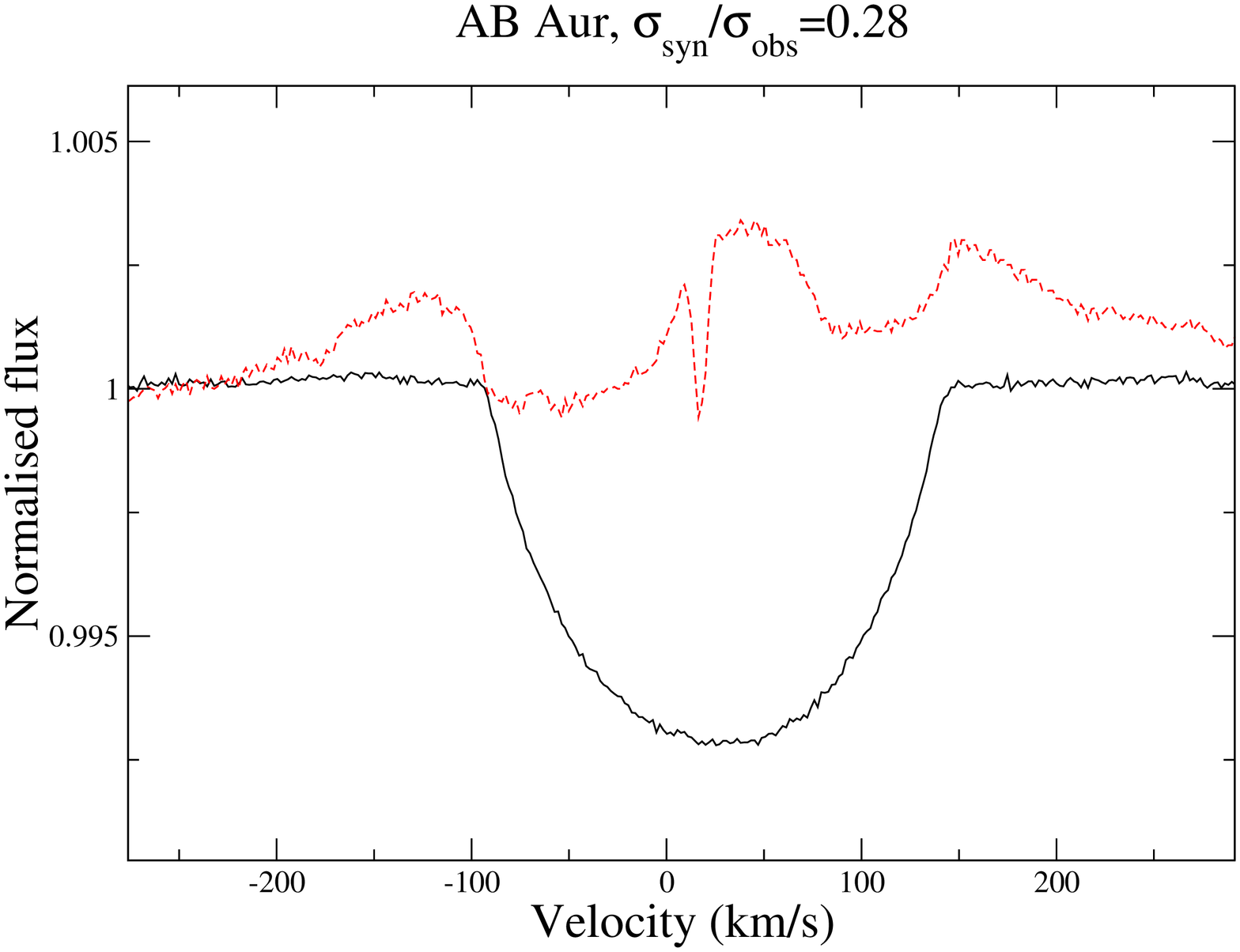}
\includegraphics[width=3cm,angle=-90]{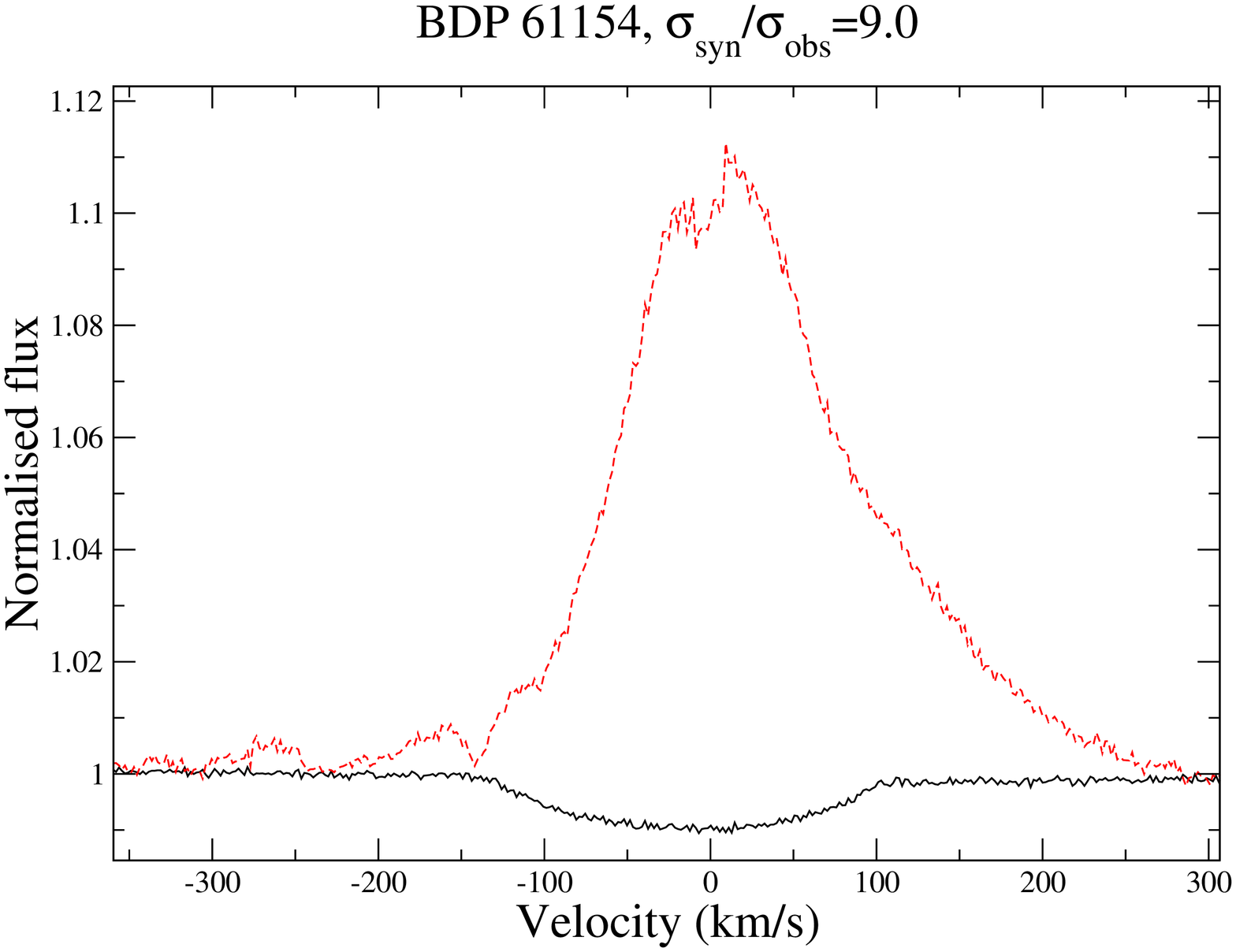}
\caption{Illustrations of observed versus synthetic LSD Stokes $I$ profiles for stars with profiles dominated by the circumstellar component. The black full line is the synthetic, and the red dashed line is observed. Upper left: HD 144668. Upper right: BF Ori. Lower left: AB Aur. Lower right: BD+61 154.}
\label{mag5}
\end{figure}

Based on the analysis of Sect.~5, we conclude that some of our observations have relatively small CS contamination, and are dominated by the photospheric component. We identified 17 stars (corresponding to 35 observations) for which this was the case; these stars are underlined in Table~\ref{bltable}. We have used these observations as a test of the accuracy of this method by comparing the longitudinal field extracted using the LSD profiles with observed Stokes $I$ profile, and those with the synthetic Stokes $I$ profile. For stars with purely photospheric spectra we expect the longitudinal field error bars to agree. (Because the value of the longitudinal field itself is determined by the details of the noise pattern, we do not expect those values to agree, except that they should have values compatible with the uncertainties.) The results of this comparison are shown in Fig.~\ref{mag3}.

\begin{figure}
\centering
\includegraphics[width=6.5cm,angle=-90]{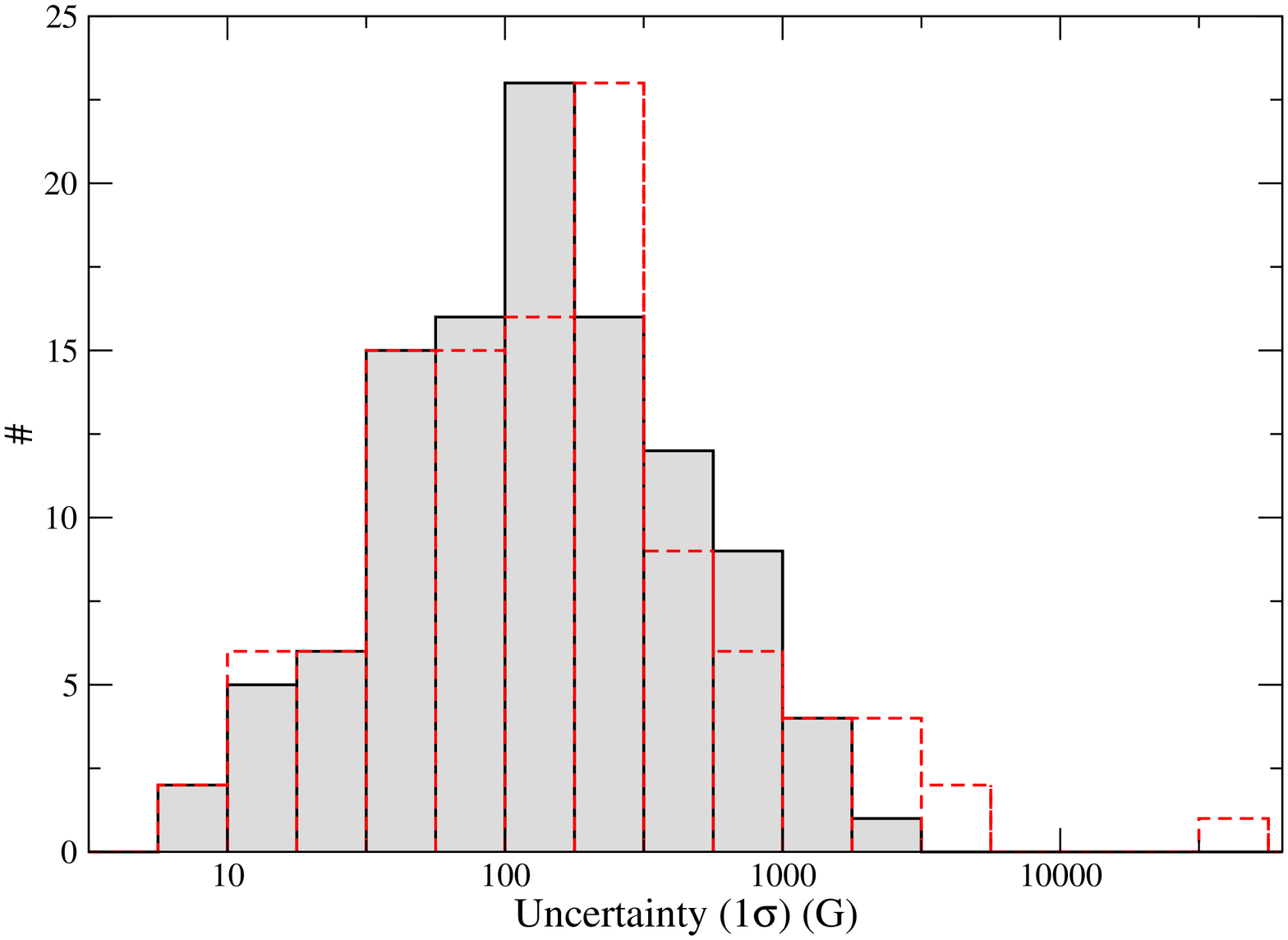}
\includegraphics[width=6.5cm,angle=-90]{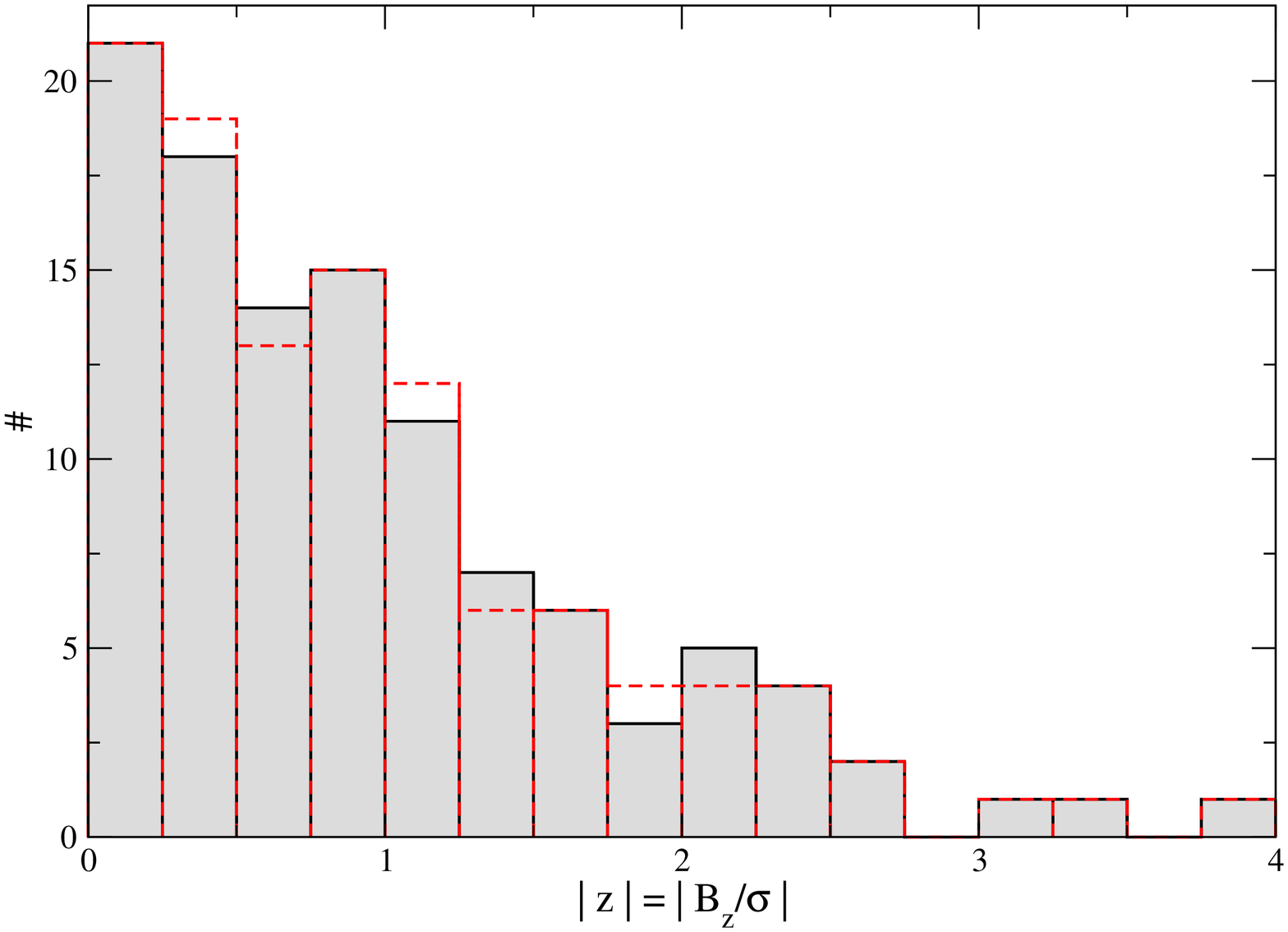}
\includegraphics[width=6.5cm,angle=-90]{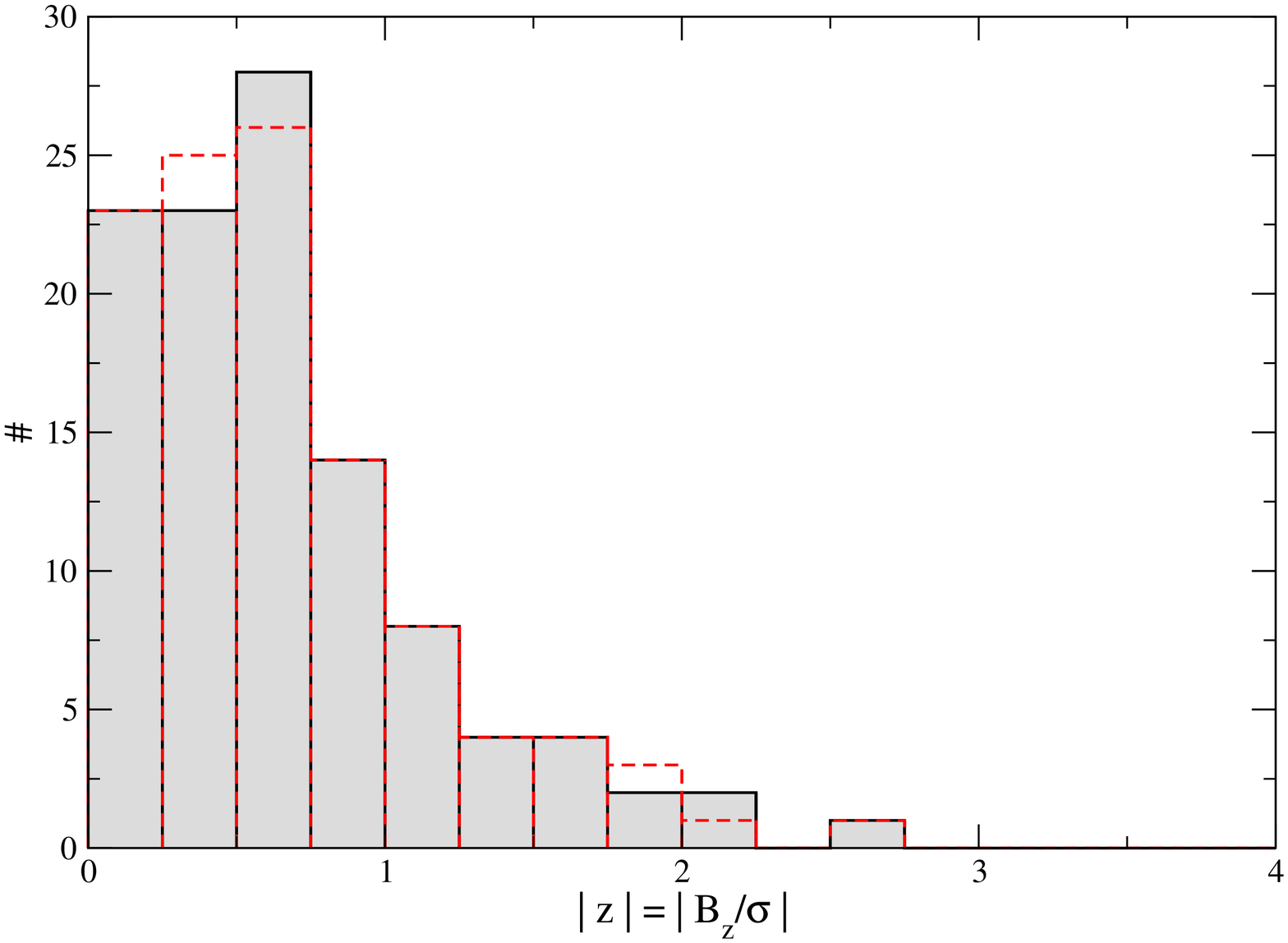}
\caption{Final results for the longitudinal field uncertainties of the program stars. {\em Upper panel -}\ Error bars from the hybrid and original LSD profiles. The black filled histogram corresponds to the hybrid profiles, while the dashed red unfilled histogram corresponds to the original profiles. {\em Middle panel -}\ Detection significance $z=|\bz/\sigma|$. Note that the three detections (i.e. $z\geq 3$) correspond to the magnetic HAeBe star LP Ori and to the suspected magnetic HAeBe star HD 35929. Both are sometimes detected in the $V$ signatures of the LSD profiles and in $\bz/\sigma$.{\em Lower panel -}\ Same as the middle but for the $N$ profiles. { Note on one side the absence of detections (i.e. $z\leq 3$), and on the other side, the numerous values with $z$ between 1 and 2, while by definition an $N$ spectrum does not contain any signal.} }
\label{mag6}
\end{figure}

The upper panel of Fig.~\ref{mag3} shows the derived value of the longitudinal field error bar using the synthetic Stokes $I$ profile ($\sigma_{\rm syn}$, on the horizontal axis) versus that derived from { the real Stokes $I$ profile ($\sigma_{\rm obs}$, on the vertical axis) of the stars with mostly photospheric spectra}. As can been seen, the correspondence between the error bars is quite close, with most points clustered tightly about the line $x=y$. The detailed agreement is summarised in the lower panel, which shows a histogram of the ratio $\sigma_{\rm syn}$/$\sigma_{\rm obs}$. The median of the distribution is 0.85, the mean is 1.2 and the standard deviation is 0.4. Most of the values of $\sigma_{\rm syn}$/$\sigma_{\rm obs}$ are clustered within $\pm 0.15$ of unity, although two significant outliers (the stars HD 34282 and HD 68695) exist at $\sim 1.5$. The dispersion results from (usually) small differences in the equivalent widths of the computed versus observed Stokes $I$ profiles. A comparison of the observed and computed Stokes $I$ profiles for 4 of the stars with mostly photospheric spectra is shown in Fig.~\ref{mag4}. The stars are typical of the sample, and illustrate the level of agreement usually achieved. Clearly some of the differences in the Stokes $I$ profiles result from small CS contributions to the real spectrum (e.g. HD~244604, lower left frame). The remainder we attribute mainly to errors in the adopted atmospheric parameters and the detailed chemistry of the star. HD 34282 and HD 68695 are both instructive in this respect: a detailed examination of their spectra reveals underabundances of the Fe peak elements that dominate their spectra, while the abundance of oxygen appears to be solar. This suggests that these two stars represent further examples of HAeBe stars with $\lambda$~Boo abundance patterns \citep{cowley10,folsom12}. In fact, one of these stars (HD 68695) was analysed by Folsom et al, who found it to exhibit clear lambda boo abundance peculiarities. From Fig.~\ref{mag3} we conclude that typically our spectrum synthesis approach is able to determine the expected longitudinal field error bar within $\pm 20$\%, although larger deviations sometimes occur for stars with strongly peculiar chemistry.

Examples of the remaining stars - those with profiles with significant CS contributions - are shown in Fig. \ref{mag5}. In these cases the magnetic diagnosis is highly uncertain, and often no reasonable diagnosis can be performed using the observed Stokes $I$ profile. { When we measure the longitudinal fields of the hybrid profiles of these objects, we find that while sometimes the error bars are reduced, sometimes they increase significantly}. Examination of individual profiles reveals that often when the error bar derived by our method is significantly larger than that from the observed profile, it is because the observed profile is in strong emission or contains strong CS absorption, artificially increasing the magnitude of the equivalent width (e.g. BDP 61 154, lower right frame of Fig. \ref{mag5}, HD 144668 and BF Ori, upper frames of Fig. \ref{mag5}). On the other hand, partial infilling of the photospheric profile as a result of CS contributions may reduce the equivalent width, thereby increasing the error bar relative to that obtained from a synthetic profile (e.g. AB Aur, lower left frame of Fig. \ref{mag5}). In some extreme cases, the infilling can produce an observed equivalent width very close to zero, resulting in a divergent longitudinal field and error bar (e.g. HD 50083, observed on 03/04/08. Interestingly, the observation obtained on 12/11/07 has much less infilling, and a much more realistic error bar).

The results we obtain from our hybrid analysis are very different from those that we would obtained using the clean masks derived for the profile analysis of Sect. 5.2. To illustrate, in Table~\ref{bltable} there are 44 measurements (corresponding to 19 stars) with "hybrid" error bars smaller than 100 G. In contrast, analysis of the "clean" profiles results in only 10 measurements smaller than 100 G. For some stars for which the raw and hybrid profiles yield quite good errors (a few 10s of G), the clean profiles produce errors of hundreds of G. This confirms our view that in many cases, the line masks required to obtain a relatively pure photospheric profile for determining e.g. \vsini\ are not suitable to obtain profiles yielding the most realistic magnetic diagnosis.

Based on these experiments and examination of individual profiles, we conclude that our hybrid approach provides a useful way of determining the longitudinal magnetic field for a large and diverse sample of HAeBe stars, and that while inherent uncertainties exist in the determination of the longitudinal field using synthetic photospheric Stokes $I$ profiles, those uncertainties (errors in atmospheric parameters, detailed chemistry) are better controlled and understood than the uncertainties associated with the determination of $\bz$ using the observed Stokes $I$ profiles. We therefore recommend the use of the hybrid determinations of $\bz$ for characterisation of the magnetic fields of individuals stars or samples of stars for which poorly understood contamination of Stokes $I$ by CS environment is a problem.

{ In order to evaluate the impact of our choices of the masks on the \bz\ and their uncertainties, we have computed new LSD profiles of different stars of our sample with various masks of different temperature, gravity and abundance. We have then computed new values of \bz\ and compared them to those of Table 4. We selected stars in our sample of various spectral types and rotation velocities, and with small CS contribution. We changed the temperature within the error bars (Table 2), the $\log g$ from 3.5 to 4.5, and the abundance at $\pm$25 \%. We find that the \bz\ vary within the error bars in all cases. The uncertainties on the \bz\ vary by a factor lower than 1\% when the $\log g$ and the abundances are varied. When we change the temperatures, the uncertainties vary from 1 \% (for large uncertainties of $\sim$800 G) to 10 \% (for uncertainties lower than $\sim$100 G).}

Of the 70 stars observed in our survey, 65 (93\%) show no direct evidence of a magnetic field. The derived characteristics of the longitudinal magnetic fields of the sample are summarised in Table~\ref{bltable}. The magnetic geometries of the detected stars, as well as interpretation of the general magnetic properties of the sample from the distributions illustrated in Fig.~\ref{mag6}, will be discussed in detail in Paper III.


\section{Conclusions}

This paper is the first of a series that presents the results of a high-resolution spectropolarimetric analysis of a sample of 70 Herbig Ae/Be stars. We carried out this analysis in order to address the problems of magnetism, angular momentum evolution and circumstellar environment during the pre-main sequence phase of intermediate-mass stars.

We obtained 132 high-resolution Stokes $I$ and $V$ spectra of 70 HAeBe stars using the instruments ESPaDOnS at CFHT and Narval at TBL. In this paper, we described the sample selection, the observations and their reduction, and the measurements that will comprise the basis of much of our following analysis. We described the determination of fundamental parameters for each target:
\begin{itemize}
\item the published effective temperatures have been verified by a visual comparison of observed with synthetic spectra. For some stars, new determinations of $T_{\rm eff}$ are given here. In the case of a few stars with high SNR observations weakly contaminated by CS material, we have re-determined their $T_{\rm eff}$ with an automatic procedure described here.
\item the luminosities have been estimated using the most reliable distance and photometric data that we could find in the literature.
\item the radius, mass and age have been determined by comparing the position of the stars in an HR diagram with PMS evolutionary tracks computed with CESAM. The ages have been measured from the birthline of \citet{behrend01}.
\end{itemize}

We discussed the Least-Squares Deconvolution method that we have applied to each of our spectra, including the careful selection, editing and tuning of the LSD line masks. We described the fitting of the LSD Stokes $I$ profiles using a multi-component model that yields the rotationally-broadened photospheric profile (providing the projected rotational velocity and radial velocity for each observation) as well as circumstellar emission and absorption components. The \vsini\ measurements are summarised in Table 2.

Finally, we detail the method that we used to confidently affirm that a star is magnetic. We diagnosed the longitudinal Zeeman effect via the measured circular polarisation inside spectral lines. In this survey, 5 (out of 70) HAeBe stars have been confirmed to be magnetic (V380 Ori and HD 72106 reported by Wade et al. 2005 ; HD 190073 by Catala et al. 2007 ; HD 200775 by Alecian et al. 2008a ; LP Ori by Petit et al. 2008). Four of them have been discovered within this program. One star (HD 35929) is reported here as a new suspected magnetic star. We also present the "hybrid" method that we have adopted in order to obtain realistic quantitative measurements of the magnetic fields of the 65 non-magnetic stars. The results are reported in Tables 3 and 4.

As an appendix, we have also provided a detailed review of each star observed.

In three forthcoming papers we will present out analysis of the rotational properties of the sample \citep[paper II;][]{paperii}, the magnetic properties of the sample (paper III ; Wade et al. in prep.), and the properties of the circumstellar environment of the Herbig Ae/Be stars (paper IV ; Alecian et al. in prep.).


 \begin{table*}
\caption{Results of the magnetic analysis of the program HAeBe stars. { The data of HD 190073, V380 ori, HD 200775 and HD 72106 have already been published and do not appear here}. Columns 1 and 2 give the name of the star and the date of the observation. Columns 3 and 4 give the limits of the integration range. The False Alarme Probability (FAP) of a Zeeman detection in the $V$ profile is indicated in column 5. { The magnetic diagnosis (ND, MD or DD) is indicated in column 6}. The $B_{\ell}$ measurement, its error ($\sigma$), and the detection significance ($B_{\ell}/\sigma$) computed using the hybrid and original profiles are given in the columns 7 to 10. The final column give the ratio of the $B_{\ell}$ errors of the hybrid over the observed solutions.}
\centering
 \label{bltable}
\begin{tabular}{lc rrcc r@{$\pm$}lr r@{$\pm$}lr c}
\hline
              &&&     &  &    & \multicolumn{3}{c}{HYBRID} &   \multicolumn{3}{c}{ORIGINAL} \\
Filename      & Date           &     Start&  End  &    FAP   & Diagnosis &     \multicolumn{2}{c}{$B_\ell\pm\sigma$}   &  $B_\ell/\sigma$  &     \multicolumn{2}{c}{$B_\ell\pm\sigma$}  &  $B_\ell/\sigma$ & $\sigma_{\rm syn}/\sigma_{\rm obs}$ \\  
           &              &     \multicolumn{2}{c}{(km/s)}       &     &      &      \multicolumn{2}{c}{(G)}     &         &       \multicolumn{2}{c}{(G)}    &         &          		   \\
\hline                                                                      
BD-06 1259                   & 12/03/09        &    -25  &    69   &     0.7990 &  ND &      62  &      87 &  0.711  &       37  &      51 &  0.711&1.71  \\
                                      & 20/02/05        &    -25  &    69   &     0.5826 &  ND &      53  &      54 &  0.986  &       31  &      31 &  0.987&1.74  \\
BD-05 1329                   &23/08/05         &  -148  &  205   &     0.6266 &  ND &     201  &     396 &  0.508  &      144  &     282 &  0.508&1.40      \\
\underline{BD-05 1324} &11/01/06       &   -59  &  118   &     0.3831 &  ND &     -51  &     111 & -0.455  &      -47  &     103 & -0.455&1.08     \\
BD+41 3731                  & 06/11/07        &  -428  &  400   &     1.0000 & ND &     2367  &    1890 &  1.252  &     3415  &    2731 &  1.250&0.69  \\
                                      & 25/08/05        &  -428  &  400   &     0.9987 &  ND &      50  &    1008 &  0.049  &       51  &    1038 &  0.049&0.97  \\
BD+46 3471                  & 25/08/05        &  -242  &  235   &     0.6788 &  ND &     -20  &     526 & -0.038  &      -45  &    1202 & -0.038&0.44  \\
BD+61 154                    & 21/02/05        &  -151  &  118   &     0.6030 &  ND &     146  &     666 &  0.219  &      -16  &      74 & -0.219&9.00  \\
                                      & 23/08/05        &  -151  &  118   &     0.2323 &  ND &     985  &     549 &  1.796  &     -177  &      98 & -1.798&5.60  \\
BD+65 1637                  & 10/06/06        &  -360  &  308   &     0.9998 &  ND &     574  &    1109 &  0.518  &    -2045  &    3955 & -0.517&0.28  \\
                                      & 24/09/09        &  -360  &  308   &     0.9997 &  ND &    -806  &     874 & -0.923  &      938  &    1023 &  0.917&0.85  \\
BD+72 1031                  &11/06/06        &   -59  &  118   &     0.9309 &  ND &     -28  &     118 & -0.234  &      -36  &     156 & -0.234&0.76     \\
                                      &11/11/07     &  -225  &  207   &     0.9810 &  ND &    1072  &    1122 &  0.955  &      412  &     431 &  0.956&2.60      \\
\underline{HD 9672}      & 24/08/05        &  -221  &  247   &     0.9509 &  ND &      27  &     111 &  0.239  &       27  &     111 &  0.239&1.00  \\
\underline{HD 17081}    & 19/02/05        &    -11  &    37   &     0.3758 &  ND &       2  &       7 &  0.268  &        2  &       8 &  0.268&0.88   \\
                                      & 20/02/05        &    -13  &    35   &     0.9972 &  ND &       0  &       7 &  0.010  &        0  &       7 &  0.010&1.00   \\
HD 31293                      & 27/11/04        &  -113  &  163   &     0.5567 &  ND &    -112  &      87 & -1.283  &      393  &     308 &  1.274&0.28  \\
                                      & 19/02/05        &  -113  &  163   &     0.8017 &  ND &     -27  &     157 & -0.172  &       31  &     179 &  0.172&0.88  \\
                                      & 21/02/05        &  -113  &  163   &     0.7151 &  ND &      51  &     103 &  0.490  &      -63  &     128 & -0.490&0.80  \\
HD 31648                      &21/02/05     &  -104  &  130   &     0.9268 &  ND &       9  &      65 &  0.134  &       10  &      77 &  0.134&0.84    \\
                                      &24/08/05     &  -104  &  130   &     0.9687 &  ND &     149  &      62 &  2.411  &      220  &      92 &  2.408&0.67    \\
\underline{HD 34282}    & 24/08/05    &  -111  &  142   &     0.3744 &  ND &    -223  &     148 & -1.505  &     -369  &     246 & -1.504&0.60    \\
HD 35187 B                   & 25/08/05   &   -85  &  139   &     0.7977 &  ND &     -94  &     101 & -0.925  &      -86  &      93 & -0.926&1.09    \\
\underline{HD 35929}    & 11/03/09    &   -53  &   95   &     0.9993 &  ND &      -8  &      35 & -0.227  &       -6  &      28 & -0.227&1.25    \\
                                      &12/11/07     &   -53  &   95   &     0.0091 &   ND &    -59  &      23 & -2.598  &      -45  &      17 & -2.598&1.35    \\
                                      &13/11/07     &   -53  &   95   &     0.9704 &   ND &   -137  &      45 & -3.007  &     -106  &      35 & -3.007&1.29    \\
                                      &20/02/09            &   -53  &   95   &     0.0002 &  MD &      33  &      17 &  1.957  &       25  &      13 &  1.957&1.31    \\
                                      &21/02/09            &   -53  &   95   &     0.0000 &   DD &     74  &      19 &  3.928  &       57  &      15 &  3.928&1.27    \\
\underline{HD 36112}    &29/11/04     &   -47  &   83   &     0.9382 &  ND &      27  &      26 &  1.034  &       26  &      25 &  1.034&1.04    \\
                                      &19/02/05     &   -47  &   83   &     0.9395 &  ND &     -35  &      33 & -1.059  &      -33  &      31 & -1.059&1.06    \\
HD 36910                      & 04/04/08    &   -81  &  143   &     0.9018 &  ND &    -141  &      85 & -1.654  &     -180  &     109 & -1.653&0.78     \\
HD 36917                      &08/11/07      &  -126  &  179   &     0.9960 &  ND &    -672  &     384 & -1.748  &     -978  &     560 & -1.745&0.69     \\
HD 36982                      &08/11/07     &   -65  &  90   &     0.9402 &  ND &    -307  &     251 & -1.224  &     -326  &     267 & -1.224&0.94     \\
                                      &09/11/07     &   -65  &  90   &     0.0059 &  ND &    -225  &     108 & -2.090  &     -227  &     109 & -2.086&0.99     \\
                                      &10/11/07     &   -65  &  90   &     1.0000 &  ND &    -10  &      79 & -0.129  &      -10  &      76 & -0.129&1.04     \\
                                      &11/11/07     &   -65  &  90   &     0.0009 &  MD &   -248  &      76 & -3.267  &     -233  &      71 & -3.265&1.07     \\
HD 37258                      &24/02/09      &  -209  &  271   &     0.6827 & ND &     -216  &     404 & -0.535  &     -196  &     366 & -0.535&1.10     \\
HD 37357                      &24/02/09      &  -127  &  170   &     0.9917 &  ND &     133  &     197 &  0.674  &      212  &     315 &  0.674&0.63     \\
HD 37806                      &24/08/05      &   -97  &  191   &     0.1909 &  ND &     239  &     179 &  1.336  &      349  &     261 &  1.334&0.69     \\
HD 38120                      &13/03/09      &   -89  &  144   &     0.3659 &  ND &     191  &     203 &  0.942  &      679  &     723 &  0.939&0.28     \\
\underline{HD 38238}    &16/03/07      &  -105  &  133   &     0.7370 &  ND &      18  &      81 &  0.221  &       20  &      90 &  0.221&0.90     \\
HD 50083                      &03/04/08     &  -279  &  278   &     0.9978 & ND &      124  &     236 &  0.525  &    10860  &   33614 &  0.323&0.01     \\
                                      &12/11/07     &  -279  &  278   &     1.0000 & ND &     -229  &     191 & -1.200  &     -304  &     254 & -1.200&0.75     \\
HD 52721                      &03/04/08      &  -237  &  280   &     0.9520 & ND &      222  &     235 &  0.943  &      212  &     225 &  0.943&1.04     \\
                                      &06/11/07     &  -237  &  280   &     1.0000 &  ND &     -22  &     204 & -0.107  &      -24  &     220 & -0.107&0.93     \\
HD 53367                      &19/02/05     &    -2  &   97   &     0.9606 &  ND &     -19  &      46 & -0.406  &      -20  &      48 & -0.406&0.96     \\
                                      &20/02/05     &    -2  &   97   &     0.9917 &  ND &      18  &      29 &  0.617  &       19  &      31 &  0.617&0.94     \\
\underline{HD 68695}    &21/02/05      &   -32  &   73   &     0.7696 &  ND &      10  &     125 &  0.078  &       15  &     188 &  0.078&0.66     \\
HD 76534 A                   &21/02/05     &   -58  &  105   &     0.9331 &  ND &    -154  &     151 & -1.019  &     -203  &     199 & -1.019&0.76     \\
HD 98922                      &20/02/05      &   -60  &   60   &     0.3729 & ND &     -144  &      71 & -2.015  &      194  &      96 &  2.012&0.74     \\
HD 114981                    & 11/01/06        &  -335  &  236   &     0.9968 &  ND &    -105  &     203 & -0.518  &     -157  &     303 & -0.518&0.67  \\
                                      & 19/02/05        &  -335  &  236   &     0.9875 &  ND &    -117  &     459 & -0.255  &     -197  &     775 & -0.255&0.59  \\
\underline{HD 135344}  & 09/01/06        &    -98  &    98   &     0.9417 &  ND &    -124  &     138 & -0.893  &     -131  &     147 & -0.893&0.94  \\
\underline{HD 139614}  & 19/02/05        &    -29  &    29   &     0.1845 &  ND &     -24  &      14 & -1.760  &      -22  &      12 & -1.760&1.17  \\
                                      & 20/02/05        &    -29  &    29   &     0.8027 &   ND &    -13  &      14 & -0.947  &      -11  &      12 & -0.947&1.17  \\
                                      & 21/02/05        &    -29  &    29   &     0.7707 &  ND &      12  &      12 &  0.998  &       11  &      11 &  0.998&1.09  \\
HD 141569                    & 06/03/07        &  -286  &  262   &     0.6638 &  ND &     -73  &     149 & -0.492  &     -591  &    1219 & -0.485&0.12  \\
                                      & 12/02/06        &  -286  &  262   &     0.8590 &  ND &     645  &     778 &  0.829  &     1672  &    2023 &  0.826&0.38  \\
 \hline													  
\end{tabular}
\end{table*}
 
 \begin{table*}
\contcaption{}
\centering
\begin{tabular}{lc rrcc r@{$\pm$}lr r@{$\pm$}lr c}
\hline
              &&&  &    &     & \multicolumn{3}{c}{HYBRID} &   \multicolumn{3}{c}{ORIGINAL} \\
Filename      & Date           &     Start&  End  &    FAP   & Diagnosis &     \multicolumn{2}{c}{$B_\ell\pm\sigma$}   &  $B_\ell/\sigma$  &     \multicolumn{2}{c}{$B_\ell\pm\sigma$}  &  $B_\ell/\sigma$ & $\sigma_{\rm syn}/\sigma_{\rm obs}$ \\  
           &              &     \multicolumn{2}{c}{(km/s)}       &     &      &      \multicolumn{2}{c}{(G)}     &         &       \multicolumn{2}{c}{(G)}    &         &          		   \\
\hline                                                                       
\underline{HD 142666}  & 19/02/05        &    -85  &    72   &     0.9918 &  ND &      28  &      78 &  0.359  &       22  &      60 &  0.359&1.30  \\
                                      & 21/02/05        &    -85  &    72   &     0.4443 &  ND &     -45  &      44 & -1.009  &      -34  &      34 & -1.010&1.29  \\
                                      & 21/05/05        &    -85  &    72   &     0.8100 &  ND &      16  &      53 &  0.298  &       12  &      39 &  0.298&1.36  \\
                                      & 21/05/05        &    -85  &    72   &     0.9985 &  ND &     -29  &      53 & -0.543  &      -21  &      39 & -0.543&1.36  \\
                                      & 22/05/05        &    -85  &    72   &     0.6419 &  ND &      54  &      62 &  0.873  &       39  &      45 &  0.873&1.38   \\
                                      & 23/05/05        &    -85  &    72   &     0.8755 &  ND &     -14  &      50 & -0.282  &      -11  &      37 & -0.282&1.35   \\
                                      & 24/05/05        &    -85  &    72   &     0.5239 &   ND &    -29  &      56 & -0.511  &      -21  &      42 & -0.511&1.33   \\
\underline{HD 144432}  &19/02/05         &    -97  &    93   &     0.7430 & ND &     -113  &      49 & -2.296  &     -101  &      44 & -2.296&1.11   \\
                                      & 20/02/05        &    -97  &    93   &     0.8485 &  ND &      -9  &      41 & -0.227  &       -8  &      36 & -0.227&1.14   \\
HD 144668                    & 23/08/05        &  -249  &  229   &     0.8869 &  ND &     299  &     145 &  2.064  &      191  &      92 &  2.066&1.58   \\
HD 145718                    & 25/08/05        &  -139  &  132   &     0.0961 &  ND &      17  &      85 &  0.204  &       15  &      72 &  0.204&1.18   \\
HD 150193                    & 23/08/05        &  -135  &  125   &     0.8757 &  ND &    -194  &     117 & -1.664  &     -382  &     230 & -1.661&0.51   \\
HD 163296                    & 21/05/05        &  -164  &  146   &     0.8992 &  ND &      41  &     106 &  0.387  &      103  &     266 &  0.387&0.40   \\
                                      & 22/05/05        &  -164  &  146   &     0.9703 &  ND &      47  &     141 &  0.333  &       59  &     176 &  0.333&0.80   \\
                                      & 23/05/05        &  -164  &  146   &     0.7855 &  ND &     138  &      96 &  1.431  &      160  &     112 &  1.431&0.86   \\
                                      & 23/05/05        &  -164  &  146   &     0.8324 &  ND &       0  &     135 & -0.001  &        0  &     131 & -0.001&1.03   \\
                                      & 24/05/05        &  -164  &  146   &     0.6017 &   ND &    109  &      96 &  1.134  &      250  &     221 &  1.132&0.43   \\
                                      & 24/05/05        &  -164  &  146   &     0.6782 &  ND &     313  &     136 &  2.292  &      523  &     229 &  2.285&0.59   \\
                                      & 24/08/05        &  -164  &  146   &     0.2564 &  ND &     202  &      93 &  2.160  &    -1835  &     998 & -1.840&0.09   \\
HD 169142                    & 19/02/05        &    -58  &    57   &     0.9872 &  ND &      12  &      28 &  0.434  &       23  &      52 &  0.434&0.54   \\
                                      & 21/02/05        &    -58  &    57   &     0.1944 &  ND &      51  &      34 &  1.521  &      100  &      66 &  1.520&0.52   \\
                                      & 21/05/05        &    -58  &    57   &     0.9848 &  ND &     -21  &      20 & -1.038  &      -40  &      38 & -1.038&0.53   \\
                                      & 23/08/05        &    -58  &    57   &     0.9943 &  ND &      28  &      12 &  2.215  &       52  &      23 &  2.214&0.52   \\
HD 174571                    & 15/04/08        &  -249  &  277   &     0.9923 &  ND &   -1733  &     687 & -2.521  &    -1825  &     724 & -2.520&0.95   \\
                                      & 16/03/07        &  -249  &  277   &     0.9983 &  ND &     213  &     544 &  0.392  &      220  &     560 &  0.392&0.97   \\
\underline{HD 176386}  & 24/08/05        &  -212  &  208   &     0.6676 &  ND &     304  &     239 &  1.272  &      309  &     243 &  1.271&0.98   \\
HD 179218                    &03/10/09         &    -67  &    98   &     0.9902 &   ND &     -1  &      39 & -0.026  &       -1  &      42 & -0.026&0.93   \\
                                      & 20/02/05        &    -67  &    98   &     0.3448 &  ND &     -97  &     114 & -0.850  &      -98  &     115 & -0.849&0.99    \\
                                      & 25/08/05        &    -67  &    98   &     0.8626 &  ND &      77  &      50 &  1.536  &       78  &      51 &  1.536&0.98    \\
\underline{HD 244314}  & 05/11/07        &    -40  &    85   &     0.9511 &  ND &     -46  &     106 & -0.436  &      -39  &      90 & -0.436&1.18    \\
\underline{HD 244604}  & 23/08/05        &    -91  &  145   &     0.8532 &  ND &     -90  &      79 & -1.138  &     -103  &      91 & -1.137&0.87    \\
HD 245185                    & 19/02/05        &  -124  &  157   &     0.7352 & ND &     -255  &     335 & -0.760  &    -1674  &    2216 & -0.755&0.15    \\
HD 249879                   & 05/04/08         &  -294  &  316   &     0.9577 & ND &     1465  &    1326 &  1.104  &      971  &     879 &  1.104&1.51    \\
HD 250550                   &07/11/07           &  -117  &   72   &     0.0256 & ND &      -60  &     249 & -0.241  &       54  &     225 &  0.241&1.11    \\
HD 259431                   &17/03/07           &   -73  &  126   &     0.7920 & ND &     -117  &     184 & -0.636  &       75  &     118 &  0.636&1.56    \\
                                     &17/03/10       &   -73  &  126   &     0.3502 &  ND &      22  &     281 &  0.078  &      -27  &     351 & -0.078&0.80    \\
                                     &24/02/09       &   -73  &  126   &     0.6835 &  ND &      93  &     197 &  0.474  &      -58  &     122 & -0.474&1.61    \\
HD 275877                   &10/12/06        &  -267  &  270   &     0.9871 & ND &       80  &     303 &  0.263  &       43  &     163 &  0.263&1.86      \\
                                     &24/09/09     &  -267  &  270   &     0.9182 &   ND &      6  &     418 &  0.014  &        5  &     346 &  0.014&1.21      \\                                                                                     
\underline{HD 278937} &20/02/05        &   -82  &  109   &     0.8934 &  ND &      74  &     119 &  0.627  &       81  &     129 &  0.627&0.92     \\
                                     &20/02/05     &   -82  &  109   &     0.7830 &  ND &    -105  &     147 & -0.717  &     -113  &     158 & -0.717&0.93     \\
                                     &21/02/05     &   -82  &  109   &     0.9398 &  ND &     -62  &     166 & -0.370  &      -67  &     180 & -0.370&0.92     \\
\underline{HD 287841} &20/02/09     &  -119  &  159   &     0.7004 &  ND &      61  &     268 &  0.229  &       55  &     242 &  0.229&1.11    \\
HD 290409                   &06/11/07     &  -299  &  301   &     0.9883 &  ND &    -939  &     980 & -0.958  &    -4140  &    4385 & -0.944&0.22    \\
\underline{HD 290500} &21/02/09     &   -73  &  131   &     0.9775 &  ND &     601  &     464 &  1.296  &      557  &     430 &  1.295&1.08    \\
HD 290770                   &24/02/09     &  -249  &  324   &     0.8890 &  ND &    1088  &     985 &  1.105  &     2264  &    2068 &  1.095&0.48    \\
HD 293782                   &10/01/06        &  -253  &  277   &     0.8452 &  ND &    1178  &     942 &  1.251  &      508  &     406 &  1.253&2.32      \\
HD 344261                   &06/11/07        &  -240  &  231   &     0.9608 & ND &     -817  &     928 & -0.880  &     -463  &     526 & -0.880&1.76      \\
                                     &23/08/05     &  -240  &  231   &     0.7389 &  ND &    -275  &     491 & -0.561  &     -162  &     289 & -0.561&1.70      \\
VV Ser                          &25/08/05        &  -100  &  201   &     0.9998 &  ND &    1138  &     485 &  2.348  &      561  &     238 &  2.355&2.04      \\
VX Cas                         &/24/08/05   &  -199  &  180   &     0.9707 &  ND &    -286  &     994 & -0.288  &     -242  &     840 & -0.288&1.18      \\
\hline													  
\end{tabular}
\end{table*}

%

\section*{Acknowledgments}

We are very grateful to O. Kochukhov, who provided his BINMAG1 code. We thank Nikolai Piskunov, the referee, which led to important improvements in the paper. EA has been supported by the Marie Curie FP6 program, and the Centre National d'Etudes Spatiales (CNES). EA has also been supported by the Ministry of Higher Education and Research (MESR) and the Ministry of Foreign and European Affairs (MAEE), via the Hubert Curien Partnership (PHC) FAST (French-Australian Science \& Technology). GAW and JDL acknowledge support from the Natural Science and Engineering Research Council of Canada (NSERC). GAW has also been supported by the DND Academic Research Programme (ARP). This research has made use of the SIMBAD database and the VizieR catalogue access tool, operated at CDS, Strasbourg (France), and of NASAs Astrophysics Data System. SM and IW have been supported by the Commonwealth of Australia under the International Science Linkages programme.

%

\nocite{*}
\bibliographystyle{mn2e}
\bibliography{haebe-data}

%

\appendix
\section{Analysis of individual stars}

This appendix describes the approach that has been followed in order to determine the \vsinis of all the stars of our sample, and the fundamental parameters (luminosity, effective temperature and surface gravity) required to estimate the masses and ages of the stars, that will be used in the statistical analysis described in paper II. The basic procedure we have followed  has been to first check that the effective temperature found in the literature corresponds to our data, and modify it if necessary. Then we have computed the LSD profiles for most of the data, { using masks of appropriate temperature and gravity for each star, and fit them (see Section 5)}. Sometimes it was necessary to add one or more Gaussian functions to the photospheric rotational velocity broadening function in order to obtain a better fit, and hence a more accurate value of the \vsini. This paper does not aim to propose a physical origin of these Gaussian functions. Most of them are assumed to have a circumstellar (CS) origin. However, a more detailed analysis of the non-photospheric spectral features observed in the spectra (and in the LSD profiles) of our sample will be presented in a forthcoming paper. This appendix summarises only the information required to fully understand the method that we applied to determine the \vsini\ of the stars.

For each of the stars we compared the observed normalised spectra with a grid of synthetic spectra, { in order to check the published values and estimate new ones if required}. As described above, these spectra assume solar chemical composition. In most cases the effective temperature could be estimated from this comparison, but not the surface gravity, due to imperfect continuum normalisation of individual echelle orders and/or circumstellar contamination. Therefore, unless specified, we used by default a surface gravity $\log g = 4.0$ (cgs) (see Sc. 4.1). In order to better understand the shape of the LSD profiles and the choice of the adopted mask, a description of the non-photospheric features has been added for each star. In these descriptions, the Balmer profiles emission types that are sometimes mentioned have been classified according to the system of \citet{beals53}.

For each one of the stars, a short discussion has also been added to support their PMS nature, and therefore to justify their membership to our sample. { The references of the photometric data and the distances used to derive the luminosity are also detailed. In the cases of stars members of the Orion OB 1 association, the distance adopted is the weighted mean of the distances of the six sub-groups described by \citet{brown94}.}

Finally, apart from LP Ori, for which we obtained more data since its magnetic detection reported in \citet{petit08}, this appendix concerns only the stars that have not been detected as magnetic. For the magnetic stars we refer the reader to the following papers: \citet{catala07,alecian08a,folsom08,alecian09b}, Alecian et al. in prep.

\subsection{BD-06 1259 (= BF Ori)}

BF Ori is a member of the subgroup c of the Ori OB 1 association \citep{warren78}, at a distance of 375~pc \citep{brown94}. It belongs to the UXOR sub-class of HAeBe objects (UXOR stars, hereinafter, Mora et al. 2004), whose the prototype is UX Ori (see Sec. \ref{sec:uxori}). These stars are strong photometric variables. For the same reasons as for UX Ori, we used the Hipparcos photometric data \citep{perryman97} at maximum brightness ($V_{\rm T}=7.81$~mag, $B_{\rm T}=7.85$~mag, in the Tycho system), and converted them to the Johnson system (see the method in Sec. \ref{sec:uxori}). We find $V = 7.85$~mag and $(B-V) = -0.028$~mag, values that have been adopted to derive the luminosity of the star. BF Ori displays strong near-IR excess, very likely due to the presence of an optically thick circumstellar accretion disk \citep{hillenbrand92}.

The spectrum of BF Ori is very complex, highly variable, and similar to other UXOR stars. In the February 2005 spectrum, we can distinguish two classes of spectral lines among the metals. The first class consists of strong and broad CS absorption features at the positions of the predicted strong photospheric lines. These lines are still observed in the March 2009 spectrum, with different shapes and increased depth. The second class of lines concerns the predicted weak lines of the spectrum, which show a photospheric component on which is superimposed a narrower circumstellar absorption. In our 2009 spectrum these lines show only photospheric components. As with other UXOR stars, these transient absorption features are assumed to come from gaseous clouds in the disk of the star (see Sec. \ref{sec:uxori}).

The Balmer lines from H$\delta$ to H$\beta$ show strong absorption components superimposed on the cores of the photospheric lines, with weak emission in the wings of the absorption component. The amplitudes of these features increase with wavelength, and their shape has changed between our two observations. H$\alpha$ is in emission with a double-peaked profile of type VI and a strong central absorption that goes below the continuum. The amplitude of the emission doubled between 2005 and 2009. The Ca~{\sc ii} K line, the He~{\sc i} lines at 5875~\AA, 6678~\AA, and 7065~\AA, and the OI 7775~\angs and O~{\sc i}~8446~\angs triplets display very strong (stronger than predicted) absorption profiles. The three IR Ca~{\sc ii} lines at 8498~\AA, 8542~\AA, and 8662~\angs (herinafter the Ca~{\sc ii} IR triplet) show strong and broad emission profiles superimposed on the three photospheric absorption lines. The Paschen lines seem also to be slightly contaminated with circumstellar emission.

\begin{figure}
\centering
\includegraphics[width=6cm,angle=90]{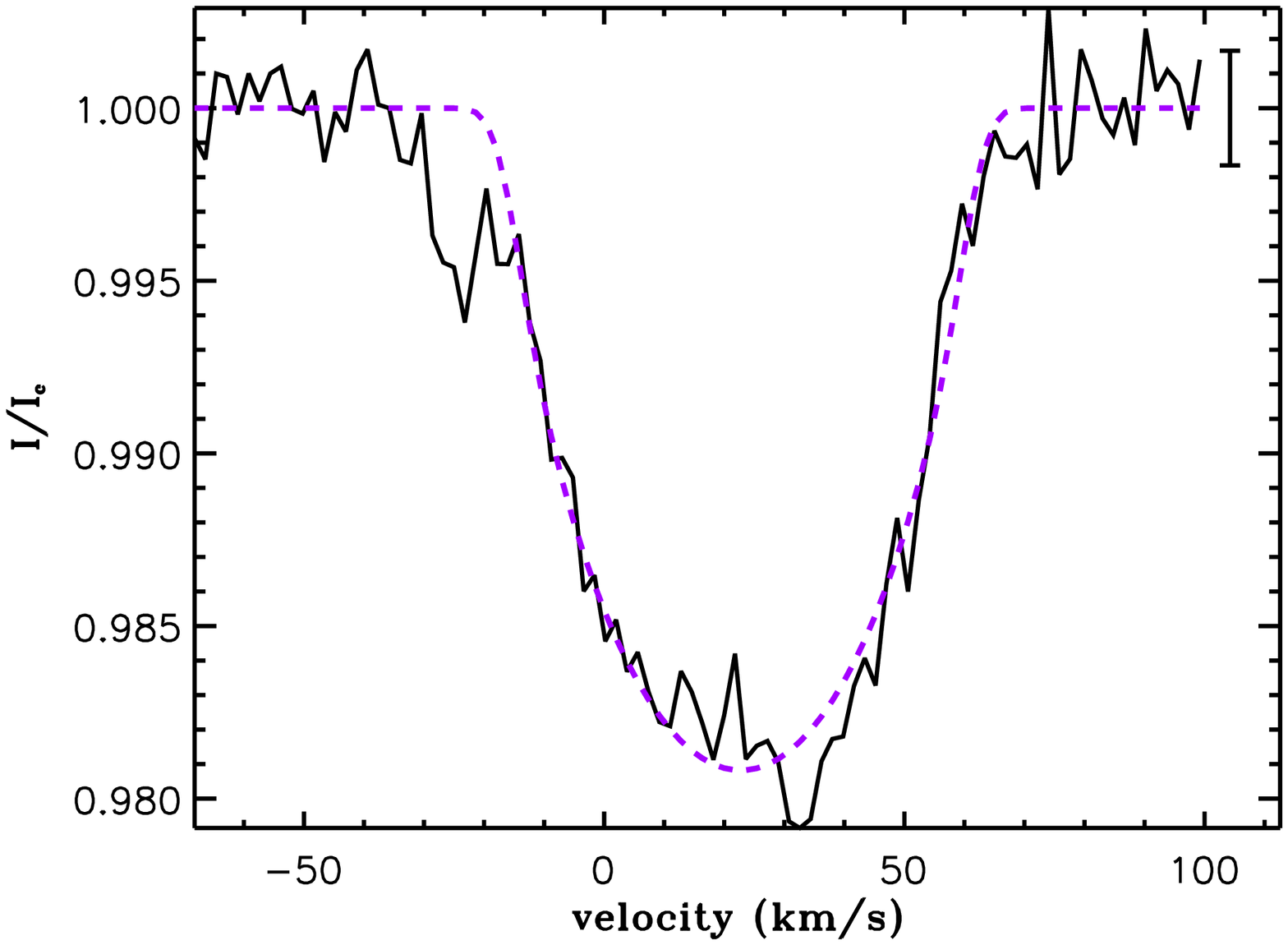}
\caption{The LSD $I$ profile of the March 2009 observation of BD-06 1259 = BF Ori (black full line), superimposed with its best fit (purple dashed-line).}
\label{fig:bfori}
\end{figure}

The wings of the Balmer lines are consistent with the temperature and surface gravity determination of \citet[][$T_{\rm eff}=8750\pm250$~K, $\log g=4.0\pm0.5$]{mora04}. We have cleaned the Kurucz mask by rejecting as many contaminated lines as possible. The resulting LSD profile for 2009 shows only a photospheric component. However the 2005 profile still shows a photospheric line with a superimposed circumstellar absorption component. We first tried to fit both observations simultaneously using the following model: a photospheric profile + a Gaussian function in absorption for the 2005 profile, and only a photospheric profile for the 2009 observation. However, we found that fitting the 2009 profile only resulted in a more accurate value of  $v \sin i$. We therefore adopted this value. The resulting fit is shown in Fig. \ref{fig:bfori}.

\subsection{BD-05 1329 (= T Ori)}

T Ori is part of the Ori OB1d cluster situated at a distance of 375~pc from the Sun \citep{brown94}. We used the photometric data of \citet{herbst99} to derive the luminosity of the star. T Ori displays a near-IR excess that could be attributed to an accretion disk \citep{hillenbrand98}.

The spectrum of T Ori is consistent with the temperature and surface gravity determination of \citet[][$T_{\rm eff} = 8500\pm300$~K, $\log g=4.2\pm0.3$]{folsom12}. The spectrum shows strong circumstellar features: the Ca~{\sc ii} K line and the O~{\sc i} 777~nm triplet display redshifted absorption. From H$\eta$ to H$\gamma$ a redshifted absorption is superimposed on the core of the line. A similar, but deeper, absorption is present in the core of H$\beta$, as well as emission in the blue wing of the CS absorption component. H$\alpha$ displays a double-peaked emission profile with a redshifted absorption component that goes below the continuum. The He~{\sc i} lines at 5875~\AA, 6678~\AA, and 7065~\AA, and the O~{\sc i}~8446~\angs triplet show inverse P Cygni profiles. Faint emission is observed in the cores of the Paschen lines. The Ca~{\sc ii}~K and metallic lines, as well as the Ca~{\sc ii} IR-triplet, do not display any CS features.

\begin{figure}
\centering
\includegraphics[width=6cm,angle=90]{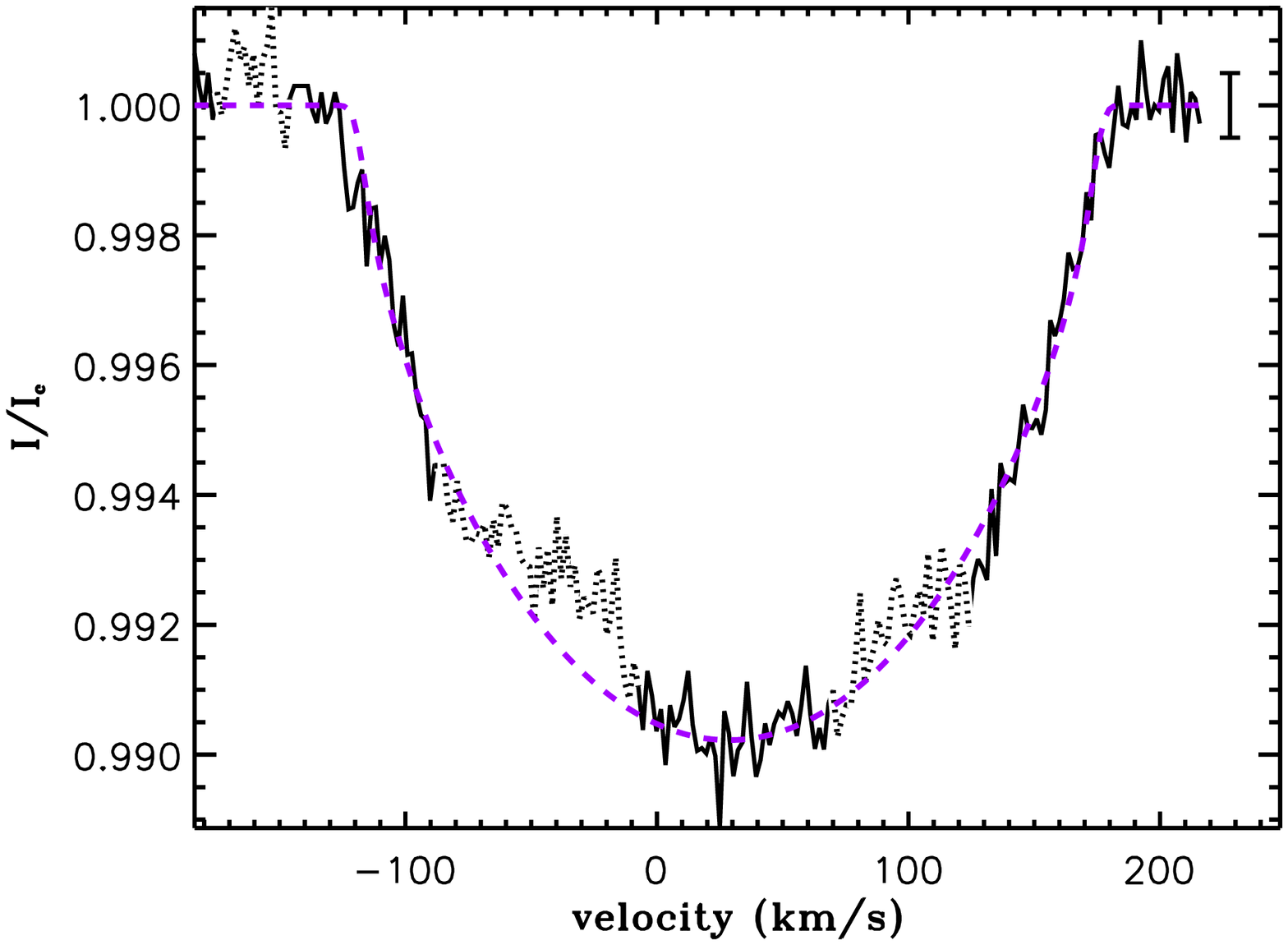}
\caption{As Fig.~A1 for BD-05 1329 (= T Ori). The dotted part of the observed profile has not been included in the fitting procedure.}
\label{fig:tori}
\end{figure}

We have cleaned the Kurucz mask in order to reject all the lines contaminated with circumstellar or interstellar features. The resulting LSD $I$ profile displays a photospheric shape slightly contaminated with CS emission. We first tried to fit the profile with a photospheric function and two emission Gaussian function. Then we fit the profile by rejecting, into the fitting procedure, the data points of the profile contaminated with emission. The last option gives a more accurate $v\sin i$ value. The result is shown in Fig. \ref{fig:tori}.

\subsection{BD-05 1324 (= NV Ori)}

\begin{figure}
\centering
\includegraphics[width=6cm,angle=90]{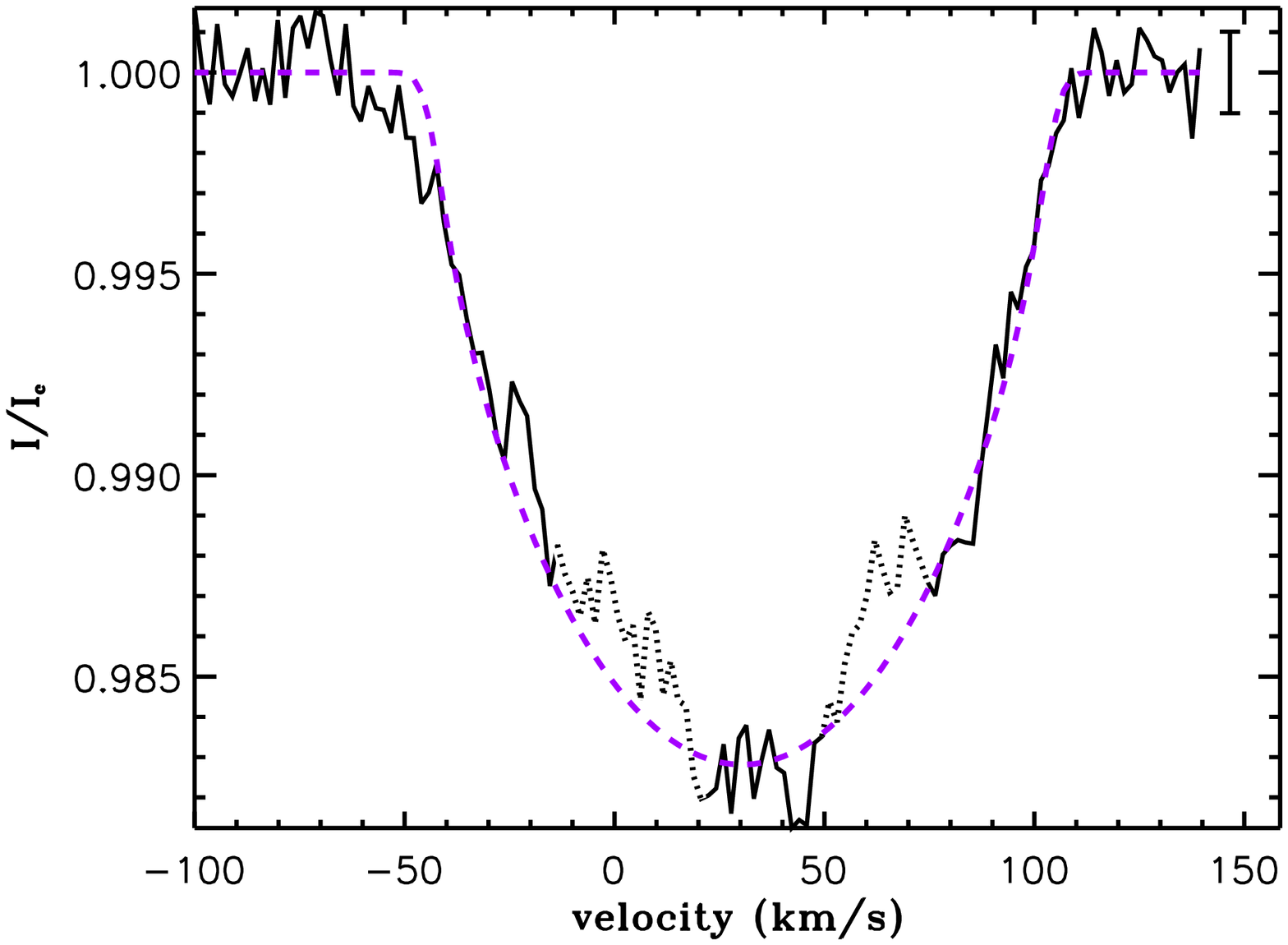}
\caption{As Fig.~A2 for BD-05 1324 (= NV Ori).}
\label{fig:nvori}
\end{figure}

NV Ori is part of the Ori OB 1 association \citep{warren77}, situated at a distance of 375~pc from the Sun \citep{brown94}. We used the photometric data ($V$ and $(B-V)$) from \citet{herbst99}, to compute the luminosity of the star. It does not seem to display infrared (IR) excess, but emission lines in its spectrum, as well as its membership to the Ori OB 1 star forming region, are good indications that the star is very young.

The spectrum of NV Ori is consistent with a $T_{\rm eff} = 6350\pm250$~K, corresponding to the spectral type determination of \citet{mora01}. It shows circumstellar emission in the core of H$\beta$, H$\gamma$, and H$\delta$. H$\alpha$ displays a strong and complex emission profile, while the O~{\sc i}~777~nm and Ca~{\sc ii} IR triplets show strong absorptions with emissions in the blue wings. The O~{\sc i} 8446~\angs absorption profile is stronger than predicted, while the He~{\sc i}~D3 line (at 5875~\AA) displays a broad absorption profile. The Ca~{\sc ii}~K, metallic and Paschen lines do not display any special features. 

The LSD $I$ profile shows a photospheric shape with broad wings, as well as a narrow and deep blueshifted CS absorption component. In order to improve the profile, we have calculated new masks by varying the cut-off of the intrinsic depths ($d_{\rm i}$) or excitation potential ($\chi_{\rm exc}$) of the lines. The best profile that we found has been computed with a mask containing only lines with $\chi_{\rm exc}>5$~eV. The resulting $I$ profile is slightly distorted in the core, which might be due to circumstellar emission. We therefore excluded the contaminated data points from the fit, and fit the profile with a photospheric function. The result is shown in Fig. \ref{fig:nvori}.

\subsection{BD+41 3731}

\begin{figure}
\centering
\includegraphics[width=6cm,angle=90]{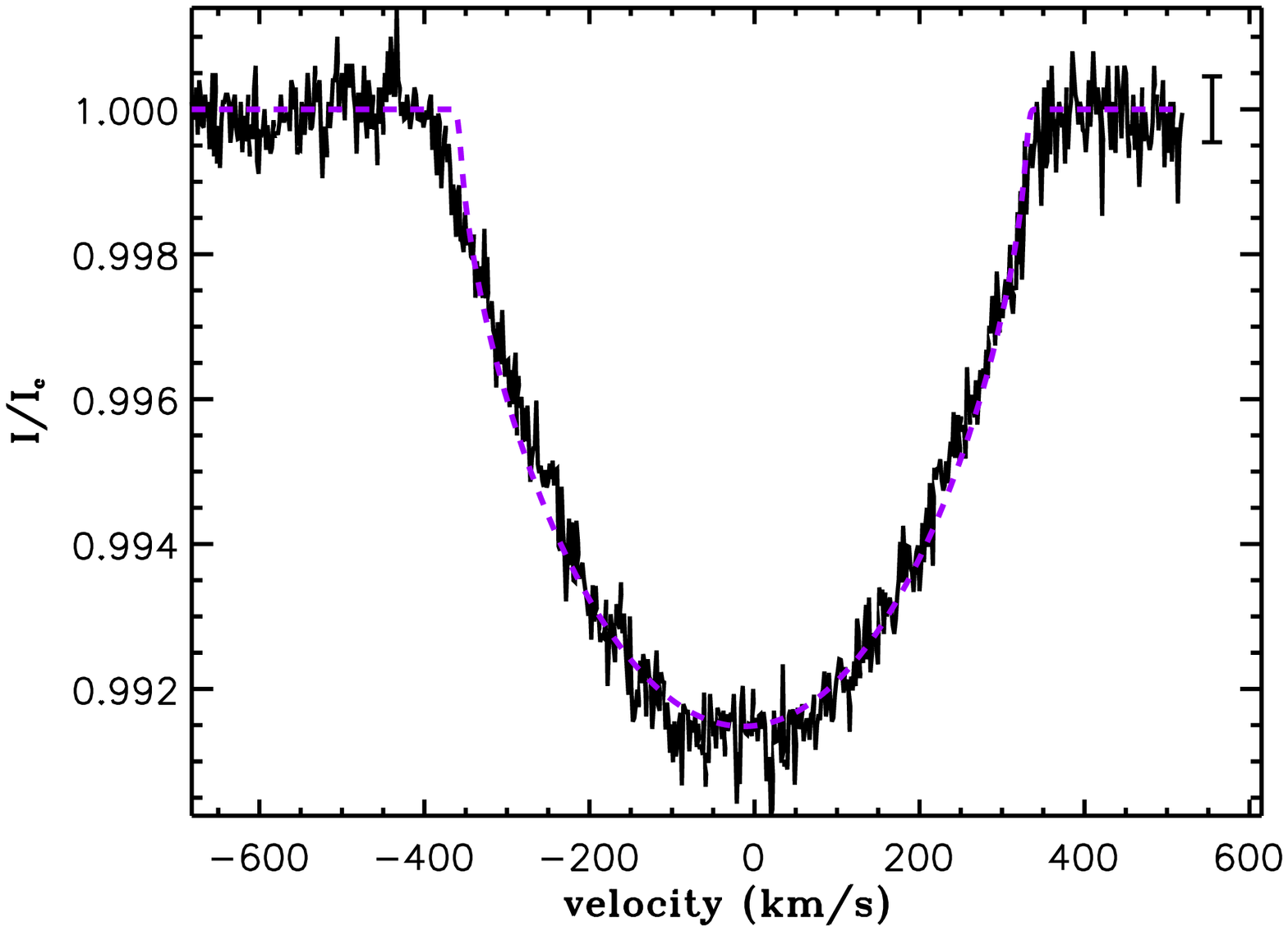}
\caption{As Fig.~A2 for the August 2005 observation of BD+41 3731.}
\label{fig:413731}
\end{figure}

BD+41 3731 is associated with the star forming region 2 Cyg and the young cluster NGC 6910 at a distance of 980~pc \citep{shevchenko91}. We used the photometric data from the Hipparcos catalogue \citep{perryman97} to compute the luminosity. This star does not seem to display a strong IR excess, and shows a spectral energy distribution similar to classical Be stars. However, its association with a star forming region, as well as to a young cluster, suggests that it is a young intermediate mass star lacking an accretion disk \citep{hillenbrand92}.

The most recent spectroscopic determination of the effective temperature and $v\sin i$ is found in \citet{finkenzeller85}. He assigns a spectral type of B2 (i.e. $T_{\rm eff} \sim 22000$~K), and $v\sin i \sim300$~\kms. Our spectra are very well fit with a non-LTE TLUSTY synthetic spectrum of $T_{\rm eff}=17000\pm1000$~K and a $v\sin i \sim345$~\kms. The difference in temperature might be due to non-LTE effects that have been taken into account in our analysis. Faint circumstellar emission is visible in the cores of the O~{\sc i}~777~nm and Ca~{\sc ii} IR triplets. The S~{\sc ii} lines (S~{\sc ii} 5201 \AA, S~{\sc ii} 5212  \AA, S~{\sc ii} 5346 \AA, and S~{\sc ii} 5429 \AA) are very weak compared to synthetic lines computed using solar abundance. They might be filled with circumstellar emission, or an indication that our adopted $T_{\rm eff}$ is too high. No other circumstellar features are observed in the spectrum. Many diffuse interstellar bands (DIB) are present and can be identified with the 1997 version\footnote{http://leonid.arc.nasa.gov/DIBcatalog.html} of the DIB catalogue \citep{jenniskens94}. 

The Kurucz mask has been cleaned by keeping lines with excitation potential greater than 1 eV, by rejecting lines contaminated with circumstellar emission, and by keeping only parts of the spectrum with high SNR and not contaminated with telluric lines. Two observations were obtained, one with ESPaDOnS in August 2005, and one with Narval in November 2007. We tried to fit both profiles separately and simultaneously, with a photospheric profile for each one of the observations. Because of its relatively low SNR (see Table 1), the Nov. 2007 observation does not improve the precision of the fitted parameters when both profiles are fitted simultaneously. Therefore the $v\sin i$ and \vrad\ values have been obtained from the fit of the Aug. 2005 profile only (Fig. \ref{fig:413731}).

\subsection{BD+46 3471}

BD+46 3471 is associated with the young cluster IC 5146 \citep{walker59} situated at a distance of 950~pc \citep{harvey08}. The luminosity has been obtained using this distance and the Hipparcos photometric data \citep{perryman97}. \citet{hillenbrand92} reports strong near-IR excess suggesting the presence of an optically thick circumstellar accretion disk.

The spectrum of BD+46 3471 is strongly contaminated with circumstellar absorption and emission lines. A blueshifted absorption and a redshifted emission are superimposed on the core of the Balmer lines from H$\zeta$ to H$\beta$, while H$\alpha$ displays a single-peaked emission line of type III with a weak absorption in the blue wing. Many metallic lines show narrow emission in the cores of their photospheric profiles. The O~{\sc i}~8446~\angs and Ca~{\sc ii} IR triplets show strong and redshifted single-peaked emission profiles. Narrow emission is detected in the cores of the Paschen lines. The Ca~{\sc ii} K  line has the central emission peak of other metal lines superimposed on the photospheric line, with a narrow, probably interstellar, absorption in the centre of the emission. The He~{\sc i}~D3 line and the O~{\sc i} 777~nm triplet display very strong absorption profiles, stronger than predicted, but the weaker O~{\sc i} lines at 615~nm have about the correct strength for solar O abundance. The few lines not contaminated with CS features are consistent with the temperature determination of \citet[][$T_{\rm eff} = 9500\pm1000$~K]{hernandez04}. 

\begin{figure}
\centering
\includegraphics[width=6cm,angle=90]{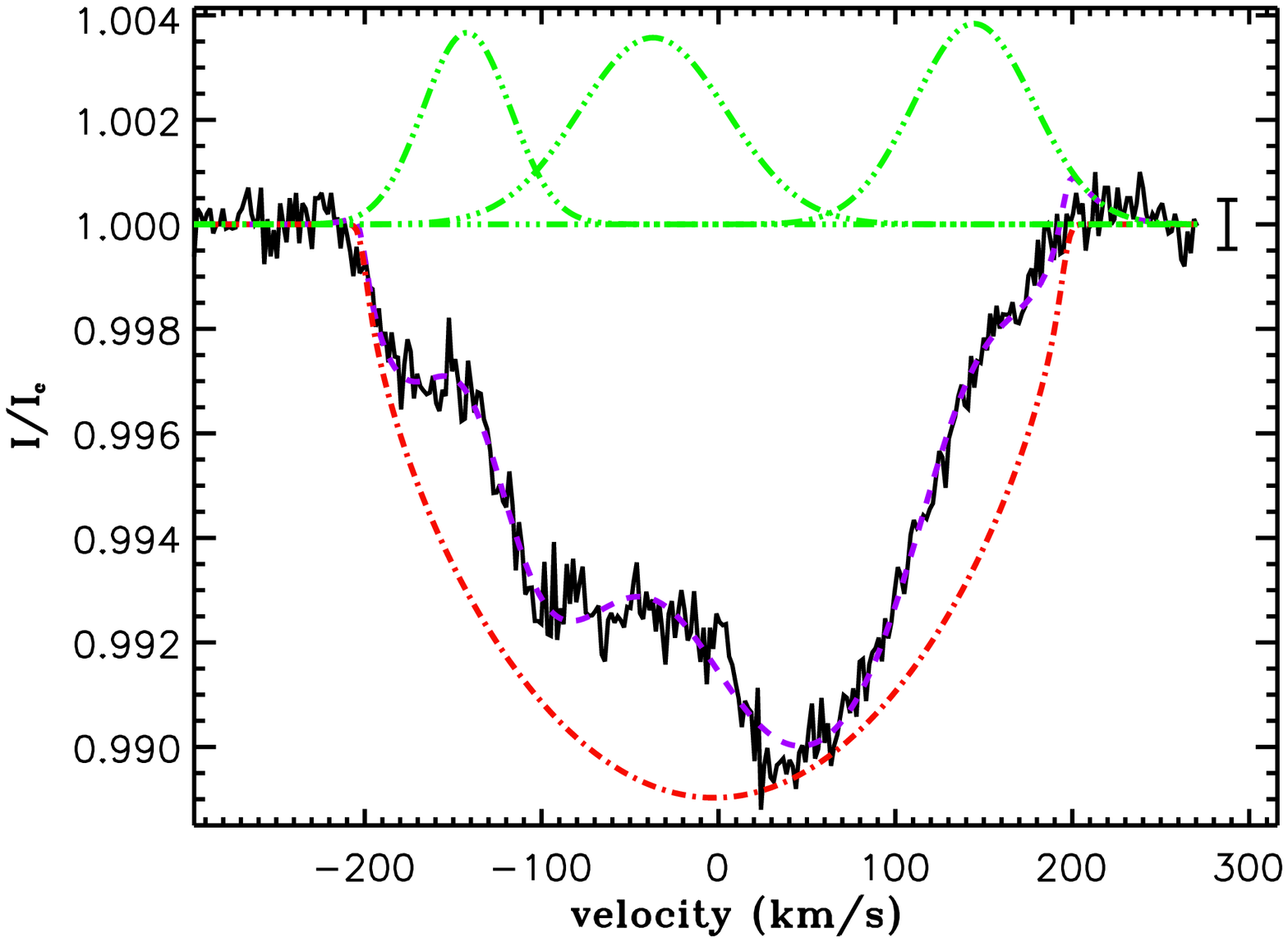}
\caption{As Fig.~A2 for BD+46 3471. The photospheric and Gaussian parts of the best fit are represented in dot-dashed and dot-dot-dot-dashed lines, respectively.}
\label{fig:463471}
\end{figure}

We tried many variations of the Kurucz mask, by doing a selection on the excitation potential and the central depth of the lines. We varied the lower value of $\chi_{\rm exc}$ between 0.5 and 10~eV, and the maximum value of the central depth between 0.5 and 0.15. In each cases the LSD profiles have the same shape and are still contaminated with emission. We therefore conclude that emission is present in all the lines of the spectrum, and we adopted a mask that has been calculated by rejecting by hand the lines visibly strongly contaminated with emission. The result gives a profile that is not strongly contaminated with emission, but that still contains few emission features.
We fit the profile using a photospheric and three Gaussian functions, which reproduces perfectly the LSD $I$ profile. The resulting fit is plotted in Fig. \ref{fig:463471}.

\subsection{BD+61 154 (= V594 Cas)}

\begin{figure}
\centering
\includegraphics[width=6cm,angle=90]{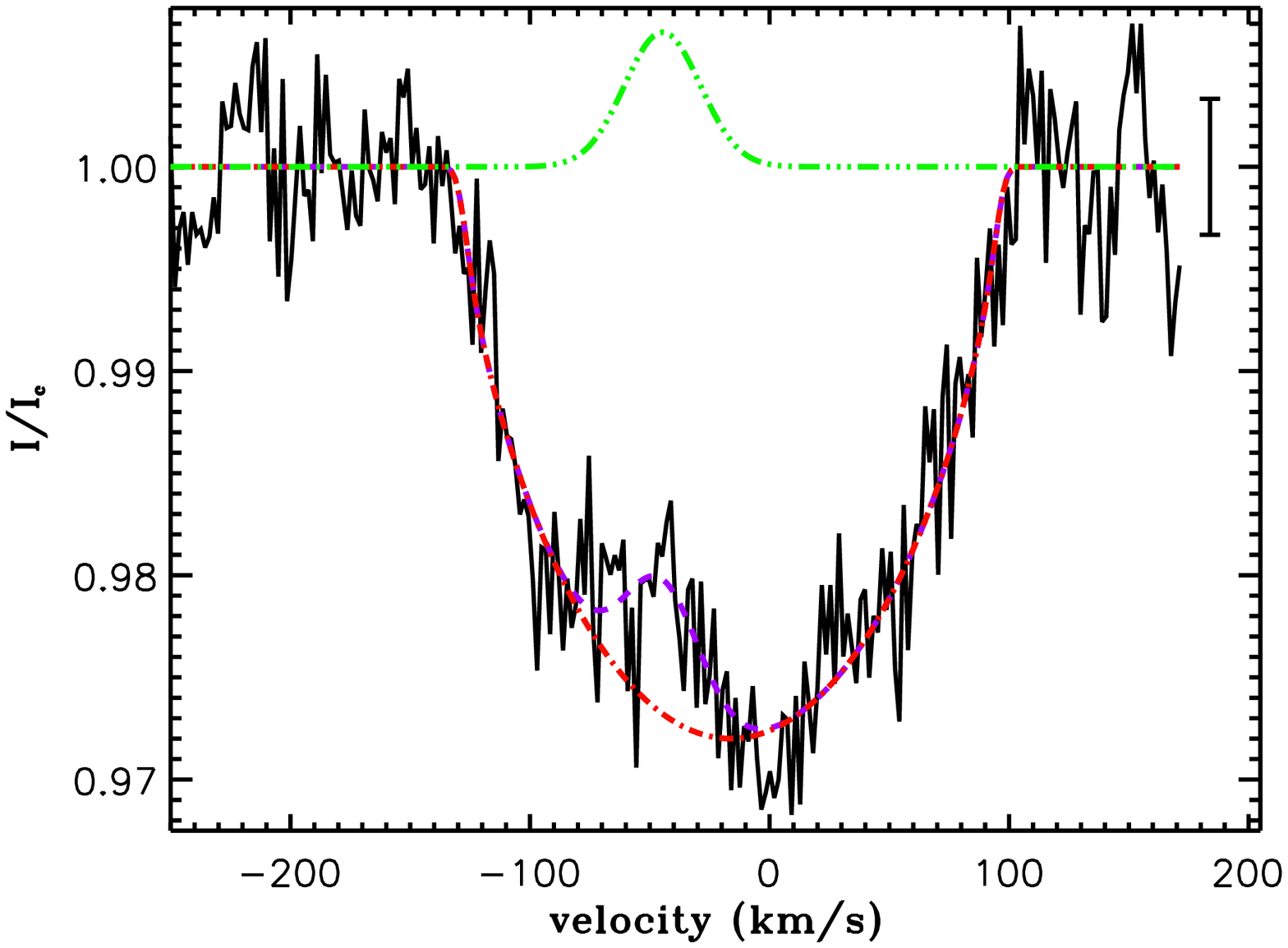}
\caption{As Fig.~A5 for the August 2005 observation of BD+61 154 (=~V594~Cas).}
\label{fig:61154}
\end{figure}

BD+61 154 is often associated with the cluster NGC 225, because its position in the sky coincides with the centre of the cluster. However \citet{kharchenko04} determined the membership probability from the proper motions and the photometry of the star, and find both values below 3\%, leading to the conclusion that this star is not member of NGC 225 and may be isolated. The Hipparcos parallax \citep{vanleeuwen07} and photometric data \citep{perryman97} have therefore been used to determine the luminosity. BD+61 154 possesses strong near-IR excess that has been attributed by \citet{hillenbrand92} to a circumstellar accretion disk.

The spectrum of BD+61 154 is strongly contaminated with circumstellar emission leaving only very few CS-free photospheric lines in the spectrum. The Ca~{\sc ii}~K line shows a double-peaked emission profile. From H$\eta$ to H$\varepsilon$, narrow emission is superimposed with the core of the photospheric profile, while P Cygni profiles of type II are observed from H$\delta$ to H$\beta$. H$\alpha$ displays a P Cygni profile of type III. The O~{\sc i}~777~nm, O~{\sc i}~8446 \angs and Ca~{\sc ii} IR triplets, as well as the Paschen lines, show single-peaked emission profiles. The He~{\sc i} lines at 5875~\AA, 6678~\AA\ and 7065~\angs display broad and asymmetric emission profiles, but He~{\sc i} 4471 and 4713 appear in absorption with about the normal strength for the photospheric temperature. Finally, many metallic lines show single-peaked emission profiles with a P Cygni blueshifted absorption component in the strongest emission lines, but a few high-excitation lines such as the Si~{\sc ii} lines at 5040 and 5056~\angs have nearly photospheric profiles.

By comparing the wings of the H$\gamma$, H$\delta$, and H$\varepsilon$ (H$\alpha$ and H$\beta$ are too strongly contaminated with CS features) with synthetic spectra, we find an effective temperature of $T_{\rm eff} = 13000\pm500$~K, slightly larger than the temperature reported in \citet{alonso09}.

We have cleaned the mask in order to reduce as much as possible the contribution of emission. The result gives a photospheric, but noisy profile with only a very faint residual emission. We fit the profile of August 2005 with a rotation function for the photospheric part, and a Gaussian function modeling the emission. We find $v\sin i=112 \pm 24$~\kms. The LSD profile of February 2005 is much noisier than the later one. Fitting this profile gives a $v\sin i$ consistent with our previous determination, but with extremely large error bars, which do not improve the precision of $v\sin i$. We therefore did not include it in this determination. The result of the fit of the Aug. 2005 profile is plotted in Fig. \ref{fig:61154}.

\subsection{BD+65 1637 (= V361 Cep)}

BD+65 1637 is associated with the reflection nebula NGC~7129 that envelops a young and compact cluster \citep{aveni72}. The luminosity has been estimated from the Hipparcos photometric data \citep{perryman97} and the photometric determination of the distance of NGC 7129 by \citet{shevchenko89}. It displays faint near-IR excess similar to classical Be stars \citep{hillenbrand92}. However its association with a reflection nebula and a young cluster suggests that the star is very young, and might be at the end of its PMS phase.

The spectrum of BD+65 1637 fits well with a non-LTE TLUSTY spectrum of $T_{\rm eff} = 18000\pm1000$~K, slightly larger than the determination of Hernandez et al. (2004), but consistent with the work of \citet{finkenzeller85}. Many iron lines appear in emission with two peaks of different amplitude, the red one being higher than the blue one. These iron emission lines contaminate some helium lines.  The other strongest He lines show a V-shape that could be the result of circumstellar contamination. The cores of the Balmer lines from H$\zeta$ to H$\beta$ show double-peaked emission lines, and H$\alpha$ is in emission with a double-peaked profile, all with a red peak higher than the blue one. The Ca~{\sc ii} IR, O~{\sc i} 777~nm, and O~{\sc i} 8446 \angs triplets, as well as the Paschen lines, also display strong double-peaked emission profiles.

\begin{figure}
\centering
\includegraphics[width=3.5cm,angle=90]{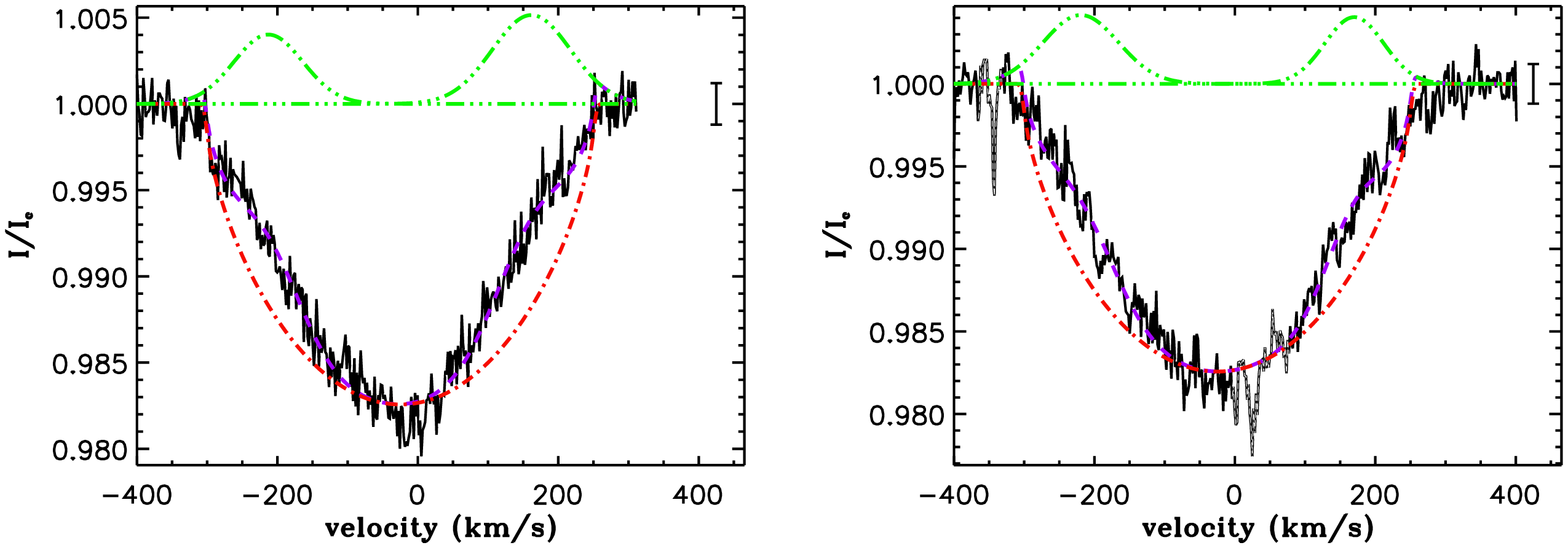}
\caption{As Fig.~A5 for the June 2006 (left) and September 2009 (right) observations of BD+65 1637 (= V361 Cep).}
\label{fig:651637}
\end{figure}

We have cleaned the mask to avoid as much as possible lines contaminated with emission. The inferred LSD $I$ profiles of both observations are in absorption, but have V-shapes, consistent with the lines of the spectrum. Taking into account the shapes of the emission features observed in the spectrum, the best and most reliable fit of the LSD $I$ profiles has been obtained with a photospheric plus two Gaussian functions. We performed a simultaneous fit of both profiles, by forcing the photospheric depth, \vsini, and \vrads to be the same for both observations. The result is shown in Fig. \ref{fig:651637}.

\subsection{BD+72 1031 (= SV Cep)}

SV Cep is associated with L1235, one of the dark clouds of the Cepheus region situated at a distance of 400~pc \citep{kun98}. SV Cep is an UXOR star, and displays strong IR excess, both suggesting that a circumstellar accretion disk is present around this star \citep{friedemann92,acke04a,juhasz07}. No Hipparcos data are available for this star, however \citet{rostopchina00} obtained UBVR photopolarimetric observations of the star. We used these data at the brightest magnitude ($V=10.48$, $(B-V)=0.344$) to derive the luminosity. 

Like other UXOR stars, SV Cep shows many redshifted or blueshifted transient absorption features in its spectrum that changed in shape and depth from June 2006 to November 2007. From H$\delta$ to H$\beta$ a strong circumstellar absorption is superimposed with the core of the Balmer lines, and in H$\beta$, emission appears in the wings of the CS absorption. The 2007 absorption components are redshifted relative to 2006. H$\alpha$ shows a strong double-peaked emission profile with both red and blue peaks of equal amplitude in 2006, while in 2007 the blue peak is stronger than the red one, and the central absorption is broader, deeper, and redshifted. Strong CS absorptions are also observed in the Ca~{\sc ii} IR- and OI 8446 \angs triplets, and in the Ca~{\sc ii} K line, as well as in many metallic lines. While these absorption features have only a single component in 2006, in 2007 they display two separate components, both redshifted compared to 2006. The He~{\sc i} lines at 5875~\AA, 6678~\AA, and 7065~\angs lines, and the O~{\sc i} 777~nm triplet display inverse P-Cygni profiles, with the absorption component being redshifted from 2006 to 2007. While the Paschen lines do not seem to be contaminated with CS features in 2006, in 2007 they display single and small emission components in the cores. The portions of the spectrum not contaminated with CS absorption is consistent with the temperature determination of \citet[][$T_{\rm eff}=9500\pm2000$~K]{hernandez04}.

\begin{figure}
\centering
\includegraphics[width=6cm,angle=90]{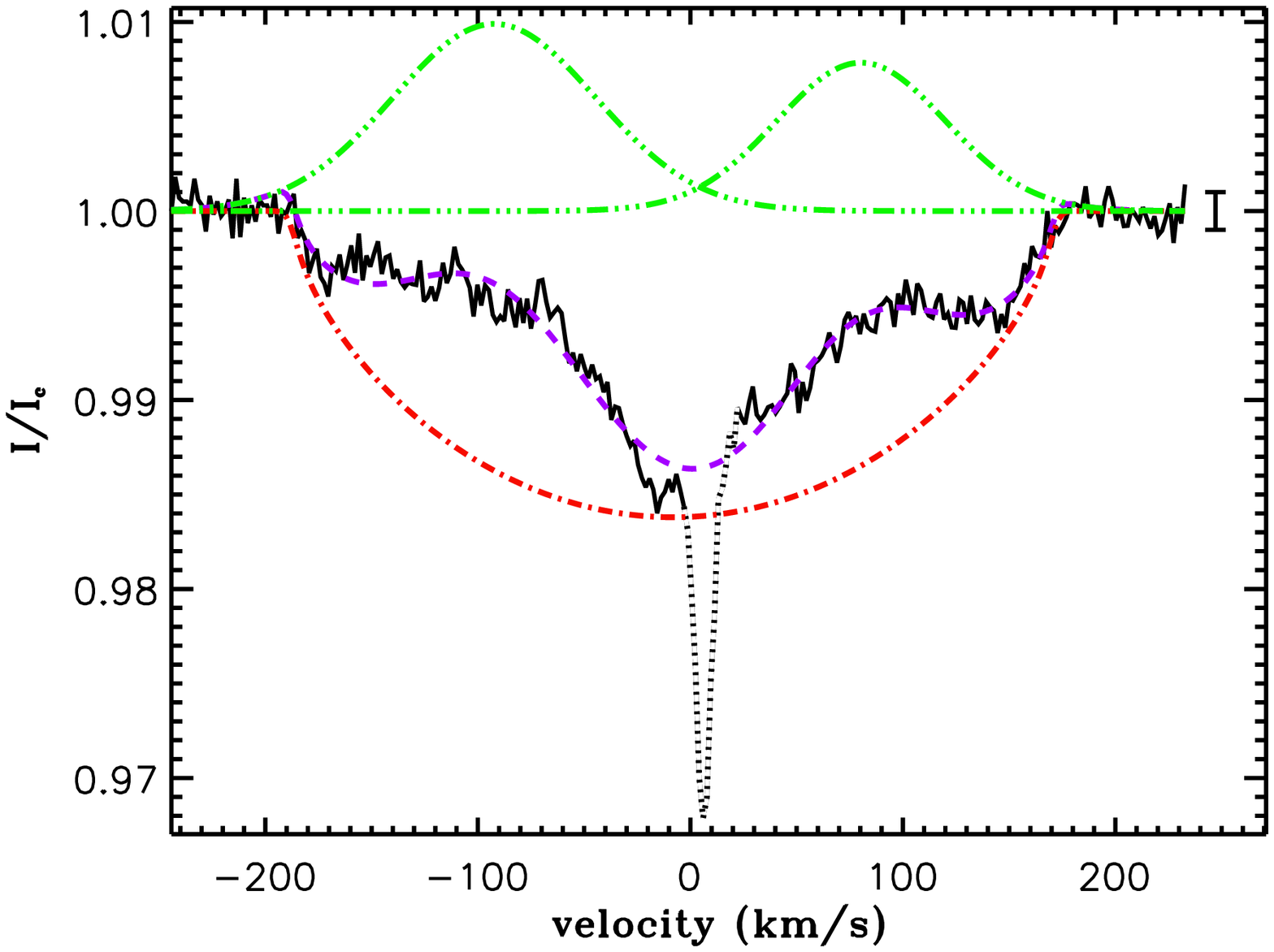}
\caption{As Fig.~A5 for the June 2006 observation of BD+72 1031 (=~SV~Cep).}
\label{fig:svcep}
\end{figure}

We tried to clean the Kurucz mask by rejecting the lines obviously contaminated with CS features. However the resulting profiles still display CS features superimposed with the photospheric profile, and could not be improved. We fitted the Jun. 2006 profile with a photospheric profile and two Gaussians modeling the emission, and by rejecting CS absorption in the core. We also fit the Nov. 2007 profile, but because of its lower SNR, the values of the fitted parameters, while consistent with the first fit, have larger error bars, and don't better constrain the model. Therefore we adopted the $v\sin i$ and \vrad\ from the fit of Jun. 2006 only (Fig.~\ref{fig:svcep}).

\subsection{HD~9672 (= 49 Cet)}

\begin{figure}
\centering
\includegraphics[width=6cm,angle=90]{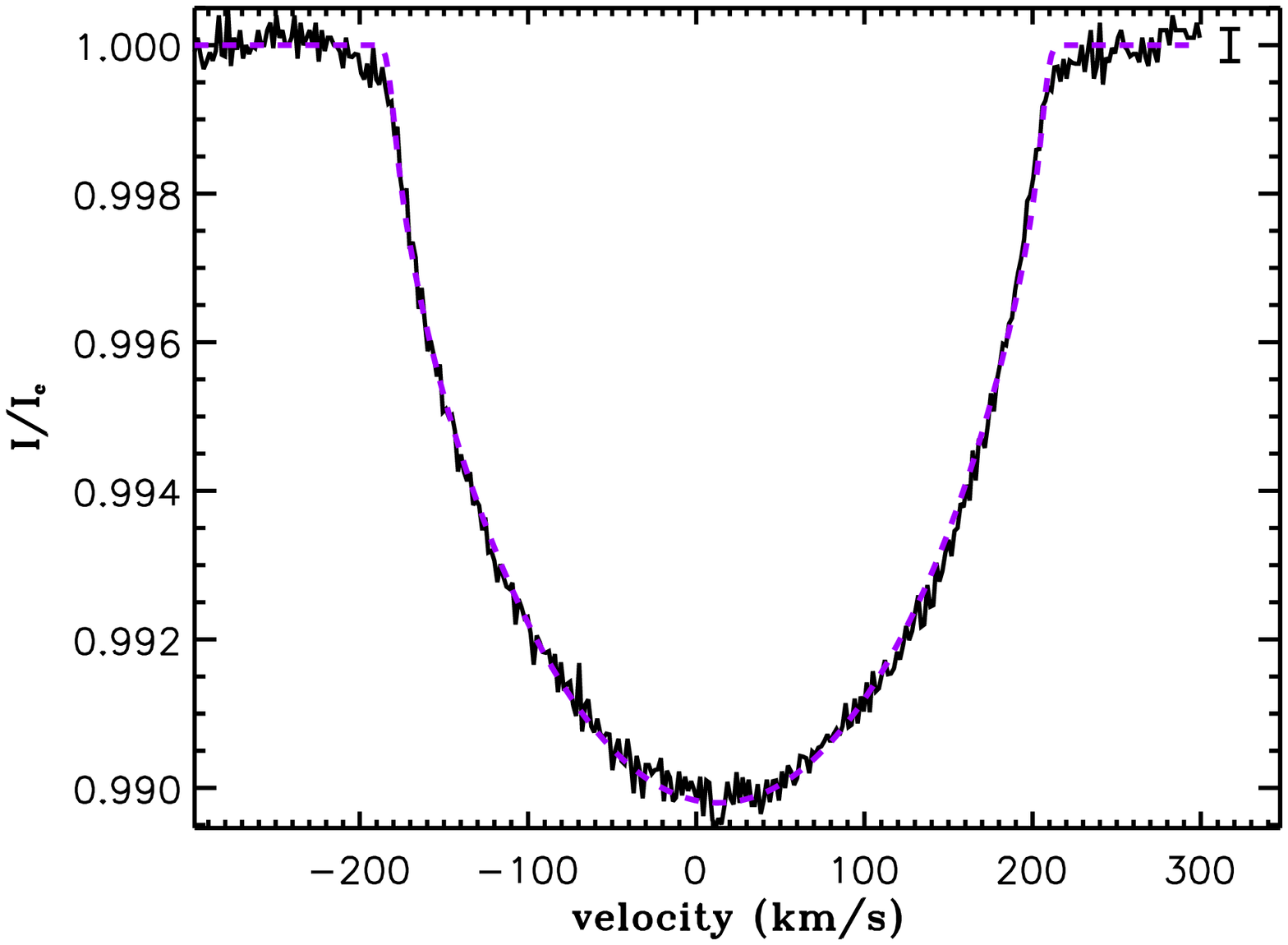}
\caption{As Fig.~A5 for HD~9672 (= 49 Cet)}
\label{fig:49cet}
\end{figure}

49 Cet is a $\beta$ Pictoris-type star having infrared excess emission attributed to circumstellar dust, but being distinguished from a Vega-type star by a much higher IR to bolometric luminosity ratio \citep{jura98}. Such a star would represent a transitional object between Herbig Ae and Vega-like stars \citep{wahhaj07}, as is confirmed by the detection of molecular gas in its close environment \citep{hughes08}. While this star cannot totally be considered to be a Herbig Ae star, we kept it in our sample because it represents an interesting and rare stellar evolutionary phase between star birth and the MS.

49 Cet is the closest star of our sample, with a distance of 61~pc \citep[Hipparcos, ][]{vanleeuwen07}. Combining both the Hipparcos distance and photometric data we derive a luminosity $\log(L/L_{\odot})=1.32$.

With our automatic procedure (see Section 4.1) we find an effective temperature of $8900\pm200$~K and $\log g=4.5$, consistent with the work of \citet{montesinos09}. It does not show any circumstellar features, therefore no special cleaning has been applied to the Kurucz mask. The LSD Stokes $I$ profile shows a clean photospheric profile, well reproduced using a photospheric function. The result of the fit is plotted in Fig. \ref{fig:49cet}.

\subsection{HD~17081 (= $\pi$ Cet)}

HD~17081 does not seem to be associated with a nebula or a star forming region. Based on its IR excess, \citet{malfait98a} has classified HD~17081 as a Vega-type star, occupying an evolutionary stage subsequent to the Herbig Ae/Be stars, but still very young, and therefore constituting an interesting object to explore in the framework of our study of rotational evolution during the PMS phase. The Hipparcos parallax \citep{vanleeuwen07} and photometric data \citep{perryman97} have been taken for the luminosity determination.

The spectrum of HD~17081 is consistent with the effective temperature and $\log g$ determination of \citet[][$12800\pm200$~K, $\log g=3.77\pm0.15$]{fossati09}. No emission is clearly observed, however, the shapes of the line profiles are distorted, possibly due to circumstellar emission or absorption, or possibly  to abundance patches at the surface of the star as proposed by \citet{fossati09}. The presence of abundance anomalies of a few elements detected in the spectrum \citep{fossati09} would support the second explanation.

\begin{figure}
\centering
\includegraphics[width=3cm,angle=90]{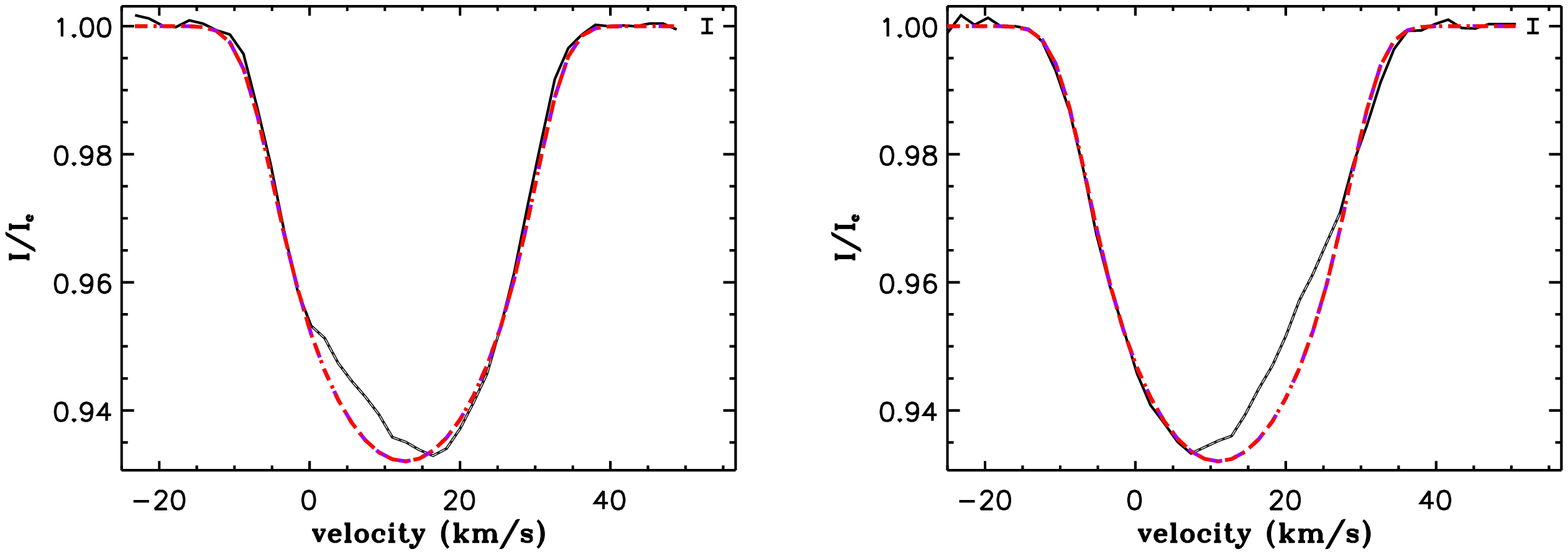}
\caption{As Fig.~A5 for the Feb. 20th (left) and Feb. 21st (right) 2005 observations of HD~17081.}
\label{fig:hd17081}
\end{figure}

The LSD profiles have been calculated with the Kuruz mask and have both been modeled simultaneously with a photospheric profile only. We tried with various techniques to model the distortion of the profiles but we could not find a solution giving consistent photospheric depths and \vsinis for both observations, as well as a better constrained value of \vsini. We therefore choose to ignore the distorted points into the fitting procedure. In this fit, the depth and the \vsinis of the photospheric profiles have been forced to be the same for both observations. We also tried to force the \vrads to be the same, but the fit was significantly less good, meaning that the radial velocity has slightly but significantly changed from the 19th  ($12.7\pm0.8$~\kms) to the 20th ($11.0\pm0.7$~\kms) February 2005. The result of the fit is shown in Fig. \ref{fig:hd17081}.

\subsection{HD~31293 (= AB Aur)}

\begin{figure}
\centering
\includegraphics[width=2.1cm,angle=90]{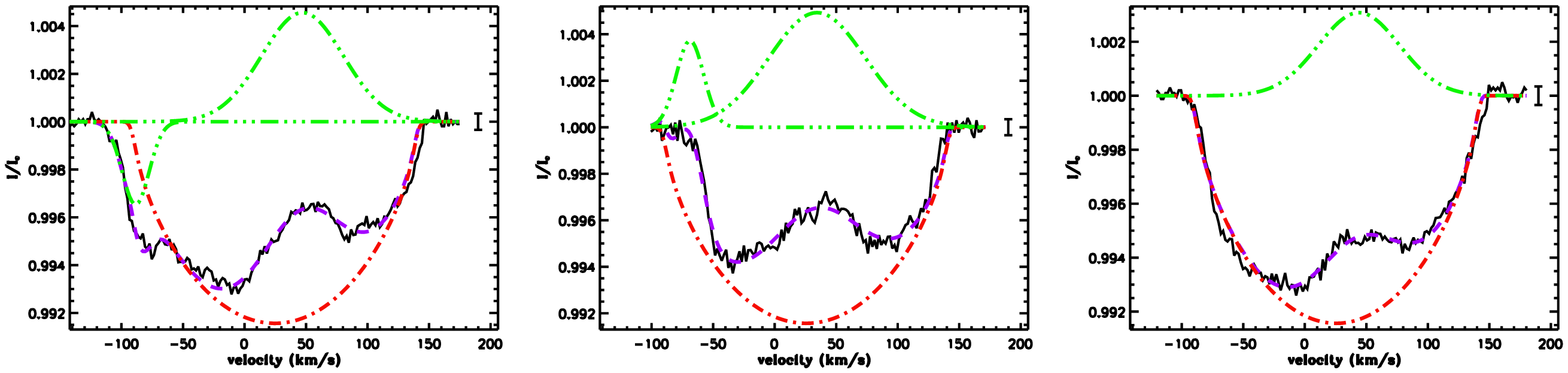}
\caption{As Fig.~A5 for the Nov. 2004 (left), Feb. 20th (middle) and Feb. 22nd (right) 2005 observations of  HD~31293 (= AB Aur).}
\label{fig:abaur}
\end{figure}

AB Aur, situated in the Taurus region at a distance of 139~pc \citep[Hipparcos, ][]{vanleeuwen07}, is associated with the dark cloud Lynds 1517, and is clearly surrounded with a bright nebula \citep{ancker98}. Hipparcos photometric data have been used for its luminosity determination. AB Aur displays near-IR excess interpreted as the presence of a circumstellar accretion disk surrounding the star \citep{hillenbrand92,malfait98a}

AB Aur is one of the best studied Herbig Ae stars with strong emission and variability in its spectrum. Rotational modulation has been detected by \citet{praderie86} and \citet{catala86b,catala93,catala99} in various UV and optical non-photospheric lines that are very likely formed in the wind. H$\alpha$ is strongly variable going from a double-peaked emission profile to a P Cygni of Type IV \citep{catala99}. In our 3 observations, H$\alpha$ displays a P Cygni profile. From H$\zeta$ to H$\gamma$ a blueshifted absorption as well as a roughly centered emission component are superimposed on the core of the Balmer lines, with increasing amplitudes with wavelength, for both absorption and emission components. A P Cygni profile of type IV is superimposed with the core of H$\beta$. Some Fe~{\sc ii} lines, including those of multiplet 42, also show P Cygni profiles, but most weaker metal lines show largely photospheric profiles, and a few, such as the O~{\sc i} lines at 615~nm are completely photospheric. The Ca~{\sc ii}~K line is contaminated with emission, while the Ca~{\sc ii} IR-, and O~{\sc i}~8446 \angs triplets, and the Paschen lines display single-peaked emission profiles. The He~{\sc i} 5878 \AA, He~{\sc i} 6678~\AA, He~{\sc i} 7065~\AA, and the O~{\sc i} 777~nm triplet show complex and variable emission profiles.

The wings of the Balmer lines are consistent with the effective temperature and $\log g$ determination of \citet[][$T_{\rm eff}=9800\pm700$~K, $\log g=3.9\pm0.3$]{folsom12}. Most of the metallic lines are shallower than predicted, suggesting that they are filled with emission. The Kurucz mask has been cleaned by keeping only lines showing similar shapes and being only slightly contaminated with CS emission. The three resulting profiles show a photospheric absorption line superimposed with a slightly redshifted and variable emission component, consistent with the photospheric profiles of the spectrum. The three profiles have been fitted simultaneously with a photospheric function and one or two Gaussians modeling the CS emission and absorption in the 2004 profile. In this fitting procedure the depth, \vsini, and \vrads of the photospheric profiles have been forced to be the same for the three observations. The result is shown in Fig. \ref{fig:abaur}.

\subsection{HD~31648 (= MWC 480)}

\begin{figure}
\centering
\includegraphics[width=2.8cm,angle=90]{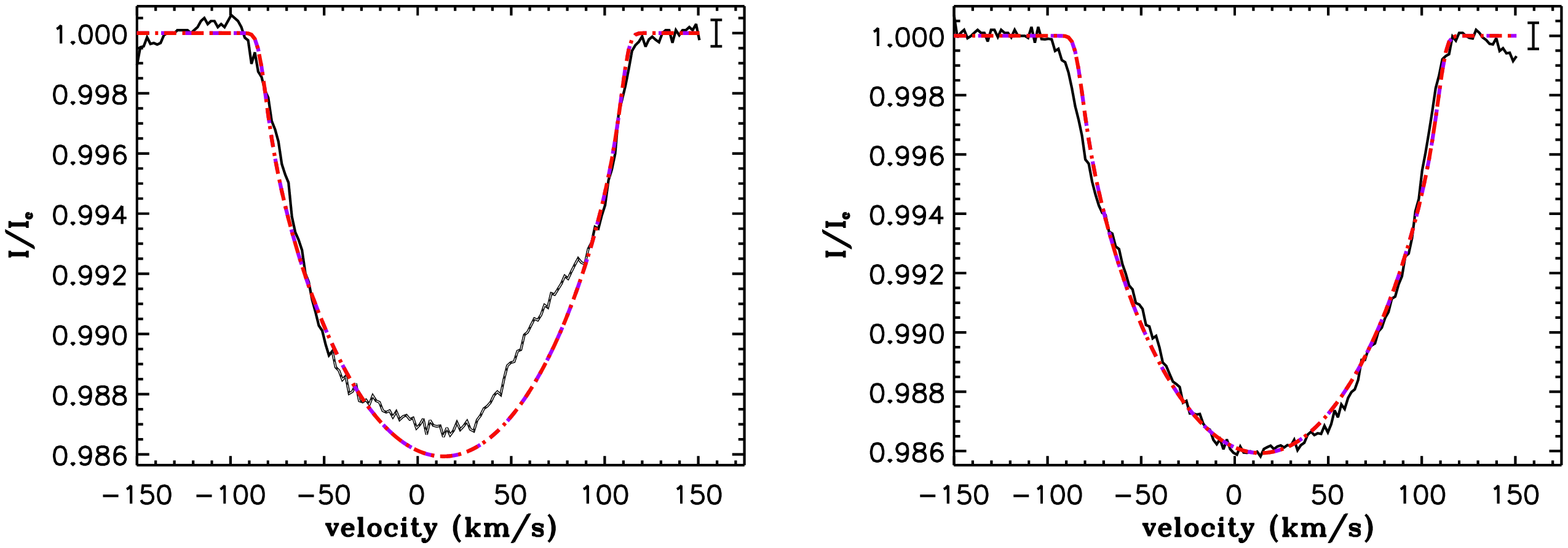}
\caption{As Fig.~A5 for the Feb. (left) and Aug. (right) 2005 observations of HD~31648.}
\label{fig:hd31648}
\end{figure}

HD~31648 lies in the star forming Taurus-Aurigae region at a distance of 137~pc \citep[Hipparcos, ][]{vanleeuwen07}. It does not show evidence of association with bright nebulae but high-resolution millimeter-wave images of continuum and molecular-line emission from dust and gas in its neighborhood revealed the presence of a proto-planetary disk \citep{mannings97}. The high IR excess due to warm dust in its close environment confirms the Herbig Ae/Be status of the star \citep{malfait98a}. We used the Hipparcos photometric data to determine its luminosity.

In the spectrum of HD~31648, the cores of the Balmer lines from H$\zeta$ to H$\gamma$ are superimposed with a blueshifted circumstellar absorption component with amplitude increasing with wavelength. A similar but stronger absorption component is superimposed on the Ca~{\sc ii} K line. H$\beta$ displays a strong blueshifted absorption component and a slightly redshifted emission in the core. H$\alpha$ has the shape of a type III P Cygni profile. Circumstellar emission and absorption are present in a few metallic lines, including the lines of multiplet 42 of Fe~{\sc ii}. The He~{\sc i} lines at 5875~\AA, 6678~\AA, and 7065~\angs lines display complex inverse P-Cygni profiles. The O~{\sc i}~777~nm triplet shows a strong (stronger than predicted), roughly centered, absorption profile with wing emissions, but the weaker O~{\sc i} lines at 615~nm are in absorption and well fit with a normal photospheric profile. The O~{\sc i}~8446 \angs triplet shows a complex emission profile, and the Ca~{\sc ii} IR-triplet displays single-peaked emission profiles, superimposed with a faint blueshifted absorption. The Paschen lines are filled with emission. Using our automatic procedure, we find that the portions of the spectrum that are free of CS features are well reproduced with $T_{\rm eff} = 8200\pm300$~K and $\log g=4.0$, consistent with the work of \citet{vieira03}.

We have cleaned the Kurucz mask keeping only lines not contaminated with CS features. The resulting profiles are slightly distorted, revealing some residual contamination by CS emission, but could not be improved. We fitted both profiles simultaneously with a photospheric function, by forcing the depth, \vsini, and \vrads of the photospheric profile to be the same for both observations, and by rejecting the distorted data points in the core of the February 2005 observation. We also tried to include a few Gaussian functions for modelling the distorted data points, instead of rejecting them, but this did not improve the fit. The result is plotted in Fig. \ref{fig:hd31648}.

\subsection{HD~34282}

\citet{merin04} carried out a thorough spectroscopic and photometric study of HD~34282 motivated by the fact that previous work placed the star below the ZAMS in the HR diagram. This star shows strong IR excess consistent with a circumstellar disk \citep{ancker98}, as well as PAH emission lines \citep{keller08}, confirming its PMS status, and meaning that this star should be situated well above the ZAMS. Their spectroscopic analysis led to an effective temperature and gravity of $T_{\rm eff} = 8625\pm200$ K and $\log g = 4.20\pm0.20$ (cgs). They also derived a sub-solar metallicity of ${\rm Fe/H} = -0.8\pm0.1$. The Balmer lines of our spectrum fits well with $T_{\rm eff} = 8750$ K and $\log g = 4.0$ consistent with their work. However using a solar metallicity, all the metallic lines are clearly underabundant as observed by \citet{merin04}.

\citet{merin04} compared the position of HD~34282 with PMS evolutionary tracks of sub-solar metallicity ($Z=0.004$) in the $\log g - \log T_{\rm eff}$ diagram, and derived the mass, age, luminosity and distance of the star that we adopted in this study.

In our data, the core of H$\alpha$ is contaminated with a double-peaked emission profile with a deep central absorption, while H$\beta$ and H$\gamma$ show a deep CS absorption centered in the core. Except for stronger absorption than predicted in the He~{\sc i} D3 line and the O~{\sc i} 777~nm triplet, no other CS manifestation is observed in the spectrum.

\begin{figure}
\centering
\includegraphics[width=6cm,angle=90]{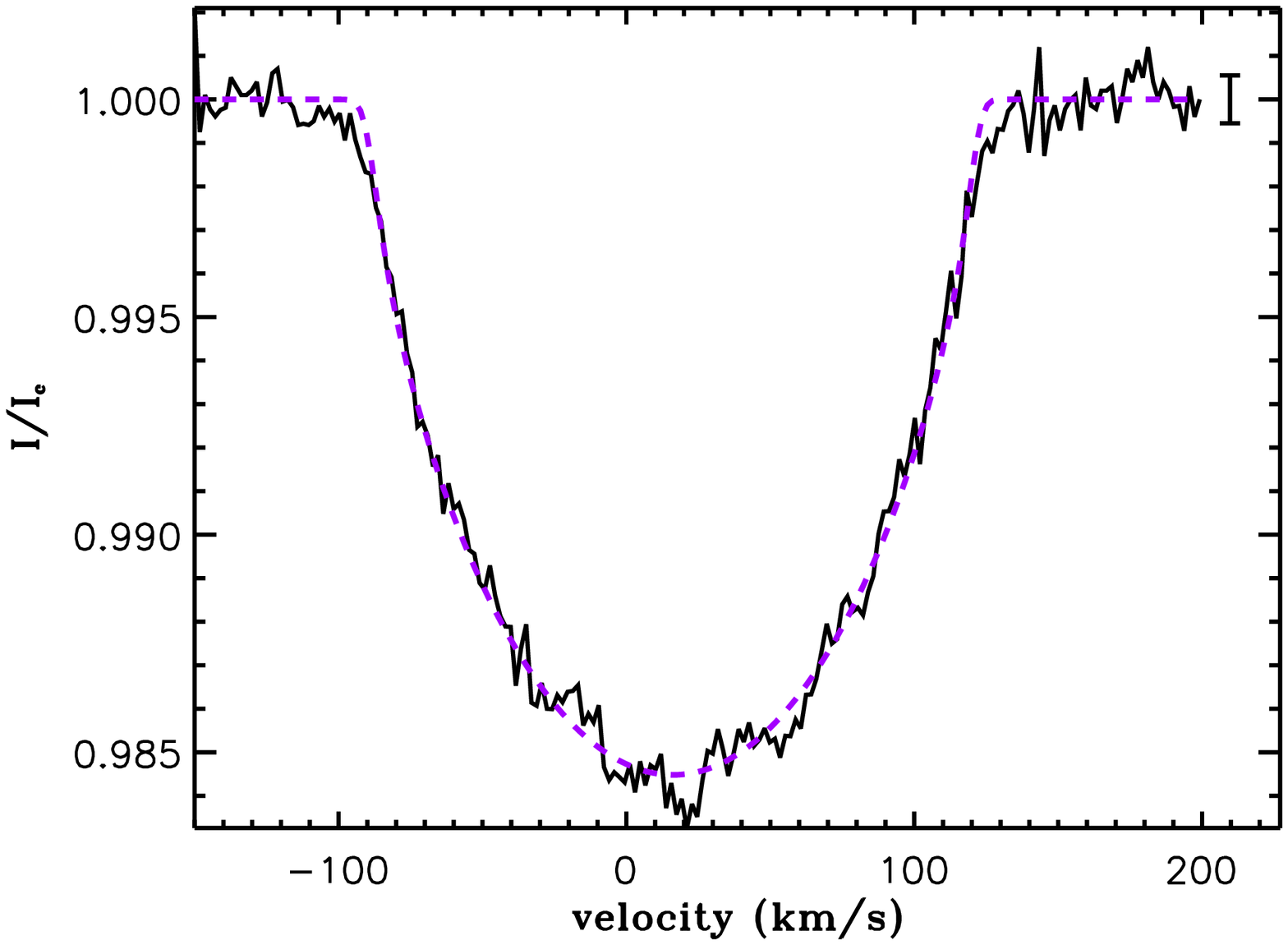}
\caption{As Fig.~A5 for HD~34282}
\label{fig:hd34282}
\end{figure}

If we use a solar metallicity and an effective temperature of 10\,000 K we are able to accurately reproduce the depth of the metallic lines of the spectrum. We therefore choose the Kurucz mask of effective temperature 10\,000~K instead of a temperature closer to the one determined by \citet{merin04}, simply to determine the rotation velocity. The result is a simple photospheric profile that has been fitted with a single photospheric function, and is shown in Fig. \ref{fig:hd34282}.

\subsection{HD~35187 B}

HD~35187 is a binary system situated at a distance of 114~pc \citep[Hipparcos,][]{vanleeuwen07} in the Taurus Auriga star forming region with a separation of 1.39 arcsec \citep{perryman97}. Hipparcos resolved the system and the Hipparcos Catalogue Double and Multiple Systems Annex (DMSA) assigned the identifier ‘B’ to the brighter component, i.e. to the more northerly component at the time of Hipparcos observations. We will follow the Hipparcos nomenclature in order to avoid confusion. Spatially resolved optical spectroscopy performed by \citet{dunkin98} revealed that HD~35187~B has a stronger H$\alpha$ emission than the HD~35187~A and displays a narrow circumstellar absorption line in the Ca~{\sc ii} K line, that is absent in the spectrum of HD~35187~A. \citet{dunkin98} derived effective temperatures of 8990 K and 7800 K for the components B and A, respectively, and suggest that only HD~35187~B is surrounded with a circumstellar disk. The spectral energy distribution analysis reveals the presence of a geometrically-flat disk, but is not able to distinguish between a circumstellar or a circumbinary disk.

On the night of August 25th 2005, the seeing was around 0.45 arcsec, allowing us to observe both components separately. However the observation of component B only is of good quality, and has been analysed. With our automatic procedure (see Section 4.1) we find an effective temperature of $8900\pm200$~K and $\log g=4.0$. It shows faint emissions in the core of H$\beta$ and H$\gamma$ roughly centred and not enough strong to fill the core. The emission profile of H$\alpha$ is of complex structure with a single emission superimposed with two faint absorption components. Apart from the He~{\sc i}~D3 line, which displays stronger absorption than predicted, no other CS features are observed in the spectrum.

\begin{figure}
\centering
\includegraphics[width=6cm,angle=90]{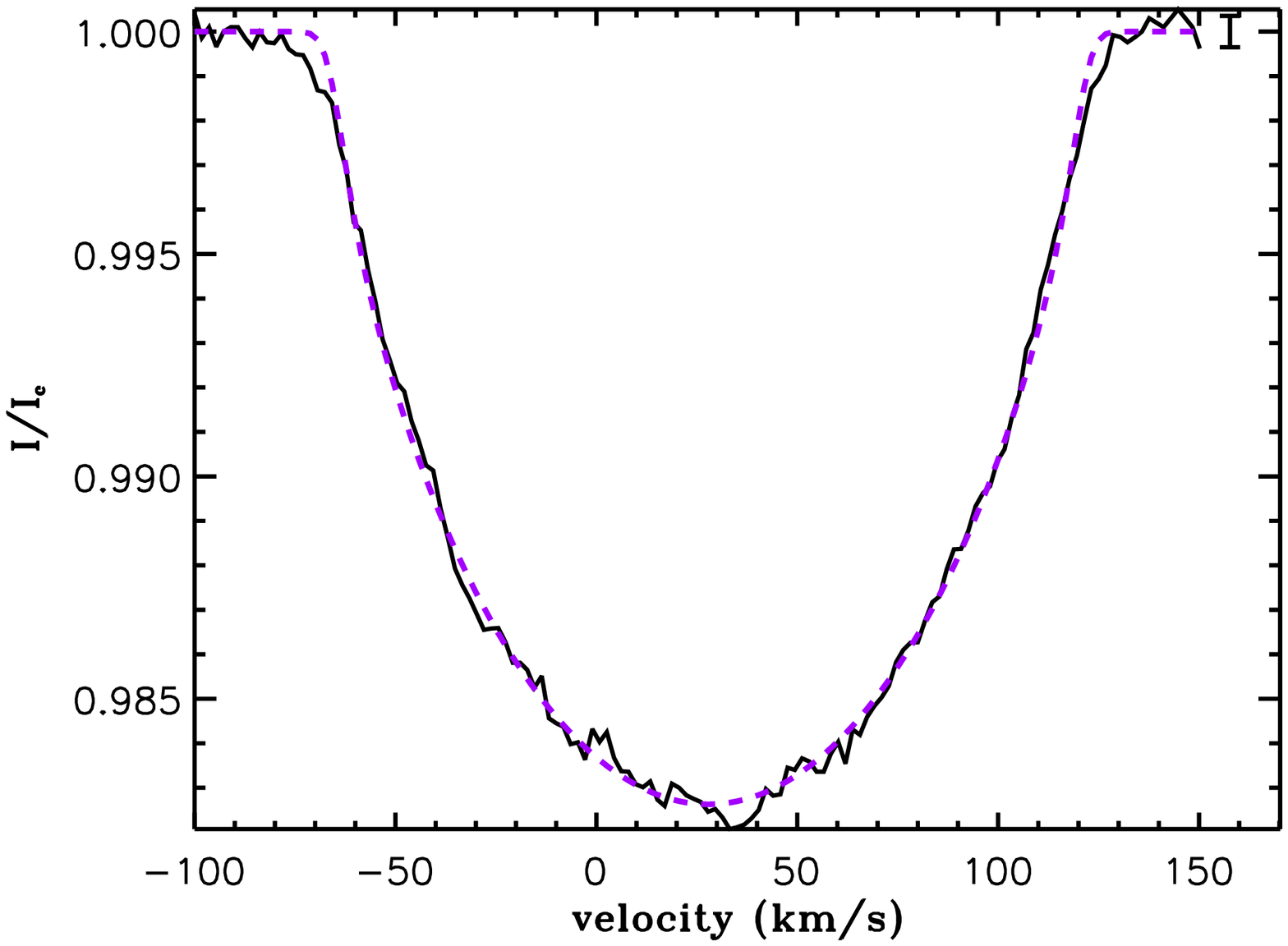}
\caption{As Fig.~A5 for HD~35187 B}
\label{fig:hd35187s}
\end{figure}

The LSD $I$ profile fits very well with a single photospheric function. The result is shown in Fig. \ref{fig:hd35187s}.

\subsection{HD~35929}

HD~35929 belongs to the Ori OB 1c association \citep{ancker98} situated at a distance of 375~pc \citep{brown94}. $\delta$ Scuti-type pulsations have been detected in the star by \citet{marconi00}. Based on pulsation and also optical and IR properties, these authors conclude that HD~35929 is probably in the PMS evolutionary stage rather than in the post-MS stage. We used the Hipparcos photometric data \citep{perryman97} to derive the luminosity.

The spectrum of HD~35929 is consistent with the effective temperature and gravity determination of \citet[][$T_{\rm eff} = 6800\pm100$~K, $\log g=3.3\pm0.1$]{miroshnichenko04}. It displays a strong and narrow single-peaked emission profile in H$\alpha$, as well as emissions with central absorption in the Ca~{\sc ii} IR-triplet. Emission also seems to be present in the core of H$\beta$ and the Ca~{\sc ii}~K line, while the O~{\sc i} 777~nm is stronger than predicted. No other CS feature is observed in the spectrum. 

\begin{figure}
\centering
\includegraphics[width=4.2cm,angle=90]{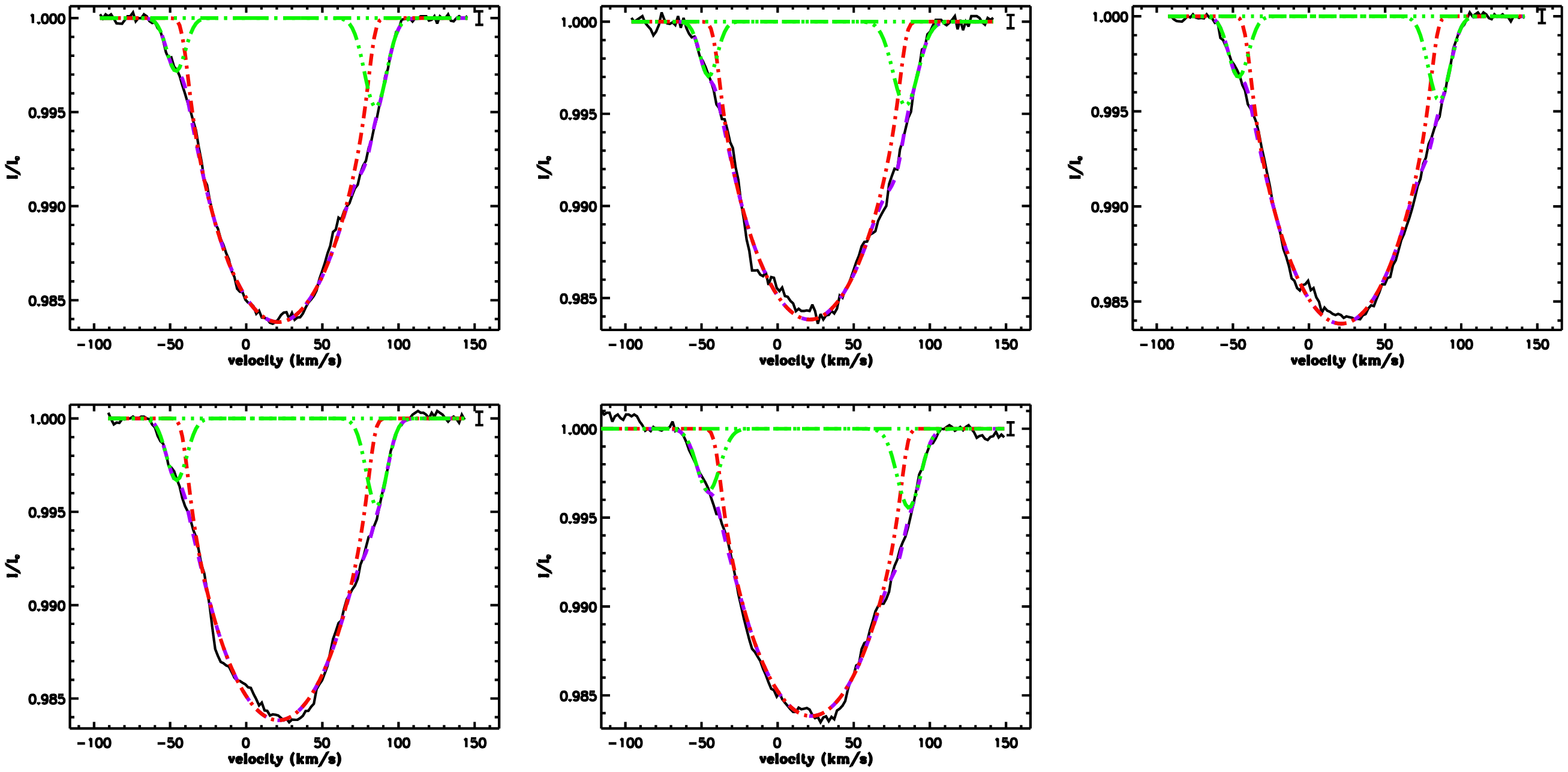}
\caption{As Fig.~A5 for the Nov. 13th, Nov. 14th 2007, Feb. 20th, Feb. 21st, and Mar. 11th 2009 (from top left to bottom right) observations of HD~35929.}
\label{fig:hd35929}
\end{figure}

We first used the Kurucz mask to compute the LSD profiles of the 5 observations. The results show $I$ profiles in absorption with broad wings. In order to get rid of these broad wings we have calculated a new mask by keeping only lines with intrinsic depth smaller than 0.5. The result shows wings without severe broadening, but now a slight distortion is detected in the profiles that could not be seen directly in the individual lines of the spectrum because of the lower SNR. We fit all the observations simultaneously with a photospheric and two Gaussian functions in absorption in both sides of each profile, by assuming that the distortions are due to CS contamination. In this fitting procedure we forced the depth, \vsini, and \vrads of the photospheric profile to be the same for the five observations. The result, shown in Fig. \ref{fig:hd35929} is satisfying. We should note however that all the Gaussians are found to be similar from one observation to another, while we would expect them to vary with time, as is observed in other HAeBe stars.

\subsection{HD~36112 (= MWC 758)}

HD~36112 is part of the Taurus-Aurigae star forming region situated at a distance of $\sim279$~pc \citep[Hipparcos, ][]{vanleeuwen07}. Strong IR excess is observed in the direction of the star that could be explained by the presence of a disk that has been directly detected with IR-interferometric data by \citet{eisner04}. We have used the Hipparcos photometric data to compute the luminosity.

\citet{beskrovnaya99} describe the features and variability observed in the spectrum of HD~36112. These are similar to those observed in our spectra. The H$\alpha$ profile shows a P Cygni shape with temporal variations in the amplitudes of the absorption and emission components. The Na D doublet also shows P Cygni profiles. Emission is detected in the He~{\sc i} lines at 5875~\AA, 6078~\AA, and 7065~\AA. Faint emission is observed in the core of H$\delta$, H$\gamma$, and H$\beta$ with amplitude increasing with wavelength. The O~{\sc i} 777~nm triplet is also strongly variable,  sometimes showing emission, sometimes absorption, while the O~{\sc i} 8446~\angs and Ca~{\sc ii} IR triplets show single-peaked emission profiles with intensity increasing from 2004 to 2005. Using our automatic procedure, we find that the portions of the spectrum that are free of CS features are well reproduced with $T_{\rm eff} = 7800\pm150$~K and $\log g=4.0$, consistent with the work of \citet{beskrovnaya99}.

\begin{figure}
\centering
\includegraphics[width=3cm,angle=90]{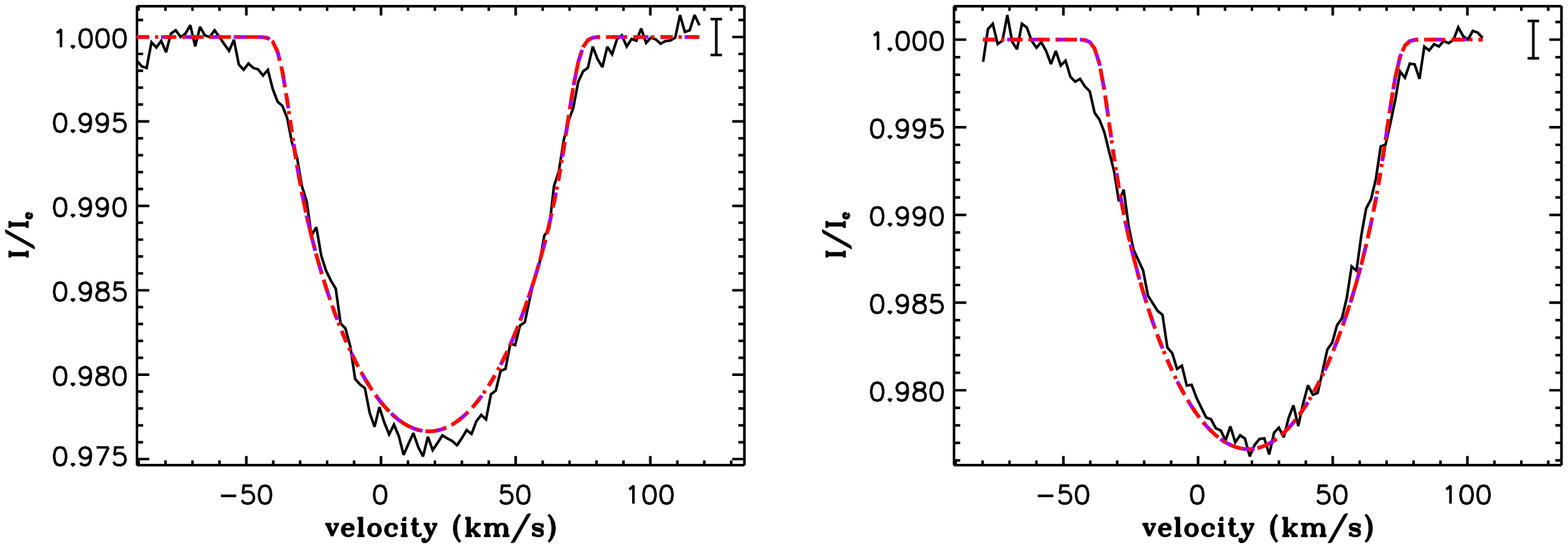}
\caption{As Fig.~A5 for the Nov. 2004 (left) and Feb. 2005 (right) observations of HD~36112.}
\label{fig:hd36112}
\end{figure}

We have cleaned the Kurucz mask by rejecting the lines strongly contaminated with emission, and used the edited mask to compute the LSD profiles. The resulting $I$ profiles show photospheric absorption lines. The core of the 2004 LSD profile is contaminated with a faint and narrow emission feature which we are not able to distinguish in the spectrum itself because of the lower SNR. We therefore tried to calculate various masks by making a selection on the intrinsic depth or excitation potential of the lines. No matter what selection is made on the intrinsic depth, the emission component is still present. However, with a mask containing only lines with $\chi_{\rm exc} > 6$~eV, the emission component disappears. We therefore chose this mask to analyse the LSD profiles. A simultaneous fit of both observations with photospheric functions of identical depth, \vsini, and \vrads gives a satisfactory result (Fig. \ref{fig:hd36112}).

\subsection{HD~36910 (= CQ Tau)}

HD~36910 seems to be part of the T-association Tau 4 \citep{kholopov59} at a distance of 113~pc \citep[Hipparcos, ][]{vanleeuwen07}. This star is strongly variable with amplitude variations greater than 1 mag, and belongs to the UXOR sub-class \citep{grinin01}. We therefore used the Hipparcos photometric data at maximum brightness ($V_{\rm T}=9.97$~mag, $B_{\rm T}=8.87$~mag, in the Tycho system), and converted them into the Johnson system with the calibration formula 1.3.20 of the Hipparcos and Tycho catalogues \citep[][p. 57]{perryman97}. We find $V = 8.77$~mag and $(B-V) = 0.94$~mag. The derived luminosity is $\log L/L_{\odot}=1.58$. Strong IR excess as well as mid-IR imaging reveal the presence of a dusty disk surrounding the star \citep{miroshnichenko99b,doucet06}.

The spectrum is consistent with the effective temperature determination of \citet[][$T_{\rm eff} = 6750\pm300$~K]{hernandez04}. A redshifted absorption component is observed in the core of the Balmer lines from H$\varepsilon$ to H$\beta$. In addition, H$\beta$ shows emission in the wings of the absorption component. H$\alpha$ displays a complex emission profile with a few redshifted absorption components. The He~{\sc i}~D3 line displays an emission profile with a superimposed redshifted absorption. The O~{\sc i} 777~nm triplet is stronger than predicted, with emission in the wings, while the O~{\sc i} 8446~\angs line displays a complex emission profile. The Ca~{\sc ii} IR-triplet seems to be contaminated with emission on the blue side of the line, and the cores of the Paschen lines are filled with emission. A faint blueshifted emission is observed in the core of the Ca~{\sc ii}~K line. The metallic lines of the spectrum seem stronger than predicted and are contaminated with CS features.

\begin{figure}
\centering
\includegraphics[width=6cm,angle=90]{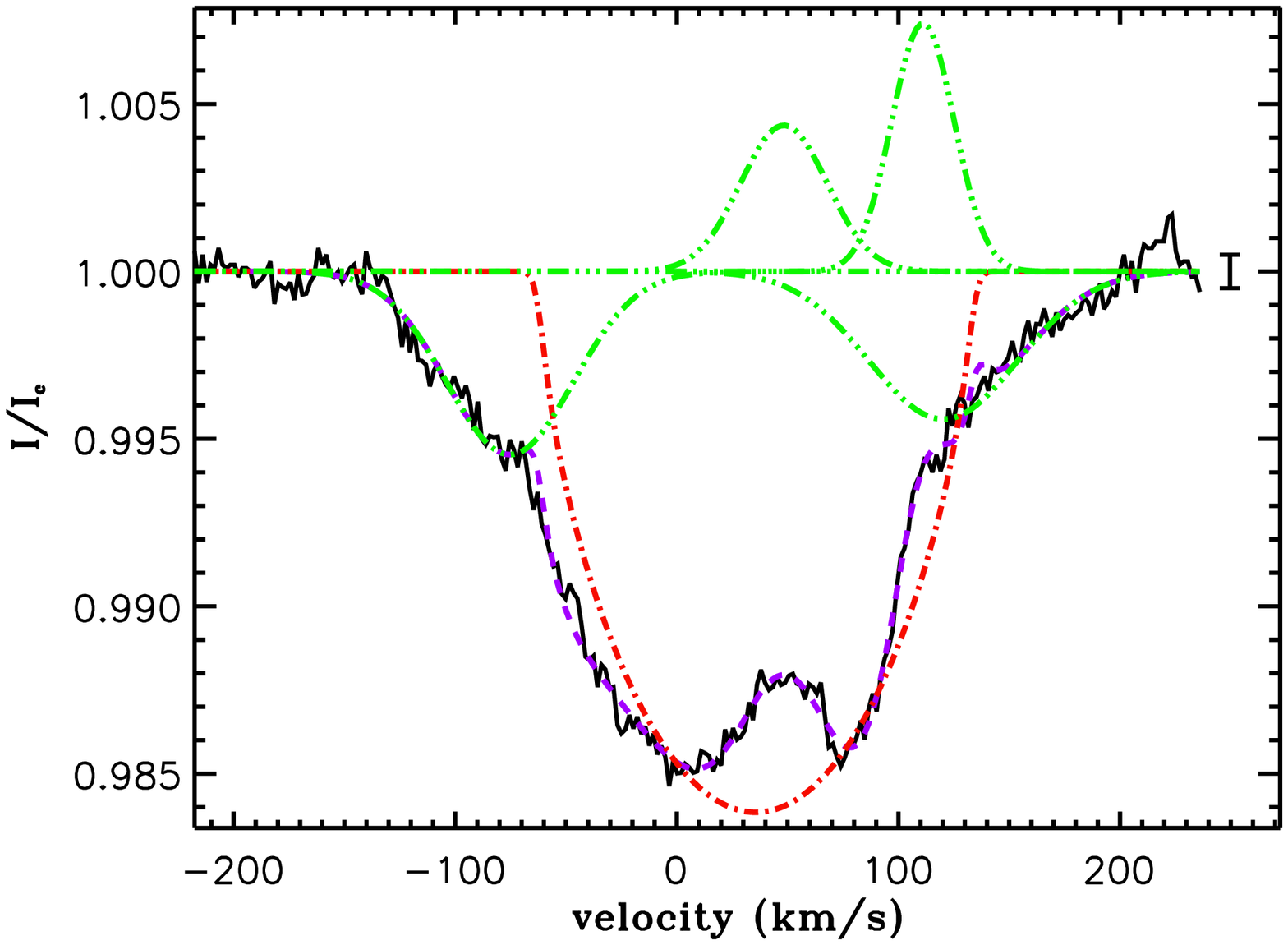}
\caption{As Fig.~A5 for HD~36910 (= CQ Tau)}
\label{fig:hd36910}
\end{figure}

We first calculated the LSD profiles without performing a special cleaning to the Kurucz mask. The result showed a very complex profile with very broad wings. We therefore tried to clean the mask by suppressing the distorted lines from the profiles, but the result was the same. We also tried to make a selection on the intrinsic line depth or on the excitation potential, but none of the different masks calculated were able to improve the LSD profile, meaning that all the lines of the spectrum are contaminated with circumstellar features. We therefore choose a mask including only lines with $\chi_{\rm exc} >  4$, because the different CS features are more easily identifiable in the resulting LSD $I$ profile. We fit the profile with a single photospheric function superimposed with four Gaussian functions modelling the CS emission and absorption components. The result is shown in Fig. \ref{fig:hd36910}. While the LSD $I$ profile seems to be dominated by CS features, the blue part (from -70 to 0~\kms) of the profile can only be reproduced with a photospheric function, which increases our confidence in the interpretation of this profile.

\subsection{HD~36917 (= V372 Ori)}

HD~36917 is part of the Ori OB Id association \citep{wolff04} situated at a distance of 375~pc \citep{brown94}. The photometric data of \citet{wolff04} have been used for the luminosity determination. The strong far-IR excess and the absence of strong near-IR excess place this object at an intermediate evolutionary state between Herbig Ae/Be and Vega-like stars \citep{manoj02}. As explained above, this object still constitutes an interesting target for our study and has been kept in the survey.

According to \citet{levato76} HD~36917 is a double-lined spectroscopic binary, which we do not confirm with our observations. The spectrum of HD~36917 is consistent with a single star of effective temperature $T_{\rm eff} = 10000\pm500$~K, in agreement with the spectral type determination of \citet{johnson65}. Many CS features are observed in its spectrum. H$\alpha$ displays a strong and narrow emission in the core. Very faint and narrow emissions, slightly blueshifted, are detected in the core of H$\beta$ and H$\gamma$. Faint emission is also present in the core of the O~{\sc i} 8446~\angs triplet, while the He~{\sc i}~D3 line and the O~{\sc i} 777~nm triplet display stronger absorption than predicted. A few metallic lines seem to be distorted, certainly due to CS features. The Ca~{\sc ii} IR-triplet shows strong double-peaked emission profiles, and the Ca~{\sc ii}~K line displays a V-shape that could be an effect of CS gas. The Paschen lines do not seem to be contaminated with CS features.

\begin{figure}
\centering
\includegraphics[width=6cm,angle=90]{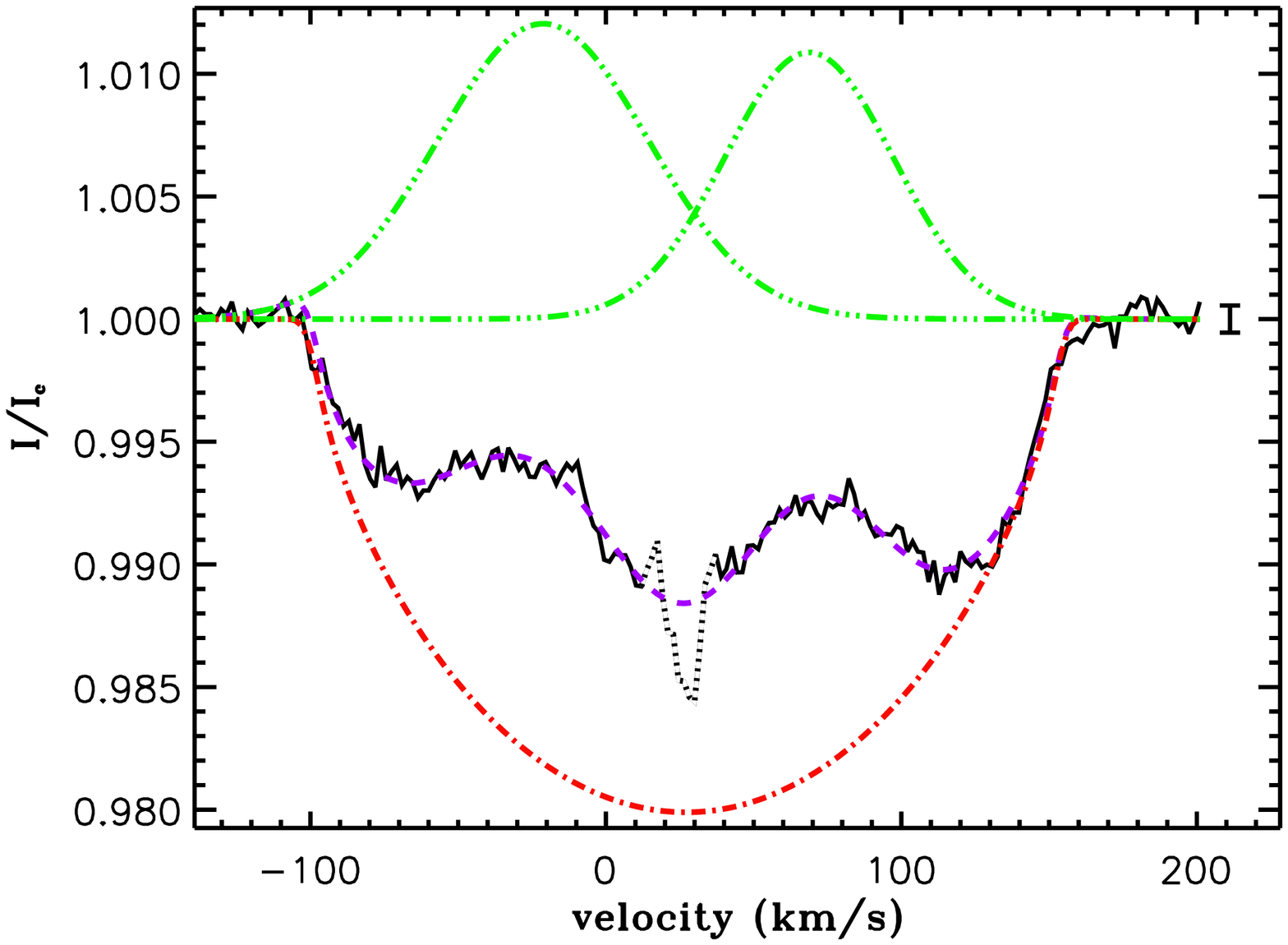}
\caption{As Fig.~A5 for HD~36917}
\label{fig:hd36917}
\end{figure}

We first calculated the LSD profiles without performing a special cleaning to the Kurucz mask. The result gives an LSD $I$ profile contaminated with CS emission. In order to avoid emission in the $I$ profile we computed various masks by making a selection on the excitation potential of the lines. We find that if all the lines of the mask have $\chi_{\rm exc} > 3$~eV, then the LSD profile is no longer contaminated with emission. We then fitted both the LSD $I$ profiles computed with the two different masks: the one resulting in a contaminated $I$ profile, and the one giving a profile free from emission. The former has been fit with a photospheric function plus two Gaussians, modeling the emission, and the latter has been fit with a single photospheric function. Both results give consistent $v\sin i$ ($127.1\pm4.6$ and $125\pm20$ \kms, respectively). The profile free from emission is noisier than the contaminated one because of the smaller number of lines taken into account in the LSD procedure, which explains the larger error bar in $v\sin i$. The contaminated profile, despite the CS contribution, gives a better constrained value of $v\sin i$, and has therefore been adopted for the determination of the final value of $v\sin i$. The fit of this profile is plotted in Fig. \ref{fig:hd36917}.

\subsection{HD~36982 (= LP Ori)}

LP Ori is part of the Ori OB 1d association at a distance of 375~pc from the Sun \citep{brown94}. As with HD~36917, \citet{manoj02} reports a strong far-IR excess and the absence of strong near-IR excess, suggesting that this star is at an intermediate evolutionary state between Herbig Ae/Be and Vega-like stars, so that this star is still an interesting target for our study. We used the photometric data of \citet{wolff04} to derive the luminosity of the star.

LP Ori was detected as an X-ray source by the Chandra Orion Ultradeep Project \citep[COUP,][]{getman05}. A magnetic field has also been detected at the surface of the star by \citet{petit08} using ESPaDOnS. \citet{petit08} performed a preliminary analysis of the magnetic properties of the star. We obtained additional Narval observations of this star in Nov. 2007. We only present here the intensity spectra obtained during that run, while a full analysis of the polarised spectra will be published in a forthcoming paper (Petit et al. in prep.).

The spectrum is well fit with a TLUSTY non-LTE spectrum of effective temperature $T_{\rm eff}=20000\pm1000$~K, consistent with the work of \citet{wolff04}. The only CS manifestations observed in the spectrum are single-peaked CS emission components in the core of the Balmer lines, a double-peaked emission profile in the core of the O~{\sc i} 8446~\angs triplet, and faint distortions in the cores of the He~{\sc i} lines at 5875~\AA, 6678~\AA, and 7065~\AA, and of the O~{\sc i} 777~nm triplet.

\begin{figure}
\centering
\includegraphics[width=6cm,angle=90]{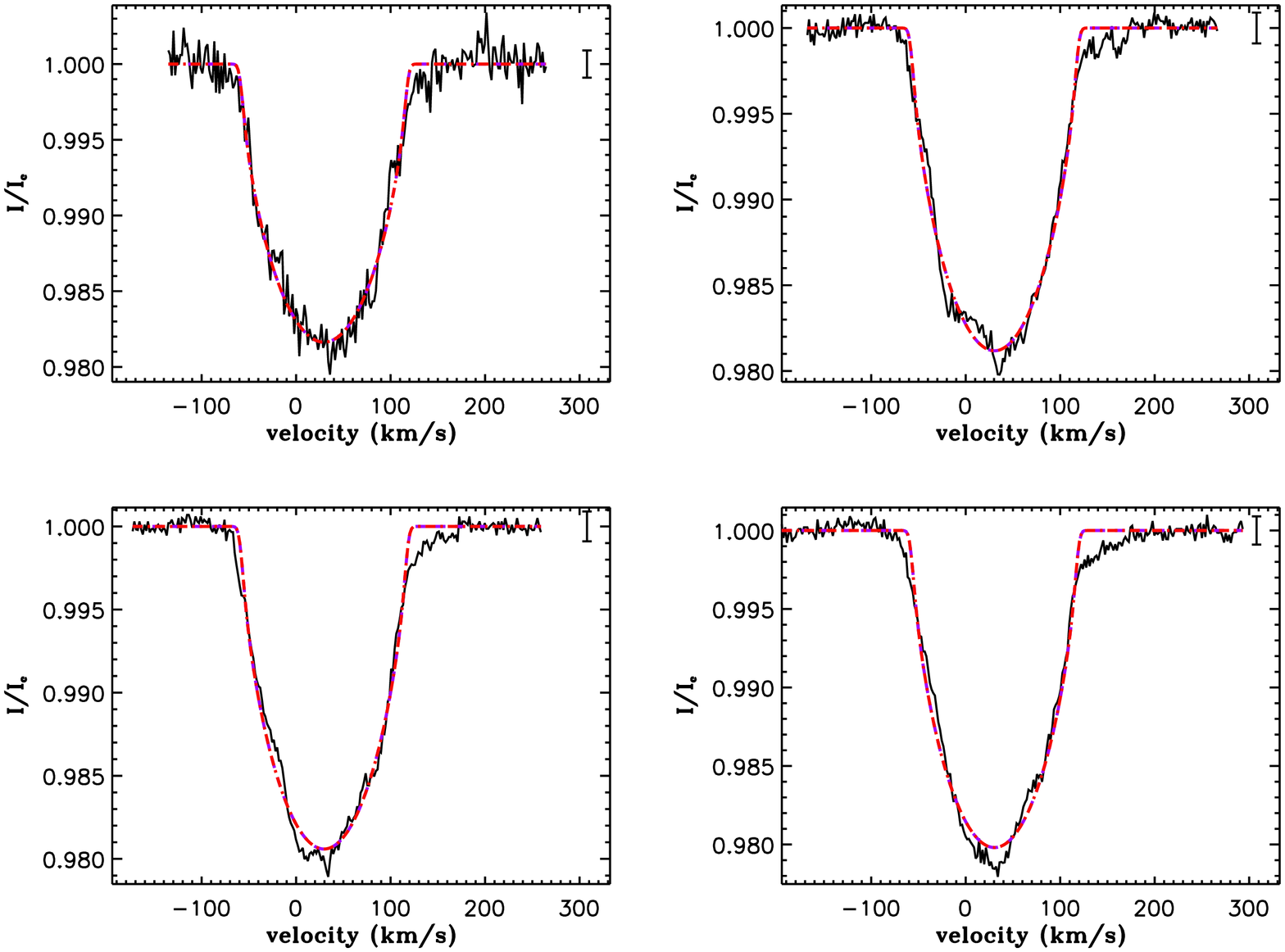}
\caption{As Fig.~A5 for the Nov. 9th, 10th, 11th and 12th 2007 (from top left to bottom right) observations of HD~36982 (= LP Ori).}
\label{fig:lpori}
\end{figure}

We have cleaned the Kurucz mask, by rejecting the strongest lines of the spectrum, as well as the lines clearly contaminated with emission. The result is a photospheric profile slightly distorted in the core, perhaps due to CS emission. We fit the four profiles simultaneously with photospheric functions only, by forcing the \vsinis and \vrads to be identical for the four profiles. The depth was left free to vary from one profile to another, because of the distortions in the core. The results of the fit are plotted in Fig. \ref{fig:lpori}.

\subsection{HD~37258 (= V586 Ori)}

\begin{figure}
\centering
\includegraphics[width=6cm,angle=90]{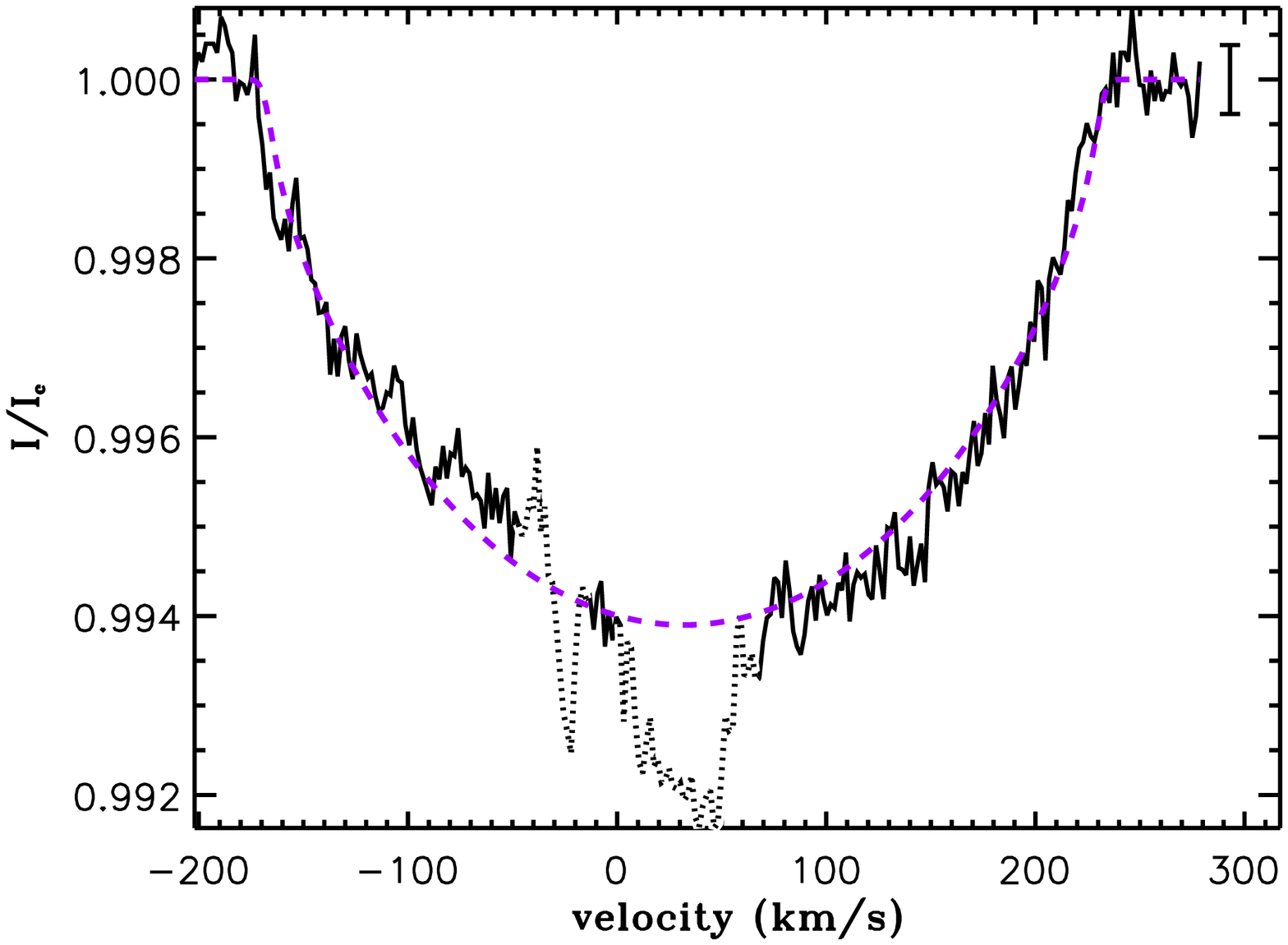}
\caption{As Fig.~A5 for HD~37258 (= V586 Ori)}
\label{fig:hd37258}
\end{figure}

HD~37258 belongs to the Ori OB Id association \citep{tian96}, at a distance of 375~pc \citep{brown94}. Photometric observations of this star revealed strong variability, with changes in brightness of up to 2 mag. These variations are assumed to be caused by circumstellar dust appearing sporadically in the line of sight \citep{pugach86}; this dust certainly belongs to the circumstellar disk revealed by the strong IR excess \citep{malfait98a}. We will therefore assume that the brightest observed magnitude is close to the real magnitude of the star. We used the photometric data of \citet{dewinter01} at the brightest magnitude ($V = 9.64$, $(B-V) = 0.140$) to compute the luminosity of the star.

The spectrum of HD~37258 is well fit with $T_{\rm eff} = 9500\pm500$~K, in agreement with the temperature determination of \citet{kovalchuk97}. The main characteristic of the spectrum is a strong contamination with redshifted CS absorption in many metallic lines, but also in the Ca~{\sc ii}~K line, in the core of the Balmer lines from H$\zeta$ to H$\beta$, in the O~{\sc i} 8446~\angs and Ca~{\sc ii} IR triplets. Emission is also present in the wings of the CS absorption in H$\beta$, and in the O~{\sc i} 8446~\angs and Ca~{\sc ii} IR triplets. H$\alpha$ displays a strong double-peaked emission with a strong central absorption that goes below the continuum. While HD 37258 has never been classified as an UXOR star, its spectral characteristics, as well as its strong photometric variability, are similar to the UXOR stars. In addition to the UXOR-type spectral features, an inverse P-Cygni profile is observed in the He~{\sc i} lines at 5875~\angs and 6678~\AA. The Paschen lines might be contaminated with circumstellar features as well.

We have cleaned the Kurucz mask to eliminate as far as possible the CS contamination. The result gives a photospheric profile, with a small residual contamination due to CS absorption, which we could not eliminate. We fit the LSD $I$ profile with a photospheric function and excluded from the fit the data points contaminated with CS features. The result is shown in Fig. \ref{fig:hd37258}.

\subsection{HD~37357}

HD~37357 belongs to the Ori OB Ic association situated at a distance of 375~pc \citep{brown94}. We used the photometric data of \citet{vieira03} to compute the luminosity of the star. The spectral energy distribution of HD~37357 displays an IR excess that is well reproduced with a double dust disk model, suggesting the presence of a protoplanetary disk surrounding the star in which the process of planet formation has already started \citep{malfait98a}.

The spectrum of HD~37357 is consistent with the temperature determination of \citet[][$T_{\rm eff} = 9250\pm500$~K]{vieira03}. Blueshifted CS absorption features are detected in a few Fe~{\sc ii} lines, as well as in the Ca~{\sc ii}~K line and in the Balmer lines from H$\delta$ to H$\beta$, the latter also showing emission in the red wing of the CS absorption feature. These CS spectral features are typical of UXOR stars, although in this object they are observed in a smaller number of lines, and with smaller depths. However, because of the absence of strong photometric variability \citep{herbst99}, we cannot class the star as an UXOR-type. These characteristics are also similar to $\beta$Pic-like objects \citep[e.g.][]{artymowicz00}. HD 37357 could therefore be in a transitional phase from UXOR-type activity to a $\beta$Pic-type star. H$\alpha$ displays a P Cygni profile of Type III. Faint emission are detected in the core of the Paschen lines. The Ca~{\sc ii} IR triplet shows single-peaked emission profiles, while the O~{\sc i}~8446~\angs triplet is filled with emission. The He~{\sc i} D3 line displays a broad double-peaked emission profile.

\begin{figure}
\centering
\includegraphics[width=6cm,angle=90]{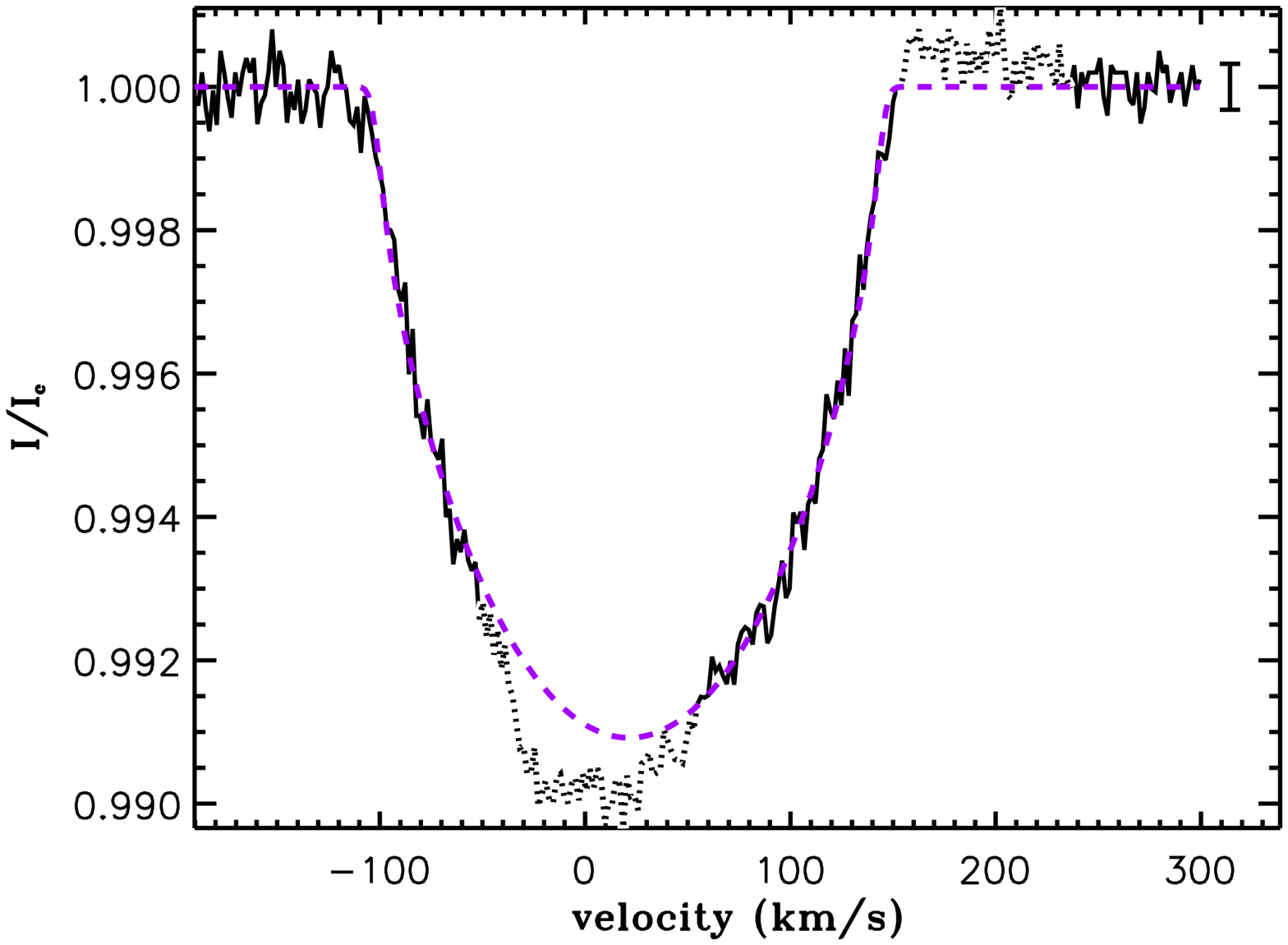}
\caption{As Fig.~A5 for HD~37357}
\label{fig:hd37357}
\end{figure}

We have cleaned the Kurucz mask K9000.40, rejecting all the lines contaminated with CS features. The result is a photospheric profile only slightly contaminated with CS absorption features, that is well fit by a photospheric function, if we exclude the contaminated data points. The profile and its best fit are plotted in Fig. \ref{fig:hd37357}.

\subsection{HD~37806 (= MWC 120)}

HD~37806 is part of the Ori OB Ib association \citep{warren78}, situated at a distance of 375~pc \citep{brown94}. We used the Hipparcos photometric data \citep{perryman97} to derive the luminosity of the star. Like HD~37357, the spectral energy distribution of HD~37806 displays near-and far-IR excess, and reveals the presence of a circumstellar disk with a gap, where planet formation could have started \citep{malfait98a}.

The spectrum is well fit with $T_{\rm eff} = 11000\pm500$~K, which is equivalent to a spectral type B8 \citep{kenyon95}. This value is inconsistent with the spectral type determination of \citet{guetter81}, but in agreement with the estimation of \citet{grady96}.

Strong blueshifted circumstellar absorption, often superimposed with still farther blueshifted emission, is present in numerous lines of the spectrum, including the strongest metallic lines, the Ca~{\sc ii} K line, and the Balmer lines from H$\theta$ to H$\beta$. H$\alpha$ displays a strong double-peaked emission with a blueshifted central absorption. The He lines at 5875~\AA, 6678~\AA, 7065~\AA, and the O~{\sc i} 777~nm triplet display strong inverse P-Cygni profiles. The O~{\sc i} 8446 \angs triplet shows a single-peaked emission, while the Ca~{\sc ii} IR-triplet show double-peaked emissions. Single-peaked emission is superimposed with the cores of the Paschen lines. Some high-excitation lines such as Si~{\sc ii} 5040 and 5056~\angs and O~{\sc i} 6156~\angs are largely or completely free of circumstellar features.

\begin{figure}
\centering
\includegraphics[width=6cm,angle=90]{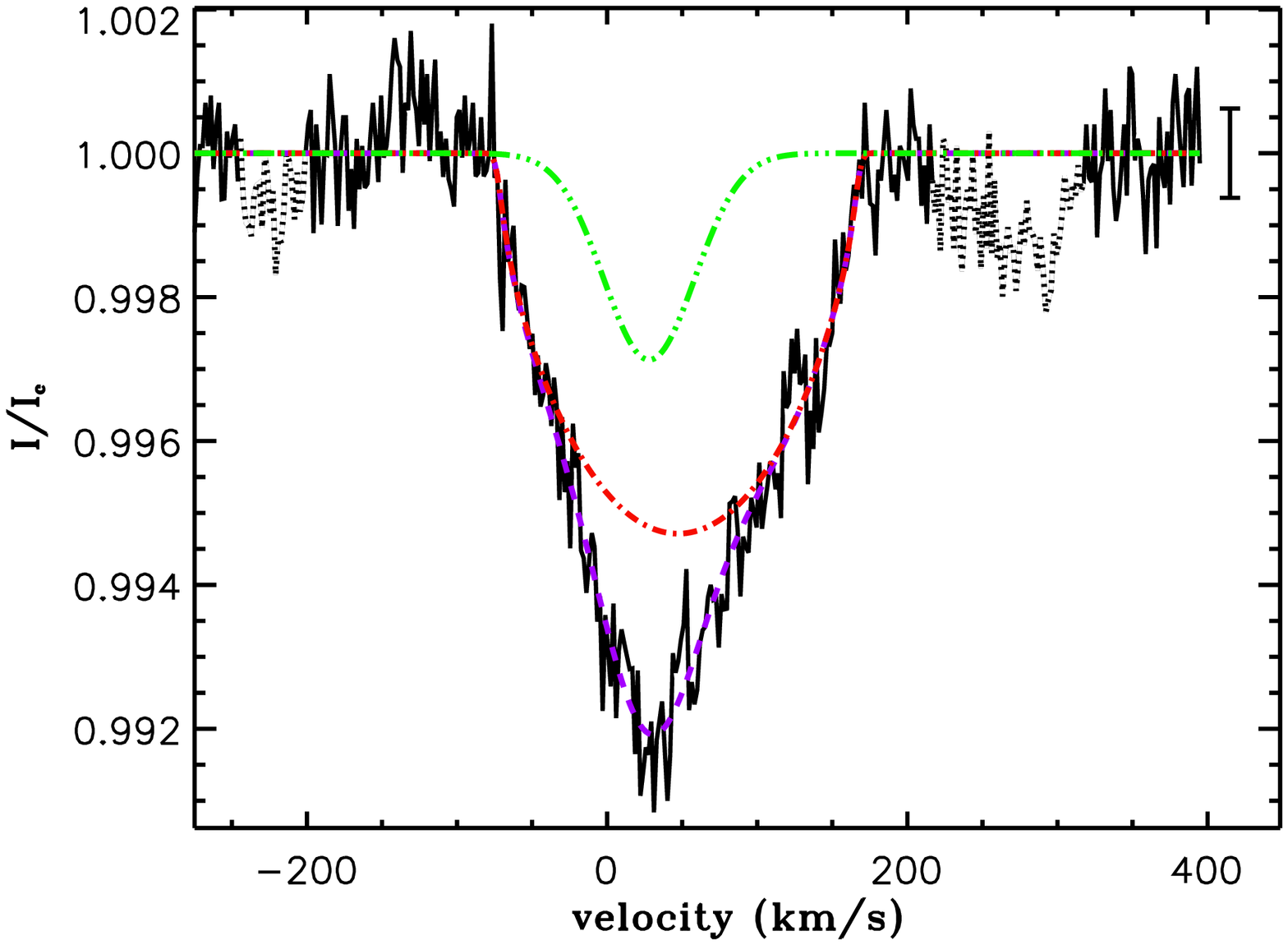}
\caption{As Fig.~A5 for HD~37806 (=MWC 120)}
\label{fig:hd37806}
\end{figure}

We have cleaned the Kurucz mask by keeping only lines that do not seem to be contaminated with CS features. The result is a photospheric profile still contaminated with a blueshifted CS absorption, which is well fit by a photospheric plus a Gaussian function. The result is plotted in Fig. \ref{fig:hd37806}.

\subsection{HD~38120}

\begin{figure}
\centering
\includegraphics[width=6cm,angle=90]{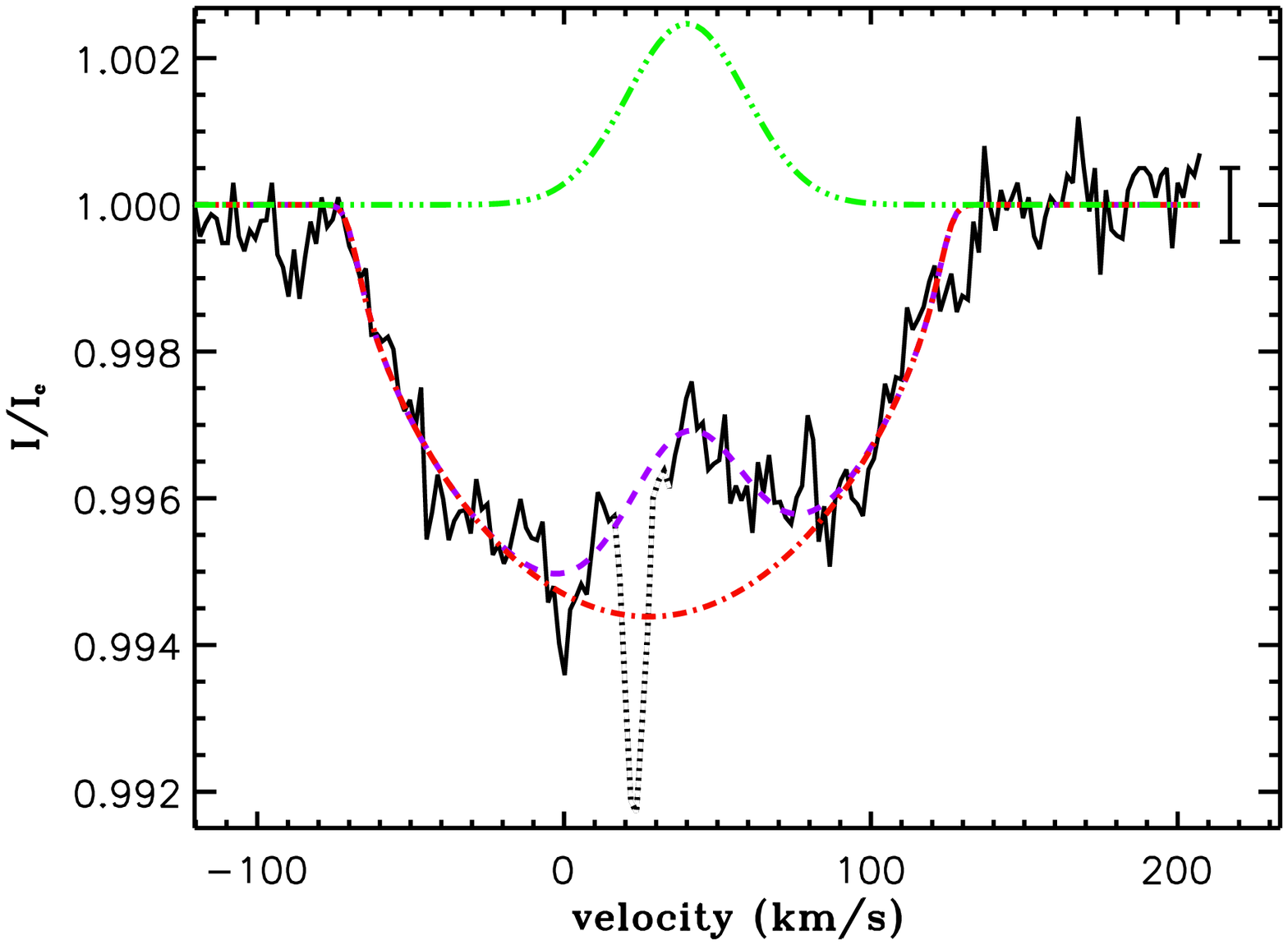}
\caption{As Fig.~A5 for HD~38120}
\label{fig:hd38120}
\end{figure}

HD~38120 belongs to the Ori OB Ic association at a distance of 375~pc \citep{brown94}. The IR and submillimetric properties of this star suggest that it could be a less-evolved object than the Vega-type stars, and is therefore very interesting for our study \citep{coulson98}. We used the Hipparcos photometric data \citep{perryman97} to derive the luminosity.

The intensity spectrum of HD~38120 is well fit with an effective temperature of $11000\pm500$~K, in agreement with the work of \citet{vieira03}. The metallic lines are very faint compared to a solar metallicity spectrum. Faint single emission is observable in the core of a few metallic lines, including the Ca~{\sc ii} K line and the Ca~{\sc ii} IR-triplet. Broad emission is observed in the He~{\sc i} lines at 5875~\AA, 6678~\AA, 7065~\AA, and in the O~{\sc i} 777~nm triplet. All Balmer lines visible in the spectrum (from H$\alpha$ to H$\theta$) have a narrow emission in the core, with strength increasing with wavelength. The Paschen lines also display narrow emission features in their cores, and the O~{\sc i} 8446 \angs triplet appears in emission.

We have calculated the LSD profiles without performing a special cleaning to the Kurucz mask. The resulting $I$ profile shows an emission component in the core, reflecting what is observed in the metallic lines. We fitted the profile with a photospheric function and a Gaussian function modeling the emission. The result is shown in Fig. \ref{fig:hd38120}.

\subsection{HD~38238 (= V351 Ori)}

\begin{figure}
\centering
\includegraphics[width=6cm,angle=90]{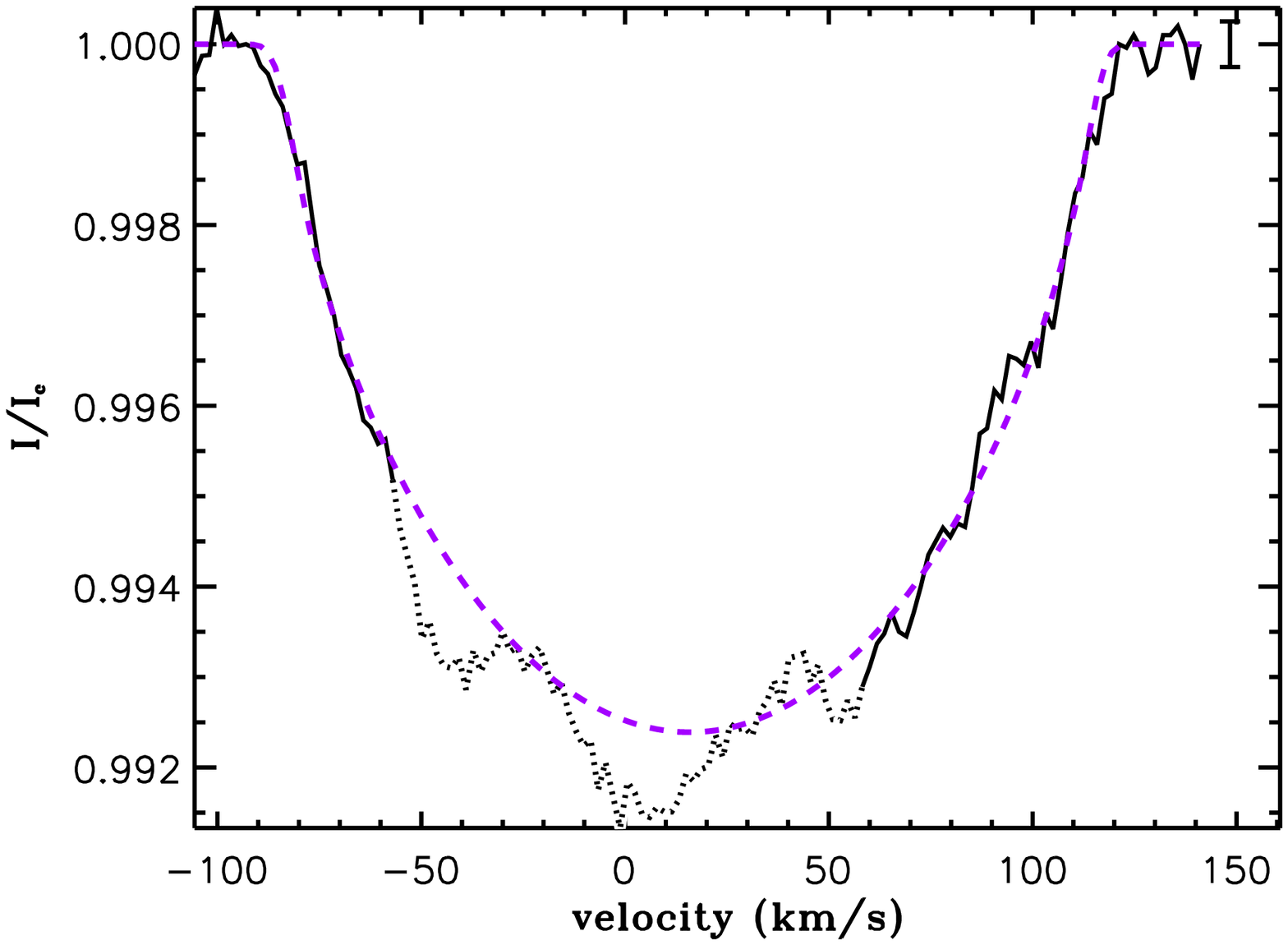}
\caption{As Fig.~A5 for HD~38238 (= V351 Ori)}
\label{fig:hd38238}
\end{figure}

HD~38238 is an irregular variable star, part of the Ori OB Id association, which is situated at a distance of 375~pc \citep{brown94}. It changes its photometric behaviour from that of a Herbig Ae star with strong photometric variations, due to extinction by circumstellar dust clouds, to that of an almost non-variable star, suggesting that HD~38238 is in the process of transition from a Herbig Ae star with strong photometric variations to a non-variable one \citep{ancker96}. We used the Hipparcos photometric data \citep{perryman97} to derive the luminosity. The IR excess observed in HD 38238 can be explained with an extended disk or dust-shell around the star \citep{ancker96}.

The spectrum of HD~38238 is well reproduced with $T_{\rm eff} = 7750\pm250$~K, consistent with the spectral type determination of \citet{ancker96} (A6-A7). No CS features are observed in the spectrum, except an inverse P Cygni of type III in the core of H$\alpha$, and perhaps faint emission in the Paschen lines.

We first calculated the LSD profiles without performing a special cleaning to the Kurucz mask. The resulting $I$ profile show broad wings. In order to avoid this, we have calculated various masks by making a selection on the intrinsic depth and excitation potential of the lines, and we find that a mask containing only lines with $d<0.6$ gives a profile with normal photospheric wings, consistent with the individual lines of the spectrum itself. The resulting LSD $I$ profile reveals distortion in the core that could be due to CS contamination. We therefore fitted the profile with a photospheric function, but excluding the contaminated data points in the core. The result is shown in Fig. \ref{fig:hd38238}.

\subsection{HD~50083 (= V742 Mon)}

HD~50083 is a candidate Herbig Ae/Be star \citep{vieira03} studied in detail by \citet{fremat06}. Using high-resolution spectroscopy, they determined the effective temperature and the surface gravity of the star ($T_{\rm eff}=20000\pm1000$~K and $\log g = 3.43\pm0.15$~(cgs)). The Hipparcos parallax \citep{vanleeuwen07} is too uncertain to be used for a distance estimation. We therefore placed the star in the $T_{\rm eff}-\log g$ diagram and compared its position with CESAM PMS evolutionary tracks to derive the mass, radius and age. We also derive a luminosity of $\log(L/L_{\odot})= 4.15\pm0.12$, and using the photometric data of \citet{vieira03}, we estimate the distance of the star to be around $1000\pm100$~pc, consistent with the new determination of the Hipparcos parallax by \citet{vanleeuwen07}. The spectral energy distribution of HD 50083 displays a weak IR excess, similar to classical Be stars \citep{sartori10}. However it might be associated with the dark nebulae LDN 1639 \citep{lynds62}, indicating that the star is young. 

The spectrum of HD~50083 is strongly contaminated with variable circumstellar features. From H$\zeta$ to H$\gamma$ double-peaked emission profiles are observed in the core of the Balmer lines, with increasing amplitude with wavelength. H$\beta$ and H$\alpha$ display single peaked emission profiles superimposed with many absorption features. Many double-peaked emission are observed in Fe~{\sc ii} lines, as well as in the Paschen lines, in the O~{\sc i}~8446 \angs and Ca~{\sc ii} IR triplets. In the He~{\sc i} lines at 5875~\angs and 6678~\AA, we observe small emission features in each side of the photospheric profile. A strong and narrow interstellar feature is only observed in the Ca~{\sc ii} K line. The O~{\sc i} 777~nm triplet displays a strong and complex emission profile.

\begin{figure}
\centering
\includegraphics[width=3cm,angle=90]{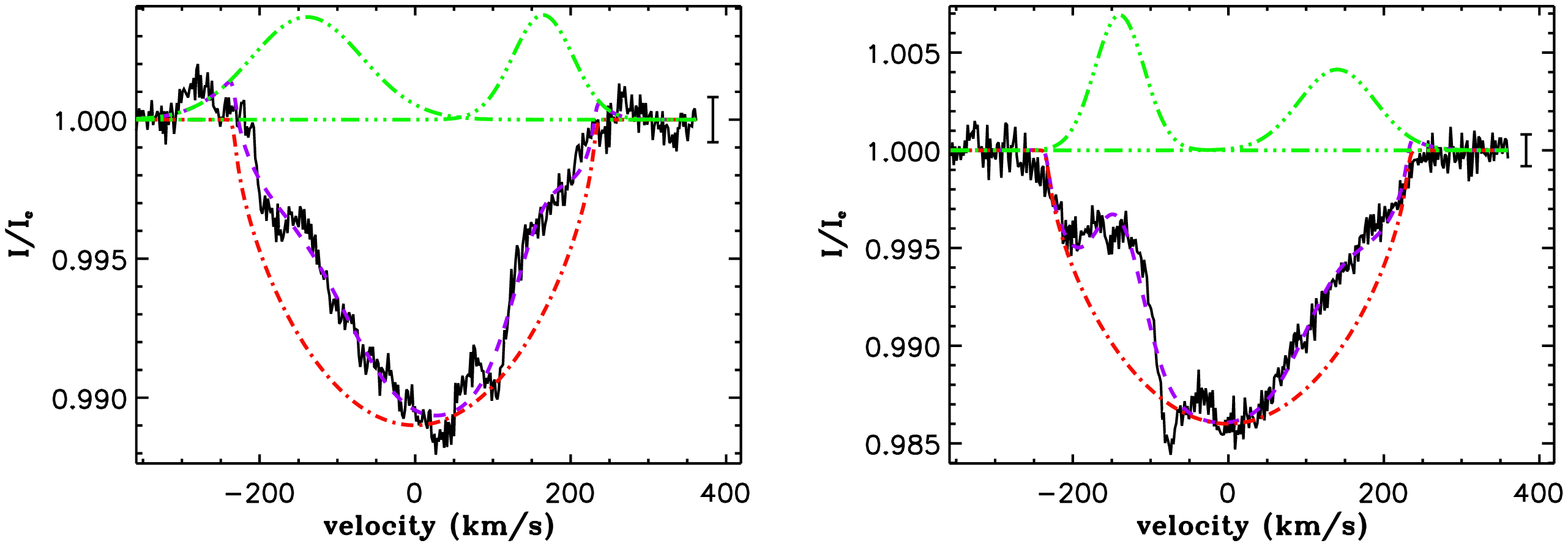}
\caption{As Fig.~A5 for the Nov. 2007 (left) and Apr. 2008 (right) observations of HD~50083.}
\label{fig:hd50083}
\end{figure}

Those spectral lines that are not contaminated with emission are well fit with a non-LTE TLUSTY model of $T_{\rm eff} = 24000$~K. This is slightly different from the determination of \citet{fremat06}, which can be explained by our inability to estimate $\log g$ with our data. We have cleaned the Kurucz mask by rejecting all spectral lines contaminated with CS features. The result is the two photospheric profiles displayed, which are still contaminated with CS emission. We fitted both profiles simultanously with a photospheric and four Gaussian functions, by forcing the \vsinis and \vrads to be the same for both observations. We were unable to find consistent results if we require the photospheric depths of both profiles to be identical. This means that the depth of the LSD profile has changed between the two observations, and therefore that additional CS features (not taken into account in our fit) may contaminate the cores of the profiles. The result of the fitting procedure is shown in Fig. \ref{fig:hd50083}.

\subsection{HD~52721 (=GU CMa)}

\begin{figure}
\centering
\includegraphics[width=6cm,angle=90]{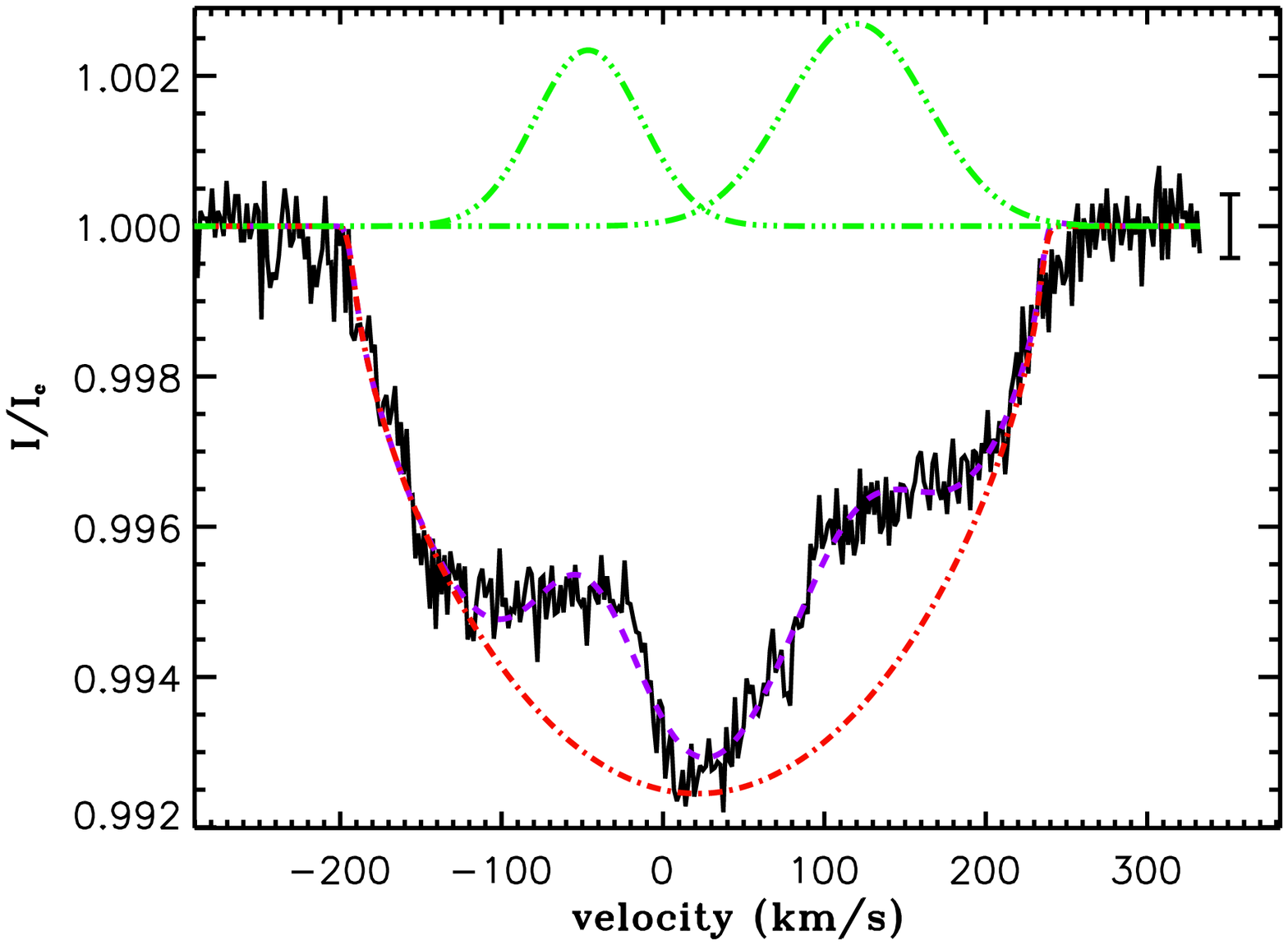}
\caption{As Fig.~A5 for the Apr. 2008 observation of  HD~52721 (=GU CMa).}
\label{fig:hd52721}
\end{figure}

HD~52721 has been carefully studied by \citet{fremat06}, who determined the effective temperature and the surface gravity of the star ($T_{\rm eff}=22500\pm2000$~K and $\log g = 3.99\pm0.20$~(cgs)). No star forming region is known to be associated with this star, and the Hipparcos parallax \citep{vanleeuwen07} is too uncertain to be used for a distance estimate. We therefore placed the star in the $T_{\rm eff}-\log g$ diagram and compared its position with CESAM PMS evolutionary tracks to derive the mass, radius and age. We also derive a luminosity $\log(L/L_{\odot})= 3.77_{-0.31}^{+0.35}$, and using the photometric data of \citet{shevchenko99}, we estimate the distance of the star to be about $670_{-110}^{+140}$~pc. Its spectral energy distribution displays a weak IR excess, similar to classical Be stars \citep{hillenbrand92}. HD~52721 illuminates the reflection nebula VdB 88 \citep{vandenbergh66}, indicating that the star must be very young.

The spectrum of HD~52721 is strongly contaminated with variable CS emission. From H$\delta$ to H$\beta$ the cores of the Balmer lines are superimposed with double-peaked emission profiles. H$\alpha$ displays a single-peaked emission line superimposed with various faint and narrow absorption. Double-peaked emission is observed in many Fe~{\sc ii} lines, as well as in the Paschen lines, in the O~{\sc i}~8446 \angs and Ca~{\sc ii} IR triplets. We observe two small emission features in the He~{\sc i} 5875~\angs line, one on each side of the photospheric profile. A strong and narrow interstellar feature is only observed in the Ca~{\sc ii} K line. The O~{\sc i} 777~nm triplet displays an emission profile superimposed with three absorption features.

The spectral lines that do not appear contaminated with CS features fit well with a non-LTE TLUSTY spectrum of $T_{\rm eff} = 26000$~K and $\log g = 4.0$, slightly different from the temperature derived by \citet{fremat06}, a difference that could be due to the NLTE effect not taken into account in the \citet{shevchenko99} analysis. We have cleaned the Kurucz mask, rejecting the lines contaminated with circumstellar features. The result gives profiles still contaminated with emission. Both observations fit well with a photospheric function plus two Gaussian functions, and give consistent results when fitted separately. However, the SNR of the Nov. 2007 profile is significantly lower than the SNR of the Apr. 2008 profile, so that the former observation does not help to better constraint the fitting parameters. Only the Apr. 2008 profile has been taken into account in the fitting procedure. The result is shown in Fig. \ref{fig:hd52721}.

\subsection{HD~53367}

HD~53367 is a candidate Herbig Be star \citep{vieira03}, and illuminates the reflection nebula VdB 93 \citep{vandenbergh66}. We used the Hipparcos parallax \citep{vanleeuwen07} and the photometric data of \citet{shevchenko99} to derive the luminosity of the star. Its spectral energy distribution is very similar to HD~52721, and for the same reasons we can argue that the star is still very young.

The spectrum of HD~53367 does not show any obvious emission in any of the Balmer lines. Hints of emission are visible in the blue side of H$\alpha$ and He lines. The spectrum is well fitted with a non-LTE TLUSTY synthetic spectrum of $T_{\rm eff}=29000\pm2000$~K, consistent with the temperature determination of \citet{shevchenko99}. The strongest lines of the spectrum do not have the shape of a photospheric function, and the fainter lines show flattened cores. 

\begin{figure}
\centering
\includegraphics[width=3cm,angle=90]{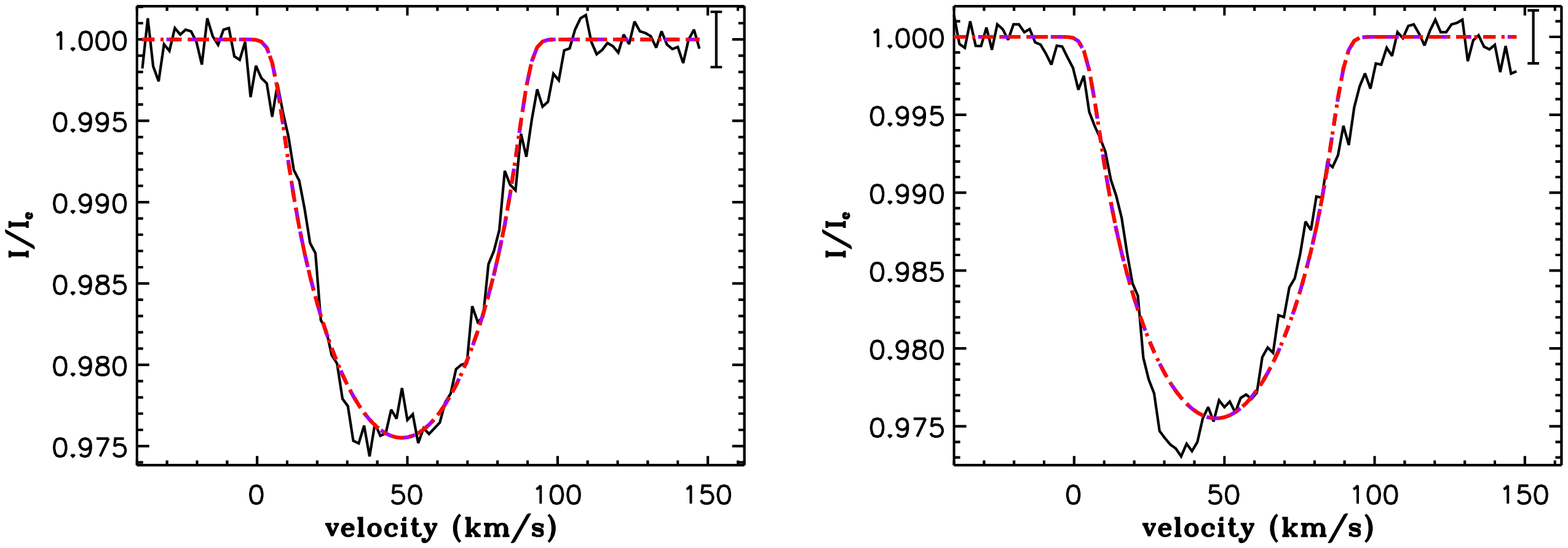}
\caption{As Fig.~A5 for the Feb. 20th (left) and 21st (right) 2005 observations of HD~53367.}
\label{fig:hd53367}
\end{figure}

We have cleaned the Kurucz mask by rejecting the strongest lines of the spectrum, in order to obtain an LSD $I$ profile as close as possible to a photospheric profile. The result is noisy but satisfying. We fitted both observations simultaneously with a single photospheric function (see Fig. \ref{fig:hd53367}).

\subsection{HD~68695}

\begin{figure}
\centering
\includegraphics[width=6cm,angle=90]{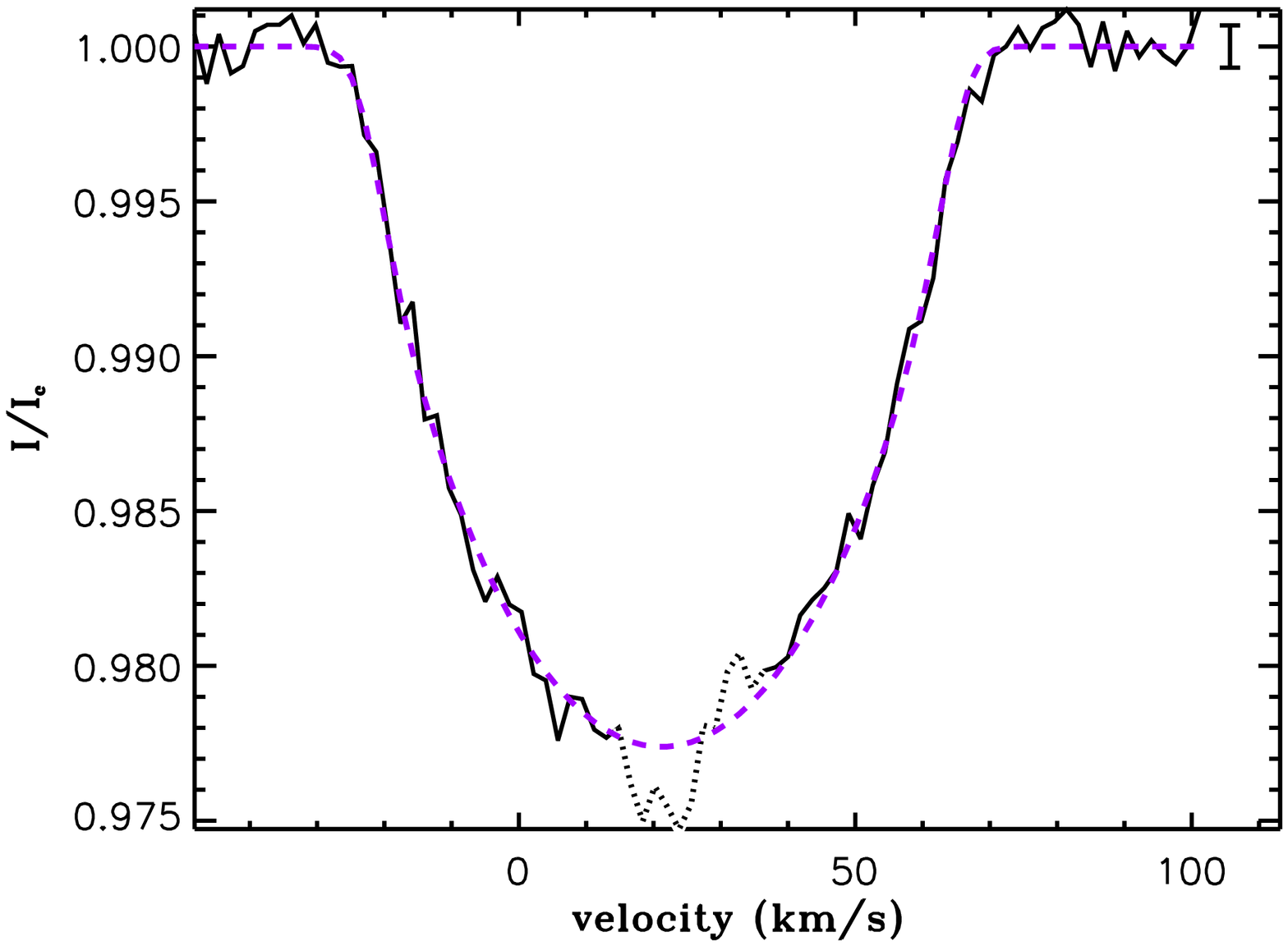}
\caption{As Fig.~A5 for HD~68695}
\label{fig:hd68695}
\end{figure}

HD~68695 is part of the Vela star cloud situated at a distance of 570~pc \citep{eggen86}. We used the photometric data of \citet{vieira03} to derive the luminosity. No spectral energy distribution has been published until now, however \citet{vieira03} reports visible and near-IR photometry of the star. By comparing the near-IR colour excesses to that of HD~38120, a Herbig Ae star of the same temperature, we find similar values, leading to the conclusion that HD~68695 may also be a less-evolved Vega-type star.

The spectrum is consistent with the temperature and gravity determination of \citet[][$T_{\rm eff} = 9500$~K, $\log g=4.3\pm0.3$]{folsom12}. The cores of the Balmer lines from H$\delta$ to H$\alpha$ are superimposed with a single-peaked emission line, with amplitude increasing with wavelength. Except for broad emission observed at the He~{\sc i} lines at 5875~\AA, 6678~\AA\ and 7065~\AA, and emission in the blue wing of the O~{\sc i} 777~nm triplet, no other CS contribution is observed in the spectrum. The abundance of C appears to be higher than the solar value, O appears approximately solar, and Ti, Cr and Fe all seem to have somewhat lower than solar abundance. This may be a young $\lambda$~Boo star as proposed by \citet{folsom12}.

We have cleaned the Kurucz mask by rejecting the lines with CS contribution. The result gives a profile in absorption, well fitted with a single photospheric function, and is plotted in Fig. \ref{fig:hd68695}

\subsection{HD~76534 A}

HD~76534 illuminates the reflection nebula VdB 24 \citep{vandenbergh66}, and is member of the Vela R2 association situated at a distance of 870~pc \citep{herbst75}. It is a visual binary star resolved by Hipparcos with a separation of 2.1 arcsec \citep{perryman97}. We observed the brighter component labeled A in the Hipparcos catalogue. We used the Hipparcos Double and Multiple System Annex (DMSA) photometric data of the component A to derive the luminosity of the star. In order to convert the magnitudes from the Tycho to the Johnson system, we used the calibration formula 1.3.20 of the Hipparcos catalogue \citep[][p.57]{perryman97}, and we find $V = 8.35$~mag and $(B-V)=0.107$~mag. Its spectral energy distribution is very similar to HD~52721, and for the same reasons we argue that the star is still very young.

The spectrum is well reproduced by a non-LTE TLUSTY synthetic spectrum of $T_{\rm eff}=18000\pm2000$~K, consistent with the temperature determination of \citet{the85a}. The cores of the Balmer lines from H$\delta$ to H$\beta$ are contaminated with double-peaked emission lines, while H$\alpha$ is totally dominated with a double-peaked emission profile. Slight emission is observed in the wings of the He~{\sc i} lines at 5875~\AA, 6678~\angs and 7065~\AA. The O~{\sc i} 777~nm and O~{\sc i} 8446 \angs triplets, and the Paschen lines display double-peaked emission profiles. The Ca~{\sc ii} IR-triplet does not seem to contribute to the emission observed in the Paschen lines. The Ca~{\sc ii} K line is only contaminated with a narrow interstellar absorption. The strongest metallic lines of the spectrum show broad wings and asymmetric shapes.

\begin{figure}
\centering
\includegraphics[width=6cm,angle=90]{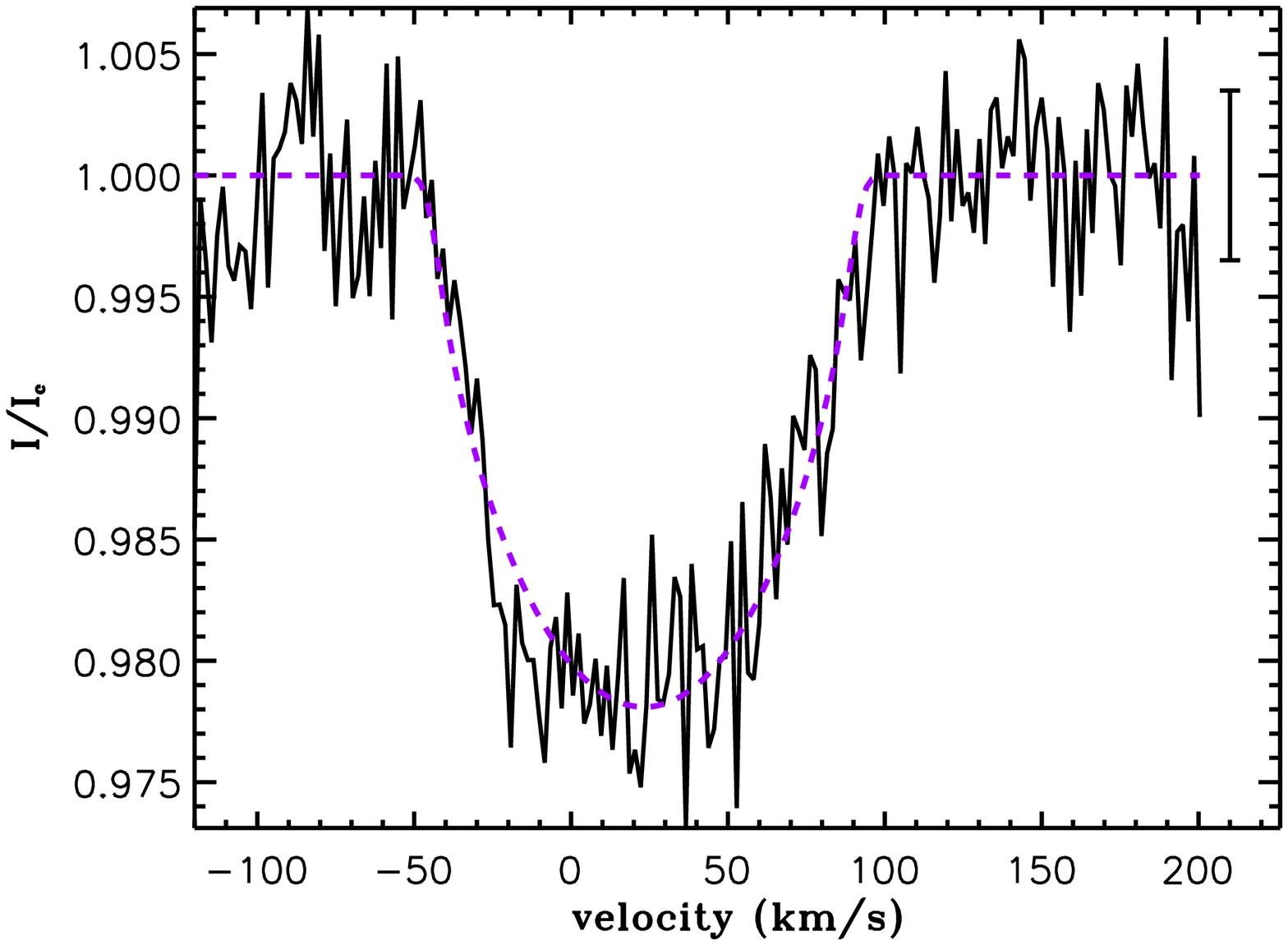}
\caption{As Fig.~A5 for HD~76534 A}
\label{fig:hd76534}
\end{figure}

We have cleaned the Kurucz mask by selecting only faint lines, with relatively symmetric shapes. The result is satisfactory, but very noisy and slightly asymmetric. It is reasonably well fit with a single photospheric function (Fig. \ref{fig:hd76534}).

\subsection{HD~98922}

HD~98922 is a Herbig Be star that does not seem to be associated with any bright nebula or star-forming region. However the large near- and far-IR excess leaves no doubt that the star is young and still surrounded with dust \citep{malfait98a}. We used the Hipparcos photometric data \citep{perryman97} and the new Hipparcos parallax \citep{vanleeuwen07} to compute the luminosity of the star. The resulting position of the star in the HR diagram is situated well above the birthline (see Fig. \ref{fig:hr}), which can be the result of either an inaccurate parallax, or an unusual star with an atypical history. The models that we use to compute the mass, radius and age could be inappropriate for this star, and have therefore not been estimated in this paper.

The spectrum is consistent with the temperature determination of \citet[][$T_{\rm eff}=10500\pm500$~K]{vieira03}. Circumstellar absorption and emission lines are present in the whole spectrum. Single peaked emission lines are observed in the core of H$\delta$ and H$\gamma$, while P Cygni profiles are superimposed with the core of the H$\beta$ and H$\alpha$ lines. Single-peaked emission lines are also observed in the core of the Paschen lines from P17 to P10. The O~{\sc i} 8446~\angs and Ca~{\sc ii} IR triplets display single-peaked emission profiles. The O~{\sc i} 777~nm triplet displays a triple-peaked emission profile. The He~{\sc i} 5875~\angs photospheric profile might be slightly distorted by CS features. The Ca~{\sc ii} K line, while very noisy, does not seem to be contaminated with CS features. Many metallic lines display double-peaked emission profiles. The lines of multiplet 42 of Fe~{\sc ii} display single-peaked emission profiles.

\begin{figure}
\centering
\includegraphics[width=6cm,angle=90]{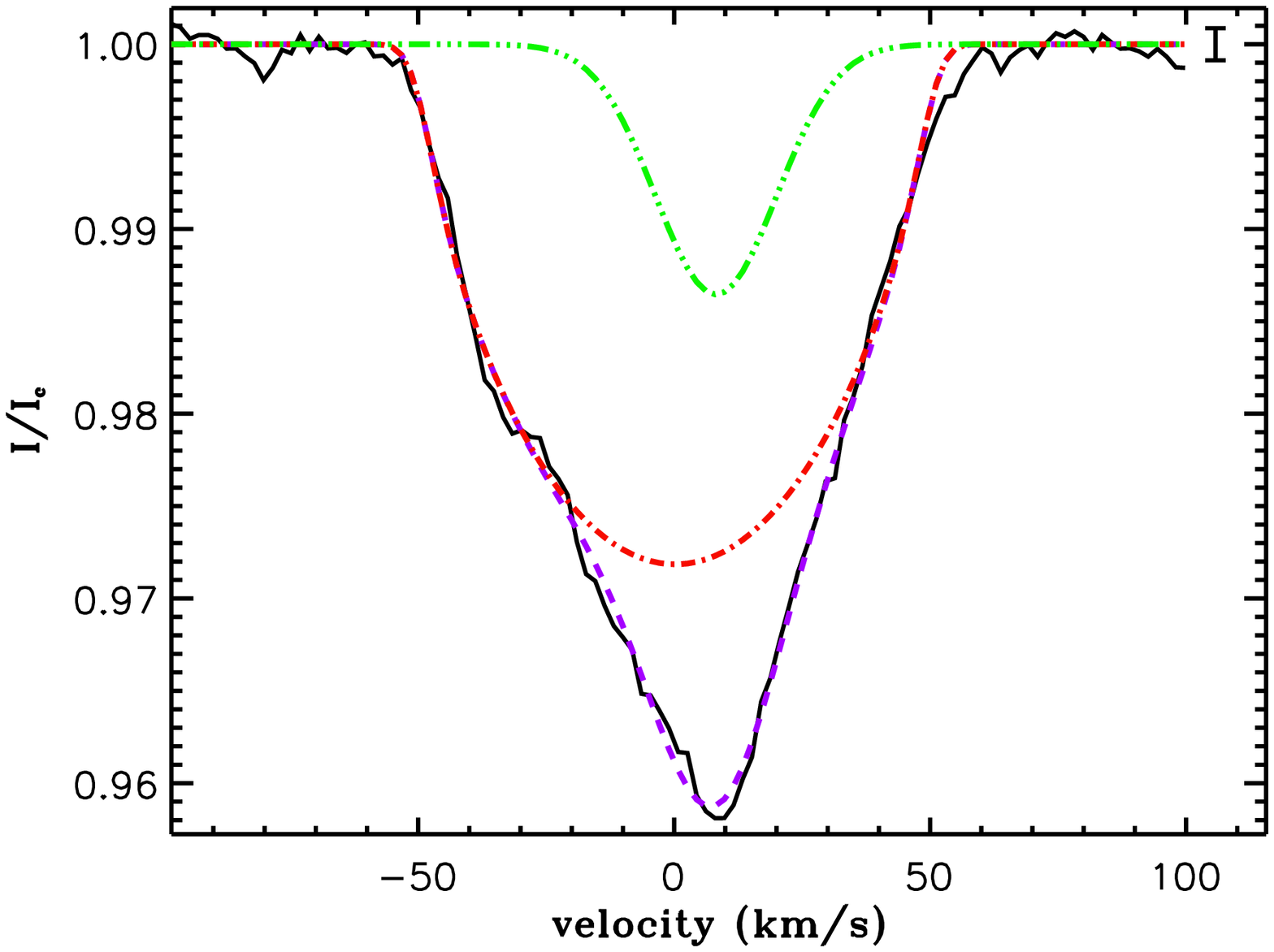}
\caption{As Fig.~A5 for HD~98922 }
\label{fig:hd98922}
\end{figure}

We have cleaned the Kurucz mask by rejecting all lines showing obvious circumstellar contamination. The resulting LSD $I$ profile still contains a circumstellar absorption contribution, which could not be eliminated. From a more careful analysis of the spectrum, we find that the photospheric lines that are not obviously contaminated with CS emission still seem to be superimposed with a narrow CS absorption, consistent with the LSD profile. We fit the LSD profile with a single photospheric function superimposed with a circumstellar Gaussian. The result is shown in Fig. \ref{fig:hd98922}. 

\subsection{HD~114981 (= V958 Cen)}

HD~114981 is a B-type star with IR excess typical of Vega-like stars \citep{mannings98}. We used the Hipparcos photometric data \citep{perryman97} and the new Hipparcos parallax \citep{vanleeuwen07} to compute the luminosity of the star \citep{vanleeuwen07}.

The spectrum of HD~114981 is well reproduced with a non-LTE TLUSTY synthetic spectrum of $T_{\rm eff}=17000\pm2000$~K, inconsistent with previous work \citep{hill70,vieira03}. This disagreement is certainly due to the lack of high-resolution spectra or to low SNR data in the earlier studies. Our $T_{\rm eff}$ is consistent with the spectral type quoted in the Michigan Catalogue of HD stars, vol. 3 \citep{houk82,houk94}. The star displays variable circumstellar contamination. Double-peaked emission lines are observed in the cores of Balmer lines from H$\delta$ to H$\beta$. H$\alpha$, the Paschen and Ca~{\sc ii} K lines, and the O~{\sc i} 777~nm and O~{\sc i} 8446 \angs triplets display double-peaked emission profiles. The Ca~{\sc ii} IR-triplet does not seem to contribute to the emission observed in the Paschen lines. The He~{\sc i} D3 line is not contaminated with CS features. Many Fe~{\sc ii} lines show double-peaked emission lines.

\begin{figure}
\centering
\includegraphics[width=3cm,angle=90]{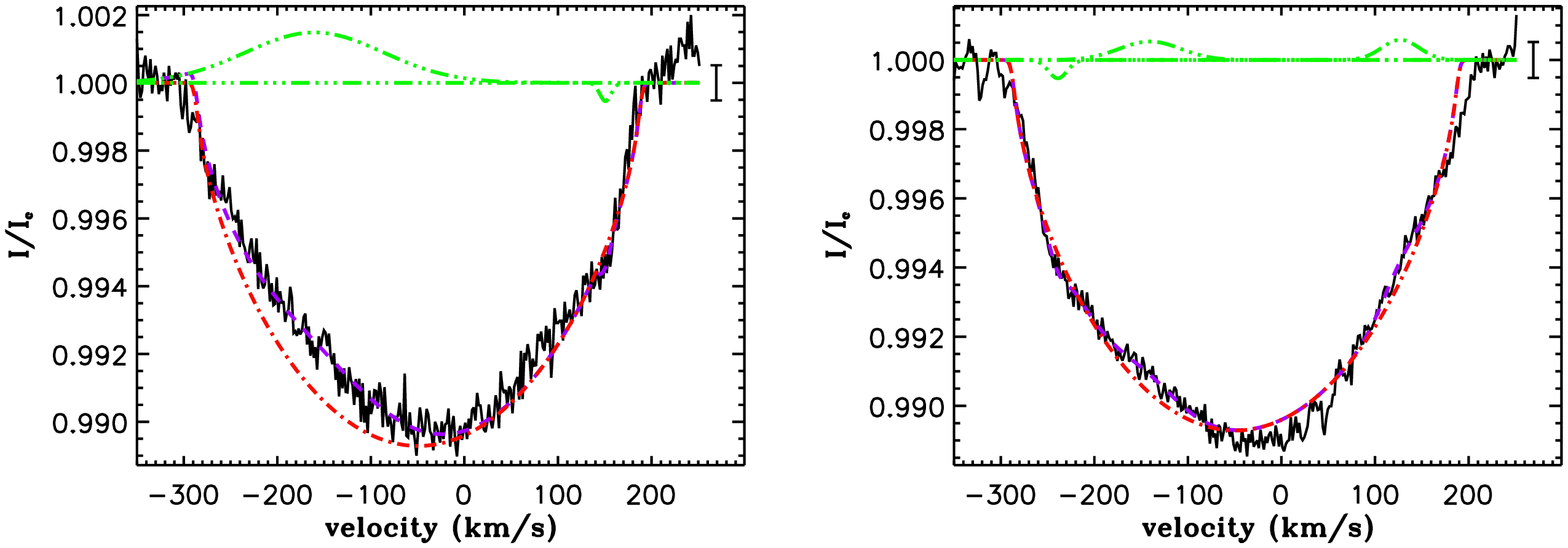}
\caption{As Fig.~A5 for the Feb. 2005 (left) and Jan. 2006 (right) observations of HD114981.}
\label{fig:hd114981}
\end{figure}

We have cleaned the Kurucz mask, by rejecting lines contaminated with emission. The resulting LSD $I$ profile is still slightly contaminated with emission, but could not be improved. We performed a simultaneous fit of both observed $I$ profiles with a single photospheric function, plus circumstellar Gaussians modelling the CS features, by forcing the depth, \vsini\ and \vrads of the photospheric profiles to be the same for both observations. The result is shown in Fig. \ref{fig:hd114981}.

\subsection{HD~135344}

HD~134344 is an F4 type star with strong near and far-IR excesses that are similar to HD~31293, and that can be well reproduced with a two-temperature dust disk model \citep{malfait98a}, confirming the Herbig Ae nature of the star. It is associated with the Upper Centaurus-Lupus star forming region \citep{dezeeuw99},  situated at a distance of 142~pc \citep{muller11}. We used the photometric data of \citet{oudmaijer92} to derive the luminosity.

The spectrum of HD~135344 is well fit with $T_{\rm eff}=6750\pm250$~K, consistent with the temperature determination of \citet{saffe08}. The core of H$\beta$ is superimposed with faint emission and a strong blueshifted absorption, while H$\alpha$ displays a P Cygni profile of type III. The cores of H$\gamma$ and H$\delta$ are also slightly contaminated with circumstellar emission. The wings of the O~{\sc i} 777~nm and O~{\sc i} 8446 \angs triplets are contaminated with emission, and the Paschen lines are filled with circumstellar emission. A broad and faint emission is observed in the He~{\sc i} D3 line. The Ca~{\sc ii} K line does not seem to be contaminated with CS features, while other metallic lines seem to be slightly distorted due to CS emission.

\begin{figure}
\centering
\includegraphics[width=6cm,angle=90]{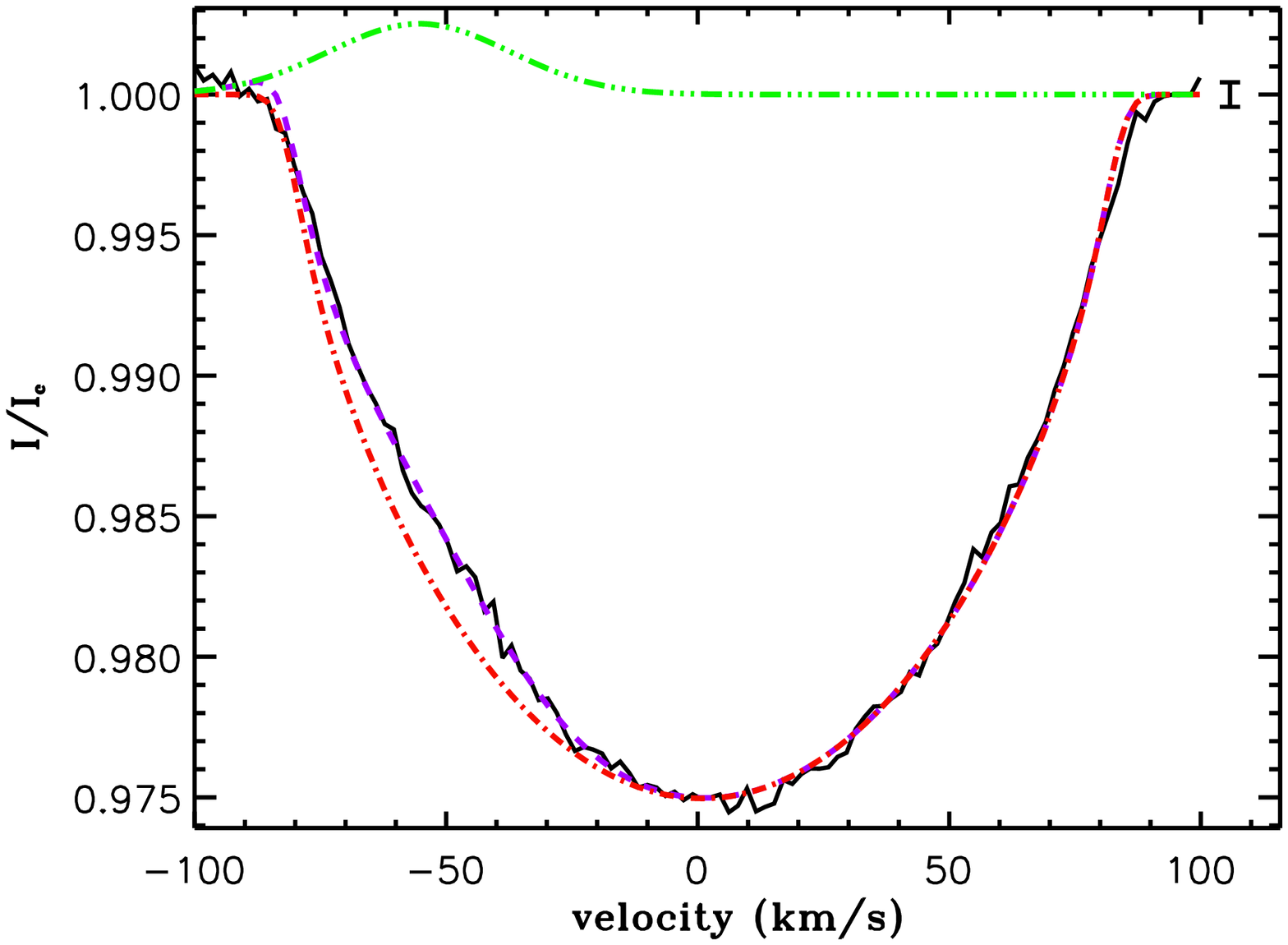}
\caption{As Fig.~A5 for HD~135344}
\label{fig:hd135344}
\end{figure}

We have calculated the LSD profiles without performing a special cleaning to the Kurucz mask. The resulting LSD $I$ profile displays a photospheric profile slightly contaminated with emission, but could not be improved. We fit the profile with a single photospheric function plus a Gaussian. The result is shown in Fig. \ref{fig:hd135344}.

\subsection{HD~139614}

\begin{figure}
\centering
\includegraphics[width=2cm,angle=90]{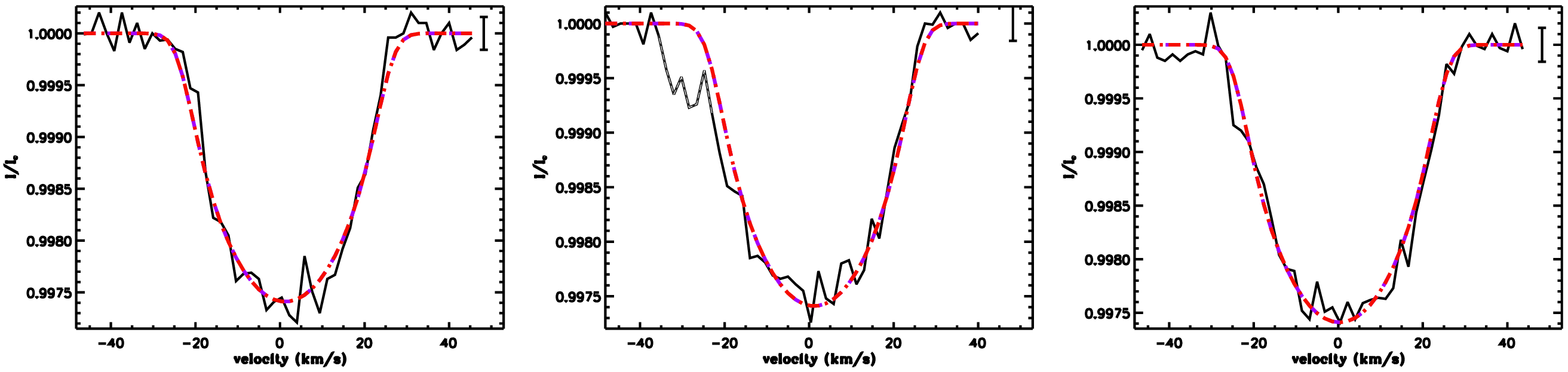}
\caption{As Fig.~A5 for the Feb. 20th (left), 21st (middle), and 22nd (right) 2005 observations of HD~139614}
\label{fig:hd139614}
\end{figure}

HD~139614 is associated with the Upper Centaurus Lupus star forming region \citep{dezeeuw99}, situated at a distance of 142~pc \citep{muller11}. We used the photometric data of \citet{vieira03} to derive the luminosity of the star. Its spectral energy distribution is  well reproduced with a two-temperature dust disk model \citep{malfait98a}, confirming the Herbig Ae nature of the star.

The spectrum is consistent with the temperature and gravity determination of \citet[][$T_{\rm eff}=7600\pm300$~K, $\log g=3.9\pm0.3$]{folsom12}. This temperature does not lead to a very exact fit to the observed metallic line spectrum, which may be due to chemical peculiarities in the atmosphere \citep[see][]{folsom12}. The spectrum is slightly contaminated with variable CS features: faint emission is observed in the core of H$\beta$, while the core of H$\alpha$ is superimposed with a strong, narrow, single-peaked emission. The He~{\sc i} D3 line displays a faint but broad emission profile. A hint of broad emission is also observed at 6878 \angs and 7065~\AA. The photospheric profile of the O~{\sc i}~777~nm triplet might also be superimposed with faint emission. No other CS contribution is observed in the spectrum.

We first calculated the LSD profiles without performing a special cleaning to the Kurucz mask. The resulting LSD $I$ profile appears in absorption, but with very large wings, of which the origin is unknown. We therefore choose to compute new masks from K8250.40 by making a selection on the central depth and the excitation potential of the lines. A mask containing lines with central depths below 0.08 provides us with LSD $I$ profiles displaying photospheric shapes. We performed a simultaneous fit of the three observations with a photospheric function, by forcing the depth, \vsini, and \vrads to be identical for the three profiles. In the profile of Feb. 20th 2005, we exclude from the fit the few points in the blue wing of the $I$ profile that suggest a small absorption component with an unknown origin. The result of the fit is shown in Fig. \ref{fig:hd139614}

\subsection{HD~141569}

HD~141569 is an isolated $\beta$ Pictoris-like star \citep{malfait98a,sahu98}, situated at a distance of 116~pc \citep[Hipparcos, ][]{vanleeuwen07}. It is the primary member of a triple system with two late-type companions orbiting at $\sim 8$ arcsec \citep{weinberger00}. The disk of the primary star has been directly observed using coronographic observations \citep{weinberger99,augereau99}. We used the Hipparcos photometric data to derive the luminosity of the primary.

The spectrum of HD~141569 is consistent with the temperature and gravity determination of \citet[][$T_{\rm eff}=9800\pm500$~K, $\log g=4.2\pm0.4$]{folsom12}. It is only very slightly contaminated with CS features. A faint double-peaked emission is present in the core of H$\beta$, and a strong double-peaked emission is superimposed on the core of H$\alpha$. Some of the metallic lines appear less deep than predicted, which could be due to CS contamination. Except for a double-peaked emission profile in the O~{\sc i}~777~nm and O~{\sc i}~8446~\angs triplets, and in a few Fe~{\sc ii} lines of the IR part of the spectrum, no other CS contribution is observed.

\begin{figure}
\centering
\includegraphics[width=3cm,angle=90]{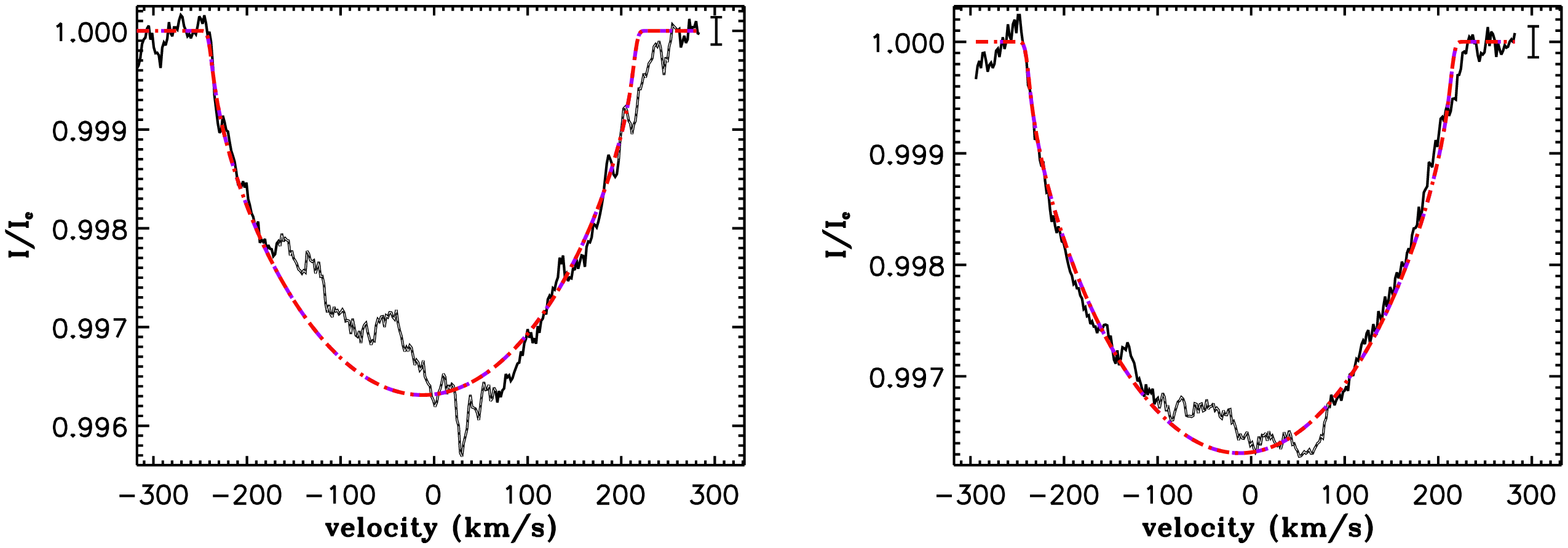}
\caption{As Fig.~A5 for the Feb. 2006 (left) and Mar. 2007 (right) observations of HD~141569. The data have been smoothed with a width equal to 4 pixels.}
\label{fig:hd141569}
\end{figure}

We first calculated the LSD profiles without performing a special cleaning to the Kurucz mask. The resulting LSD $I$ profiles display photospheric profiles contaminated with emission. In order to reduce the contamination, we have cleaned the mask by rejecting, as far as possible, lines that seem to be contaminated with emission. The resulting $I$ profiles look better but are still slightly contaminated. We therefore choose to fit both profiles simultaneously with a single photospheric function, by rejecting the data points contaminated with CS emission, and by forcing the photospheric depth, \vsini, and \vrads to be the same for both observations. The result is shown in Fig. \ref{fig:hd141569}.

\subsection{HD~142666 (= V1026 Sco)}

HD~142666 could be associated with the Upper Scorpius star forming region at a distance of 145~pc \citep{preibisch08}. We used the photometric data of \citet{vieira03} to derive the luminosity. Its spectral energy distribution is  well reproduced with a two-temperature dust disk model \citep{malfait98a}, confirming the Herbig Ae nature of the star.

Using our automatic procedure, we find that the spectrum of HD~142666 is well reproduced with $T_{\rm eff}=7900\pm200$~K and $\log g=4.0$, consistent with the work of \citet{vieira03}. The spectrum is contaminated with variable CS features. Redshifted circumstellar absorption components are observed in the core of the Ca~{\sc ii} K and Balmer lines from H$\zeta$ to H$\beta$. In addition, emission is present in the wings of the CS absorption component in H$\beta$. The core of H$\alpha$ is superimposed with an inverse P Cygni profile. The He~{\sc i}~D3 line displays a strong and broad absorption component, while the O~{\sc i}~777~nm triplet is stronger than predicted. No other circumstellar features are observed in the spectrum.

\begin{figure}
\centering
\includegraphics[width=2cm,angle=90]{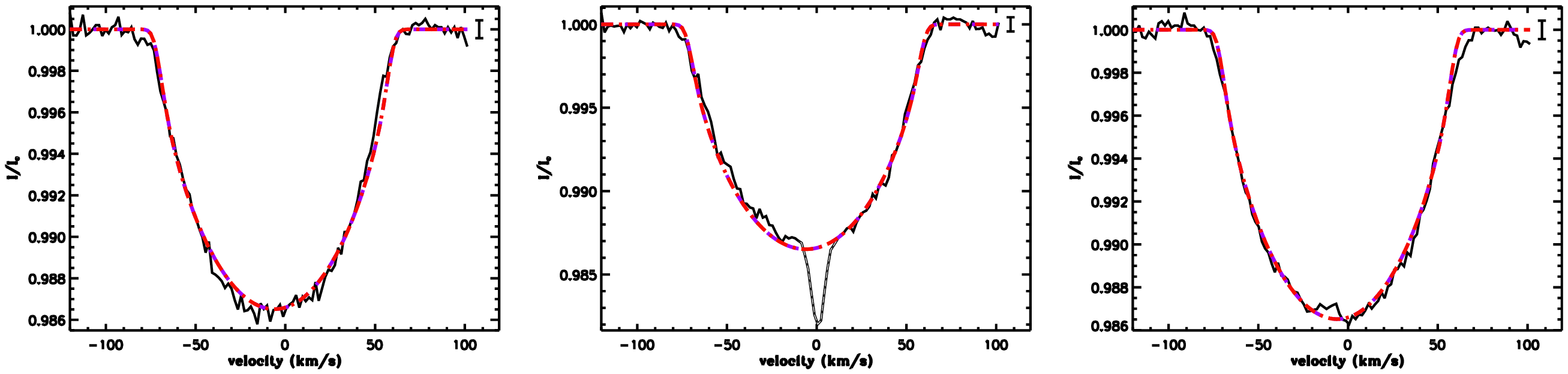}
\caption{As Fig.~A5 for the Feb. 20th (left), May 23rd 07:50 (middle) and May 25th (right) 2005 observations of HD~142666}
\label{fig:hd142666}
\end{figure}

We first calculated the LSD profiles without performing a special cleaning to the Kurucz mask. The resulting LSD $I$ profiles display broad wings that could not be correctly reproduced with a photospheric function. We therefore computed various other masks by making a selection on the central depth and excitation potential of the individual lines in the mask K8000.40. The mask including only lines with central depths of less than 0.5 gives profiles with photospheric-like shapes. The profiles of Feb 21st, May 21st, and May 23rd 2005 display asymmetric shapes. We first fit all the observations simultaneously with a single photospheric function, but the result was not satisfactory. We then fit simultaneously the symmetric profiles of Feb. 19th, May 22nd, and May 24th 2005 only, leading to a better solution. The result is shown in Fig. \ref{fig:hd142666}.

\subsection{HD~144432}

\begin{figure}
\centering
\includegraphics[width=3cm,angle=90]{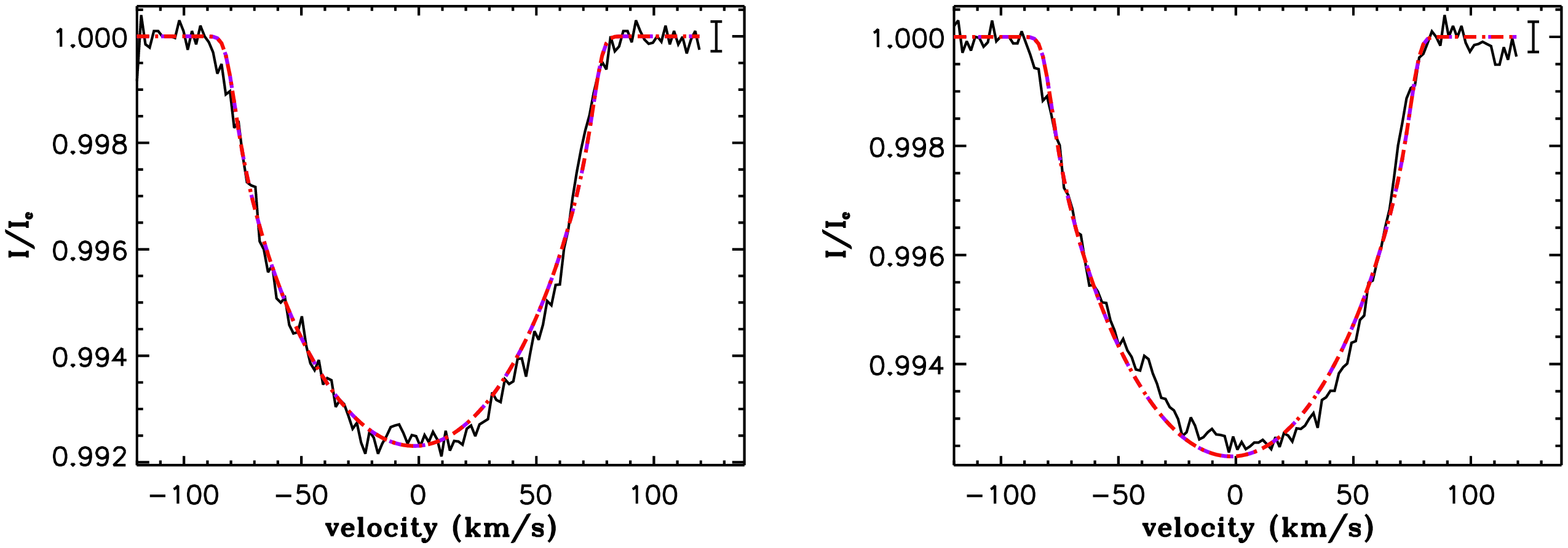}
\caption{As Fig.~A5 for the Feb. 20th (left) and 21st (right) observations of HD~144432.}
\label{fig:hd144432}
\end{figure}

Combining the new Hipparcos data reduction \citep{vanleeuwen07} and the membership criteria of \citep{dezeeuw99}, we find that HD~144432 is a member of the Upper Scorpius OB association, situated at a distance of 145~pc \citep{preibisch08}. We used the Hipparcos photometric data \citep{perryman97} to derive the luminosity of the star. The spectral energy distribution, as well as the the optical photometric and spectroscopic properties of HD~144432 strongly suggest that it is surrounded by a disk, confirming the Herbig Ae nature of the star.

Using our automatic procedure, we find that the spectrum of HD~144432 is well reproduced with $T_{\rm eff}=7500\pm300$~K and $\log g=3.5$, consistent with the work of \citet{vieira03}. Its spectrum is slightly contaminated with variable CS features. Blueshifted circumstellar absorption components are observed in the core of the Balmer lines from H$\delta$ to H$\beta$. In addition, emission is observed in the wings of the CS absorption component of H$\beta$. H$\alpha$ displays a complex P Cygni profile, with two blueshifted absorption components. Faint emission is observed in the core of the Ca~{\sc ii} K line. The He~{\sc i} D3 line displays a faint and broad emission profile. Emission is observed in the blue wing of the O~{\sc i} 777~nm triplet, as well as in the blue and red wings of the O~{\sc i} 8446 \angs line. The Ca~{\sc ii} IR-triplet displays double-peaked emission profiles. No other CS features are observed in the spectrum.

We first calculated the LSD profiles without performing a special cleaning to the Kurucz mask. The resulting LSD $I$ profile displays an absorption profile with broad wings, not well reproduced by a photospheric function. We have therefore calculated many masks from K7500.40 by making a selection on the central depth or excitation potential of the lines. We find that a mask containing lines with central depths of less than 0.4 leads to photospheric profiles. We fit both observations simultaneously, by forcing the photospheric depths, \vsini\ and \vrads to be identical for both observations. The result is shown in Fig. \ref{fig:hd144432}

\subsection{HD~144668 (= HR 5999)}

Combining the Hipparcos data reduction \citep{perryman97} and the membership criteria of \citep{dezeeuw99}, we find that HD~144668 is part of the Upper Centaurus Lupus OB association, situated at a distance of 142~pc \citep{muller11}. This distance is consistent with the Hipparcos parallax, and will be used in the following. HD 144668 illuminates the reflection nebula B149 \citep{bernes77,magakian03} situated in the molecular cloud SL~14 \citep{sandqvist76}. We used the Hipparcos photometric data \citep{perryman97} to derive the luminosity of the star. Its spectral energy distribution is  well reproduced with a two-temperature dust disk model \citep{malfait98a}, a disk that has also been detected by \citet{preibisch06} using mid-infrared long-baseline interferometry.

Using our automatic procedure, we find that the spectrum of HD~144668 is well reproduced with $T_{\rm eff}=8200\pm200$~K and $\log g=3.5$, consistent with the work of \citet{tjinadjie89}. The Balmer lines from H$\zeta$ to H$\beta$ are superimposed with a centered CS absorption. In addition, emission is observed in the wings of the CS absorption of H$\beta$. H$\alpha$ displays a double-peaked emission profile. Strong circumstellar absorption is also observed in some Fe~{\sc ii} lines and in the Ca~{\sc ii} K line, as well as  in the O~{\sc i} 8446 \angs, and in Ca~{\sc ii} IR triplets. Emission is also observed in the wings of the Ca~{\sc ii} IR lines. The O~{\sc i} 777~nm triplet displays a very deep absorption profile with weak emission in the blue wing. The He~{\sc i} 5875~\angs line shows an inverse P Cygni profile. The Paschen lines do not seem to be contaminated with CS features.

\begin{figure}
\centering
\includegraphics[width=6cm,angle=90]{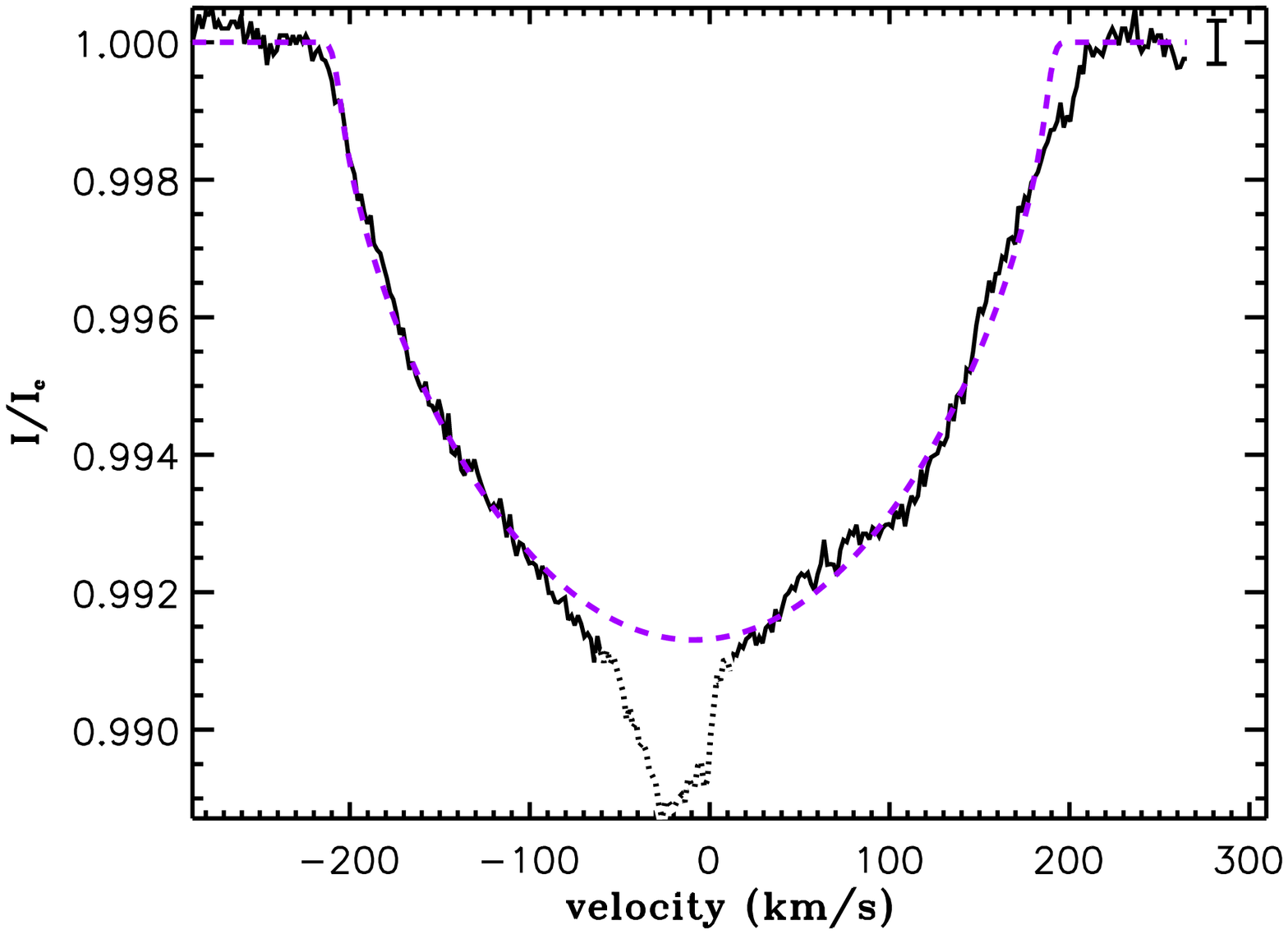}
\caption{As Fig.~A5 for HD~144668}
\label{fig:hd144668}
\end{figure}

We first cleaned the Kurucz mask by rejecting lines contaminated with CS features. The resulting LSD $I$ profile displays broad wings and a small CS contamination in the core. In order to improve this profile we have calculated various masks from the cleaned mask, by making a selection on the central depth or the excitation potential of the lines. The best result was obtained with a mask containing lines with central depths of less than 0.7. The result still shows weak CS contamination but could not be improved. It is well fit with a photospheric function, by excluding from the fit the data points contaminated with CS features. The result is shown in Fig. \ref{fig:hd144668}.

\subsection{HD~145718 (V718 Sco)}

HD~145718 is part of the Upper Scorpius OB association, situated at a distance of 145~pc \citep{preibisch08}. This distance is consistent with the new Hipparcos parallax determination of \citet{vanleeuwen07}, and will be used in the following. We used the Hipparcos photometric data \citep{perryman97} to derive the luminosity. IR, optical and submillimetric observations strongly suggest the presence of a circumstellar disk around the star \citep{dent05,guimaraes06}.

Using our automatic procedure, we find that the spectrum of HD~145718 is well reproduced with $T_{\rm eff}=8100\pm200$~K and $\log g=4.0$, consistent with the work of \citet{vieira03}. The core of the Balmer lines, from H$\delta$ to H$\beta$, are superimposed on redshifted CS absorption. Emission is present in the wings of the CS absorption component of H$\beta$. The core of H$\alpha$ is superimposed on an inverse P Cygni profile with two redshifted absorption components (instead of one). The He~{\sc i} D3 line displays a broad and asymmetric absorption profile. Due to strong non-LTE effects, the O~{\sc i} 777~nm triplet is stronger than predicted. The cores of a few metal lines seem to be distorted due to CS features. No other CS contribution is observed.

\begin{figure}
\centering
\includegraphics[width=6cm,angle=90]{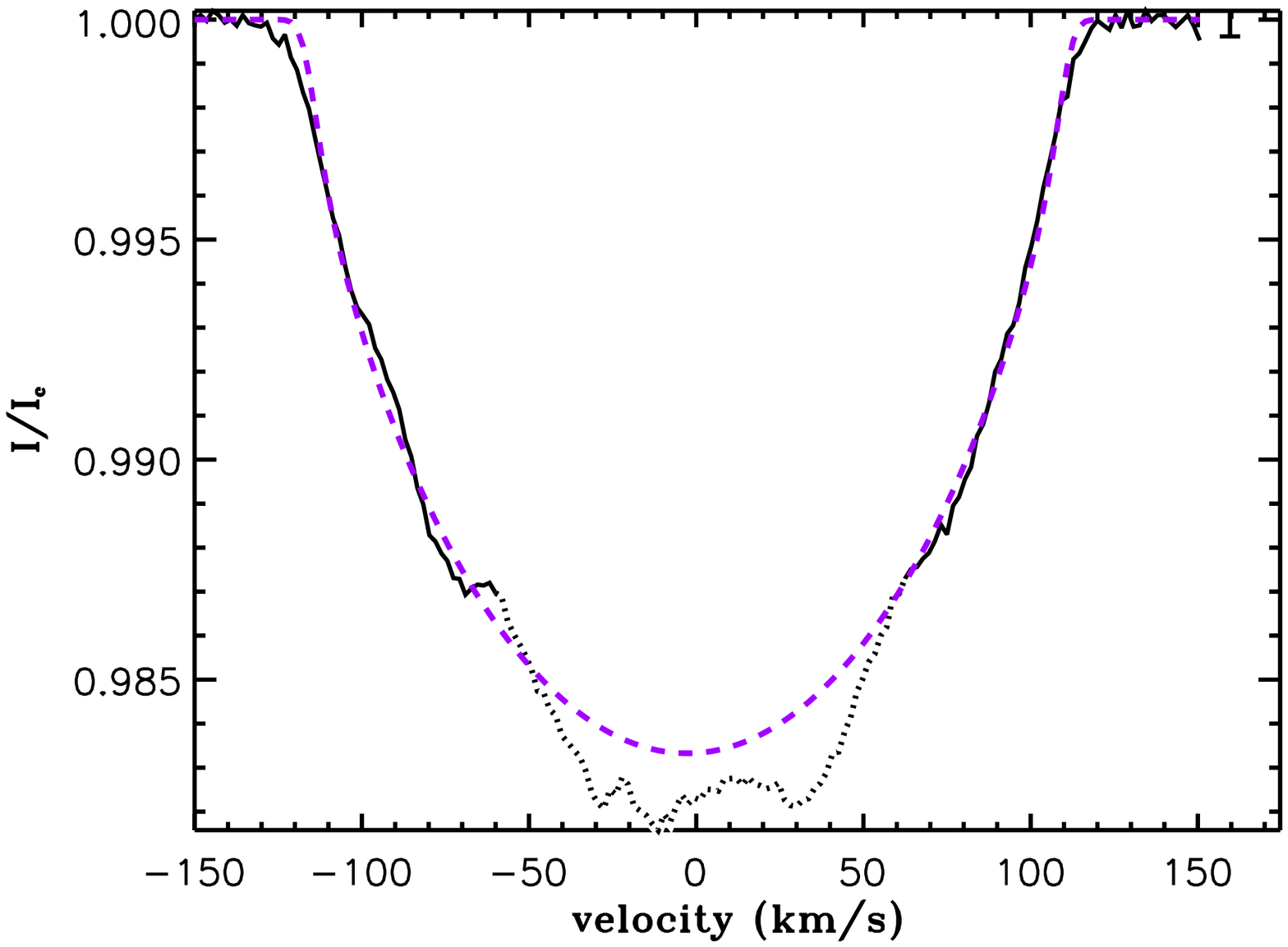}
\caption{As Fig.~A5 for HD~145718}
\label{fig:hd145718}
\end{figure}

We first calculated the LSD profiles without performing a special cleaning to the Kurucz mask. The resulting $I$ profile displays a photospheric shape with a slight distortion in the core due to CS contribution. We tried to improve the fit by calculating new masks by making a selection on the central depth and the excitation potential of the lines. We find that a mask containing only lines with an excitation potential greater than 8~eV does not show CS contamination, but is much noisier than the first computed $I$ profile. We fit both computed profiles with a single photospheric function. We excluded from the fit the data points contaminated with CS features in the first LSD $I$ profile that we computed. Both results give consistent \vsini, but the second profile, due to a lower SNR, is less accurate. We therefore determined \vsini\ from the first profile that we computed. The result of the fitting is shown in Fig. \ref{fig:hd145718}.

\subsection{HD~150193 (V2307 Oph)}

HD~150193 is part of the Upper Scorpius OB association situated at a distance of 145~pc \citep{preibisch08}. This distance is consistent with the new Hipparcos parallax determination of \citep{vanleeuwen07}, and will be used in the following. We used the Hipparcos photometric data \citep{perryman97} to derive the luminosity. Its IR spectral energy distribution is  well reproduced with a two-temperature dust disk model \citep{malfait98a}, describing a disk that has been spatially resolved by \citep{eisner09} using near-IR interferometry.

The stellar spectrum is well reproduced with an effective temperature of $T_{\rm eff}=9500\pm500$~K, consistent with the work of \citet{hernandez05}. Blueshifted CS absorption components are superimposed on the core of the Balmer lines from H$\zeta$ to H$\beta$, with increasing depth with wavelength. Emission is observed in the wings of the CS absorption of H$\beta$. H$\alpha$ displays a P Cygni profile of type III. A strong absorption component with emission in the wings is also observed in the Ca~{\sc ii} K line. The He~{\sc i} lines at 5875~\AA, 6678~\angs and 7065~\angs display broad emission profiles. The lines of the multiplet 42 of Fe~{\sc ii} have blueshifted CS absorption superimposed on centered emission profiles. The O~{\sc i} 777~nm triplet shows emission in the blue wing. The O~{\sc i} 8446~\angs triplet is filled with emission, while the Ca~{\sc ii} IR-triplet displays single-peaked lines. The core of the Paschen lines seem to be filled with emission. The cores of a few other metallic lines might be superimposed on a faint emission component.

\begin{figure}
\centering
\includegraphics[width=6cm,angle=90]{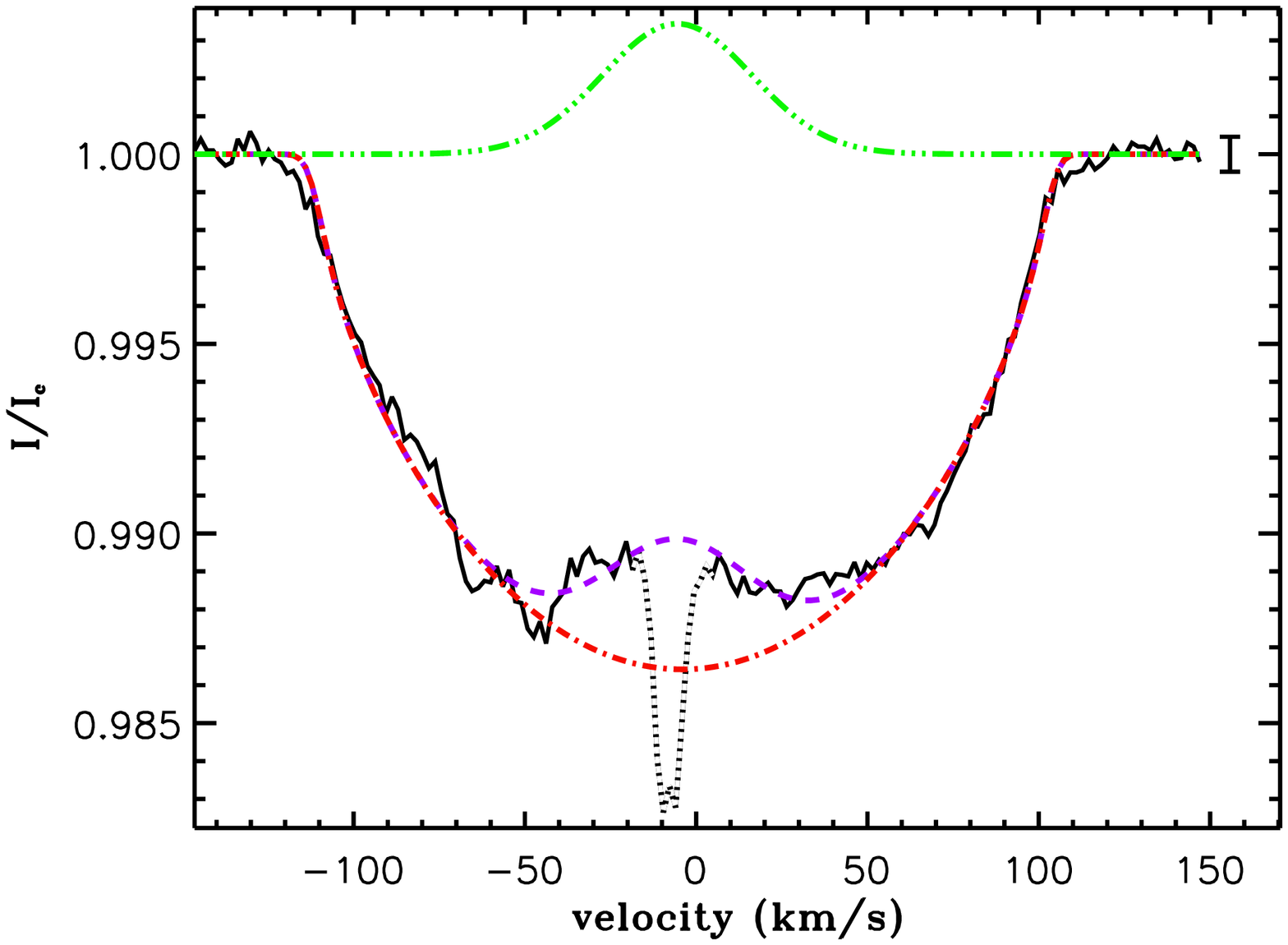}
\caption{As Fig.~A5 for HD~150193}
\label{fig:hd150193}
\end{figure}

We have calculated the LSD profiles without performing a special cleaning to the Kurucz mask. The resulting $I$ profile displays a photospheric profile still slightly contaminated with CS features, but could not be improved. It is well fit with a single photospheric profile together with a Gaussian that models the emission. The small absorption component in the core has been rejected from the fit. The result is shown in Fig. \ref{fig:hd150193}.

\subsection{HD~152404 (= AK Sco)}

HD~152404 is a binary system, part of the the Upper Centaurus Lupus OB association, situated at a distance of 103~pc \citep[Hipparcos,]{vanleeuwen07}. A strong IR excess is recorded in the direction of the system, confirming its PMS nature \citep{andersen89}.

Our observations confirm the SB2 nature of the system, with two components of similar temperature. We use a modified version of the BINMAG1 code of O. Kochukhov (private communication), which computes the composite spectrum of a binary star, to fit the observed spectrum. This code takes as input two synthetic spectra of different effective temperatures and gravities, each corresponding to one of the two components. The code convolves the synthetic spectra with instrumental, turbulent and rotational broadening profiles and combines them according to the radii ratio of the components specified by the user, and the flux ratio at the considered wavelength given by the atmosphere models, to produce the spectrum of the binary star. The individual synthetic spectra have been calculated in the local thermodynamic equilibrium (LTE) approximation, using the code SYNTH of Piskunov (1992). SYNTH requires, as input, atmosphere models obtained using the ATLAS~9 code (Kurucz 1993) and a list of spectral line data obtained from the VALD database\footnote{http://ams.astro.univie.ac.at/$\sim$vald/} (Vienna Atomic Line Data base). Our observations are consistent with an effective temperature of $6500\pm100$~K for both components \citep{alencar03} and a ratio of radii $R_{\rm P}/R_{\rm S}=1.3\pm0.15$. We used the new Hipparcos parallax \citep{vanleeuwen07} and the photometric data \citep{perryman97} to compute the luminosity of the system, then we derived the luminosity of the two components: $L_{\rm P}/L_{\odot}=0.94\pm0.21$, and $L_{\rm S}/L_{\odot}=0.71\pm0.21$.

A redshifted CS absorption component is superimposed on the core of the Balmer lines from H$\varepsilon$ to H$\beta$. H$\alpha$ displays an inverse P Cygni profile of type III. Faint emission is observed in the core of the Ca~{\sc ii}~K line. Strong CS absorption components are observed in the He~{\sc i} lines at 5875~\AA, 6678~\angs and 7065~\AA, as well as in the O~{\sc i}~777~nm triplet. No other CS feature is observed.

\begin{figure}
\centering
\includegraphics[width=6cm,angle=90]{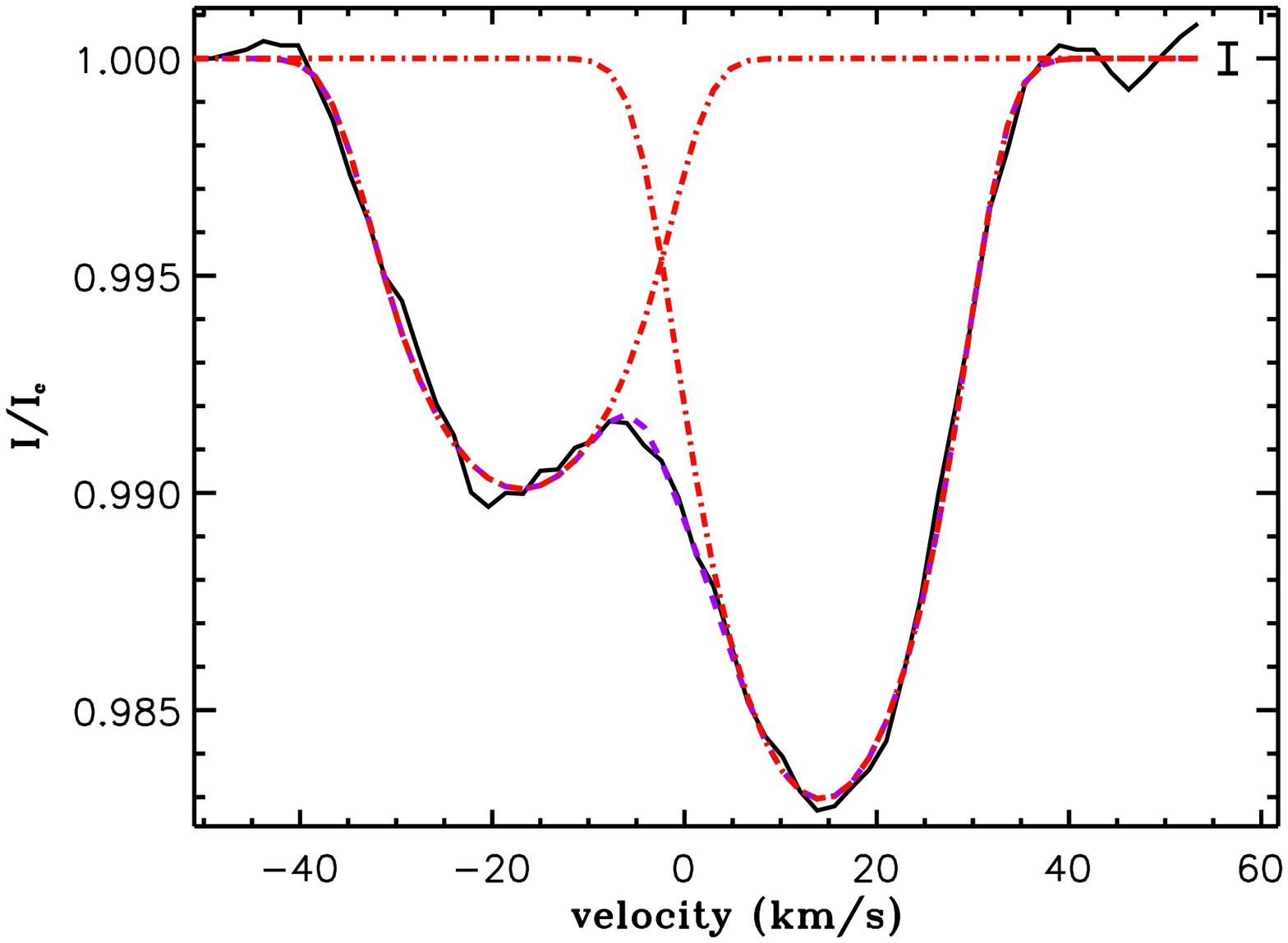}
\caption{As Fig.~A5 for AK Sco}
\label{fig:aksco}
\end{figure}

We first calculated the LSD profiles without performing a special cleaning to the Kurucz mask. The result shows the photospheric composite profile of the binary with very broad wings. We have therefore calculated various masks by making a selection on the central depth and excitation potential of the lines. We find that a mask containing only lines with central depths of less than 0.30 gives a satisfactory result: an LSD profile with photospheric wings. In order to determine the \vsinis and \vrad\ of the two components, we performed a least-square fit to the composite $I$ profile of the binary. Each profile is fitted with the normalised sum of two photospheric functions. We adopted an isotropic macroturbulent velocity of 2 \kms\ in order to fit the wings of the LSD I profiles. The free parameters of the fitting procedure are the centroids, depths and $v\sin i$ of both components. The result is shown in Fig. \ref{fig:aksco}. As both stars have the same temperature, the ratio of the equivalent widths of the two components measured on the LSD profile is equal to the luminosity ratio $L_{\rm P}/L_{\rm S}$ \citep{alecian08a}. From our fit to the composite profile we derive a luminosity ratio $L_{\rm P}/L_{\rm S}=1.7$, consistent with our radii ratio determined above from the fit of the spectrum ($R_{\rm P}/R_{\rm S}=1.3\pm0.15$).

\subsection{HD~163296}

\begin{figure}
\centering
\includegraphics[width=6cm,angle=90]{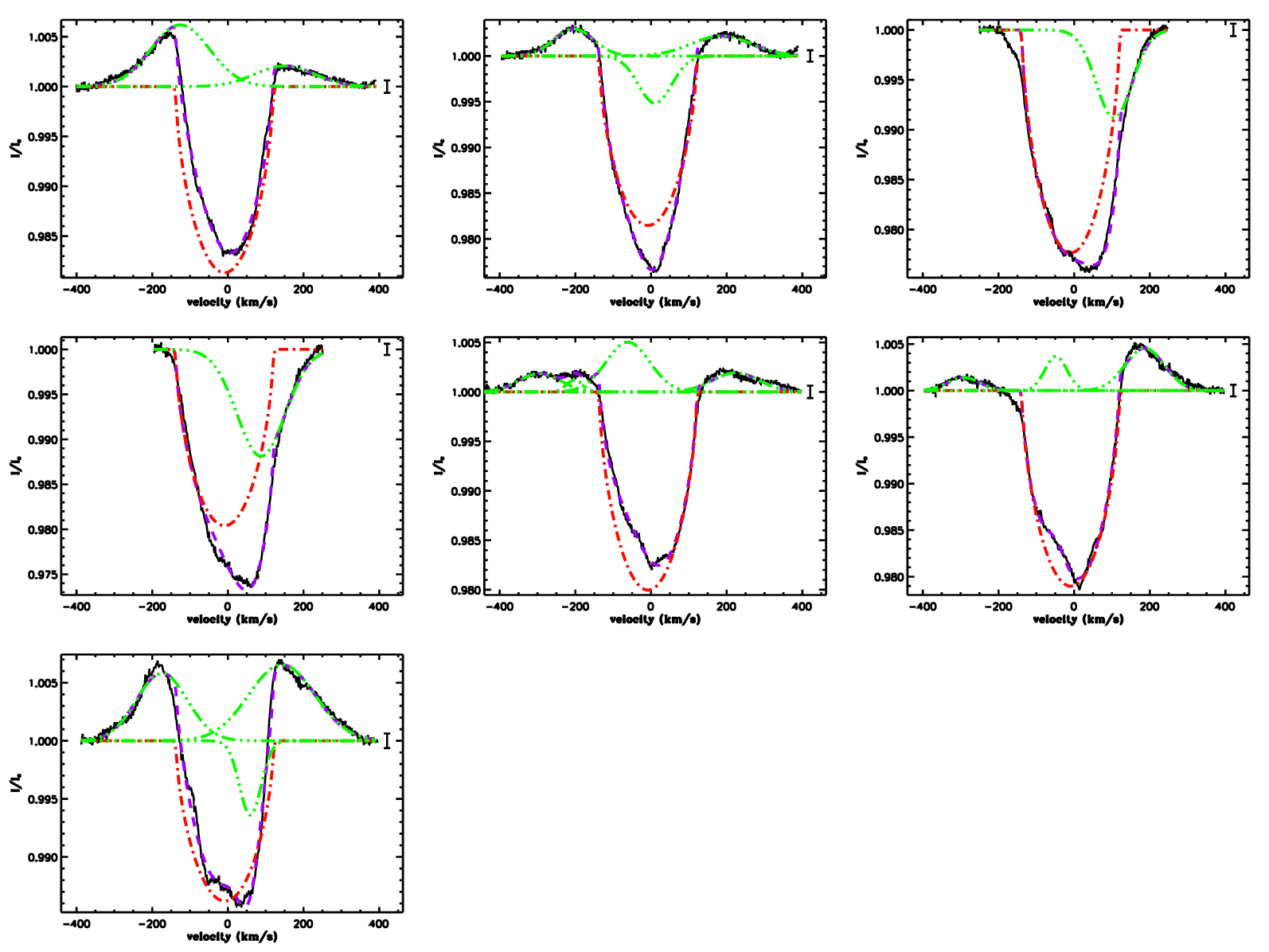}
\caption{As Fig.~A5 for the May 22nd, 23rd, 24th 09:49, 24th 14:48, 25th 09:52, 25th 14:53, and Aug. 25th 2005 observations (from top left to bottom right) of HD~163296.}
\label{fig:hd163296}
\end{figure}

HD~163296 is an isolated Herbig Ae star situated at a distance of 119~pc \citep{vanleeuwen07}, not associated with any star forming regions, and not illuminating a bright nebulosity. However the star is associated with Herbig-Haro objects \citep{grady00a}. It displays a large IR excess well reproduced with a two-temperature disk, perhaps reflecting the presence of a gap inside the accretion disk \citep{malfait98a}. The presence and the geometry of the disk has been confirmed by \citet{grady00a} using coronagraphic imaging with the Space Telescope Imaging Spectrograph (STIS) on board the Hubble Space Telescope (HST). All these characteristics leave no doubt that HD~163296 belongs to the Herbig Ae/Be class of objects. We used the Hipparcos photometric data to compute the luminosity of the star.

HD~163296 displays circumstellar activity that could be of magnetic origin, such as rotational modulation in non-photospheric lines \citep{catala89}, X-ray emission \citep{gunther09}, and emission lines of highly ionised species \citep[e.g. O VI, ]{deleuil05}. The spectrum displays strong circumstellar emission. H$\alpha$ is strongly variable, going from a double-peaked emission line profile to a P Cygni of type IV profile \citep{the85a,catala89}. In our spectra of May 2005, H$\alpha$ appears single-peaked with variable blueshifted absorption components, while in August 2005, a strong P Cygni of type IV profile is observed, confirming the high variability of H$\alpha$. H$\beta$ displays variable emission and blueshifted absorption components. Faint emission is observed in the cores of H$\gamma$ and H$\delta$. Fe~{\sc ii} lines, including those of multiplet 42, the O~{\sc i} 8446~\angs and Ca~{\sc ii} IR triplets display single-peaked emission profiles. Faint blueshifted emission is also present in the core of the Paschen lines. The Ca~{\sc ii} K line displays a strong and complex absorption profile. The He~{\sc i} lines at 5875~\AA, 6678~\AA, and 7065~\angs show very broad double-peaked emission profiles, with a stronger blue peak with respect to the red one. The portions of the spectrum not contaminated with CS features is consistent with temperature and gravity determination of \citet[][$T_{\rm eff}=9200\pm300$~K, $\log g=4.2\pm0.3$]{folsom12}.

We first cleaned the mask in order to reject the lines contaminated with CS features. The resulting LSD $I$ profiles still show contamination. We therefore tried to make a selection on the central depth of the lines and the excitation potential, but we could not improve the profiles. We therefore fit the 7 observations at once, with single photospheric functions superimposed with Gaussians modelling the emission or absorption features where necessary, and by forcing the \vsinis and \vrad\ to to be identical for all the observations. The photospheric depths of the profiles can vary from one observation to the other, because of the variable CS contamination. The result is shown in Fig. \ref{fig:hd163296}.

\subsection{HD~169142}

\begin{figure}
\centering
\includegraphics[width=6cm,angle=90]{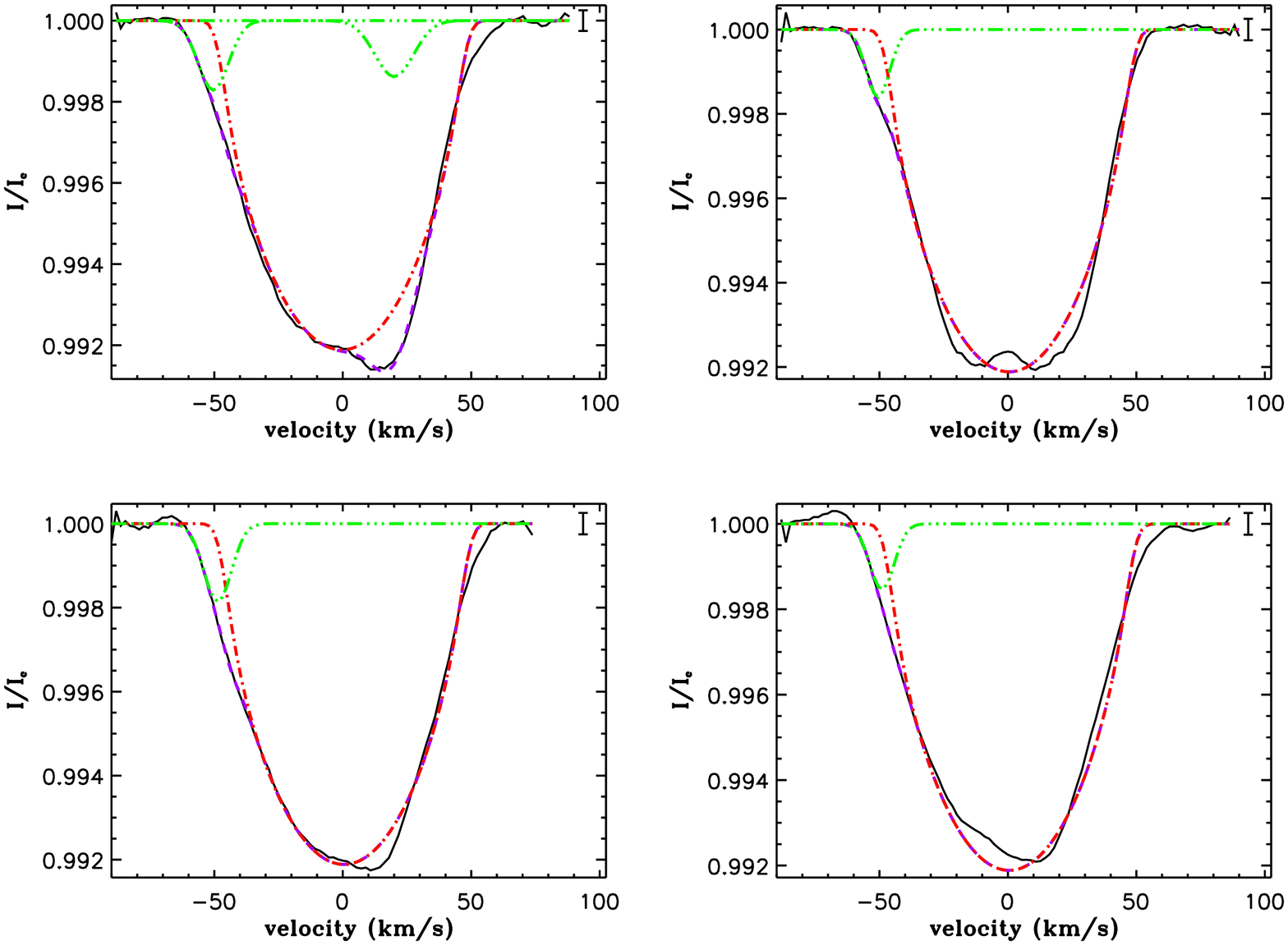}
\caption{As Fig.~A5 for the Feb. 20th, 22nd, May 22nd, and Aug. 24th 2005 observations (from top left to bottom right) of HD~169142}
\label{fig:hd169142}
\end{figure}

HD~169142 is an isolated Herbig Ae star not associated with a star forming region, nor illuminating a bright nebulosity. No Hipparcos data has been obtained, which makes a precise determination of its distance from the Sun very difficult. We therefore used the photometric distance derived by \citet{sylvester96}, and the photometric data of \citet{vieira03} to compute the luminosity. The spectral energy distribution in the near- and far-IR can be well reproduced with a two-temperature dust disk model, like many other Herbig Ae stars \citep{malfait98a}.

The spectrum of HD~169142 is consistent with the temperature and gravity determination of \citet[][$T_{\rm eff}=7500\pm200$~K]{folsom12}. It displays few CS features. Faint emission is observed in the core of H$\beta$. H$\alpha$ displays a single-peaked emission profile. The He~{\sc i} D3 line shows a double-peaked emission profile. The O~{\sc i} 777~nm triplet is slightly stronger than predicted, with weak emission in the blue wing. No other CS contribution is observed in the spectrum.

We first calculated the LSD profiles without performing a special cleaning to the Kurucz mask. The resulting LSD $I$ profiles are absorption profiles slightly distorted due to circumstellar contribution, and have broad wings, not easy to fit with a photospheric function. Therefore we have calculated various masks from by making a selection on the depth and excitation potential. We find that a mask containing lines with central depths of less than 0.6 gives a good compromise between the SNR of the resulting $I$ profile and its shape. The $I$ profiles are still slightly contaminated with CS features, but could not be improved. We fit the four observations simultaneously, with a single photospheric function and several Gaussians modelling the CS contribution. In this fit we force the photospheric depth, \vsini, and \vrads to be the same for the four observations. The result is shown in Fig.~\ref{fig:hd169142}.

\subsection{HD~174571 (=MWC 610)}

HD~174571 is often classified as a classical Be star. However its near-IR excess is large and more similar to those observed among the Herbig Ae/Be class of objects \citep{vieira03}. No star forming regions or nebulosity are known to be associated with the star, and the Hipparcos parallax \citep{perryman97} is too uncertain to constrain the distance of the star. We therefore could not estimate the luminosity directly using photometric data. Instead, we used the well-constrained effective temperature and surface gravity of \citet[][$T_{\rm eff}=21000\pm1500$~K and $\log g=4.00\pm0.10$~(cgs)]{fremat06} to place the star in the $T_{\rm eff}-\log g$ diagram and derive the mass, radius, and age. We then derived a luminosity $\log L/L_{\odot}=3.58\pm0.21$, and using the photometric data of \citet{vieira03}, we estimated the distance of the star at $540_{-70}^{+80}$~pc.

The spectrum of HD~174571 is well reproduced with a non-LTE synthetic spectrum of effective temperature $T_{\rm eff}=22000$~K, consistent with the work of \citet{fremat06}. Variable CS contributions are observed in the spectra. Double-peaked emission features are superimposed on the core of the Balmer lines from H$\delta$ to H$\beta$ with increasing strength with wavelength. H$\alpha$ displays a double-peaked emission profile. The strongest He lines of the spectrum are slightly distorted, which could be due to CS material. Many double-peaked emission profiles are observed in the spectrum, which could come from CS Fe~{\sc ii}. The O~{\sc i} 777~nm triplet presents emission in the wings. The O~{\sc i} 8446~\angs triplet displays a double-peaked emission profile. Double-peaked emission is also observed in the cores of the Paschen lines. While the Ca~{\sc ii}~K line is only contaminated with a strong and narrow interstellar (IS) absorption component, the Ca~{\sc ii} IR triplet seems to display single-peaked emission.

\begin{figure}
\centering
\includegraphics[width=3cm,angle=90]{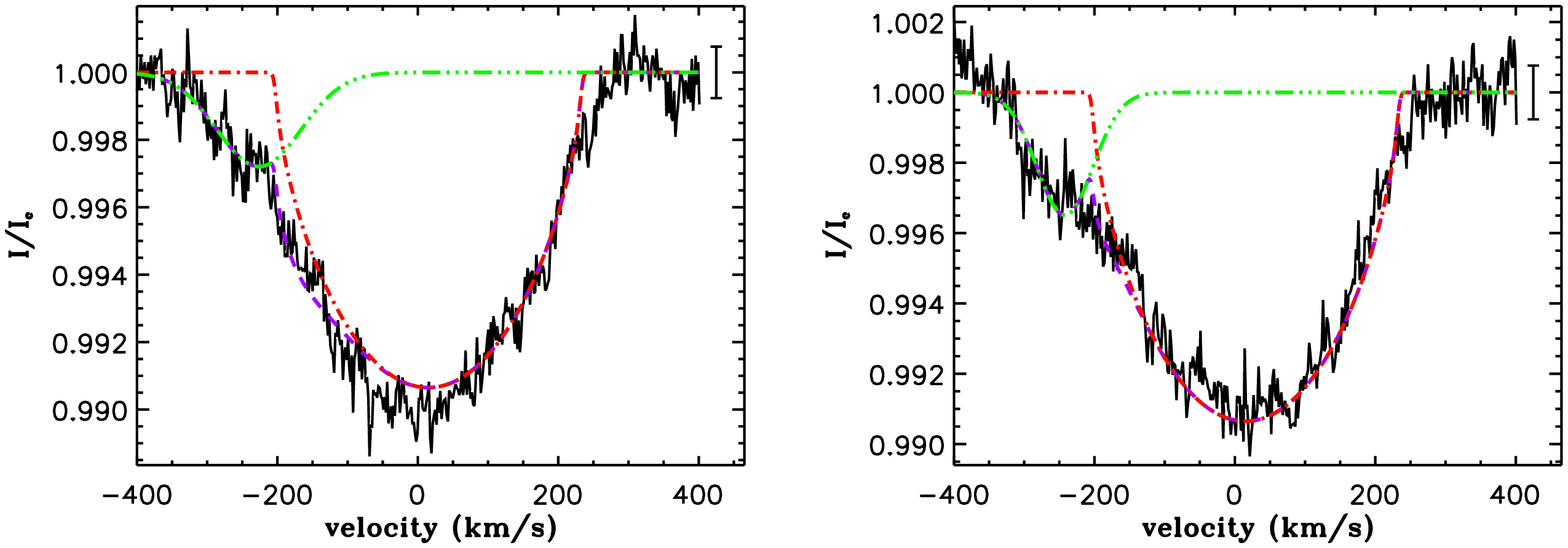}
\caption{As Fig.~A5 for the Mar 2007 (left) and Apr. 2008 (right) observations of HD~174571}
\label{fig:hd174571}
\end{figure}

We have cleaned the Kurucz mask in order to eliminate the CS contribution. The resulting LSD $I$ profiles do not look like photospheric profiles, certainly due to CS contamination, and could not be improved. We fit both observations simultaneously, with a single photospheric function, and Gaussians modelling the CS features. The result is shown in Fig. \ref{fig:hd174571}.

\subsection{HD~176386}

HD~176386 is located in the R Coronae Australis star-forming region \citep{fernandez08}, at $\sim1$ arcminute from TY CRa. We used the new Hipparcos parallax \citep{vanleeuwen07} and photometric data \citep{perryman97} to derive the luminosity of the star. HD~176386 lacks strong near-IR emission; however its mid-IR energy distribution reveals the presence of an extended emission region and a large-scale structure, as well as silicate emission distributed in a disk \citep{prusti94,boersma09}. Ongoing accretion has been reported \citep{grady93}. This star is therefore very probably still in the PMS phase of stellar evolution.

Using our automatic procedure, we find that the spectrum of HD~176386 is well reproduced with $T_{\rm eff}=11500\pm350$~K and $\log g=4.5$, consistent with the work of \citet{paunzen01}. The He~{\sc i} lines at 5875~\angs and 6678~\angs display V-shaped profiles that could be due to CS emission. The O~{\sc i} 777~nm triplet is stronger than predicted. No other CS manifestation is observed in the spectrum.

\begin{figure}
\centering
\includegraphics[width=6cm,angle=90]{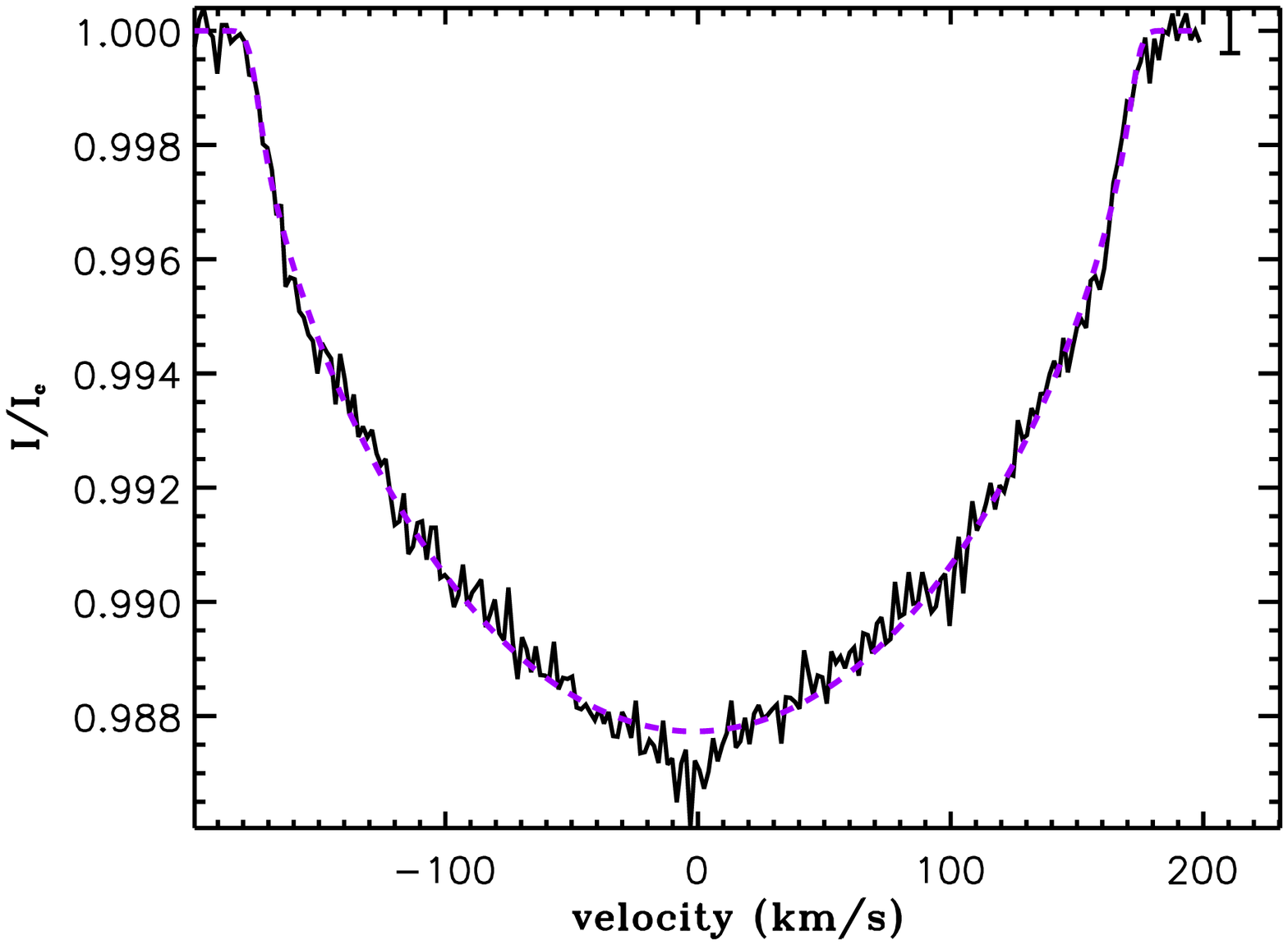}
\caption{As Fig.~A5 for HD~176386}
\label{fig:hd176386}
\end{figure}

We have calculated the LSD profiles without performing a special cleaning to the Kurucz mask. The resulting LSD $I$ profile displays a photospheric shape well fitted with a single photospheric function. The result is plotted in Fig. \ref{fig:hd176386}

\subsection{HD~179218}

HD~179218 is an isolated Herbig Ae star displaying strong IR excess that is well reproduced with a two-temperature dust-disk model \citep{malfait98a}. We used the new Hipparcos parallax \citep{vanleeuwen07} and photometric data \citep{perryman97} to derive the luminosity of the star.

The spectrum of HD~179218 is consistent with the temperature and gravity determination of \citet[][$T_{\rm eff}=10500$~K, $\log g=3.9\pm0.2$]{folsom12}. The observed Fe~{\sc ii} lines seem systematically fainter than the computed lines, revealing possible circumstellar contamination. Faint emission is observed in the core of the Balmer lines from H$\delta$ to H$\beta$, while H$\alpha$ displays a strong single-peaked emission superimposed on the core of the photospheric profile. The Ca~{\sc ii} K line, and the O~{\sc i} 777~nm and O~{\sc i} 8446 \angs triplets display absorption profiles that are stronger than predicted. The Paschen lines and the Ca~{\sc ii} IR triplet do not seem to be contaminated with circumstellar features. The He~{\sc i} lines at 5875~\angs and 6678~\angs display inverse P-Cygni profiles.

\begin{figure}
\centering
\includegraphics[width=2cm,angle=90]{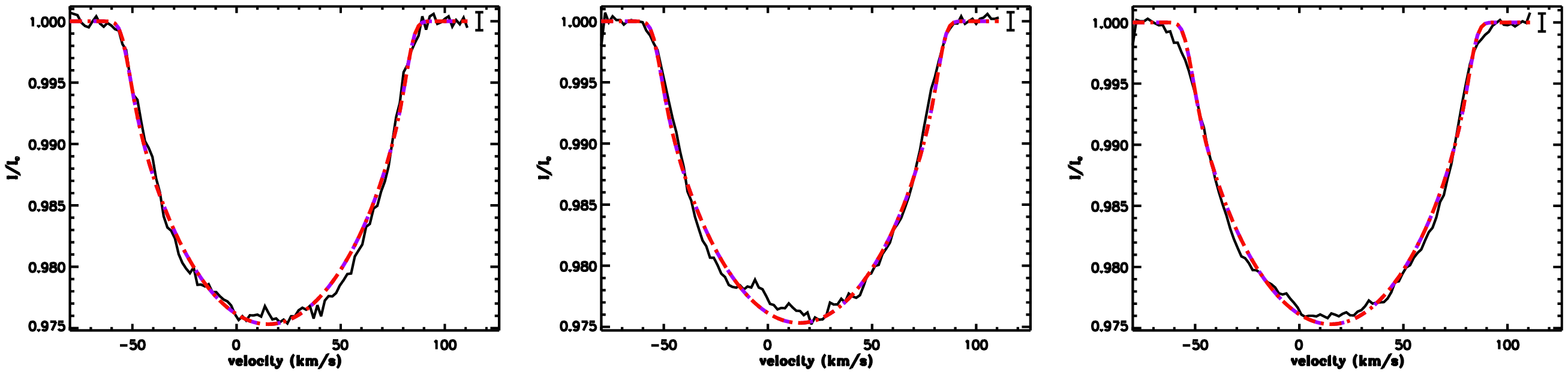}
\caption{As Fig.~A5 for the Feb. 2005 (left), Aug. (middle) 2005, and Oct. 2009 (right) observations of HD~179218.}
\label{fig:hd179218}
\end{figure}

We have cleaned the Kurucz mask by removing the lines obviously contaminated with CS features, then calculated the LSD profiles for the three observations. The three profiles display photospheric shapes. We fit them simultaneously with photospheric functions, by forcing the photospheric parameters (depth, $v\sin i$, $v_{\rm rad}$) to be the same for each one of the observations. The resulting fit is plotted in Fig. \ref{fig:hd179218}.

\subsection{HD~203024}

HD~203024 is a Herbig Ae star belonging to the Cepheus R2 association, situated at a distance of 420~pc \citep{kun00}. The spectral energy distribution shows a slope in the IRAS wavelength region that is typical of optically thick disks surrounding Herbig Ae/Be stars \citep{kun00}. 

\citet{corporon99} reports the detection of the Li~{\sc i} line at 6707 \AA, and conclude that a low-mass star companion might be present close to the Herbig star. In our spectra we detect the spectral lines of a secondary star at almost the same radial velocity as the primary. The primary is a fast rotator, with $v\sin i=162$~\kms; the secondary rotates more slowly with $v\sin i=57$~\kms. We used the same method as for HD~152404 (= AK~Sco) to compute the synthetic composite spectrum of this binary star.  We find that our spectra are well reproduced with effective temperatures of 9250 K and 6500 K for the primary (P) and secondary (S), respectively, with a ratio of radii $R_{\rm P}/R_{\rm S}$ equal to $1.5$. We also estimate the ratio of the luminosity to be $L_{\rm P}/L_{\rm S} = 9.2$. We used the SIMBAD photometric data to derive the luminosity of the system: $\log(L/L_{\odot})=1.93$, and then the luminosity of each component: $\log(L_{\rm P}/L_{\odot})=1.88$ and $\log(L_{\rm S}/L_{\odot})=0.93$.

The temperature that we derive for the secondary is too high to display a detectable Li~{\sc i} line at 6707 \angs in the spectrum. We confirm the presence of a line at about 6708 \angs but it is not clear whether it is of photospheric origin. As the spectra that we obtained are relatively noisy compared to the depth of the lines from the secondary, we tried to change the effective temperature of the secondary. A temperature as low as 5000 K could still fit the spectrum reasonably well, but even at this temperature we do not fit the feature observed at 6708 \AA. The feature observed at 6708 \AA\ does not appear to belong to the secondary component that we detect, and its origin is therefore not understood.

The spectrum displays little circumstellar contamination. The blue part of the Ca~{\sc ii} K line is superimposed on a strong and complex CS absorption, while the O~{\sc i} 777~nm and Ca~{\sc ii} IR triplets are stronger than predicted by the synthetic spectrum. Faint emission is superimposed on the blue side of the core of H$\alpha$. A faint absorption is visible in the He~{\sc i} D3 line, while no photospheric feature is predicted by the synthetic spectrum. No other evidence of CS material is observed in the spectrum.

\begin{figure}
\centering
\includegraphics[width=3cm,angle=90]{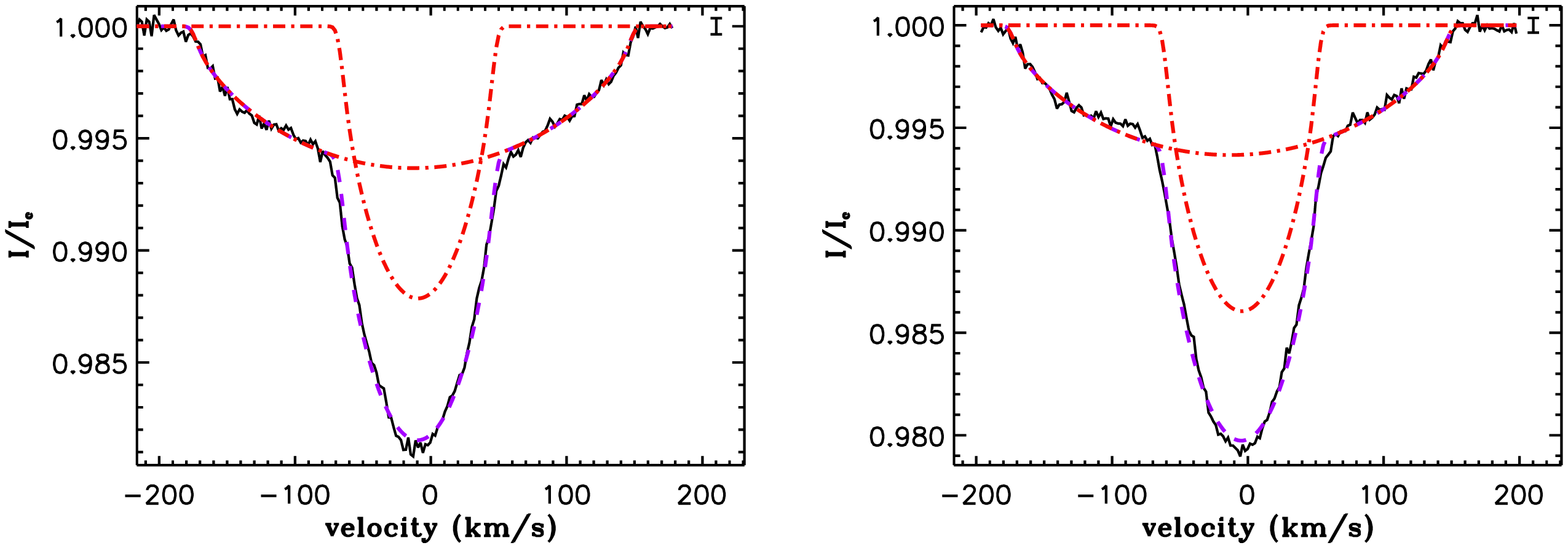}
\caption{As Fig.~A5 for the Aug. 2005 (left) and Nov. 2007 (right) observations of HD~203024.}
\label{fig:hd203024}
\end{figure}

We have cleaned the mask in order to reject the lines contaminated with CS features to compute the synthetic profiles of the binary star. The resulting LSD profiles show both components of the system at almost the same radial velocity. While the secondary radial velocity changed slightly from August 2005 to November 2007, that of the primary did not change. Due to the different temperatures of the stars, and the fact that the LSD profiles have been computed with a line list appropriate for the primary, the ratio of the LSD profiles of both components cannot be used to derive the luminosity ratio (as in HD~200775, e.g., Alecian et al. 2008). However the fit of the combined profile can give us the radial velocities and the \vsinis of both stars. We used the same method as for AK~Sco to the fit the LSD profiles of both observations, by fixing a macroturbulent velocity of 2~\kms, and by forcing the \vsinis of both components, as well as the photospheric depth and the radial velocity of the primay to be identical for both observations. We could not find a satisfactory fit of both profiles by forcing the photospheric depth of the secondary to be identical for both observations, which means that its depth has varied, but this is not yet understood. The result is presented in Fig. \ref{fig:hd203024}.

\subsection{HD~216629 (=IL Cep)}

\begin{figure}
\centering
\includegraphics[width=2.2cm,angle=90]{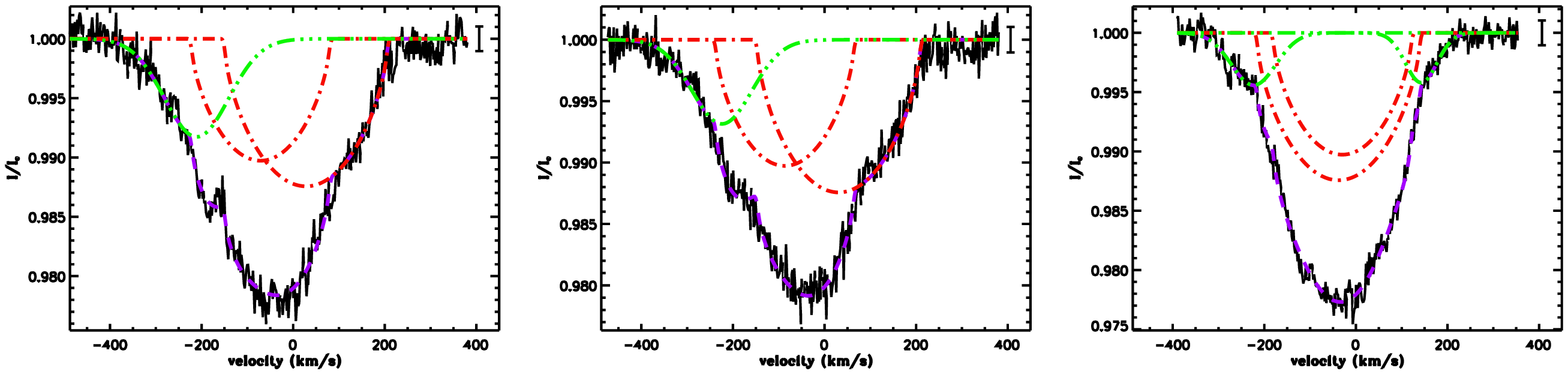}
\caption{As Fig.~A5 for the Jun. 2006 (left), Dec. 2006 (middle), and Nov. 2007 (right) observations of HD~216629 (= IL Cep).}
\label{fig:ilcep}
\end{figure}

\citet{jordi96} re-analysed the membership of the Cepheus OB3 association, and found that IL~Cep is part of the youngest subgroup (b) of the region. The Hipparcos parallax \citep{perryman97} is not well enough constrained to be useful. We therefore used the \citet{crawford70} distance modulus determination ($d=720_{-150}^{190}$~pc) and the Hipparcos photometric data to derive the luminosity of the star.  The resulting position of the star in the HR diagram is well below the ZAMS, which may mean that the luminosity determination is wrong, and therefore IL~Cep is not a member of Cep OB3.

The spectrum of IL~Cep is well fit with an effective temperature $T_{\rm eff}=19000$~K, consistent with the work of \citet{finkenzeller85}. \citet{wheelwright10} find that IL Cep is a double-lined spectroscopic binary, using spectroastrometric methods. They attempted a separation of the unresolved spectrum into two composite spectra, and find that the secondary has a spectral type B4, while the primary is a B3 star. Indications of binarity can be found in our spectra, as explained below. As both stars have different temperatures they should have different luminosity. The quality of our spectra is not sufficient to attempt a determination of the luminosity ratio. For this reason, and also because the position of the binary in the HR diagram is well below the ZAMS (see Fig. \ref{fig:hr}), no mass, radius, nor age of the components have been estimated.

Circumstellar features are observed in the spectrum, and vary from one observation to the other. The cores of H$\gamma$ and H$\delta$ are filled with CS emission. A double-peaked emission line is superimposed on the core of H$\beta$, while H$\alpha$ displays a single-peaked emission profile. Many Fe~{\sc ii} lines (including those of multiplet 42), the Paschen lines, and the O~{\sc i} 777~nm, O~{\sc i} 8446 \AA, and Ca~{\sc ii} IR triplets display double-peaked emission profiles. A narrow and deep absorption line is observed in the Ca~{\sc ii} K line. The broadening and the depth of the He~{\sc i} lines vary from one observation to another, which may be due to the spectroscopic companion mentioned by \citet{wheelwright10}. 

We have cleaned the mask in order to reject the lines contaminated with CS features to compute the synthetic profiles of the binary star. The profiles of June and December 2006 look similar, revealing a broad and complex photospheric profile, that could be formed by an SB2, superimposed with a broad blue wing. The profile of November 2007 is different showing a single photospheric profile, with slightly less broad wings, and with a photospheric depth much deeper than in 2006. This observation could have been obtained during a spectroscopic conjunction, so that the observed profile could be the result of the superposition of the two components at a similar radial velocity, and therefore consistent with an SB2 profile. We fit the three observations simultaneously with the photospheric function of a double star (as in HD~152404 = AK~Sco), and many Gaussian functions to reproduce the wings. In this fitting procedure the photospheric depths and \vsinis of both components were forced to be identical for both observations, while the \vrads were free to vary from one observation to another. The radial velocities of the two components are not significantly different between June 2006 and Dec. 2006, suggesting that both observations may have been obtained at a similar orbital phase, while they are definitely different in 2007. The resulting fit is shown in Fig. \ref{fig:ilcep}.

The success of this fitting procedure, the fact that in 2007 both radial velocities are found to be very similar, and that the depth of the combined profile is much larger than in the other observations, confirm the SB2 nature of IL~Cep. The combination of large \vsinis for both components (180 and 150~\kms), the small velocity separation between the two components of the spectral lines, and the relatively low SNR of our data, explain our inability to detect the SB2 nature of IL Cep within the spectrum itself, and illustrate the capability of the LSD procedure to provide more information on the object than the analysis of the individual spectral lines.

\subsection{HD~244314 (=V1409 Ori)}

\begin{figure}
\centering
\includegraphics[width=6cm,angle=90]{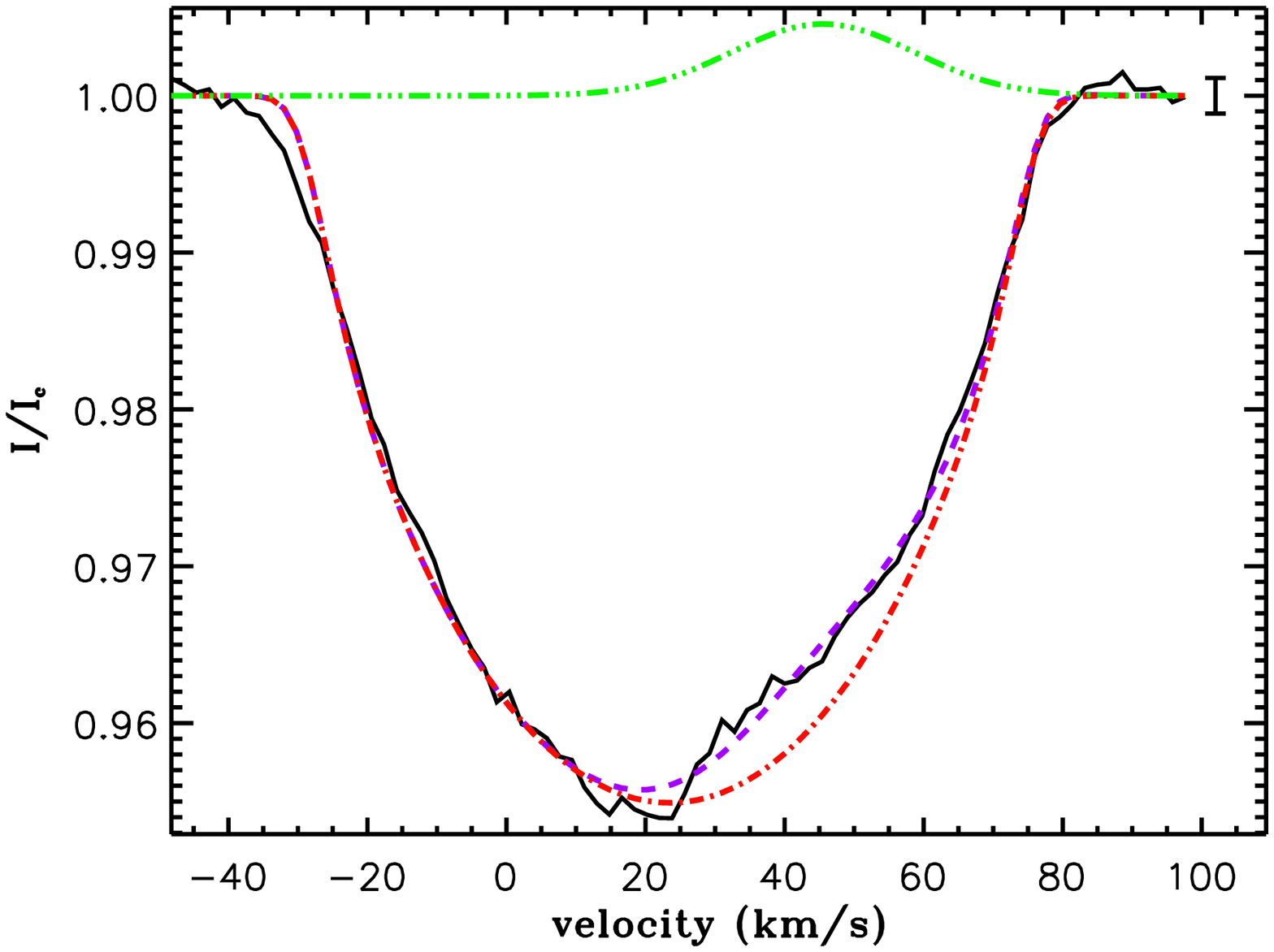}
\caption{As Fig.~A5 for HD~244314}
\label{fig:hd244314}
\end{figure}

HD~244314 is part of the Orion star forming region situated at a distance of 375~pc \citep{brown94}. We used the \citet{miroshnichenko99a} photometric data ($V_{\rm 0}$ and $E(B-V)$) to derive the luminosity of the star. \citet{miroshnichenko99a} detected near-IR excess, and classified the star as a PMS candidate.

The spectrum of HD~244314 is consistent with the effective temperature determination of \citet[][$T_{\rm eff} = 9250\pm500$~K]{vieira03}, and displays little CS contamination. The cores of H$\gamma$ and H$\delta$ are filled with emission and superimposed on a narrow blueshifted CS absorption line. The core of H$\beta$ is sumperimposed on a P~Cygni profile. H$\alpha$ displays a P~Cygni profile of type II. The core of the Ca~{\sc ii} K photospheric line is filled with emission and superimposed on a narrow and deep IS absorption line. The wings of the Fe~{\sc ii} multiplet 42 lines are in emission. The He~{\sc i} lines at 5875~\AA, 6678~\angs and 7065~\angs show broad and double-peaked emission profiles. The wings of the O~{\sc i} 777~nm and O~{\sc i} 8446 triplets are in emission. The core of the Paschen lines are filled with emission, and the Ca~{\sc ii} IR-triplet displays single-peaked emission lines.

We have calculated the LSD profiles without performing a special cleaning to the Kurucz mask. The result shows a photospheric profile slightly distorted, which we believe to be due to circumstellar contamination. We fit the profile with a photospheric function and one Gaussian function. The result is shown in Fig. \ref{fig:hd244314}.

\subsection{HD~244604 (=V1410 Ori)}

\begin{figure}
\centering
\includegraphics[width=6cm,angle=90]{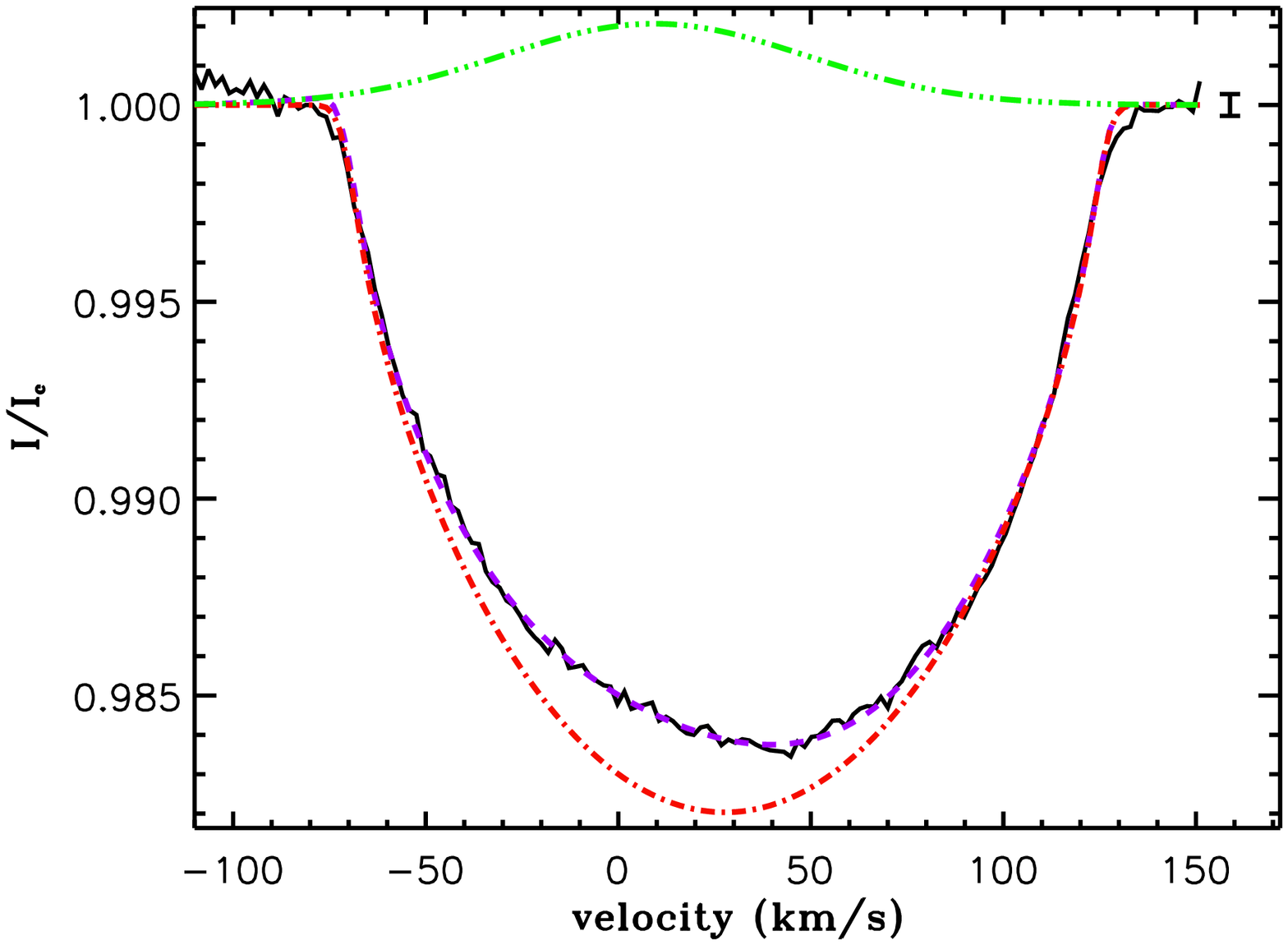}
\caption{As Fig.~A5 for HD~244604}
\label{fig:hd244604}
\end{figure}

HD~244604 is part of the Orion star forming region situated at a distance of 375~pc \citep{brown94}. We used the photometric data of \citet{dewinter01} to derive the luminosity of the star. The spectral energy distribution displays an IR excess well reproduced using a model of a two-temperature dust disk, as is the case for most HAeBe stars \citep{malfait98a}.

Using our automatic procedure, we find that the spectrum of HD~244604 is well reproduced with  $T_{\rm eff} = 8200\pm200$~K and $\log g=4.0$, consistent with the work of \citet{miroshnichenko99a}.  The cores of the Balmer lines from H$\zeta$ to H$\gamma$ are filled with emission. The core of the H$\beta$ profile is superimposed on a single-peaked, slightly redshifted, emission, as well as a blueshifted absorption. H$\alpha$ displays a P Cygni profile of type II. A narrow, faint and blueshifted CS absorption line is superimposed on the photospheric profile of the Ca~{\sc ii} K line. Faint emission is observed in the core of the photospheric profiles of the lines of multiplet 42 of Fe~{\sc ii}, and in the O~{\sc i} 8446 \angs triplet. Broad and faint emission are observed in the He~{\sc i} lines 5875~\AA, 6678~\angs and 7065~\AA, where neither absorption nor emission is predicted by the synthetic spectrum. Faint emission is observed in the wings of the O~{\sc i} 777~nm triplet. The cores of the Paschen lines are filled with emission, and the Ca~{\sc ii} IR-triplet displays strong and single-peaked emission profiles.

We have cleaned the mask in order to reject the lines contaminated with CS features, to compute the LSD profiles. The result shows a slightly distorted single photospheric profile, certainly due to CS contamination, but could not be improved. We fit the profile with a photospheric function and a Gaussian function. The result is shown in Fig. \ref{fig:hd244604}.

\subsection{HD~245185 (=V1271 Ori)}

HD~245185 is part of the $\lambda$ Ori star-forming region \citep{murdin77} situated at a distance of 450~pc \citep{dolan01}. We used the Tycho-2 photometric data \citep{hog00}, and converted them into the Johnson system using the calibration formula 1.3.20 of the Hipparcos and Tycho catalogues \citep[][p. 57]{perryman97}, to derive the luminosity of the star. The spectral energy distribution displays IR excess that is well reproduced using a two-temperature dust disk model, similar to most HAeBe stars \citep{miroshnichenko99b}.

The spectrum is consistent with the temperature and gravity determination of \citet[][$T_{\rm eff}=9500\pm750$~K]{folsom12}. All the metallic lines of the spectrum, except the O~{\sc i} and Si~{\sc ii} lines between 6000 and 6500 \AA, are much weaker than predicted by the synthetic spectrum. This might be due to CS extinction, but it could also be due chemical peculiarities consistent with the work of \citet{folsom12} that report $\lambda$ Boo peculiarities. Many circumstellar features are observed in the spectrum. The cores of the metallic lines, including Ca~{\sc ii} K, are superimposed on narrow emission features. The cores of the Balmer lines from H$\zeta$ to H$\beta$ are also superimposed on emission lines, of increasing amplitude with wavelength. H$\alpha$ displays a double-peaked emission line. The He~{\sc i} lines at 5875~\AA, 6678~\AA, and 7065~\angs show broad and double-peaked emission profiles. The cores of the Paschen lines are filled with emission, and the O~{\sc i} 8446 and Ca~{\sc ii} IR-triplets show small P~Cygni profiles. The wings of the O~{\sc i} 777~nm triplet are in emission.

\begin{figure}
\centering
\includegraphics[width=6cm,angle=90]{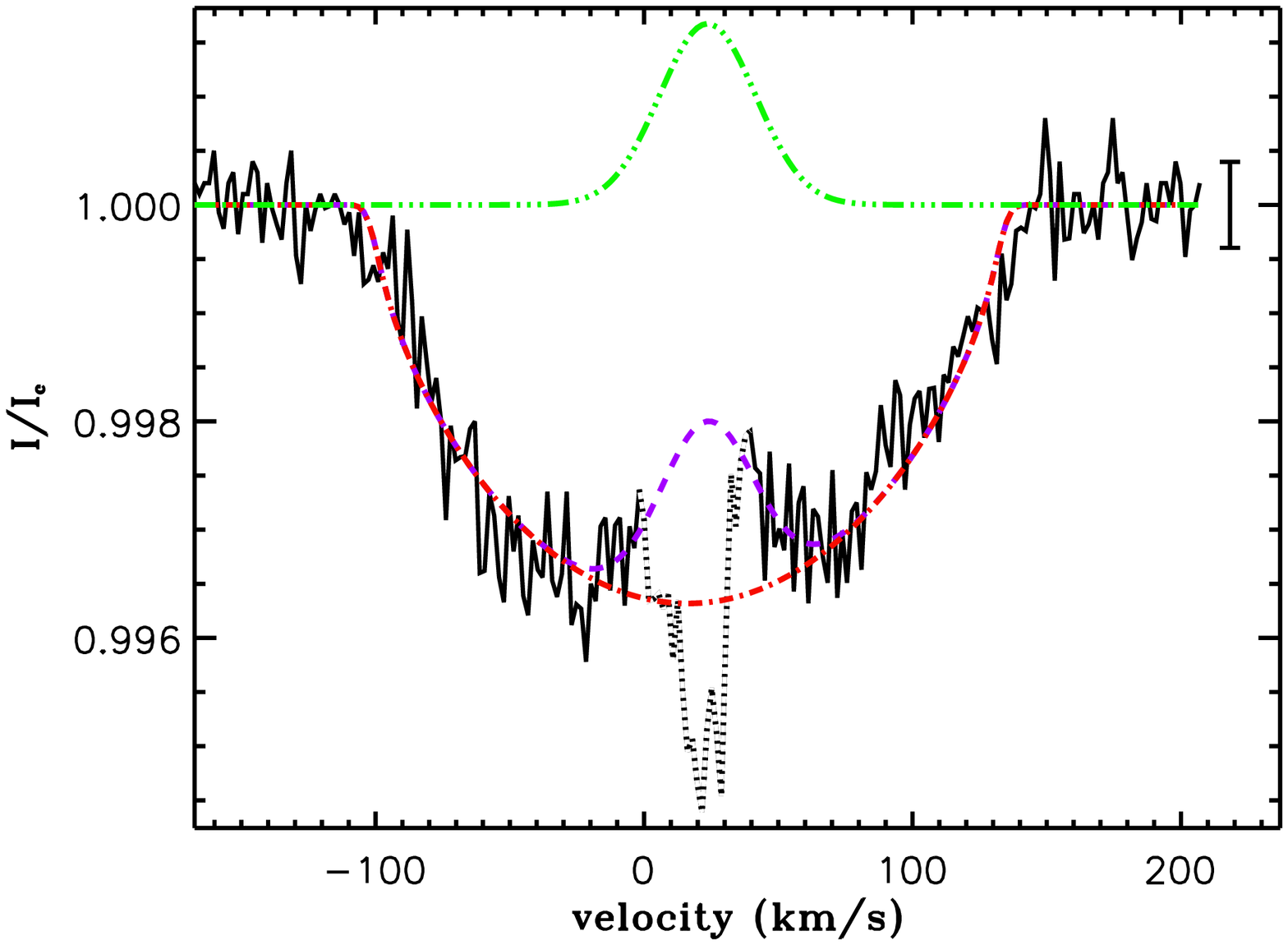}
\caption{As Fig.~A5 for HD~245185}
\label{fig:hd245185}
\end{figure}

We tried to clean the the Kurucz mask, but could not get rid of the emission in the core of the LSD profile. We therefore decided to use the full mask to obtain a more accurate value of the \vsini, and fit the LSD profile with a photospheric function and a Gaussian. The result is shown in Fig. \ref{fig:hd245185}

\subsection{HD~249879}

\begin{figure}
\centering
\includegraphics[width=6cm,angle=90]{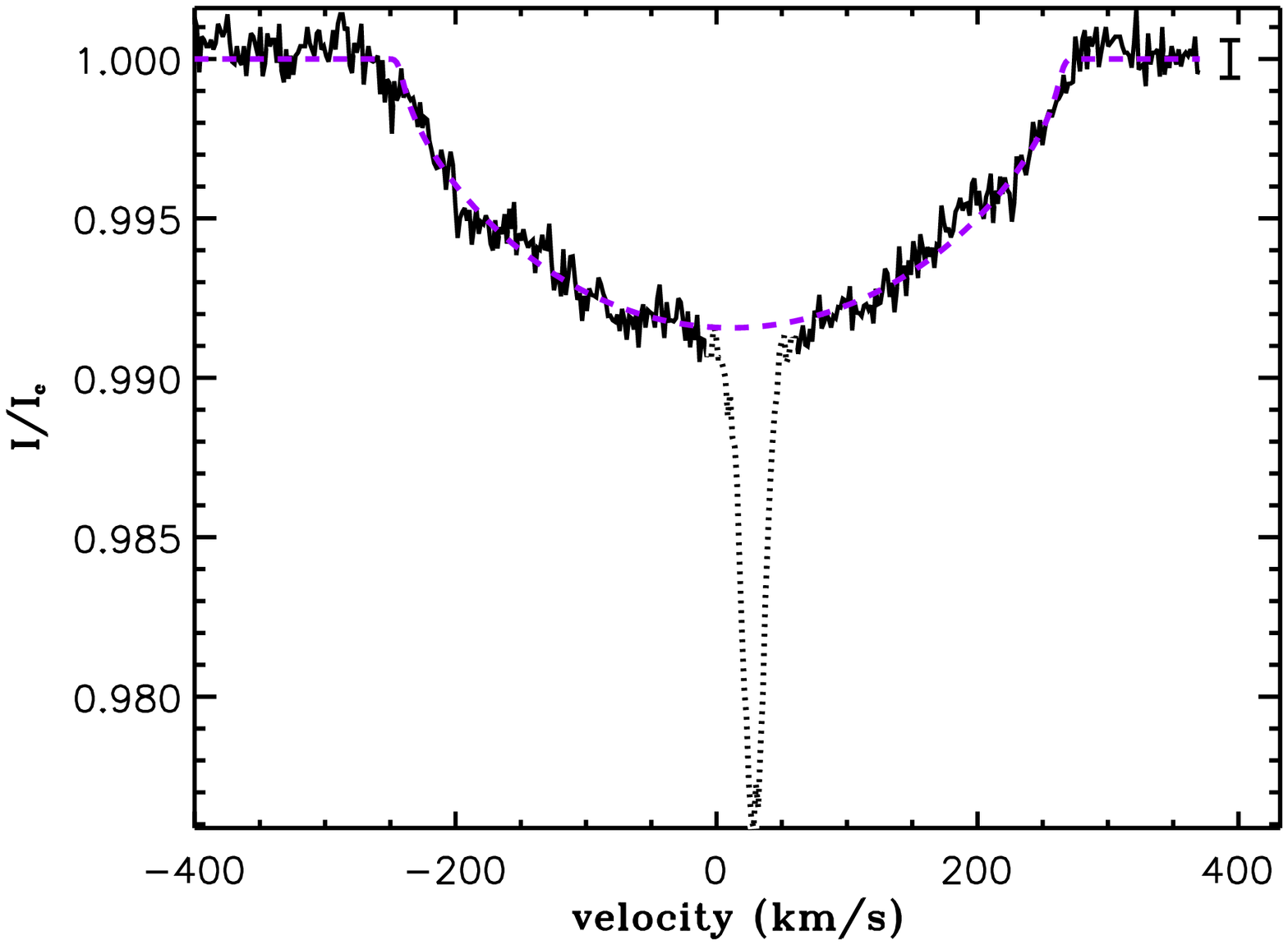}
\caption{As Fig.~A5 for HD~249879}
\label{fig:hd249879}
\end{figure}

\citet{vieira03} proposes that HD~249879 is part of the Gemini OB1 star association, situated at a distance $2000$~pc \citep{vieira03}. We used the photometric data of \citet{vieira03} to derive the luminosity of the star. No spectral energy distribution has yet been published, but the strong IR excess, as well as the presence of H$\alpha$ emission led \citet{gregorio92} to classify the star as an Herbig Ae/Be.

The metallic lines and the Balmer lines of the spectrum are well fit with an effective temperature $T_{\rm eff} = 9000\pm1000$~K, which is inconsistent with the unique temperature determination available in the literature of 12000 K \citep{vieira03}. We suggest that the combination of a large \vsinis ($\sim 250$~\kms) \citet{vieira03} and their low SNR ($\sim 80$) meant that they were not able to resolve the metallic lines. In contrast, we can easily detect and resolve the metallic lines with our higher SNR spectrum ($\sim 550$), and therefore get a better estimate of the effective temperature of the star. The spectrum shows only a few CS features. The Ca~{\sc ii} K line is much stronger than predicted. Weak emission is observed in the cores of H$\alpha$, the O~{\sc i} 777~nm triplet, and the Paschen line P12. The O~{\sc i} 8446 \angs triplet seems to show double-peaked emission, while the Ca~{\sc ii} IR triplet might display an inverse P Cygni profile.

We have cleaned the mask in order to reject the lines contaminated with CS features, to compute the LSD profiles. The result is a photospheric profile with a strong and narrow absorption line superimposed on the core. Its origin is unknown, but a careful examination of the spectrum confirms that the metallic lines are indeed superimposed on narrow absorption. We therefore decided to ignore this absorption in the fitting procedure, using only a photospheric function. The result is shown in Fig. \ref{fig:hd249879}.

\subsection{HD~250550}

HD~250550 is located centrally in a small obscuring cloud, and is also the central object of a bright arc of nebulosity \citep{herbig60}. No reliable distance or parallax determination were obtained for this star. We are therefore unable to determine its luminosity. Based on its spectral energy distribution, HD~250550 has been classified as a group I object by \citet{hillenbrand92}; these are objects that are surrounded with flat optically thick accretion disks.

The spectrum of HD~250550 is highly contaminated with CS features, which makes temperature determination difficult. The few portions of the spectrum not contaminated with CS features is well fit with $T_{\rm eff}=12000\pm1500$ ~K, consistent with the work of \citet{hernandez04}. The strongest metallic lines show emission profiles, sometimes with a P Cygni absorption component. When no emission is observed in these lines, a strong redshifted CS absorption is superimposed on the photospheric profiles. A strong blueshifted CS absorption is superimposed on the core of the Balmer lines, from H$\theta$ to H$\gamma$, of increasing strength with wavelength. In H$\gamma$ and H$\delta$, a redshifted emission is also observed in the core of the lines. The core of H$\beta$ is superimposed on a P Cygni profile with a saturated blueshifted absorption. H$\alpha$ displays a P~Cygni profile of type IV. The Ca~{\sc ii} K line shows a strong highly blueshifted absorption line, as well as centered IS absorption, and emission in the red wing of the profile. The He~{\sc i} lines at 5875~\AA, 6678~\AA, and 7064 \angs are much stronger than predicted, and display  V-like shapes. The O~{\sc i} 777~nm triplet shows a P Cygni profile. The core of the Paschen lines are superimposed on emission, while the O~{\sc i} 8446~\angs and Ca~{\sc ii} IR triplets show very strong single-peaked emission profiles.

\begin{figure}
\centering
\includegraphics[width=6cm,angle=90]{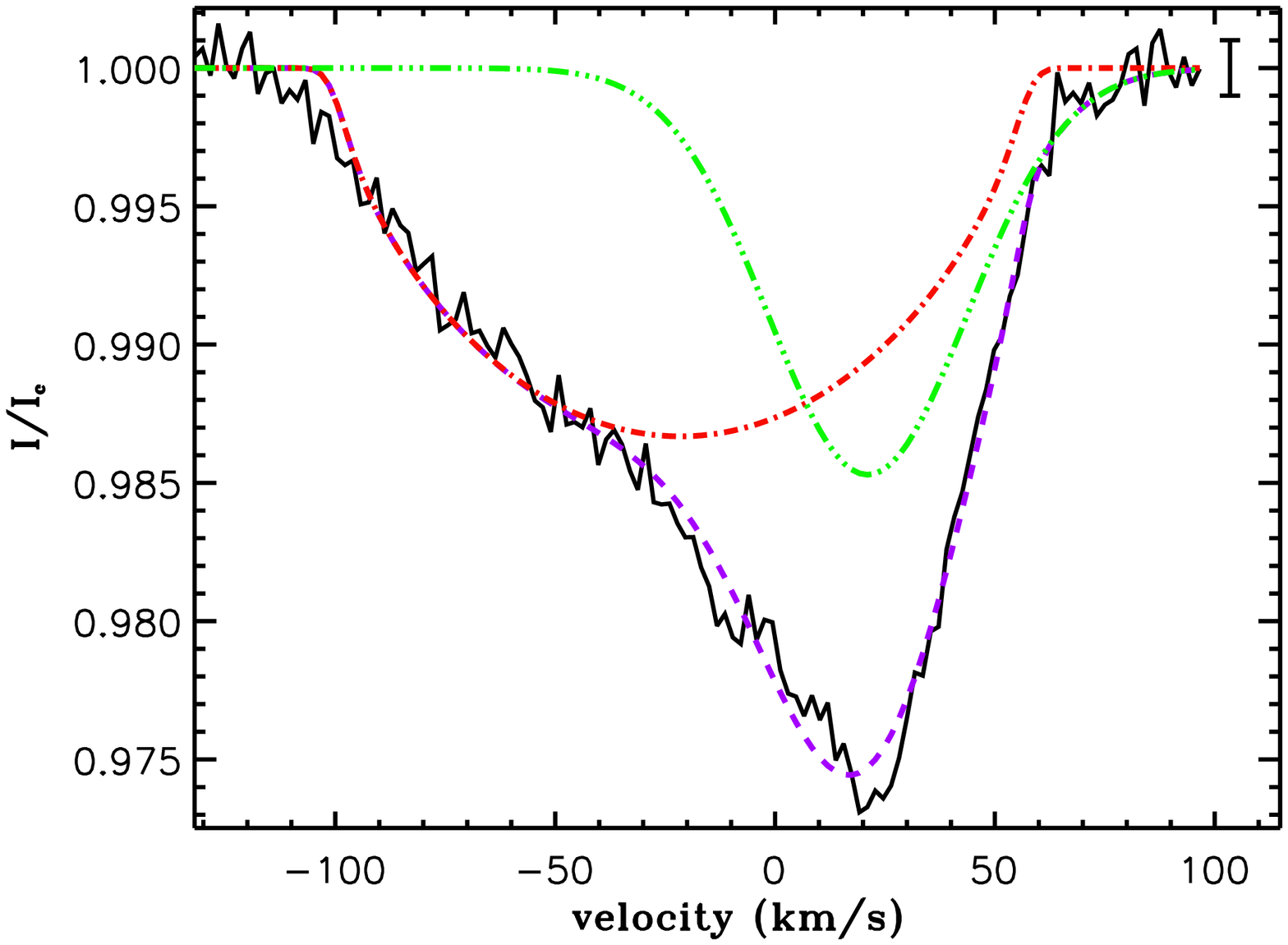}
\caption{As Fig.~A5 for HD~250550}
\label{fig:hd250550}
\end{figure}

We have cleaned the mask in order to reject the lines contaminated with CS features. The resulting LSD $I$ profile shows a photospheric profile superimposed to a CS redshifted absorption line, but could not be improved. We therefore fit the LSD $I$ profile with a photospheric function and a Gaussian function. The result is shown in Fig. \ref{fig:hd250550}.

\subsection{HD~259431 (=V700 Mon)}

HD~259431 is associated with the reflection nebula VdB 82, a member of the association of reflection nebulae Mon R1 \citep{vdb66}. Mon R1 is part of the same cloud complex as NGC 2264 \citep{herbst82}, situated at a distance of 660~pc \citep{kharchenko05}, which disagrees with the Hipparcos parallax of HD~259431 \citep[$\pi=3.45\pm1.41$, ][]{perryman97}. The new Hipparcos parallax of \citet{vanleeuwen07} ($\pi=5.78\pm1.22$) is even larger than the previous determination, and still inconsistent with an association of this star with Mon R1. Furthermore this recent determination of the parallax is also inconsistent with our temperature determination for the star; in the HR diagram, the star falls well below the ZAMS. Because of the reflection nebulosity, we suspect that  the Hipparcos data might not be suitable for parallax determination, and we adopt a distance of 660~pc. We used the Hipparcos photometric data \citep{perryman97} to derive the luminosity.  The spectral energy distribution as well as mid-IR interferometric observations are well reproduced with an optically thick gaseous disk \citep{kraus08}.

The spectrum of HD~259431 is strongly contaminated with variable CS features. The few metallic lines that are not contaminated are well fit with an effective temperature $T_{\rm eff}=14000\pm1000$~K, consistent with the work of \citet{hernandez04}. The cores of the Balmer lines, from H$\iota$ to H$\gamma$, are superimposed on a double-peaked emission profile, with a narrow central absorption that is deeper than the emission. In H$\delta$ and H$\gamma$ narrow blueshifted absorption components appear in the blue part of the CS emission. H$\beta$ displays a double-peaked emission profile, with a strong central absorption that goes below the continuum, and a blueshifted narrow absorption superimposed on the blue wing of the emission. H$\alpha$ shows a strong double-peaked emission profile. The Ca~{\sc ii} K lines shows a broad emission superimposed on a narrow absorption that could be of IS origin. Most of the metallic lines display broad double-peaked emission profiles, with a relatively faint central absorption. The central absorption is much stronger in the lines of multiplet 42 of Fe~{\sc ii}. The red part of the He~{\sc i} lines at 5875~\angs and 6678~\angs are greatly changed between March 2007 and February 2009: the profile of 2007 is much broader on the red part, while the blue wing stayed unchanged. The red part of these He lines in 2009 is consistent with the predicted photospheric profile. While the O~{\sc i} 777~nm triplet shows a single-peaked emission profile, the Paschen lines and the O~{\sc i} 8446 \angs and Ca~{\sc ii} IR triplets display strong double-peaked emission profiles, with central absorptions that are faint and narrow in the O~{\sc i} triplet and in the Paschen lines, but stronger and broader in the Ca~{\sc ii} triplet.

\begin{figure}
\centering
\includegraphics[width=6cm,angle=90]{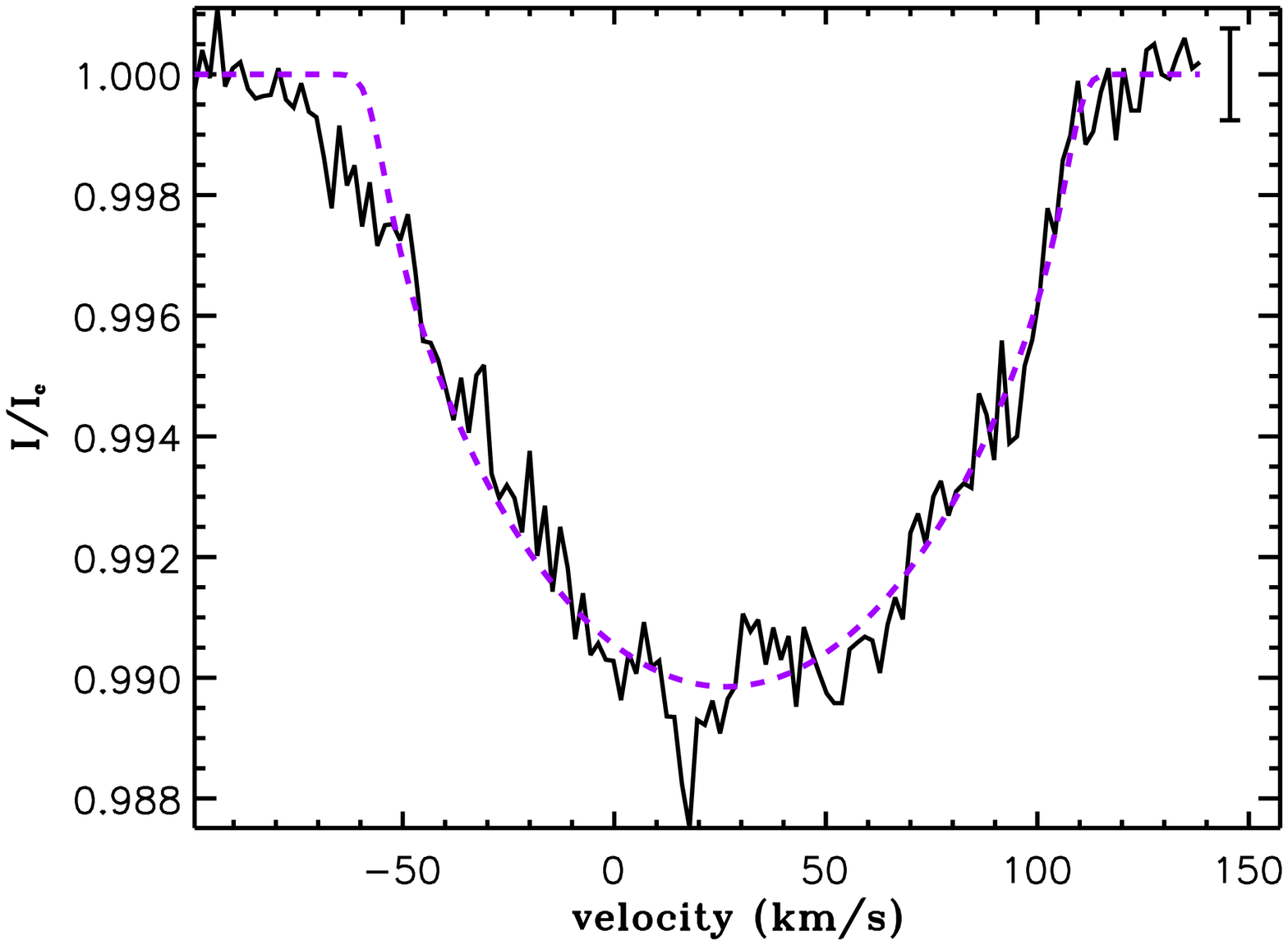}
\caption{As Fig.~A5 for the 2009 observation of HD~259431.}
\label{fig:hd259431}
\end{figure}

We have cleaned the mask in order to reject the lines contaminated with CS features. The resulting three LSD $I$ profiles display photospheric shape, and can be fitted with a photospheric function, giving consistent \vsini. However, the profiles from 2007 and 2010 are much noisier than that of 2009, so that a combination of the three profiles did not improve the \vsinis determination of the star compared to determination from the 2009 profile alone. We therefore decided to fit only the profile of 2009 with a single photospheric function. The result is shown in Fig. \ref{fig:hd259431}.

\subsection{HD~275877 (=XY Per)}

XY Per illuminates the bright nebulosity VdB 24 \citep{vdb66}. It is a strong visible and near-IR photometric variable star \citep[e.g.][]{oudmaijer01}, and belongs to the UXOR class of stars \citep{mora04}, whose the prototype is HD 293782 = UX Ori (Sec. A61). For the same reasons as in the case of UX Ori, we used the photometric data of \citet{oudmaijer01} at the brightest magnitude, together with the Hipparcos parallax \citep{perryman97}, to determine the luminosity of the star. We did not use the new Hipparcos parallax of \citet{vanleeuwen07} as it is totally inconsistent with the old one, and with the effective temperature of the star. The near-IR excess observed in XY Per is typical of the Herbig Ae/Be class of objects \citep{eiroa01}.

The spectrum of XY Per is strongly contaminated with CS features, as is common for UXOR-type stars. Uncontaminated, lines and the wings of the Balmer lines are consistent with $T_{\rm eff}=9000\pm500$~K, corresponding to the spectral classification of \citet{mora01}. Most of the metallic lines, including the Ca~{\sc ii} K line, as well as the core of the Balmer lines are superimposed on strong CS absorption. Emission is also observed in the wings of the CS absorption in the core of H$\beta$. H$\alpha$ displays a double-peaked emission profile, with a deep narrow central absorption that goes below the continuum. The He~{\sc i} lines at 5875~\AA, 6678~\AA\ and 7065~\AA, and the O~{\sc i} 777~nm triplet show inverse-P Cygni profiles, with weak blue emission, and a strong central absorption. The O~{\sc i} 8446 \angs triplet is also much stronger than predicted. The Ca~{\sc ii} IR triplet shows double-peaked emission lines, with very strong and narrow central absorptions. The core of the Paschen lines may be contaminated with CS emission.

\begin{figure}
\centering
\includegraphics[width=6cm,angle=90]{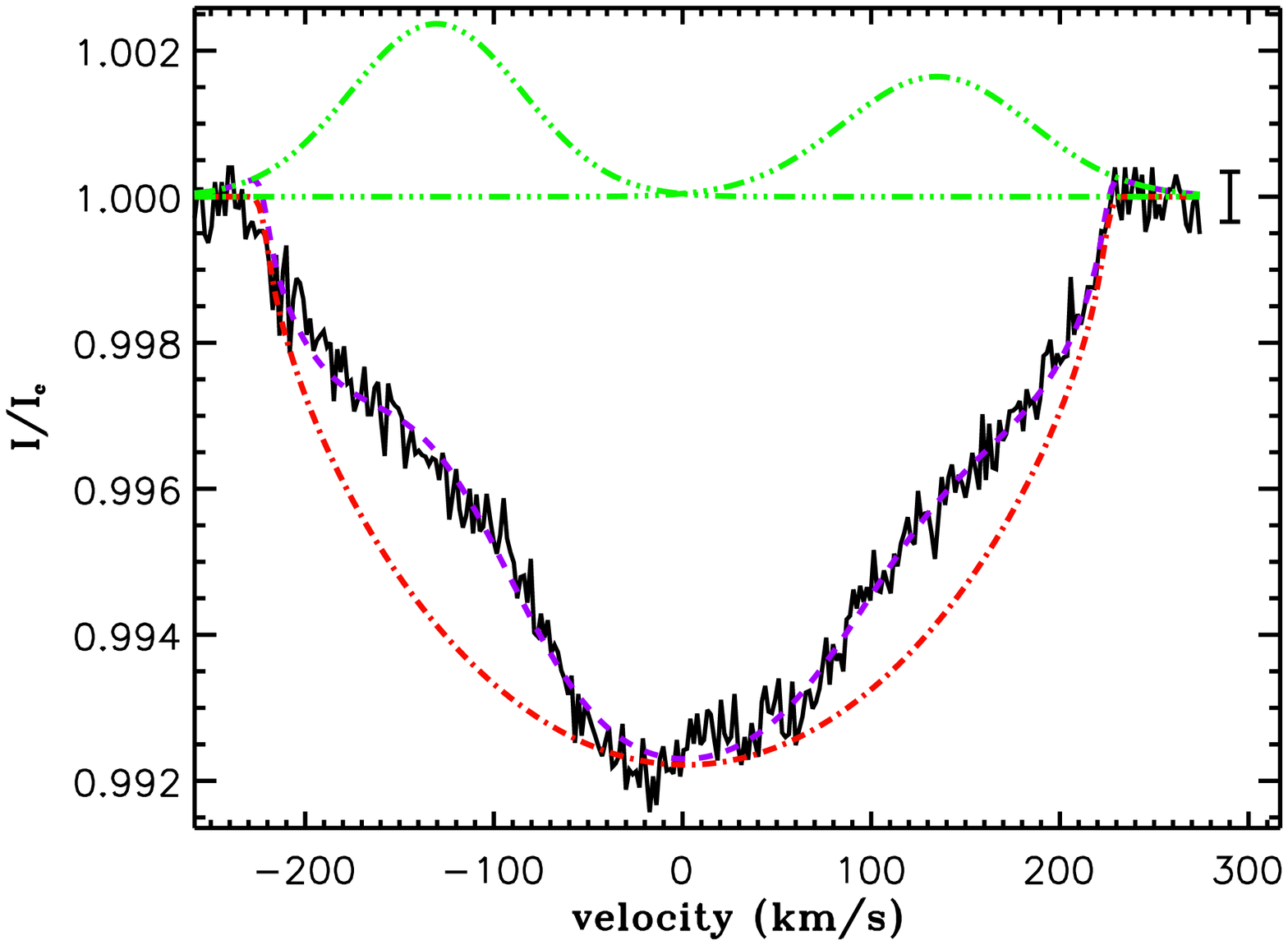}
\caption{As Fig.~A5 for the 2006 observation of XY Per.}
\label{fig:xyper}
\end{figure}

We have cleaned the mask in order to reject the lines contaminated with CS features. The resulting LSD $I$ profiles shows photospheric shapes still slightly contaminated with emission on both the blue and red sides. After a careful examination of the spectrum, the few lines not obviously contaminated with CS features are found to have a V shape, consistent with the LSD $I$ profiles. Both profiles are well fitted with a photospheric plus Gaussian functions, giving consistent \vsini. However, because the profile from 2009 is relatively noisy, including it in the fit together with the profile from 2006 did not improve the \vsinis determination of the star. We therefore decided to only fit the profile of 2006 with a single photospheric function and two Gaussian functions. The result is shown in Fig. \ref{fig:xyper}.

\subsection{HD~278937 (= IP Per)}

According to its Galactic coordinates and its proper motions \citep[from the Tycho-2 catalogue,][]{hog98,hog00}, IP Per is very likely associated with the Per OB 2 association situated at a distance of $\sim320$~pc \citep{dezeeuw99}. IP Per is a photometric variable; the variations are assumed to be caused by moving circumstellar material which shadows the star when it crosses the line of sight \citep{miroshnichenko01}. Therefore, the minimum magnitude is very likely close to the true magnitude of the star. We used the photometric data of \citet{miroshnichenko01} at the brightest magnitude to derive the luminosity of the star. A large IR excess is clearly seen in the spectral energy distribution, which can be reproduced with a two-temperature disk model, as in other HAeBe stars \citep{miroshnichenko01}.

The spectrum of IP Per is consistent with the temperature and gravity determination of \citet[][$T_{\rm eff}=8500$~K, $\log g=4.1\pm0.2$]{folsom12}. Only few CS features are observed in its spectrum. Weak emission fills the core of H$\gamma$, while the core of H$\beta$ is superimposed on a faint single-peaked emission. H$\alpha$ displays a single-peaked emission profile of type VI. The He~{\sc i} 5875~\angs line shows a faint inverse-P Cygni profile in the observation of Feb. 21st 2005, while the He~{\sc i} lines at 5875~\angs and 6678~\angs show broad double-peaked emission profiles in the observations of Feb. 20th 2005. The O~{\sc i} 777~nm triplet is stronger than predicted, and emission is observed in the wings of the profile in the observations of Feb. 20th.

\begin{figure}
\centering
\includegraphics[width=2.3cm,angle=90]{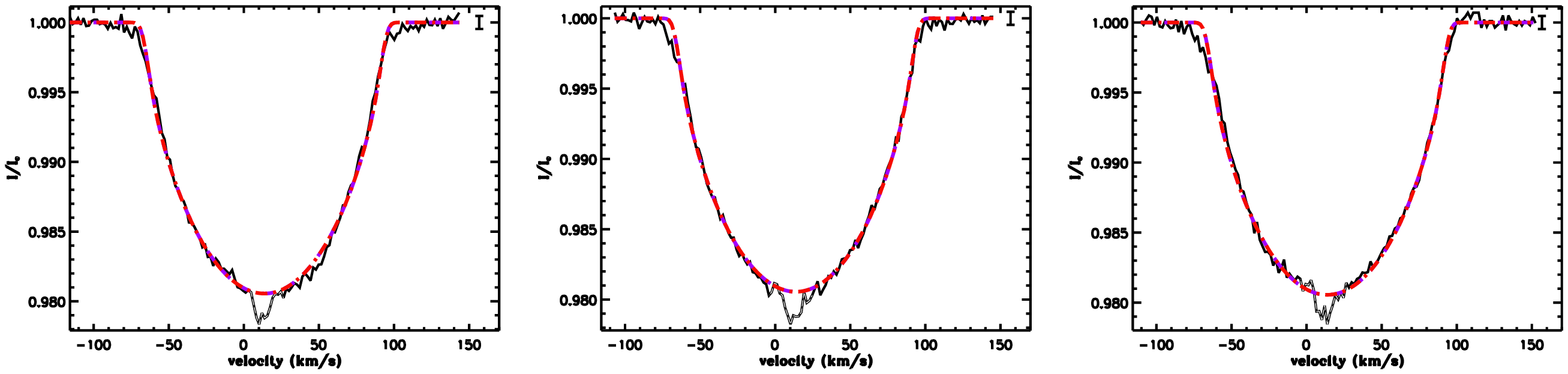}
\caption{As Fig.~A5 for the Feb. 21st 06:26 (left), 21st 07:50 (middle), and 22nd (right) 2005 observations of IP Per.}
\label{fig:ipper}
\end{figure}

We have calculated the LSD profiles without performing a special cleaning to the Kurucz mask. The three LSD $I$ profiles display generally photospheric shapes, superimposed, in the core, on a faint and narrow absorption of unknown origin. A careful look at the spectra leads to the conclusion that this absorption could be present in all the metallic lines, but is lost in the noise, and thus cannot be corrected by cleaning the mask. We therefore fit the three profiles, simultaneously, with a photospheric function, but excluding the data points inside the cores. The result is shown in Fig. \ref{fig:ipper}.

\subsection{HD~287823}

From its Galactic coordinates \citep{perryman97}, HD~287823 might be part of the Orion OB1 association, situated at a distance of $\sim375$~pc \citep{blaauw64,brown94}. The strong near-IR excess observed in the direction of HD~287823 reveals the presence of an optically thick inner disk around the star, as for many HAeBe stars \citep{hernandez05}.

In our spectrum, we detect the spectral lines of a companion of almost the same projected rotational velocity, with a large enough radial velocity separation from the primary that the two spectra are well separated.  The lines are sharp enough that we needed to include macroturbulent velocities which we set to 2 \kms. We find that our spectra are well reproduced with effective temperatures of 10000 K and 7000 K for the primary (P) and secondary (S) respectively, and with a ratio of radii $R_{\rm P}/R_{\rm S}$ equal to $1.5$, leading to a ratio of the luminosities: $L_{\rm P}/L_{\rm S} = 9.4$. We used the Hipparcos photometric data \citep{perryman97} to derive the luminosity of the system, $\log(L/L_{\odot})=1.83$, then the luminosity of each component: $\log(L_{\rm P}/L_{\odot})=1.79$ and $\log(L_{\rm S}/L_{\odot})=0.82$.

\citet{doering09}, from IRAS observations, report the detection of an IR companion at a distance of 2.4 arcsec. The ESPaDOnS aperture being 1.6 arcsec, either the IR companion is the same as the spectroscopic companion observed in our spectrum, which means that the secondary has moved closer to the primary between the IRAS observations and ours, or the IR companion is not the same as the spectroscopic companion, and HD~287823 would therefore be a triple system.

\begin{figure}
\centering
\includegraphics[width=6cm,angle=90]{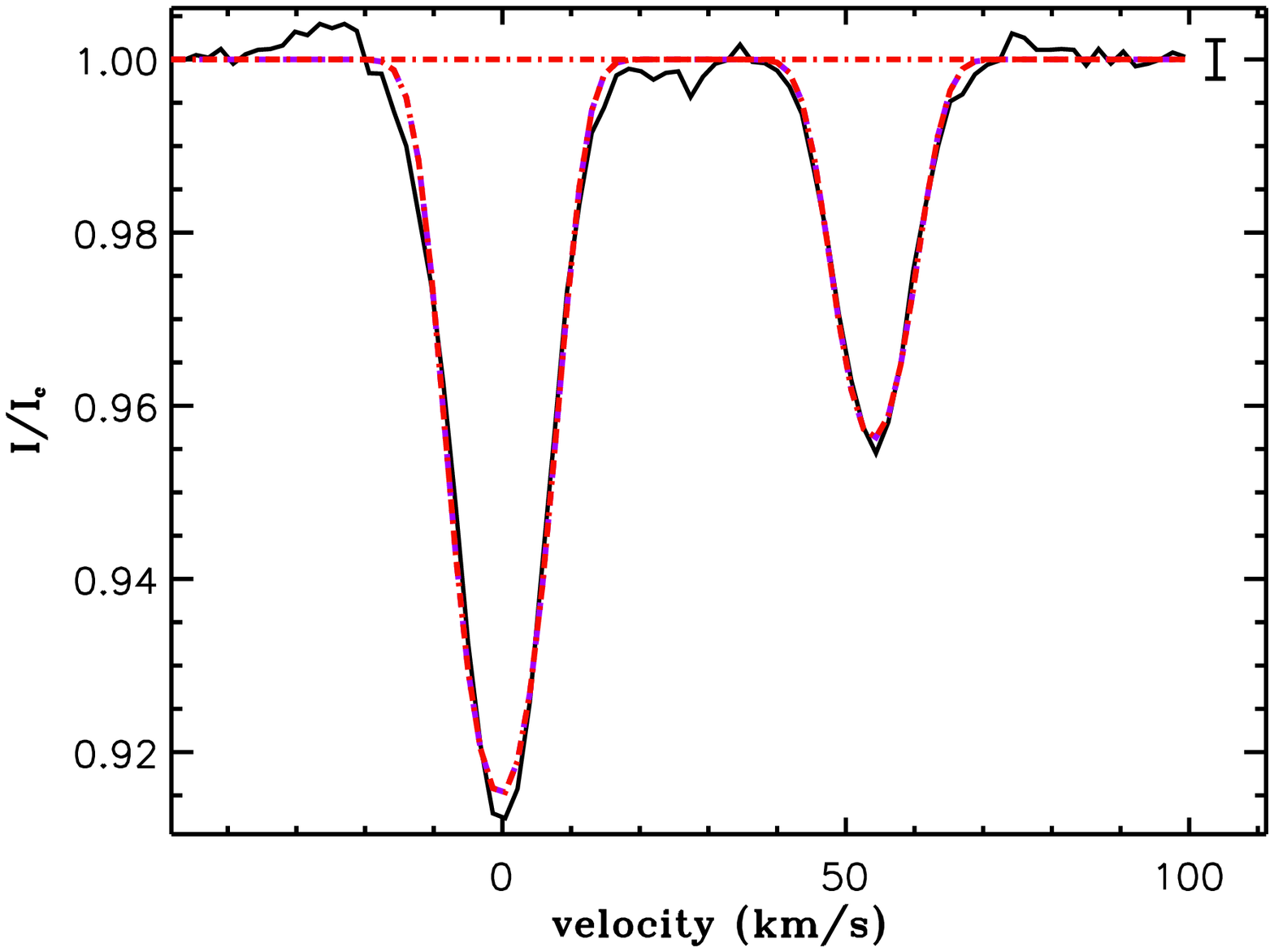}
\caption{As Fig.~A5 for HD~287823}
\label{fig:hd287823}
\end{figure}

Circumstellar features are only observed in few lines: weak emission is present in the core of H$\alpha$ and H$\beta$, while the wings of the O~{\sc i} 8446 \angs and Ca~{\sc ii} IR triplets seem to be in emission. We used the Kurucz mask of 10\,000~K, more suitable for the primary that dominates the flux of the system, to calculate the LSD profile of the binary. While the line ratio of the LSD profile of the primary with respect to the secondary cannot be used for a determination of the luminosity ratio, the \vsinis and \vrads of both components can still be retrieved from the fit of the LSD profile of the binary. We used the same method as for HD~152404 = AK~Sco, to fit the profile of the binary, and the result is shown in Fig. \ref{fig:hd287823}.

\subsection{HD~287841 (= V346 Ori)}

HD~287841 is part of the Orion OB 1a association \citep{hernandez05}, situated at a distance of 375~pc \citep{brown94}. We used the Hipparcos photometric data \citep{perryman97} to derive the luminosity of the star. The spectral energy distribution in the near-IR is consistent with the presence of an optically thick inner disk around the star, as is found for most HAeBe stars \citep{hernandez05}.

The spectrum of HD~287841 is consistent with the temperature and gravity determination of \citet[][$T_{\rm eff}=7550\pm250$~K, $\log g=3.5\pm0.4$]{bernabei09}. CS features are only observed in few spectral lines. The core of H$\beta$ is contaminated with weak emission, while the core of H$\alpha$ is superimposed on a single-peaked emission line. The O~{\sc i} 777~nm triplet is stronger than predicted, and emission might be present in the cores of the Paschen lines.

\begin{figure}
\centering
\includegraphics[width=6cm,angle=90]{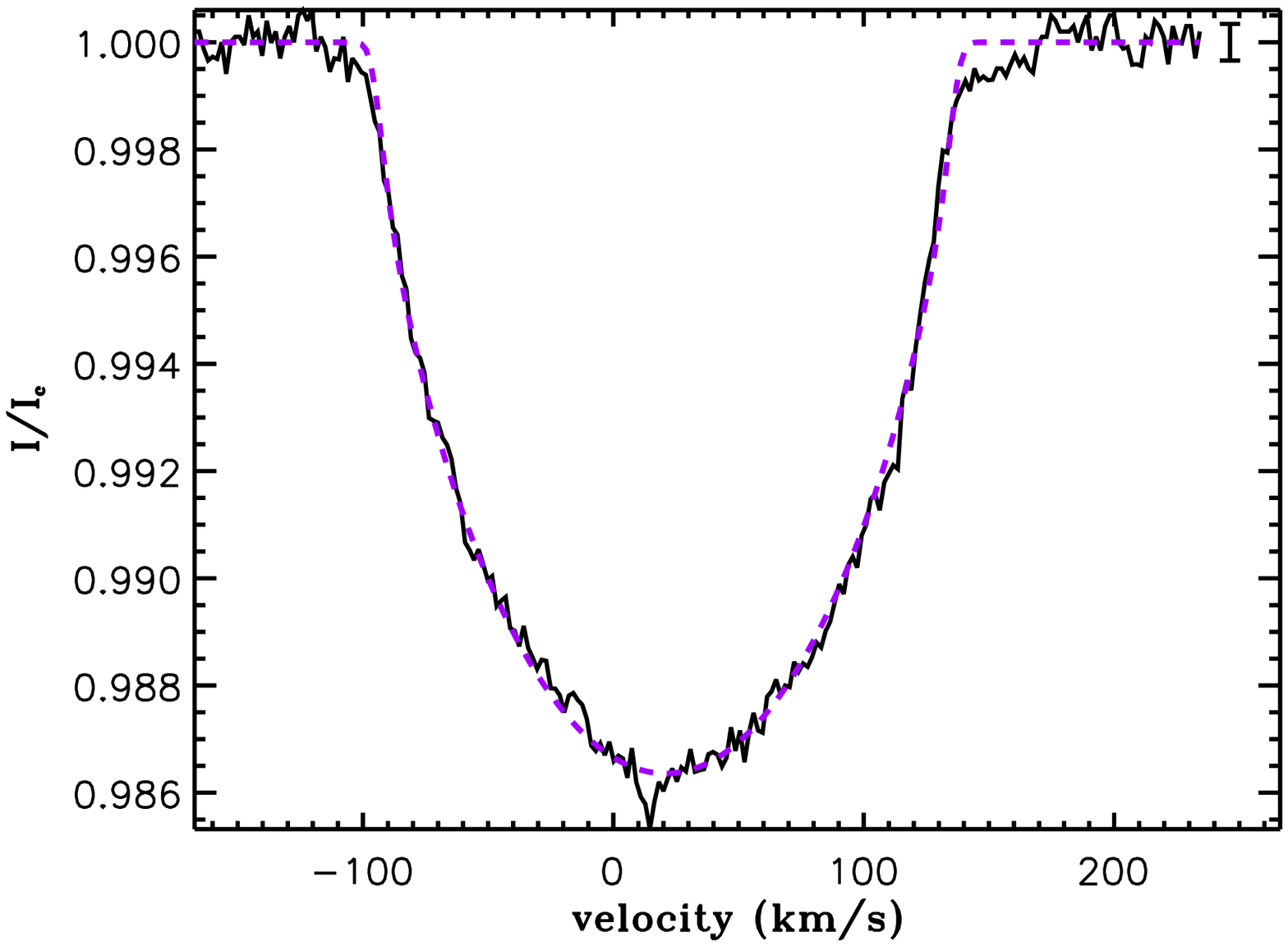}
\caption{As Fig.~A5 for HD~287841}
\label{fig:hd287841}
\end{figure}

We have calculated the LSD profiles without performing a special cleaning to the Kurucz mask. The LSD $I$ profile displays a photospheric shape that fits very well with a photospheric function. The result is shown in Fig. \ref{fig:hd287841}.

\subsection{HD~290409}

HD~290409 is part of the Orion OB 1 star forming region \citep{vieira03}, situated at a distance $\sim375$~pc \citep{brown94}. The near-IR excess observed in the direction of the star is similar to that observed for other Herbig Ae/Be stars \citep{doering09}.

Near-IR observations of HD~290409 reveal variability that could be due to a companion \citep{doering09}. In our spectrum we distinguish two components, one with an effective temperature around $9000\pm500$~K that provides a fit to the Balmer lines, and a second component that accounts for the metallic lines, at a temperature around 5000~K. By applying the LSD method with a mask for 9000 K, we obtain the LSD $I$ profile of a binary star with a rapidly rotating primary, and a secondary rotating more slowly, consistent with the observed spectrum. We used the same method as in the case of HD~152404 = AK~Sco to compute the synthetic composite spectrum of a binary star.  We find that our spectrum is well reproduced with effective temperatures of 9000 K and 5000 K for the primary (P) and secondary (S), respectively, and with a ratio of radii $R_{\rm P}/R_{\rm S}$ equal to $1.5$. We also estimate the ratio of the luminosity: $L_{\rm P}/L_{\rm S} = 23.6$. We used the photometric data of \citet{vieira03} to derive the luminosity of the system: $\log(L/L_{\odot})=1.32$, and then the luminosity of each component: $\log(L_{\rm P}/L_{\odot})=1.30$ and $\log(L_{\rm S}/L_{\odot})=-0.07$. The secondary is therefore a PMS solar-type star, and does not belong to the HAeBe-class.

Few CS features are observed in the spectrum. The cores of the Balmer lines, from H$\varepsilon$ to H$\beta$, are superimposed on a redshifted CS absorption of increasing depth with wavelength. In addition, emission is observed in the wings of the CS absorption in H$\beta$. H$\alpha$ displays a complex emission profile. The O~{\sc i} 777~nm triplet is stronger than predicted, and the cores of the Paschen lines seem to be superimposed on faint and narrow single-peaked emission lines.

\begin{figure}
\centering
\includegraphics[width=6cm,angle=90]{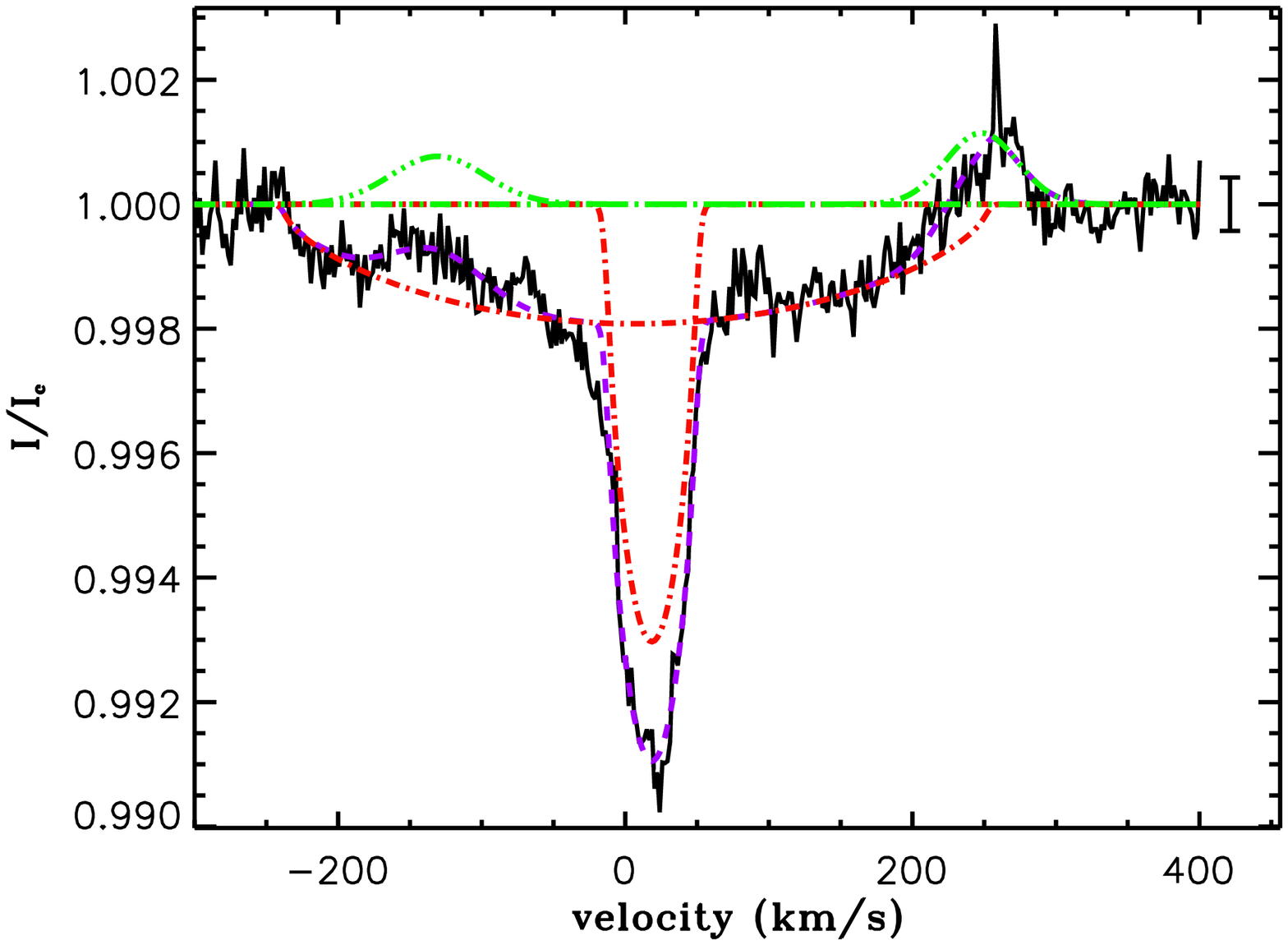}
\caption{As Fig.~A5 for HD~290409}
\label{fig:hd290409}
\end{figure}

We used the Kurucz mask of 9\,000~K, more suitable for the dominant primary, to compute the LSD profiles of the binary. The resulting $I$ profile shows photospheric shapes for both stars, superimposed on weak emission. We used the same method as AK~Sco to fit the LSD profile of the binary, but we also added two Gaussian functions in order to fit the CS emission. The result is shown in Fig. \ref{fig:hd290409}. We tried to eliminate the CS contamination by cleaning the mask, and fit the $I$ profile with only the sum of two photospheric functions, but the results were much noisier and did not improve the determination of the \vsinis of the two stars.

\subsection{HD~290500}

HD~290500 is part of the Orion OB 1 star forming region \citep{vieira03}, situated at a distance $\sim375$~pc \citep{brown94}. We used the photometric data of \citet{guetter79} to derive the luminosity of the star. The near-IR excess observed in the direction of the star is similar to that observed around other Herbig Ae/Be stars \citep{doering09}.

The metallic lines of the spectrum are well fit with an effective temperature $T_{\rm eff}=9000\pm500$~K, consistent with the spectral type of the Henry Draper Extension Charts (HDEC) catalogue \citep{nesterov95}. The spectrum of HD 290500 is only slightly contaminated with CS features. The core of H$\beta$ is superimposed on a P Cygni profile. H$\alpha$ displays a double-peaked emission profile of type VI, with a narrow and strong central absorption. The O~{\sc i} 777~nm triplet is stronger than predicted, and the He~{\sc i} lines at 5875~\angs and 6678~\angs show inverse-P Cygni profiles. The spectrum longward of 8000 \angs is too noisy to be analysed.

\begin{figure}
\centering
\includegraphics[width=6cm,angle=90]{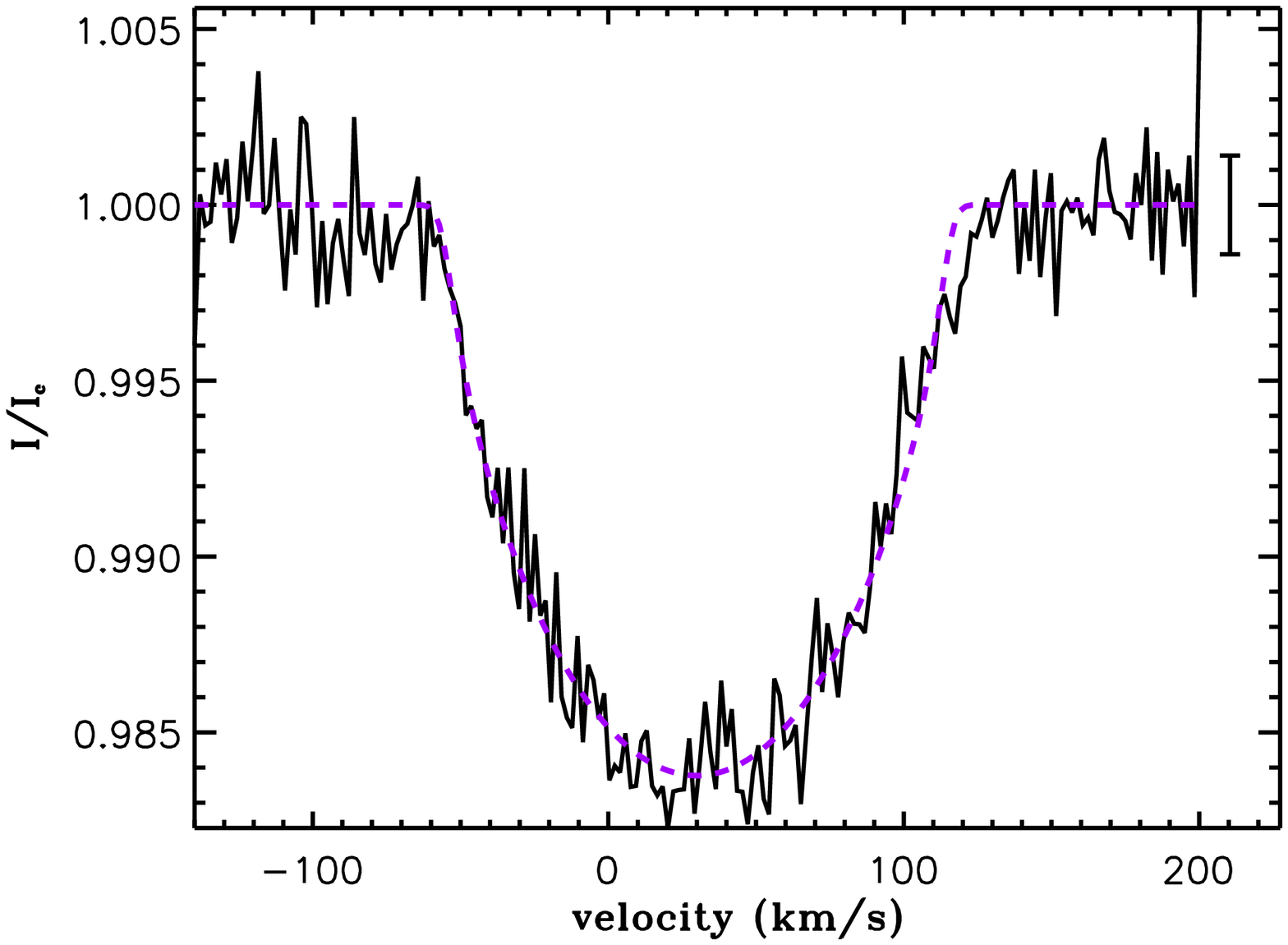}
\caption{As Fig.~A5 for HD~290500}
\label{fig:hd290500}
\end{figure}

We have cleaned the mask in order to reject the lines contaminated with CS features. The LSD resulting $I$ profile displays a photospheric shape well fitted with a photospheric function. The result is shown in Fig. \ref{fig:hd290500}.

\subsection{HD~290770}

HD~290770 is part of the belt of Orion OB1 \citep{guetter76}, situated at a distance of 375~pc \citep{brown94}. We used the photometric data of \citet{vieira03} to derive the luminosity of the star. The near-IR excess observed in the direction of the star is similar to that observed for other Herbig Ae/Be stars \citep{doering09}.

The spectrum of HD 290770 is almost without photospheric lines. This could be due to the combination of a high temperature and a large \vsini. While the Balmer lines seem to be consistent with an effective temperature $T_{\rm eff}=10000$~K, the few photospheric lines that do not seem to be contaminated with CS features, and that are useful for temperature determination, such as the Ca~{\sc ii}~K and He~{\sc i} 4026 \angs lines, are strongly in favour of an effective temperature $T_{\rm eff}=12000$~K. We therefore adopted a temperature of $11000\pm1000$~K, consistent with the temperature determination of \citet{vieira03}.

Few circumstellar features are observed in the spectrum of HD 290770. The cores of the Balmer lines, from H$\theta$ to H$\beta$, are superimposed on a single-peaked emission line of increasing amplitude with wavelength. From H$\varepsilon$ to H$\beta$ a blueshifted absorption component appears in the core of the profiles. H$\alpha$ displays a single-peaked emission profile of type VI, superimposed on two weak absorption components in the blue wing of the emission profile. The He~{\sc i} 4471 \angs line is contaminated with emission, while the He~{\sc i} lines at 5875~\AA, 6678~\AA, and 7065~\angs show broad double-peaked emission profiles. The O~{\sc i} 777~nm triplet is stronger than predicted. The O~{\sc i} 8846 \angs triplet is in emission, and single-peaked emission lines are superimposed on the core of the Paschen lines.

\begin{figure}
\centering
\includegraphics[width=6cm,angle=90]{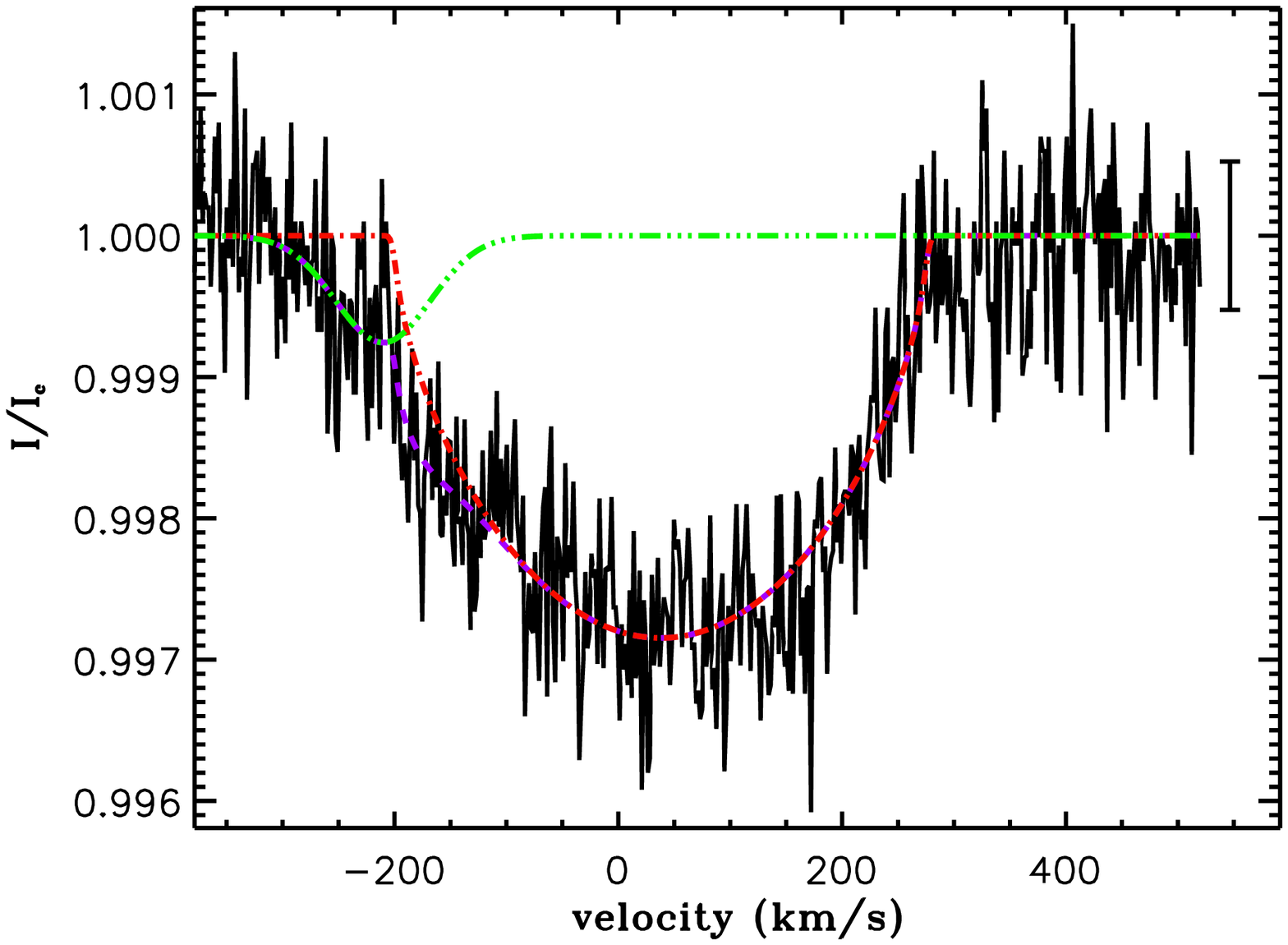}
\caption{As Fig.~A5 for HD~290770}
\label{fig:hd290770}
\end{figure}

We have cleaned the mask in order to reject the lines contaminated with CS features. The resulting LSD $I$ profile displays a photospheric shape with weak CS contamination in the blue wing. We fit the profile with a photospheric function and a Gaussian function. The result is shown in Fig. \ref{fig:hd290770}.

\subsection{HD~293782 ( = UX Ori)}\label{sec:uxori}

UX Ori is part of the Orion OB1d association, situated at a distance of 375~pc \citep{brown94}. UX Ori is the prototype star of the UXOR-type objects, which are found to be strongly irregular photometric variables. The Hipparcos magnitudes \citep{perryman97}, in the Tycho $V_{\rm T}$ system, vary from 8.6~mag to 11.0~mag. Corresponding to the PMS nature of HAeBe stars, this variability is assumed to be caused by moving circumstellar material, shadowing the star when it crosses the line of sight \citep{mora02}. Therefore the brightest magnitude observed by Hipparcos is very likely close to the true magnitude of the star. In order to convert the magnitudes ($V_{\rm T}=8.6$~mag, $B_{\rm T}=9.7$~mag) from the Tycho to the Johnson system, we used the calibration formula 1.3.20 of the Hipparcos and Tycho catalogues \citep[][p.57]{perryman97}, and we find $V = 8.53$~mag and $(B-V)=0.615$~mag. 

The spectrum of UX Ori is very specific to the UXOR-type stars, and is strongly contaminated with transient circumstellar absorption and emission features in the Balmer lines and also in the metallic lines. Most of the photospheric profiles of the metal lines are contaminated with strong redshifted absorption, sometimes superimposed on blueshifted emission. \citet{mora02} argue that these absorption lines come from large bodies in the circumstellar disk of these stars. H$\alpha$ shows a double-peaked emission profile of type VI, with a redshifted central absorption that goes below the continuum. The cores of the other Balmer lines, from H$\varepsilon$ to H$\beta$, are superimposed on redshifted absorption and blueshifted emission, both of which increase as the wavelength increases. The Ca~{\sc ii} K line is much stronger than predicted, and displays a V-shape. The He~{\sc i} lines at 5875~\angs and 6678~\angs show inverse P Cygni profiles. The O~{\sc i} 777~nm and O~{\sc i} 8446~\angs triplets show also inverse P Cygni profiles, but with very weak blue emission, and very strong red absorption. The Ca~{\sc ii} IR triplet also displays P Cygni profiles, while faint blueshifted emission is present in the cores of the Paschen lines. The wings of the Balmer lines and those metallic lines which are not contaminated with emission are consistent with an the temperature and gravity determination of \citet[][$T_{\rm eff}=9250\pm500$, $\log g=4.0$]{mora02}. 

\begin{figure}
\centering
\includegraphics[width=6cm,angle=90]{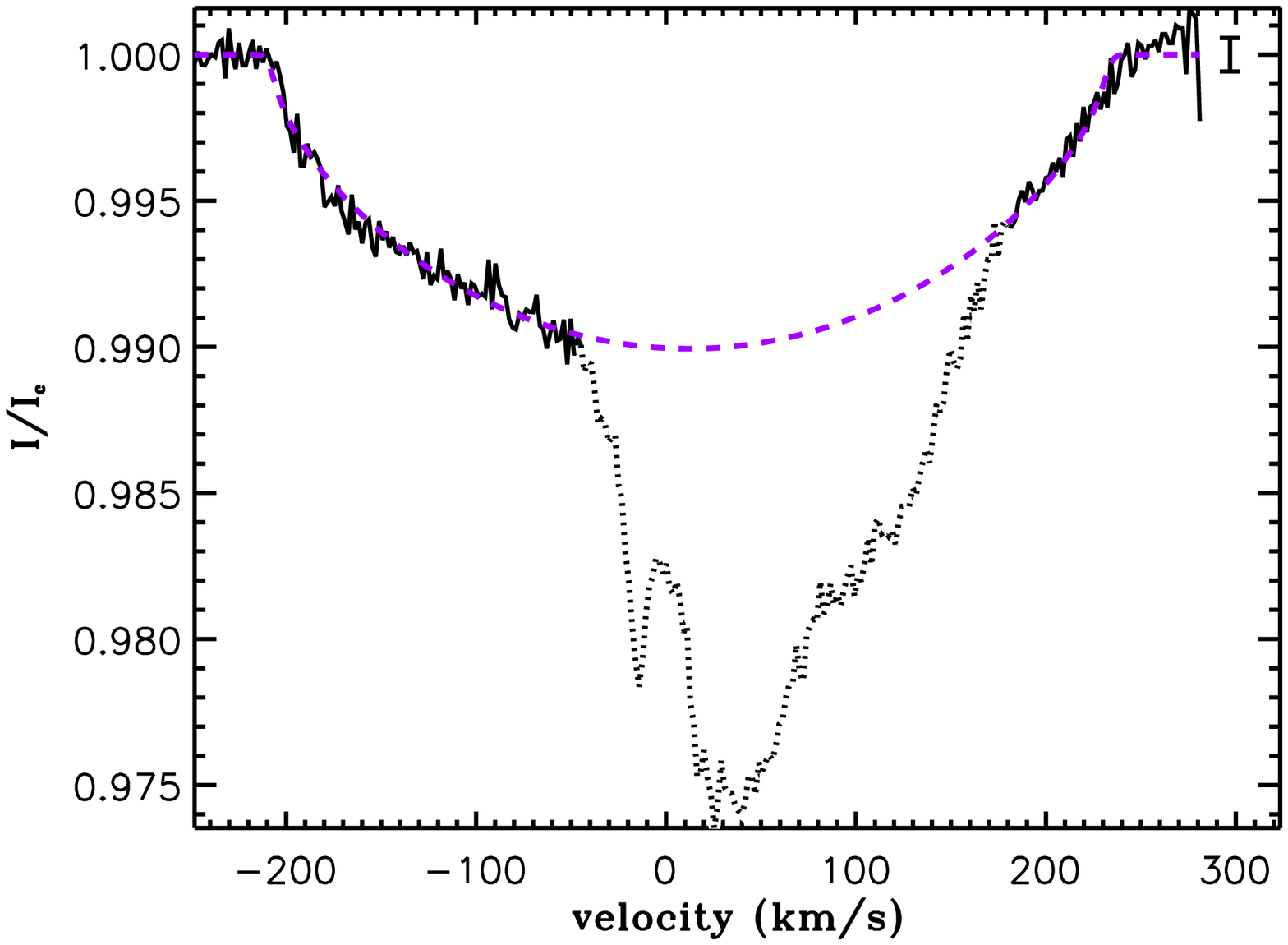}
\caption{As Fig.~A5 for HD~293782 (=UX Ori)}
\label{fig:uxori}
\end{figure}

We first calculated the LSD profiles without performing a special cleaning to the Kurucz mask, resulting in a strongly contaminated $I$ profile. We have cleaned the mask in many ways, without improvement of the $I$ profile. We therefore choose to fit the contaminated $I$ profile by rejecting from the fit the contaminated part of the profile. The result is shown in Fig. \ref{fig:uxori}.

\subsection{HD~344361 (= WW Vul)}

WW Vul is an isolated Herbig Ae star, at a distance of $\sim700$~pc, as estimated by \citet{montesinos09}. WW Vul is an UXOR star \citep{mora04}, a class of stars that are known to be strong photometric variable. Strong variability was confirmed by \citet{herbst99}, who find an amplitude of variation going up to 2.15~mag for WW Vul. We used the photometric data of these authors to compute the luminosity of the star.

WW Vul displays CS contamination in the spectral lines, similar to other UXOR stars. The lines not contaminated with CS features are consistent with the temperature determination of \citet[][$T_{\rm eff}=9500$ K]{mora04}. Most of the photospheric lines, including Ca~{\sc ii} K, and the Balmer lines from H$\eta$ to H$\beta$, are superimposed on strong redshifted absorption features, and sometimes show emission as well. H$\alpha$ displays a double-peaked emission profile of type VI, with a redshifted central absorption that goes below the continuum. The He~{\sc i} lines at 5875~\AA, 6678~\AA, and 7065~\AA, as well as the O~{\sc i} 8446 and Ca~{\sc ii} IR triplets show inverse P Cygni profiles. The O~{\sc i} 777~nm triplet is much stronger than predicted, and the cores of the Paschen lines are filled with emission.

\begin{figure}
\centering
\includegraphics[width=6cm,angle=90]{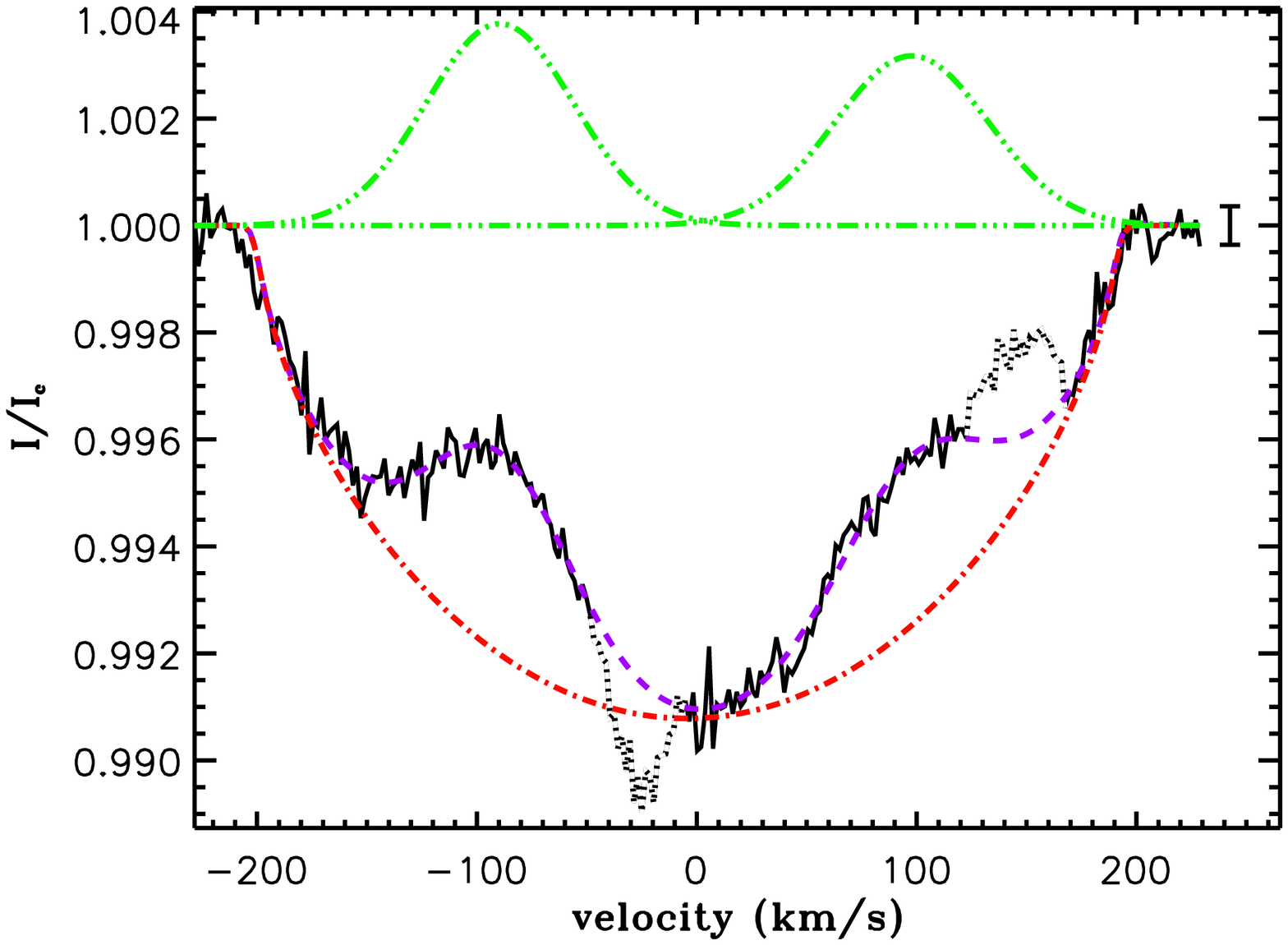}
\caption{As Fig.~A5 for the 2005 observation of HD~344361 (= WW Vul).}
\label{fig:wwvul}
\end{figure}

We have calculated the LSD profiles without performing a special cleaning to the Kurucz mask. Both profiles display photospheric shapes with CS contamination, and can both be fitted with a photospheric function and Gaussian functions, giving consistent \vsini. However, because the profile of 2007 is much noisier than the earlier observation, inclusion of these data did not improve the \vsinis determination of the star, when combined with the profile of 2005. We therefore decide to fit only the profile from 2005 with a single photospheric function and two Gaussian functions. The result is shown in Fig. \ref{fig:wwvul}.

\subsection{LkH$\alpha$ 215}

LkH$\alpha$ 215 is associated with the reflection nebulae NGC 2245 \citep{magakian03}, situated at a distance of 900~pc \citep{oliver96}. We used the photometric data of \citep{herbst99} to derive the luminosity of the stars. LkH$\alpha$ 215 possesses strong near-IR excess that has been attributed by \citet{hillenbrand92} to a circumstellar accretion disk.

The spectrum of LkH$\alpha$~215 contains two kind of lines: (i) broad, very shallow and rare lines, always superimposed on the second kind of lines ; (ii) sharp and very deep lines, that have shifted in radial velocity between the 2008 and the 2009 observations. Circumstellar emission in many metallic lines as well as a large \vsinis are assumed to be at the origin of the first kind of lines. The second kind of lines is assumed to come from a slow rotator with a high metallicity. In spite of the peculiarity of this spectrum, as well as its low SNR, we attempted to determine the effective temperature of both stars. From the wings of the Balmer lines as well as the presence of few He~{\sc i} lines, we find that a temperature around 14000 K fits the lines of the fast rotator satisfactorily, which we assume to be the primary component. The lines of the slow rotator, the secondary component, seem to be produced by a photosphere at 14000 K, as well. However these lines are much deeper than those predicted by a solar composition synthetic spectrum. Varying the gravity and the effective temperature within reasonable ranges does not improve the fit. A strong metallicity is therefore more likely to be the cause of the large depth of these lines. While our temperature determination is highly uncertain, the SB2 nature of LkH$\alpha$~215 appears firmly established, and we will therefore consider it as a binary in the following.

Due to the uncertain determination of the temperature of both stars, and the low SNR of our data, we are not able to constrain usefully the luminosity ratio of the two components. As both stars seem to have similar temperatures, and are very young, it is reasonable to assume that they have a common origin and a similar age, and therefore similar luminosities. The luminosity ratio could therefore be close to 1.

The spectrum of LkH$\alpha$ 215 is contaminated with circumstellar features. As mentioned above, we suspect that the absence of many metallic lines, as well as the faintness of others, are not only due to a large \vsini, but also to contamination by CS emission. We also observe emission in the cores of the Balmer lines from H$\varepsilon$ to H$\gamma$. H$\beta$ displays a double-peaked emission profile of type VI with a central absorption that goes well below the continuum. H$\alpha$ shows a double-peaked emission profile of type V \citep[according to ][]{beals53}, with a redshifted central absorption that reaches the continuum. The O~{\sc i} 8446 \angs triplet displays a double-peaked emission profile, while the Ca~{\sc ii} IR-triplet shows a single-peaked emission line. The Paschen lines seem to be filled with emission, and superimposed on deep and narrow central absorption.

\begin{figure}
\centering
\includegraphics[width=4cm]{FIGURES/I_lkha215.ps}
\hfill
\includegraphics[width=4cm,angle=90]{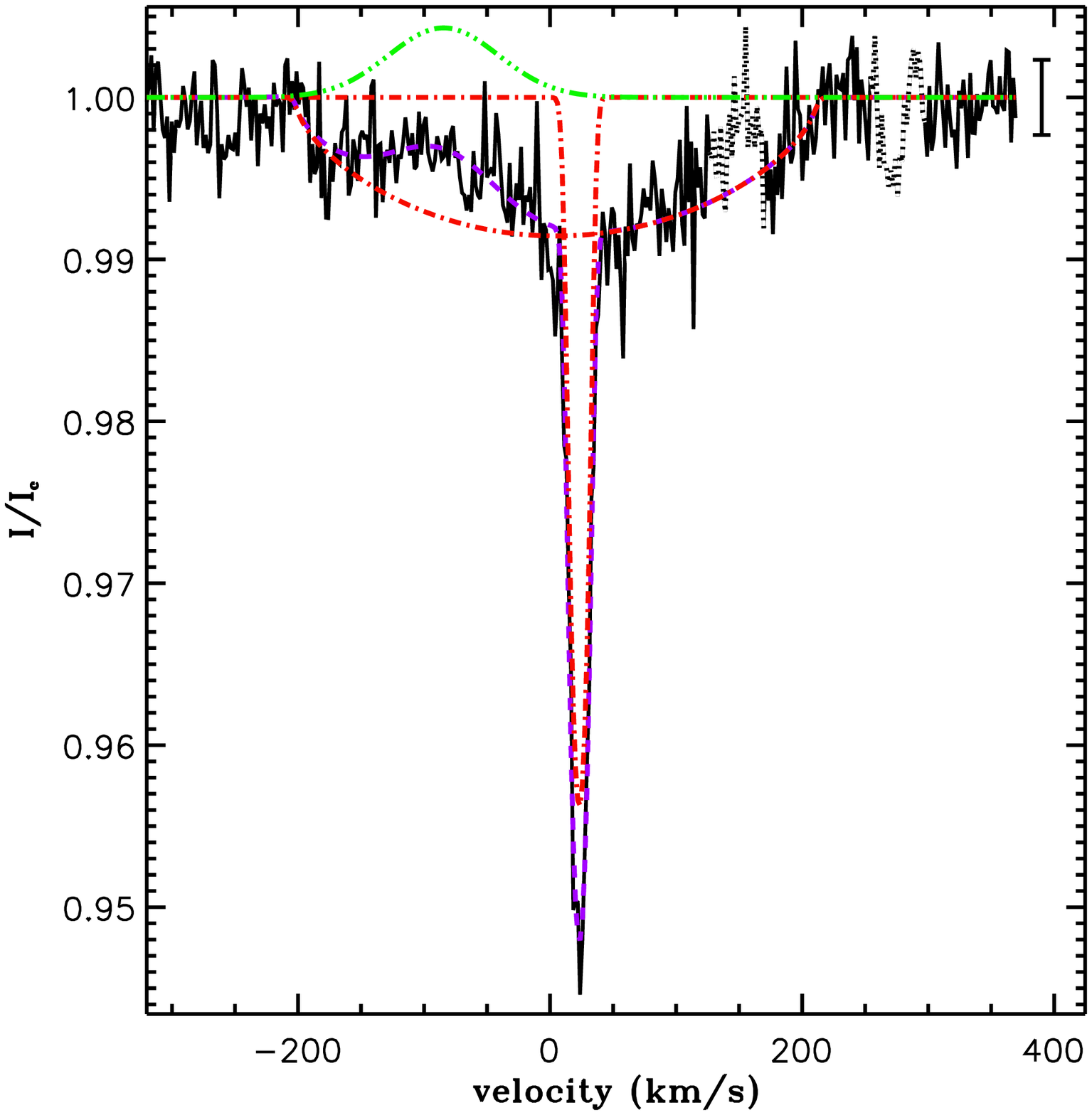}
\caption{Left: LSD $I$ profiles of LkH$\alpha$~215 observed on Apr. 15th 2008 (black full line) and  Mar. 11th 2009 (red dashed-line). Note the shift in radial velocity of the secondary. Right: As Fig.~A5 for the Mar. 11th 2009 observation of LkH$\alpha$~215.}
\label{fig:lkha215}
\end{figure}

We have cleaned the mask in order to reject the lines contaminated with CS features, to compute the LSD profiles of the binary. The $I$ profile of 2009 is much less noisy than the profile of 2008. The result is that while both components are visible in the profile of 2009, only the secondary component is visible in the profile of 2008 (Fig. \ref{fig:lkha215} left). In Fig. \ref{fig:lkha215} (left), we show the shift in radial velocity of the secondary component from 2008 (black full line) to 2009 (red dashed line). Due to the low SNR of the profile of 2008, we did not use it to determine the \vsinis of the primary, but we nevertheless fit it to determine the radial velocity of the secondary component in 2008. The \vsinis of both components as well as the radial velocity of the primary and the radial velocity of the secondary in 2009 have been determined from the fit of the 2009's profile. While we tried to eliminate the lines contaminated with emission, we still have a small contamination in the 2009 profile that we could not improve. We therefore fit the profile with a photospheric function for a binary (as in HD 152404 = AK~Sco) and a Gaussian function. The result is shown in Fig. \ref{fig:lkha215}.

\subsection{MWC 1080}

The MWC 1080 system is a small stellar group embedded within the dark cloud LDN1238. Using the local standard of rest velocity of this dark cloud ($v_{\rm LSR}=-30.3$~\kms) determined by \citet{levreault85}, and the revised prescription for calculating kinematic distances of \citet{reid09} to estimate the distance of the system at about 2300 pc. The most luminous star, MWC 1080 (V628 Cas), has been classified as a B0e star with a flat optically thick circumstellar disk \citep{wang08}.

The ESPaDOnS spectrum of MWC 1080 is totally contaminated with emission, sometimes superimposed on circumstellar absorption. No photospheric lines is detected in the spectrum. All the Balmer lines display P Cygni profiles of type IV, except H$\alpha$ which shows a P Cygni line of type III. We are therefore not able to give an estimate of the effective temperature of the star, or to confirm the spectral type of \citet{hillenbrand92}.  In the absence of photospheric lines $v\sin i$ can not be determined.

\subsection{VV Ser}

VV Ser is located in the Serpens molecular cloud, situated at a distance of 260~pc \citep{straizys96}. We used the photometric data of \citet{herbst99} to compute the luminosity of the star. Based on its spectral energy distribution, VV Ser has been classified as a group I object by \citet{hillenbrand92}, one of a class of objects that are surrounded with flat optically thick accretion disk.

The spectrum of VV Ser is strongly contaminated with circumstellar features, and has been clasified as an UXOR-type star \citep{pontoppidan07}. CS absorption components and sometimes emission are superimposed on the photospheric profiles of many metallic lines (including Ca~{\sc ii} K). The cores of the Balmer lines from H$\eta$ to H$\gamma$ are also superimposed on blueshifted CS absorption, with emission on the blue side of the CS absorption. Both emission and absorption increase with increasing wavelength. H$\beta$ displays a double-peaked emission profile of type VI, with a slightly blueshifted central absorption that goes well below the continuum. H$\alpha$ shows a double-peaked emission profile of type V, with a slightly blueshifted  central absorption that almost reaches the continuum. The He~{\sc i} lines at 5875~\AA, 6678~\AA, and 7065~\angs display inverse P Cygni profiles with very strong absorption. The O~{\sc i} 777~nm triplet is much stronger than predicted, while the O~{\sc i} 8446 \angs triplet has a double-peaked emission profile. Double-peaked emission profiles are also superimposed on the cores of the Paschen lines, while the Ca~{\sc ii} IR-triplet displays double-peaked emission profiles with very strong central absorption. The wings of the photospheric profiles are consistent with the effective temperature determination of \citet[][$T_{\rm eff}=14000\pm2000$]{hernandez04}. 

\begin{figure}
\centering
\includegraphics[width=6cm,angle=90]{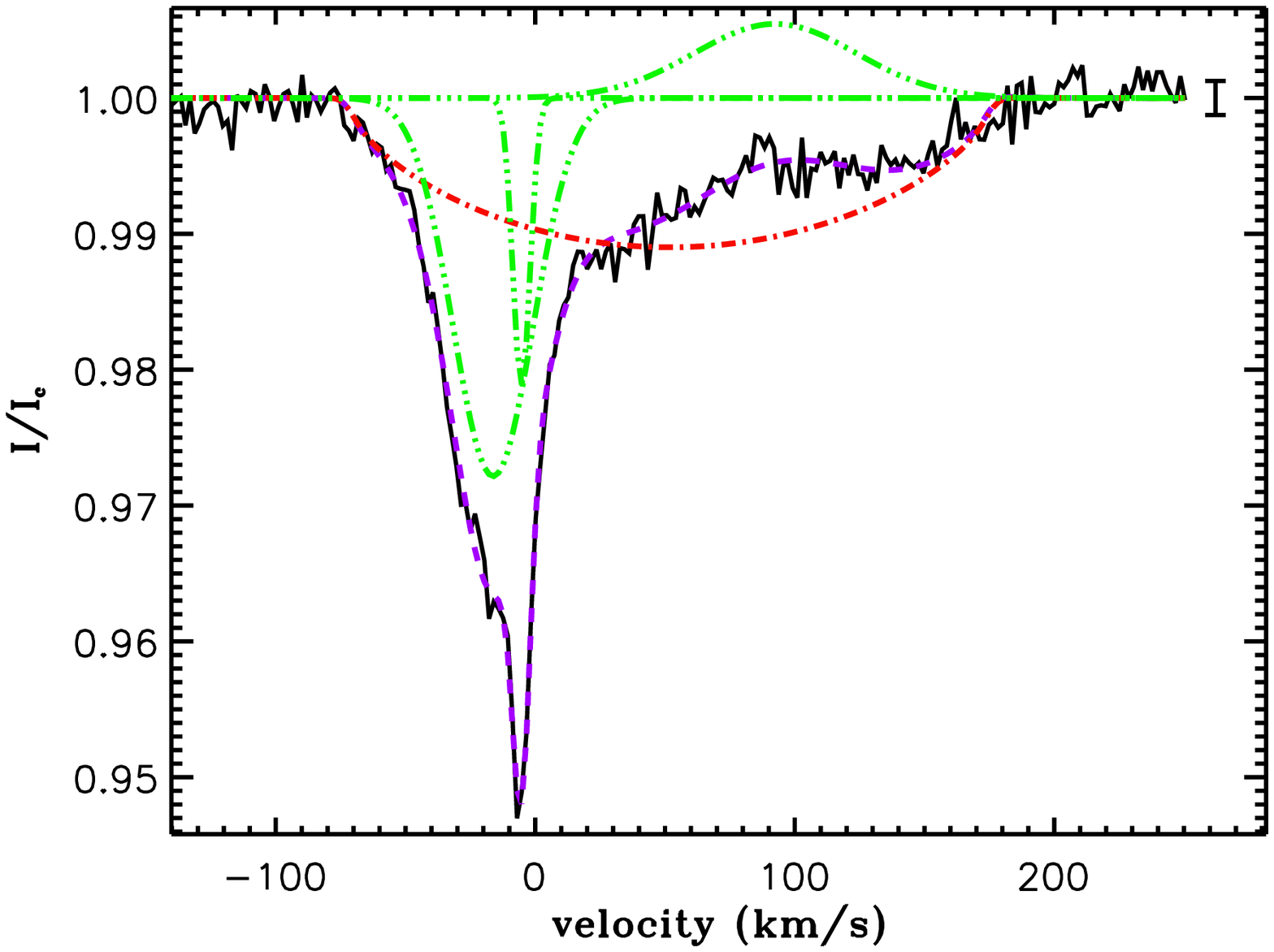}
\caption{As Fig.~A5 for VV Ser.}
\label{fig:vvser}
\end{figure}

We have calculated the LSD profiles without performing a special cleaning to the Kurucz mask. The resulting LSD $I$ profile is contaminated with circumstellar features. We tried to clean the mask, without improving the $I$ profile. We therefore fit the contaminated $I$ profile with a photospheric function and 3 Gaussian functions modelling the circumstellar components. The result is shown in Fig. \ref{fig:vvser}.

\subsection{VX Cas}

VX Cas is an UX Orionis star situated at a distance of 620~pc \citep{montesinos09}. We used the photometric data of \citet{herbst99} to determine the luminosity of the star. VX Cas possesses a strong IR excess due to thermal emission from its circumstellar dust \citep{shakhovskoi03}.

The spectrum of VX Cas is well fit with anconsistent with the effective temperature of \citet[][$T_{\rm eff}=9500\pm1500$~K]{hernandez04}. Most of the metallic lines (including Ca~{\sc ii} K)  are superimposed on blueshifted CS absorption components. Blueshifted CS absorption is also observed in the core of the Balmer lines from H$\varepsilon$ to H$\beta$. In addition, emission is present in the wings of the CS absorption in H$\gamma$ and H$\beta$. H$\alpha$ displays a double-peaked emission profile of type VI, with a slightly blueshifted central absorption that reaches the continuum. The He~{\sc i} lines at 5875~\angs and 6678~\AA, as well as the O~{\sc i} 777~nm triplet, are much broader and much deeper than predicted. Although the spectrum above 8000 \AA\ is quite noisy, it appears that the Ca~{\sc ii} IR triplet may be in emission, and emission may fill the core of the Paschen lines.

\begin{figure}
\centering
\includegraphics[width=6cm,angle=90]{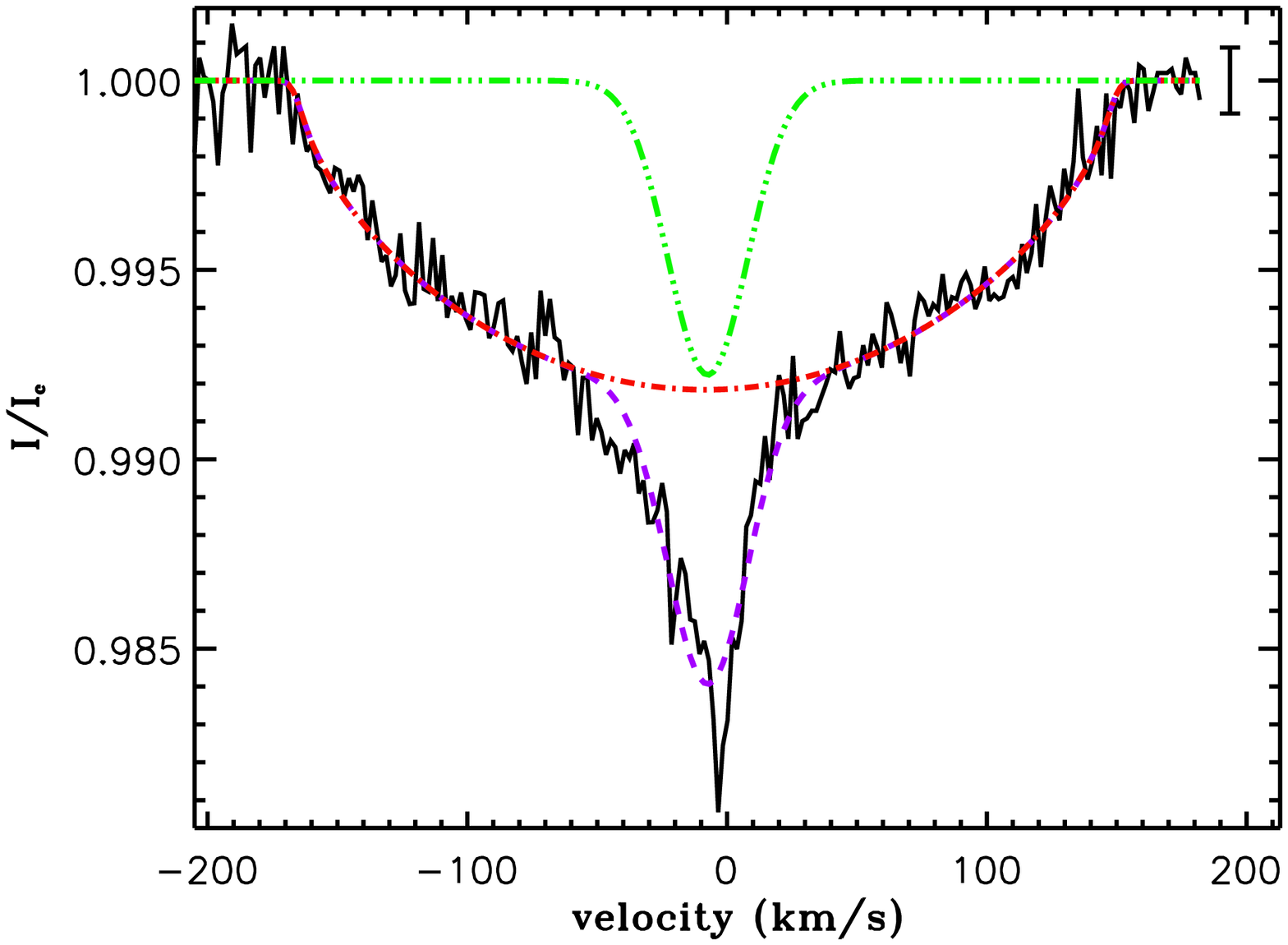}
\caption{As Fig.~A5 for VX Cas.}
\label{fig:vxcas}
\end{figure}

We have calculated the LSD profiles without performing a special cleaning to the Kurucz mask. The resulting LSD $I$ displays a photospheric profile superimposed on a narrow central circumstellar absorption. We tried to clean the mask by rejecting, as far as possible, the contaminated lines. The result gives a profile which is still contaminated with the CS absorption, and with a much lower SNR. We therefore decided to fit the profile with the highest SNR, with a photospheric function plus a Gaussian function. The result is shown in Fig. \ref{fig:vxcas}.

\end{document}